%% file: sneutrinoL-LSP-V60.tex
\documentclass[a4paper,11pt]{article}
\usepackage{jheppub}
\usepackage{soul}
\usepackage{tabularx}
\newcolumntype{C}{>{\centering\arraybackslash}X}
\input{defsV2}
\allowdisplaybreaks

\title{Searching for left sneutrino LSP at the LHC}

\author[a,b]{Pradipta Ghosh,}  
\author[c,d]{I\~naki Lara,}
\author[e,f]{Daniel~E.~L\'opez-Fogliani,}
\author[c,d]{Carlos~Mu\~noz,} 
\author[g]{and Roberto~Ruiz~de~Austri} 
  \affiliation[a]{Department of Physics, Vidyasagar College, 39, Sankar Ghose Lane, Kolkata 700006, India}
    \affiliation[b]{Department of Physics, Indian Institute of Technology Delhi,
Hauz Khas, New Delhi-110016, India}
    \affiliation[c]{Departamento de F\'{\i}sica Te\'{o}rica, Universidad Aut\'{o}noma de Madrid,
Campus de Cantoblanco, 28049 Madrid, Spain}
\affiliation[d]{Instituto de F\'{\i}sica Te\'{o}rica UAM-CSIC, 
  Campus de Cantoblanco, 28049 Madrid, Spain}
  \affiliation[e]{Instituto de F\'isica de Buenos Aires UBA \& CONICET, Departamento de F\'isica, Facultad de Ciencia Exactas y Naturales, Universidad de Buenos Aires,
1428 Buenos Aires, Argentina}
\affiliation[f]{Pontificia Universidad Cat\'olica Argentina, 1107 Buenos Aires, Argentina}
  \affiliation[g]{Instituto de F\'{\i}sica Corpuscular CSIC--UV, 
c/ Catedr\'atico Jos\'e Beltr\'an 2, 46980 Paterna,
\\
 Valencia, Spain}
\emailAdd{tphyspg@gmail.com}
\emailAdd{inaki.lara@csic.es}
\emailAdd{daniel.lopez@df.uba.ar}
\emailAdd{c.munoz@uam.es}
\emailAdd{rruiz@ific.uv.es}

\abstract{We analyze relevant signals 
expected at the LHC for a left sneutrino as the lightest supersymmetric particle (LSP).
The discussion is carried out in the `$\mu$ from $\nu$' supersymmetric standard model ($\mn$), where the presence of $R$-parity breaking couplings involving right-handed neutrinos solves the $\mu$ problem and reproduces neutrino data. 
The sneutrinos are pair produced via a virtual $W$, $Z$ or $\gamma$ in the $s$ channel.
From the prompt decay of a pair of left sneutrinos LSPs of any family, a significant diphoton 
signal plus missing transverse energy (MET) from neutrinos can be present
in the mass range 118--132 GeV, with 13 TeV center-of-mass energy and an integrated luminosity of 100 fb$^{-1}$.  
In addition, in the case of a pair of tau left sneutrinos LSPs, given the large value of the tau Yukawa coupling 
diphoton plus leptons and/or multileptons can appear.
We find that the number of expected events for the multilepton signal, 
together with properly adopted search strategies,
is sufficient to give a significant evidence for 
a sneutrino of 
mass in the range 130--310 GeV, even with the integrated luminosity of 20 fb$^{-1}$.
In the case of the signal producing diphoton plus leptons, an integrated luminosity of 
100 fb$^{-1}$ is needed to give a significant evidence in the
mass range 95--145 GeV.
Finally, 
we discuss briefly the presence of displaced vertices and the associated range of masses.}

\preprint{\begin{flushright}
          \hspace*{3cm} %CPHT-x, LPT-Orsay-16-x, 
 IFT-UAM/CSIC-17-062
 \\
FTUAM-17-11

            \end{flushright}}

\date{\today}

\begin{document}
\maketitle
%%%%%%%%%%%%%%%%%%%%%%%%%%%%%%%%%%%%%%%%%%%%%%%%%%%%%%%%%%%%%%%%%%%%%
%%%%%%%%%%%%%%%%%%%%%%%%%%%%%%%%%%%%%%%%%%%%%%%%%%%%%%%%%%%%%%%%%%%%%
%%%%%%%%%%%%%%%%%%%%---------INTRODUCTION---------%%%%%%%%%%%%%%%%%%%
%%%%%%%%%%%%%%%%%%%%%%%%%%%%%%%%%%%%%%%%%%%%%%%%%%%%%%%%%%%%%%%%%%%%%
%%%%%%%%%%%%%%%%%%%%%%%%%%%%%%%%%%%%%%%%%%%%%%%%%%%%%%%%%%%%%%%%%%%%%
\section{Introduction}
\label{Introduction} 
%%%%%%%%%%%%%%%%%%%%%%%%%%%%%%%%%%%%%%%%%%%%%%%%%%%%%%%%%%%%%%
%%%%%%%%%%%%%%%%%%%%%%%%%%%%%%%%%%%%%%%%%%%%%%%%%%%%%%%%%%%%%%%%

% WE USE SARAH (SPHENO, ETC) WITH ITS CONVENTIONS, EXCEPT FOR THE GAUGE COUPLINGS AND THE YUKAWA MATRICES FOR QUARKS AND LEPTONS THAT ARE TRANSPOSED IN SARAH

% DISCUSS IN THE INTRODUCTION THAT THE LH SNEUTRINO CANNOT BE THE LSP IN THE MSSM/NMSSM, BECAUSE IT WOULD BE STABLE AND THEREFORE THE DARK MATTER, AND THIS IS EXCLUDED BY DIRECT DETECTION EXPERIMENTS SINCE THE SCATTERING CROSS SECTION WOULD BE TOO LARGE BECAUSE OF THE EXCHANGE OF Z.

In supersymmetry (SUSY),
the `$\mu$ from $\nu$' supersymmetric standard model ($\mu\nu$SSM~\cite{LopezFogliani:2005yw,Escudero:2008jg}, see Refs.~\cite{Munoz:2009an,Munoz:2016vaa} for reviews)
% ($\mn$)~\cite{LopezFogliani:2005yw, Escudero:2008jg} 
is a natural extension of the 
minimal supersymmetric standard model (MSSM, see Ref.~\cite{Martin:1997ns} for a review), since only trilinear couplings involving right-handed neutrino superfields, $\hat \nu^c_i$ with $i=1,2,3$, are 
%necessary to be added to its 
added to the superpotential. Thus,
in addition to the usual Dirac Yukawa couplings for neutrinos $Y^{\nu}_{ij} \, \hat H_u\, \hat L_i \, \hat \nu^c_j$, other two types of couplings can be present by gauge invariance solving
%the MSSM superpotential.
%Besides, these couplings 
%The latter solve 
crucial problems of the MSSM.
In particular, the couplings between the three families of right-handed neutrino and Higgs superfields,
$\lambda_{i} \, \hat \nu^c_i\,\hat H_d \hat H_u$,
%, $i=1,2,3$,
generate an effective $\mu$ term solving the so-called
$\mu$ problem~\cite{Kim:1983dt}.
This occurs when the SUSY partners of the right-handed neutrinos,
the right sneutrinos 
%of the three families 
$\tilde \nu_{iR}$, develop vacuum expectation values (VEVs)
after the successful electroweak symmetry breaking (EWSB), with the result
$\mu^{\text{eff}}=\la_i \langle \widetilde \nu_{iR}\rangle$*.
%, solving the so-called $\mu$ problem~\cite{Kim:1983dt}.
%Besides, by the same argument,
Besides, the couplings among right-handed neutrino superfields,
$\frac{1}{3}\kappa{_{ijk}} \hat \nu^c_i\hat \nu^c_j\hat \nu^c_k$,
generate effective Majorana masses for right-handed neutrinos
%the right-handed neutrinos 
of the order of the EWSB scale, 
$\left(m^{\text{eff}}_{\mathcal M}\right)_{ij}={2}\kappa_{ijk}\langle \widetilde \nu_{kR}\rangle$*,
instrumental 
in solving the $\nu$ problem, i.e. the generation of neutrino masses 
and mixing in SUSY. The solution is obtained through a generalized electroweak-scale seesaw mechanism, involving also the neutralinos, that can accommodate the correct neutrino
data with
$Y^{\nu}_{ij} \lsim 10^{-6}$~\cite{LopezFogliani:2005yw,Escudero:2008jg,Ghosh:2008yh,Bartl:2009an,Fidalgo:2009dm,Ghosh:2010zi} (see Refs.~\cite{LopezFogliani:2010bf,Ghosh:2010ig} for reviews).
%solving the so-called $\nu$ problem~\cite{Munoz:2016vaa}.

Both types of couplings discussed above, determined by
$\lambda_i$ and $\kappa_{ijk}$, break explicitly $R$ parity ($R_p$).
Nevertheless,
in the limit $Y^\nu_{ij} \to 0$, $\hat \nu^c_i$ can be identified as
pure singlet superfields without lepton number and
$R_p$ is not broken.
Therefore, $Y^\nu_{ij}$ are the parameters determining 
the violation of $R$ parity ($\rpv$), and as a consequence such violation is small
in the $\mn$.
%with
%the Dirac Yukawa couplings for neutrinos $Y^{\nu}_{ij} \, \hat H_u\, \hat L_i \, \hat \nu^c_j$
%determining the size of the breaking, i.e. $R_p$ is conserved in the limit
%$Y^{\nu}_{ij}\rightarrow 0$.
As is well known, 
% In supersymmetric models where $R$ parity ($R_p$) is conserved, 
% the lightest supersymmetric particle (LSP) is stable, implying that it has to be neutral, since
% otherwise it would bind with nuclei and would be excluded from unsuccessful searches for
% exotic heavy isotopes (see e.g. Ref.~\cite{Kudo:2001ie} and references therein).
% As a consequence, direct searches for supersymmetry (SUSY) at the LHC are based mainly on the presence of missing energy in the final states.
%However, 
in models with $\rpv$ the LSP\footnote{The notion of LSP is in fact misleading in the context of $\rpv$ models, since
SUSY and non-SUSY states are mixed. Nevertheless, for dominant SUSY composition of the
lightest eigenstate, to keep this nomenclature, as we will do in what follows, is reasonable.} is not stable, decaying into
standard model (SM) particles, and basically all SUSY particles (sparticles) are potential candidates 
for LSPs, not only the neutral ones as in $R_p$ conserving models where they are stable and therefore contribute to the dark matter.
This means that in the $\mn$,  
%are potential candidates 
%for LSPs: 
squarks, gluinos,
%charged 
sleptons\footnote{In what follows, the notation sleptons/leptons will be used for the charged sleptons/leptons, and sneutrinos/neutrinos for the neutral sleptons/leptons. 
%(and leptons for the  charged leptons).
}, sneutrinos, neutralinos and charginos, are potential candidates for LSPs.
%, in the $\mn$ all sparticles are candidates for the LSP, as discussed above. 
Therefore, an analysis of the LHC phenomenology associated to each candidate is crucial to test the model.

In this work we start with the systematic analysis of relevant signals 
expected at the LHC for LSP candidates in the $\mn$. As a first candidate we will concentrate on the SUSY partner of the left-handed neutrino, {\it the left sneutrino}, studying in particular its dominant pair 
production channels and decays.\footnote{
For previous analyses in the literature studying other possible signals of 
the $\mn$ at colliders, mainly through light singlet scalars and neutralinos, see Refs.~\cite{Bartl:2009an,Bandyopadhyay:2010cu,Fidalgo:2011ky,Ghosh:2012pq,Ghosh:2014rha,Ghosh:2014ida}. 
%\footnote{
Also, an extension of the $\mn$ and its associated phenomenology was discussed in 
Ref.~\cite{Fidalgo:2011tm} in the context of an extra $U(1)$ gauge symmetry.} 
% Long decay chains in this scenario of left sneutrino LSP might
% give rise to other potentially interesting signals, but that analysis is beyond the scope of this work and we plan to cover it in a forthcoming 
% publication~\cite{sneutrinolongchains}.
{It is worth noticing here that
in $R_p$ conserving models where the left sneutrino LSP is stable and  therefore contributes to thermal dark matter~\cite{Ibanez:1983kw,Hagelin:1984wv}, is ruled out by direct detection experiments~\cite{Falk:1994es,Arina:2007tm}. For proposals to revive it through the breaking of lepton number or inspired in extra dimensions/gauge mediation, see e.g. Refs.~\cite{Hall:1997ah}
and~\cite{Chala:2017jgg}, respectively.}
% As mentioned in the introduction, this problem is not present in 
% $\rpv$ models since the LSP is not stable.}
If the left sneutrino is not the LSP, then the invisible width of the $Z$ puts a lower limit on its mass of about 45 GeV. Also, under the assumption of gaugino and sfermion mass universality at the GUT scale in the MSSM, searches for gauginos and sleptons give rise to a lower limit of about 94 GeV~\cite{Abdallah:2003xe}. 

Related to what was discussed before, although the LSP is not stable
in the $\mn$,
%It is worth noticing here that 
SUSY candidates for dark matter exist in models with $\rpv$.
This is in particular the case of the gravitino~\cite{Borgani:1996ag,Takayama:2000uz}. Although it decays into SM particles
as any other LSP, its lifetime can be longer than the age of the Universe since the
decay width is suppressed both by the inverse of the Planck mass and by the $\rpv$ 
%$R_p$ breaking 
parameters. The latter are 
%expected to be 
very small in the $\mn$, since they are set by the neutrino Yukawa 
couplings $Y^{\nu}_{ij} \lsim 10^{-6}$.
% Yukawa couplings.
Searches for $\mn$ gravitino dark matter\footnote{Concerning other cosmological issues in the $\mn$,
in Ref.~\cite{Chung:2010cd} the generation of the baryon asymmetry of the universe was analysed in the model, with the interesting result that electroweak baryogenesis can be realised.} in Fermi-LAT data through gamma-ray lines have been
carried out in Refs.~\cite{Choi:2009ng,GomezVargas:2011ph,Albert:2014hwa,Gomez-Vargas:2016ocf}, obtaining stringent constraints on the gravitino mass and the lifetime.
It is worth noticing that since 
%Under this assumption of gravitino dark matter, 
%and therefore 
the gravitino is assumed to be the LSP in this framework, each candidate for LSP mentioned above would in fact be the next-to-LSP (NLSP).
%, since the gravitino would be the LSP.
Nevertheless, the analysis of their phenomenology at the LHC is not altered, since they also decay
into ordinary particles using the same channels as if they were the LSP.
Thus the results of this work can also be applied to the case of a left sneutrino NLSP, with the gravitino as the LSP.

The paper is organized as follows.
In Section~\ref{themodel}, the main characteristics of the $\mn$ useful for our computation are briefly discussed.
In Section~\ref{masses}, the spectrum of the model is analyzed, paying special attention
to the neutral fermion mass matrix which determines neutrino masses and mixing.
%\R{me falta por arreglar este parrafo}.
%We start with a review of the $\mn$ in Section~\ref{Section:munuSSM}.
%, introducing the notation and our assumptions.
In Section~\ref{Section:Sneutrino-LSP}, we analyze in detail how the left sneutrino can become
the LSP in some regions of the parameter space of the model, defining at the same time several
interesting benchmark points (BPs). In particular, we study points with a left sneutrino LSP of the first two families, and separately points with a left sneutrino LSP of the third family. The different decay modes of the left sneutrino, depending on
its nature, scalar or pseudoscalar, are discussed Section~\ref{decaymodes}. 
% In the case of the pseudoscalar state its decay is mainly into neutrinos or 
% leptons, with the later case via tau sneutrino due to its large Yukawa coupling.
% Unlike the pseudoscalar state, the scalar one can have a sizable decay into photons, in a way not very different from the Higgs.
% Also lepton decays are possible for the scalar tau sneutrino.
In Section~\ref{Section:detectionSneutrino-LSP},
we study the dominant pair production channels of sneutrinos at the LHC, as well as the signals.
% First, we will see that the production can occur via a $Z$ channel giving rise
% to a pair of scalar and pseudoscalar left sneutrinos.
% Another important source of sneutrino LSP is the direct production of the left 
% slepton
% NLSP. This occurs via a $W^{\pm}$ decaying into the NLSP and a sneutrino, with the NLSP dominantly decaying into another sneutrino plus a very soft $W^{\pm}$ subproduct.
These can consist of a diphoton plus missing transverse energy (from neutrinos), a diphoton plus leptons, and multileptons.
%  is obtained in the case of sneutrinos LSP of any generation, whereas a multilepton state and/or a diphoton state plus leptons arise via the production of a pair of
%sneutrinos LSP of the third family. 
For the regions of the parameter space analyzed, we compute the number of expected events for the signals. Given properly modified search techniques, it is sufficient to give a significant evidence with 13 TeV center-of-mass energy using the current or future integrated luminosity, for a sneutrino mass in the range $95-310$ GeV.
Our conclusions and prospect for future studies of displaced vertices are presented in Section~\ref{Section:Summary-Conclusion}.
Finally, a plethora of useful formulae are given in the Appendices. 
In Appendix~\ref{Section:munuSSM}, the superpotential and the associated soft terms of the $\mn$ are briefly reviewed and discussed.
%the main characteristics of the $\mn$ are reviewed.
In Appendix~\ref{masasm}, the scalar and fermion mass matrices of the model are shown.
Finally, in Appendix~\ref{Section:Coupling}, the relevant interactions for the decays of the left sneutrino are obtained.

\section{The $\mn$}
% and neutrino physics}
\label{themodel}

The couplings of the superpotential relevant for this work are given by
\bea
W = &&
\epsilon_{ab} \left(
Y^e_{ij} \, \hat H_d^a\, \hat L^b_i \, \hat e_j^c +
Y^d_{ij} \, \delta_{\alpha\beta}\, \hat H_d^a\, \hat Q^{b}_{i\alpha} \, \hat d_{j\beta}^{c} 
+
Y^u_{ij} \, \delta_{\alpha\beta}\, \hat H_u^b\, \hat Q^{a}_{i\alpha} \, \hat u_{j\beta}^{c}
\right)
\nonumber\\
% &+&
% \epsilon_{ab} Y^{\nu}_{ij} \, \hat H_u^b\, \hat L^a_i \, \hat \nu^c_j -
&+&   
\epsilon_{ab} \left(
Y^{\nu}_{ij} \, \hat H_u^b\, \hat L^a_i \, \hat \nu^c_j 
-
%\epsilon_{ab}
\lambda_{i} \, \hat \nu^c_i\, \hat H_u^b \hat H_d^a
\right)
+
\frac{1}{3}
\kappa{_{ijk}} 
\hat \nu^c_i\hat \nu^c_j\hat \nu^c_k
% \nonumber\\
% &+&   
% \epsilon_{ab} \left(\lambda_{ijk} \hat L_i^a \hat L_j^b \hat e^c_k 
% + 
% \delta_{\alpha\beta} \lambda'_{ijk} \hat L_i^a \hat Q_j^{b \alpha} \hat d^{c \beta}_k \right)
\ ,
\label{superpotential2}
\eea
as discussed in the Introduction and Appendix~\ref{Section:munuSSM}.
Together with the corresponding soft SUSY-breaking terms, they give rise to the following
tree-level neutral scalar potential:
% $V^{(0)}$
% receives the $F$ and $D$ term contributions
% in addition to terms from $\mathcal{L}_{\text{soft}}$ in Eq.~(\ref{2:Vsoft}), and is therefore given by~\cite{LopezFogliani:2005yw, Escudero:2008jg}:
\begin{equation}
V^{(0)} = V_{\text{soft}} + V_F  +  V_D\ , 
\label{finalpotential}
\end{equation}
with
\bea
V_{\text{soft}}  =&&
%&=& 
\left(
T^{\nu}_{ij} \, H_u^0\,  \widetilde \nu_{iL} \, \widetilde \nu_{jR}^* 
- T^{\lambda}_{i} \, \widetilde \nu_{iR}^*\, H_d^0  H_u^0
+ \frac{1}{3} T^{\kappa}_{ijk} \, \widetilde \nu_{iR}^* \widetilde \nu_{jR}^* 
\widetilde \nu_{kR}^*\
%T_{\nu_{ij}}H^0_{u}\tilde{\nu}_{i}\tilde{\nu}^{c}_{j}
%m_{\tilde{\nu}_{i}}^{2}\tilde{\nu}_{i}\tilde{\nu}_{i}^{*}
%-T_{\lambda_{i}}\tilde{\nu}^c_{i}H^0_{d}H^0_{u}
% +
% \frac{1}{3} {T_{\kappa_{ijk}}\tilde{\nu}^c_{i}\tilde{\nu}^c_{j}\tilde{\nu}^c_{k}}
+
\text{h.c.} \right)
\nonumber\\
&+&
%&&+
\left(m_{\widetilde{L}_L}^2\right)_{ij} \widetilde{\nu}_{iL}^* \widetilde\nu_{jL}
% \widetilde{L}_{Li}^{a^{^*}}  \widetilde{L}^a_{Lj} 
+
\left(m_{\widetilde{\nu}_R}^2\right)_{ij} \widetilde{\nu}_{iR}^* \widetilde\nu_{jR} +
m_{H_d}^2 {H^0_d}^* H^0_d + m_{H_u}^2 {H^0_u}^* H^0_u
% m_{H_1}^{2}H^0_1{^*}H^0_{1}+m_{H_2}^{2}H^0_{2}H_{2}^{0*}
% m_{\tilde{L}_{ij} }^2 \, \tilde{\nu}_{Li} \, \tilde{\nu}_{Lj}^* +
%m_{\tilde{\nu}_{ij}}^{2}\tilde{\nu}_{Ri}\tilde{\nu}^{*}_{Rj}
\ ,
\label{akappa}
\\
\nonumber
\\
%\end{eqnarray}
%with
%$a_{\nu_{ij}}\equiv (A_\nu Y_\nu)_{ij}$, 
%$a_{\lambda_i}\equiv (A_\lambda\lambda)_i$,  
%$a_{\kappa_{ijk}}\equiv (A_\kappa \kappa)_{ijk}$,
%Besides the potential receives the $D$ and $F$-term
%contributions
%\begin{equation}
%\end{equation}
%with $G^2\equiv g_{1}^{2}+g_{2}^{2}$, and
%
%\begin{eqnarray}
V_{F}  =&&
 \lambda_{j}\lambda_{j}^{*}H^0_{d}H_d^0{^{^*}}H^0_{u}H_u^0{^{^*}}
 +
\lambda_{i}\lambda_{j}^{*}\tilde{\nu}^{*}_{iR}\tilde{\nu}_{jR}H^0_{d}H_d^0{^*}
 +
\lambda_{i}\lambda_{j}^*
\tilde{\nu}^{*}_{iR}\tilde{\nu}_{jR}  H^0_{u}H_u^0{^*}   
\nonumber\\                                              
&+&
\kappa_{ijk}\kappa_{ljm}^{*}\tilde{\nu}^*_{iR}\tilde{\nu}_{lR}
                                   \tilde{\nu}^*_{kR}\tilde{\nu}_{mR}
%  \nonumber\\
%   &-& 
- \left(\kappa_{ijk}\lambda_{j}^{*}\tilde{\nu}^{*}_{iR}\tilde{\nu}^{*}_{kR} H_d^{0*}H_u^{0*}                                      
 -Y^{\nu}_{ij}\kappa_{ljk}^{*}\tilde{\nu}_{iL}\tilde{\nu}_{lR}\tilde{\nu}_{kR}H^0_{u}
 \right.
 \nonumber\\
 &+&
 \left.
 Y^{\nu}_{ij}\lambda_{j}^{*}\tilde{\nu}_{iL} H_d^{0*}H_{u}^{0*}H^0_{u}
% \nonumber \\
% &+&
+{Y^{\nu}_{ij}}^{*}\lambda_{k} \tilde{\nu}_{iL}^{*}\tilde{\nu}_{jR}\tilde{\nu}_{kR}^* H^0_{d}
 + \text{h.c.}\right) 
\nonumber \\
 &+& 
Y^{\nu}_{ij}{Y^{\nu}_{ik}}^* \tilde{\nu}^{*}_{jR}
\tilde{\nu}_{kR}H^0_{u}H_u^0{^*}                                                
 +
Y^{\nu}_{ij}{Y^{\nu}_{lk}}^*\tilde{\nu}_{iL}\tilde{\nu}_{lL}^{*}\tilde{\nu}_{jR}^{*}
                                  \tilde{\nu}_{kR}  
 +
Y^{\nu}_{ji}{Y^{\nu}_{ki}}^*\tilde{\nu}_{jL}\tilde{\nu}_{kL}^* H^0_{u}H_u^{0*}\, ,
\\
\nonumber
\\
V_D  =&&
\frac{1}{8}\left(g^{2}+g'^{2}\right)\left(\widetilde\nu_{iL}\widetilde{\nu}_{iL}^* 
+H^0_d {H^0_d}^* - H^0_u {H^0_u}^* \right)^{2}\, .
\label{dterms}
\eea
The electroweak gauge couplings are estimated at the $m_Z$ scale by
$e=g\sin\theta_W=g'\cos\theta_W$.
Since only dimensionless trilinear couplings are present in the 
superpotential, the EWSB is determined by the
soft terms of the scalar potential. Thus all known particle physics phenomenology can be reproduced in the $\mn$ with one scale, the about 1 TeV scale of the soft terms, avoiding the introduction of
`ad-hoc' high-energy scales.
% Upon EWSB, not only Higgses but also left and right sneutrinos acquire VEVs, and fields with the same color, electric charge and spin mix giving rise to a rich phenomenology.
With the choice of CP conservation,\footnote{The $\mu\nu$SSM with spontaneous 
CP violation was studied in Ref.~\cite{Fidalgo:2009dm}.} 
one can define the neutral scalars as
\bea
%H_d^0 &=& \frac{1}{\sqrt 2} (H_{dR}+ i H_{dI}) + v_d\ ,
H_d^0 &=& \frac{1}{\sqrt 2} \left(H_{d}^\mathcal{R} + v_d + i\ H_{d}^\mathcal{I}\right)\ ,
\\
\label{vevd}
\nonumber
\\
%H^0_u &=& \frac{1}{\sqrt 2} (H_{uR} +i H_{uI}) + v_u\ ,  
H^0_u &=& \frac{1}{\sqrt 2} \left(H_{u}^\mathcal{R}  + v_u +i\ H_{u}^\mathcal{I}\right)\ ,  
\\
\label{vevu}
\nonumber 
\\
\widetilde{\nu}_{iR} &=&
%     \frac{1}{\sqrt 2} (\widetilde{\nu}^c_{iR}+ i \widetilde{\nu}^c_{iI})+\nu_i^c,  
      \frac{1}{\sqrt 2} \left(\widetilde{\nu}^{\mathcal{R}}_{iR}+ v_{iR} + i\ \widetilde{\nu}^{\mathcal{I}}_{iR}\right),  
\\
\label{vevnuc}
\nonumber
\\
% \widetilde{\nu}_i &=& \frac{1}{\sqrt 2} (\widetilde{\nu}_{iR}  +i \widetilde{\nu}_{iI})   + \nu_i\ ,
  \widetilde{\nu}_{iL} &=& \frac{1}{\sqrt 2} \left(\widetilde{\nu}_{iL}^\mathcal{R} 
  + v_{iL} +i\ \widetilde{\nu}_{iL}^\mathcal{I}\right)\ ,
\label{vevnu}
\eea
in such a way that after the EWSB they develop the real VEVs 
\bea
\langle H_d^0 \rangle = \frac{v_d}{\sqrt 2}\ , \, \quad 
\langle H_u^0 \rangle = \frac{v_u}{\sqrt 2}\ , \,
\quad 
\langle \widetilde \nu_{iR}\rangle = \frac{v_{iR}}{\sqrt 2}\ , \,  \quad
\langle \widetilde \nu_{iL} \rangle = \frac{v_{iL}}{\sqrt 2}
\ .
\label{vevs}
\eea
The eight minimization conditions 
with respect to $v_d$, $v_u$,
$v_{iR}$ and $v_{iL}$
can then be written as
%%%%%%%%%%%%%%%%%%%%%%%%%%%%%%%%%%%%%%%%%%%%%%%%%%%%%%%%%%%%%%%%
{\small
\bea
 m_{H_{d}}^{2}
%v_{d}
& =& 
%\frac{1}{2}\left\{
-\frac{1}{8}\left(g^{2}+g'^{2}\right)\left(v_{iL}v_{iL}+v_{d}^{2}-v_{u}^{2}\right)
%v_{d}
 - \frac{1}{2}\lambda_i\lambda_{j}v_{iR} v_{jR} 
 - \frac{1}{2}\lambda_{i}\lambda_{i}
%v_{d}
v_{u}^{2}
%\right.
\nonumber\\ 
  &&
%\left.  
 +v_{iR}\tan\beta\left(\frac{1}{\sqrt 2} T^{\lambda}_i 
%v_{u}
%v_{d}
   + \frac{1}{2}\lambda_{j}\kappa_{ijk}
%v_{\nu_i^c} 
v_{kR} 
%v_{u}
\right)
% \nonumber\\ 
%   &&
   + Y^{\nu}_{ij} \frac{v_{iL}}{2v_d}
\left(\lambda_k v_{kR} v_{jR} +
%   + Y_{\nu_{ij}}
\lambda_{j} 
%v_{\nu_{i}}
v_{u}^2\right)
%\right\}
- \frac{\sqrt 2}{v_d}V^{(n)}_{v_d},
%- \frac{V^{(n)}_{v_d}}{v_d/\sqrt 2}
   \label{tadpoles1}
\\ 
\nonumber
\\
m_{H_{u}}^{2}
%v_{u} 
& =& \frac{1}{8}\left(g^{2}+g'^{2}\right)\left(v_{iL}v_{iL}+v_{d}^{2}-v_{u}^{2}\right)
%v_{u}
 - \frac{1}{2}\lambda_{i}\lambda_j
%v_{u}
v_{iR}v_{jR}
-\frac{1}{2}\lambda_{j}\lambda_{j}v_{d}^2
\nonumber\\
&&
+ \lambda_j Y^{\nu}_{ij}v_{iL}v_{d} 
%v_u 
   - \frac{1}{2}Y^{\nu}_{ij}Y^{\nu}_{ik}
%v_{u}
v_{kR}v_{jR}
- \frac{1}{2}Y^{\nu}_{ij}Y^{\nu}_{kj}v_{iL}v_{kL} 
   \nonumber \\ 
&&  
+ \frac{v_{iR}}{\mathrm{tan} \beta}\left(\frac{1}{\sqrt 2} T^{\lambda}_i
+\frac{1}{2}\lambda_{j}\kappa_{ijk}v_{kR}\right)
-\frac{v_{iL}}{v_u}\left(\frac{1}{\sqrt 2} T^{\nu}_{ij}v_{jR}+
\frac{1}{2}Y^{\nu}_{ij}\kappa_{ljk}v_{lR}v_{kR}\right)
%  - \frac{\sqrt 2 V^{(n)}_{v_u}}{v_u}, 
\nonumber\\
&&
- \frac{\sqrt 2}{v_u} V^{(n)}_{v_u}\ , 
   \label{tadpoles2}
\\ 
\nonumber\\
  (m^2_{\widetilde{\nu}_R})_{ij}v_{jR} &=&    
   \frac{1}{\sqrt 2} \left(-T^{\nu}_{ji}v_{jL}v_{u}+ T^{\lambda}_iv_u v_d - T^{\kappa}_{ijk}v_{jR}v_{kR}\right) 
   -\frac{1}{2} \lambda_i\lambda_{j}\left(v_{u}^{2}+v_{d}^{2}\right)v_{jR} 
   + \lambda_{j}\kappa_{ijk}v_{d}v_{u}v_{kR}
   \nonumber \\ 
  &&
   - \kappa_{lim}\kappa_{ljk}v_{mR}v_{jR}v_{kR}
   + \frac{1}{2}Y^{\nu}_{ji}\lambda_{k}v_{jL}v_{kR}v_{d}
   + \frac{1}{2}Y^{\nu}_{kj}\lambda_{i}v_{d}v_{kL}v_{jR}
- Y^{\nu}_{jk}\kappa_{ikl}v_{u}v_{jL}v_{lR}
   \nonumber \\ 
  &&
    - \frac{1}{2}Y^{\nu}_{ji}Y^{\nu}_{lk}v_{jL}v_{lL}v_{kR}
    - \frac{1}{2}Y^{\nu}_{ki}Y^{\nu}_{kj}v_{u}^{2}v_{jR}\
    -\ V^{(n)}_{v_{iR}}\ ,  
     \label{tadpoles3}
     \\
    \nonumber\\
 (m^2_{\widetilde{L}_L})_{ij}v_{jL}& =&  -\frac{1}{8}\left(g^{2}+g'^{2}\right)
\left(v_{jL}v_{jL}+v_{d}^{2}-v_{u}^{2}\right)v_{iL}
-\frac{1}{\sqrt 2} T^{\nu}_{ij}v_u v_{jR}
   +\frac{1}{2}Y^{\nu}_{ij}\lambda_{k}v_{d}v_{jR}v_{kR}
     \nonumber \\
%\label{tadpo}
 &&
     + \frac{1}{2}Y^{\nu}_{ij}\lambda_{j}v_{u}^{2}v_{d}
%     \nonumber \\
% &&
     - \frac{1}{2}Y^{\nu}_{il}\kappa_{ljk}v_{u}v_{jR}v_{kR}
 - \frac{1}{2}Y^{\nu}_{ij} Y^{\nu}_{lk}v_{lL}v_{jR}v_{kR}
   - \frac{1}{2}Y^{\nu}_{ik}Y^{\nu}_{jk}v_{u}^{2}v_{jL}
\nonumber\\
&&
 - V^{(n)}_{v_{iL}}\ ,
\label{tadpoles4}
\eea}
%%%%%%%%%%%%%%%%%%%%%%%%%%%%%%%%%%%%%%%%%%%%%%%%%%%%%%%%%%%%%%%%
%%%%%%%%%%%%%%%%%%%%%%%%%%%%%%%%%%%%%%%%%%%%%%%%%%%%%%%%%%%%%%%%
where
$\tan\beta\equiv\frac{v_u}{v_d}$, 
$V^{(n)}_{x}\equiv \partial V^{(n)}/\partial x$ with
$x=v_d,\,v_u,\,v_{iR},\,v_{iL}$, and $V^{(n)}$ represents the $n$--loop
radiative correction to the potential, $V=V^{(0)} + V^{(n)}$.
%where the tree-level neutral scalar potential $V^{(0)}$ can be found in Appendix~\ref{Section:munuSSM}. 
%In our computation
%we use {\tt SPheno} {{v}}3.3.6~\cite{Porod:2003um,Porod:2011nf} with the complete 2--loop radiative corrections (which are crucial to reproduce the Higgs mass) generated by SARAH~
%{\bf{[If needed add Staub:2013tta]}}\cite{Staub:2008uz,Staub:2011dp}, except for the computation of the charged scalar masses where only the 1-loop correction is implemented.
%The latter approximation introduces in our BPs small deviations of about 1--4\% in the mass of the relevant lightest charged scalar.
The scale at which the EWSB conditions are imposed is 
$M_{EWSB} = \sqrt{m_{\tilde t_l} m_{\tilde t_h}}$, 
where $m_{\tilde t_l}$ and $m_{\tilde t_h}$ 
correspond to the lightest and heaviest stop mass eigenvalues, respectively, measured at $M_{EWSB}$.

% \vspace{0.25cm}

% \noindent
% {\bf The parameter space}

%\noindent 
%We can see from Eqs.~(\ref{akappa})-(\ref{dterms}) 

The free parameters in the neutral scalar sector of the $\mn$
at the low scale $M_{EWSB}$
are therefore: 
$\lambda_i$, $\kappa_{ijk}$, $Y^{\nu}_{ij}$, $m_{H_{d}}^{2}$, $m_{H_{u}}^{2}$,  
$(m_{\widetilde{\nu}_R}^2)_{ij}$,
$(m_{\widetilde{L}_L}^2)_{ij}$, 
%$m^2_{\widetilde{\nu}_{ij}^{c}}$, 
%$m^2_{\widetilde{L}_{ij}}$,
$T^{\lambda}_i$, $T^{\kappa}_{ijk}$ and $T^{\nu}_{ij}$.
From the minimization conditions 
%of Eqs.~(\ref{tadpoles1})--(\ref{tadpoles4}), 
we can eliminate
the 
%low-energy 
soft masses $m_{H_{d}}^{2}$, $m_{H_{u}}^{2}$, $m^2_{\widetilde{\nu}_{iR}}$ and
$m^2_{\widetilde{L}_{iL}}$ in favor
of the VEVs, assuming 
%$\vd$, $\vu$, $v_{iR}$, and $v_{iL}$. 
%For that, 
in the case of the sleptons 
%we have to take into account the arguments above Eq.~(\ref{tyukawa}) for the use of
diagonal sfermion mass matrices.
%, as in the previous subsection,
%$m_{\widetilde{\phi}_{ij}}^2=\delta_{ij} m_{\widetilde{\phi}_{i}}^2$,
% the minimization conditions (\ref{tadpoles3}) and (\ref{tadpoles4}) turn out to be simplified, and we can eliminate
% Since strong upper bounds upon the intergenerational scalar mixing 
% exist, we will assume that 
% such mixings are negligible, and therefore that the sfermion soft mass 
% matrices are diagonal in the flavour space.
%This occurs for example in several string compactifications as a consequence of 
%having diagonal
%Kahler metrics, or when the dilaton is the source of SUSY breaking \cite{bim}. 
%Thus using the above minimization conditions, 
%it 
%is operationally easier 
% $m^2_{\widetilde{\nu}_{iR}}$ and
% $m^2_{\widetilde{L}_{iL}}$
%$\mHd$, $\mHu$, $m_{\widetilde{L}_{i}}$, and $m_{\widetilde{\nu}_{i}^{c}}$,
% in favor of the VEVs.
%, $\vd$, $\vu$, $v_{\nu^c_i}$, and $v_{\nu_i}$.
In addition, using $\tanb$ and the SM Higgs VEV,
$v/\sqrt 2= \sqrt 2 m_Z/\sqrt{g^2 + g'^2}\approx$ 174 GeV,
%$v= 2 M_Z/\sqrt{g^2 + g'^2}\approx 246$ GeV,
%$\frac{v^2}{2}= \frac{2 M_Z^2}{(g^2 + g'^2)}\approx$ (174 GeV)$^2$,
%$v^2/2= 2 M_Z^2/(g^2 + g'^2)\approx$ (174 GeV)$^2$,
%$v^2= 4 M_Z^2/(g^2 + g'^2)\approx$ (246 GeV)$^2$,
% (CHECK THIS VALUE),
%where $G^2\equiv g_1^2 + g_2^2$,
%$\tanb$, and $v_{\vi}$, 
we can determine the SUSY Higgs 
VEVs, $v_d/\sqrt 2$ and $v_u/\sqrt 2$, through $v^2 = v_d^2 + v_u^2 + \sum_i v^2_{iL}$. 
Since $v_{iL} \ll v_d, v_u$, we obtain 
%we can define the value of $\tanb$ as usual,
%$\tan\beta=\frac{v_u}{v_d}$, and therefore
$v_d\approx v/\sqrt{\tan^2\beta+1}$.
Assuming that all the soft trilinear parameters are proportional to the Yukawa couplings
\bea
T^{e}_{ij} &=& A^{e}_{ij} Y^{e}_{ij}\ , \;\;
T^{d}_{ij} = A^{d}_{ij} Y^{d}_{ij}\ , \;\;
T^{u}_{ij} = A^{u}_{ij} Y^{u}_{ij}\ ,
\label{tyukawa}
\\
T^{\nu}_{ij} &=& A^{\nu}_{ij} Y^{\nu}_{ij}\ , \;\;
T^{\lambda}_i= A^{\lambda}_i\lambda_i\ , \;\;
T^{\kappa}_{ijk}= A^{\kappa}_{ijk} \kappa_{ijk}\ ,
\label{tmunu}
\eea
where the summation convention on repeated indexes does not apply,
we are then left with  
the following set of variables as independent parameters in the neutral scalar sector:
\bea
\lambda_i, \, \kappa_{ijk},\, Y^{\nu}_{ij}, \tan\beta, \, v_{iL}, \, v_{iR}, \, A^{\lambda}_i, \, A^{\kappa}_{ijk}, \, A^{\nu}_{ij}\ ,
\label{freeparameters}
\eea
The rest of soft parameters of the model, namely the following gaugino masses,
%$M_{1,2,3}$, 
scalar masses, 
%$m^2_{\tilde Q_i,\tilde u^c_i,\tilde d^c_i,\tilde e^c_i}$,
and trilinear parameters:
%, $A_{u_{ij},d_{ij},e_{ij}}$, 
%
\bea
M_1, \, M_2,\, M_3, \, m_{\tilde Q_{iL}},\, 
m_{\tilde u_{iR}}, \, m_{\tilde d_{iR}}, \,
m_{\tilde e_{iR}}, \,
A^u_{ij}, \, A^d_{ij}, \, A^e_{ij}\ ,
\label{freeparameterssoft}
\eea
are also
taken as free parameters and specified at low scale.
%where the summation convention on repeated indexes does not apply for this case.
It is worth remarking nevertheless that,
%, as discussed in Subsection~\ref{masses},
to reproduce neutrino data, one has to impose extra constraints on the parameters of the model.
Using the simplified formula of 
Eq.~(\ref{Limit no mixing Higgsinos gauginos}) below, one can trivially see
that the parameters $Y^{\nu}_{i}$, $\lambda_i$, $\kappa$, $v_{R}$, $v_{iL}$ and $M_{1,2}$ must be constrained in order to obtain the experimentally probed neutrino masses and mixing angles.
% Using the simplified formula 
% (\ref{Limit no mixing Higgsinos gauginos}), we already see
% that once fixed for example $Y_{\nu_{i}}$, $\kappa$ and $v_{\nu^c}$, the other parameters
% $v_{\nu_i}$, and $M_{1,2}$ are constrained in order to reproduce neutrino masses and mixing angles.

% \vspace{0.25cm}

% \noindent
% {\bf Effective terms}

% \noindent 
After the successful EWSB, several crucial terms are effectively generated in the $\mn$. Note
%As we can easily see 
from Eq.~(\ref{tadpoles3}) that
the VEVs of the right sneutrinos
%, $v_{iR}$, 
are naturally of the order of the EWSB scale
\beq
\frac{v_{iR}}{\sqrt 2}\approx 1\ \text{TeV}
%m^{\text{eff}}_{\mathcal D}
\ ,
\label{vevR}
\eeq 
implying that the $\mu$ problem of the MSSM~\cite{Kim:1983dt} is solved thanks to the presence of the 
$5^{\rm th}$ term in the superpotential above,
%of Eq.~(\ref{superpotential2}),
%(\ref{superpotential}).
which generates 
%After the successful EWSB, it generates 
an effective $\mu$ term with 
%%%%%%%%%%%%%%%%%%%%%%%%%%%%%%%%%%%%%%%%%%%%%%%%%%%%%%%%%%%%%%
%%%%%%%%%%%%%%%%%%%%%%%%%%%%%%%%%%%%%%%%%%%%%%%%%%%%%%%%%%%%%%%%
\beq
\mu^{\text{eff}}=\la_i \frac{v_{iR}}{\sqrt 2}\ .
\label{muterm}
\eeq
In addition, the $6^{\rm th}$ term in the superpotential generates effective Majorana masses for the right-handed neutrinos 
%of that order,$2\kappa_{ijk}v_{\nu^c_k}$,
%
\beq
\left(m^{\text{eff}}_{\mathcal M}\right)_{ij}={2}\kappa_{ijk}\frac{v_{kR}}{\sqrt 2}\ ,
\label{majorana}
\eeq
and, as a consequence, we can implement naturally a (generalized) electroweak-scale seesaw
in the $\mn$ which includes the neutralinos, asking for 
neutrino Yukawa 
% couplings of the order of the electron Yukawa coupling or smaller,
% $Y_{\nu_{ij}} \sim 10^{-6} - 10^{-7}$ \cite{LopezFogliani:2005yw,
% Escudero:2008jg,Ghosh:2008yh,Bartl:2009an,Fidalgo:2009dm,Ghosh:2010zi},
couplings of the order of the electron Yukawa coupling or smaller 
(see the first two terms of Eqs.~(\ref{Limit no mixing Higgsinos gauginos}) and (\ref{Limit no mixing Higgsinos gauginos2}) below)~\cite{LopezFogliani:2005yw,
Escudero:2008jg,Ghosh:2008yh,Bartl:2009an,Fidalgo:2009dm,Ghosh:2010zi}:
\beq
Y^{\nu}_{ij} \lsim 10^{-6}\ .
\label{yukawas}
\eeq
% i.e. we work with Dirac masses for neutrinos
% $m_D\sim Y_{\nu}v_u\lsim 10^{-4}$ GeV. 
This means that we work with Dirac masses for neutrinos of the order of
\beq
\left(m^{\text{eff}}_{\mathcal D}\right)_{ij}=Y^{\nu}_{ij} \frac{v_u}{\sqrt 2}
%m_D\sim Y^{\nu}v_2
\lsim 10^{-4}
\ \text{GeV}\ . 
\label{dirac}
\eeq
and that no `ad hoc' high-energy scales (larger than a TeV) are necessary to
reproduce experimentally consistent neutrino masses.
It is worth pointing out in this context that
%$\rpv$ is small in the $\mn$ since, as explained below Eq.~(\ref{superpotential}),
%it is determined by the value of
%$Y^\nu$.
the VEVs 
of the left sneutrinos are much smaller than the other VEVs. 
%in Eq.~(\ref{vevs}).
This is because of
the small value of $Y^\nu$.
% in the $\mn$.
%, unlike the other VEVs which are of order the electroweak scale.
We can see in this respect that in Eq.~(\ref{tadpoles4}),
$v_{iL}\to 0$ as $Y^{\nu}_{ij}\to 0$.
%to fulfill
%$\partial V^{(0)}/\partial v_{\nu_i} \approx 0$. 
% It is then easy to estimate the values of these VEVs as $v_{\nu}\lsim m_D$~\cite{LopezFogliani:2005yw}.
It is then easy to estimate the values of 
%left sneutrino 
VEVs as
$v_{iL}\lsim m^{\text{eff}}_{\mathcal D}$~\cite{LopezFogliani:2005yw}, thus:
\beq
\frac{v_{iL}}{\sqrt 2}\lsim 10^{-4}
\ \text{GeV}
%m^{\text{eff}}_{\mathcal D}
\ .
\label{vevdirac}
\eeq
This result allows that the 
%generalized 
seesaw of the $\mn$
%, which include the neutralinos, 
works properly, since
the third term $\sim$
$v_L^2/M$
in Eqs.~(\ref{Limit no mixing Higgsinos gauginos}) and (\ref{Limit no mixing Higgsinos gauginos2}) below, is of the same order as the first two.
Finally, the $4^{\rm th}$ term in the 
superpotential 
%of Eq.~(\ref{superpotential2}) 
generates effective bilinear $\rpv$ couplings 
%%%%%%%%%%%%%%%%%%%%%%%%%%%%%%%%%%%%%%%%%%%%%%%%%%%%%%%%%%%%%%
%%%%%%%%%%%%%%%%%%%%%%%%%%%%%%%%%%%%%%%%%%%%%%%%%%%%%%%%%%%%%%%%
\beq
\epsilon_i^{\text{eff}}=Y^{\nu}_{ij}\frac{v_{jR}}{\sqrt 2}\ ,\label{bilinearterm}
\eeq
%%%%%%%%%%%%%%%%%%%%%%%%%%%%%%%%%%%%%%%%%%%%%%%%%%%%%%%%%%%%%%
%%%%%%%%%%%%%%%%%%%%%%%%%%%%%%%%%%%%%%%%%%%%%%%%%%%%%%%%%%%%%%%%
%
as those constituting the bilinear $R$-parity
violating model (BRpV, see Ref.~\cite{Barbier:2004ez} for a review).

%\vspace{0.25cm}

%\noindent 
Recapitulating, the superpotential of the $\mn$ 
%Eq.~(\ref{superpotential2}) 
serves both the purposes of 
solving the $\mu$ problem and generating non-zero neutrino masses and mixing solving the
$\nu$ problem.
As a consequence of the new terms introduced in the superpotential to solve these challenges, 
$R_p$ is explicitly broken with its breaking controlled by the small Yukawa couplings for neutrinos, 
i.e. $R_p$ 
%$R$-parity 
is restored for $Y^{\nu}_{ij}\to 0$.

\section{The spectrum of the model}

\label{masses}

% The main characteristics of the $\mn$ are reviewed in Appendix~\ref{Section:munuSSM}.
% There, the superpotential and the associated soft terms are discussed, as well as the scalar potential, the parameter space, and the relevant effective terms appearing after the EWSB.
Similar to the MSSM, where the couplings and Higgs VEVs determine the mixing of
Bino, Wino and Higgsinos, producing the four neutralino states, 
the new couplings and sneutrino VEVs 
%(see Eq.~(\ref{vevs})) 
in the $\mn$ induce new mixing of states~\cite{LopezFogliani:2005yw, Escudero:2008jg}.
Summarizing, there are ten neutral fermions (neutralinos-neutrinos),
five charged fermions (charginos-leptons),
eight neutral scalars and seven neutral pseudoscalars (Higgses-sneutrinos),
and seven charged scalars (charged Higgses-sleptons).
The associated mass matrices were studied in 
Refs.~\cite{Escudero:2008jg,Bartl:2009an}, and can be found 
%In this work we write them
%\footnote{{We follow the convention of using for the eigenstates the 
%names of the detected particles: neutrinos, leptons, Higgses.}} 
in our Appendix~\ref{masasm}.

%The scalar mass matrices are written in  Appendix~\ref{masasm}.1.
Concerning the neutral scalars, 
the right and left sneutrino VEVs lead to mixing of the neutral Higgses with the
sneutrinos in the scalar potential, giving rise to $8\times 8$ (`Higgs') mass matrices for scalar and pseudoscalar states.
%in Appendices~\ref{SubAppendix-scalarmasses} and~\ref{SubAppendix:NeutrealScalar}, respectively. 
%They are written
%(before and) after
%replacing the values of the soft masses with the corresponding VEVs obtained %through the minimization conditions of 
%Eqs.~(\ref{tadpoles1})--(\ref{tadpoles4}), assuming that 
%the sfermion soft mass matrices are diagonal in flavor space.
Note that after rotating away the
pseudoscalar would be Goldstone boson, we are left with seven 
pseudoscalar
states. 
%It is also worth noticinghere that 
The $5\times 5$ Higgs-right sneutrino submatrix is almost decoupled from
the $3\times 3$ left sneutrino submatrix,
%\footnote{Thus although the LSP is in fact the lightest mass eigenstate of the whole matrix, the composition of the sneutrino will dominate over the others. Also, the notion of LSP is misleading in the context of $\rpv$ models, since SUSY and non-SUSY states are mixed. Nevertheless, for dominant SUSY composition of the lightest eigenstate to keep this nomenclature as we will do in what follows, seems reasonable.}
since the mixing occurs through terms
proportional to $Y^{\nu}_{ij}$ or $v_{iL}$ 
(see Eqs.~(\ref{prueba5})--(\ref{prueba6}) and~(\ref{pruebas6})--(\ref{pruebas8})), 
% and these quantities are very small, $Y_{\nu_{ij}} \lsim 10^{-6}$ and
% $v_{\nu_i}\lsim 10^{-4}$ GeV, as discussed in 
% Subsection~\ref{minimization}.
and these quantities are very small in order to satisfy neutrino data, as shown in 
Eqs.~(\ref{yukawas}) and~(\ref{vevdirac}).
Besides, the former $5\times 5$ submatrix is of the
next-to-MSSM (NMSSM, see Ref. \cite{Ellwanger:2009dp} for a review) type, apart from the small corrections proportional to 
$Y^{\nu}_{ij}$, and the fact that in the NMSSM there is only one singlet.

The charged scalars have
%, we write in the Appendix
%\ref{SubAppendix:chargedScalar} the 
a $8\times 8$ (`charged Higgs') mass matrix.
Similar to the 
%neutral scalar 
Higgs mass matrices where some sectors are decoupled, the $2\times 2$ 
charged Higgs submatrix is decoupled from the $6\times 6$ slepton 
submatrix (see Eqs.~(\ref{pruebass11})--(\ref{olvidado})). 
In addition, the right sleptons are decoupled from the left ones,
%and ~(\ref{pruebass6}), 
since the mixing terms are suppressed by the electron-type Yukawa couplings
or $v_{iL}$ (see Eq.~(\ref{pruebass5})).

%Finally, let us remark that
The squark mass matrices,
%are written in Appendices~\ref{squarks} and~\ref{squarks2}.
when compared to the MSSM/NMSSM case, maintain their structure essentially unaffected, provided that one uses the effective $\mu$ term of Eq.~(\ref{muterm}), and neglects
the terms proportional to small parameters such as $Y^{\nu}_{ij}$, $v_{iL}$.
% and $\lambda'_{ijk}$.

Concerning the fermion mass matrices,
% are written in Appendix~\ref{masasm}.2.
%\ref{masasm}.
the neutral one 
%mass matrix 
%is shown there 
%in Appendix~\ref{nfms} and 
will be discussed below in the context of neutrino physics, since it is crucial for determining neutrino masses and mixing.
For the charged fermions,
the MSSM charginos mix with the leptons in the $\mn$ giving rise to a
$5\times 5$ (`lepton') mass matrix.
%shown in the Appendix.
%~\ref{cfms}.
Nevertheless, the $2\times 2$ chargino submatrix is basically decoupled from the
$3\times 3$ lepton submatrix, since the off-diagonal entries are supressed 
by $Y^{\nu}_{ij}$, $Y^{e}_{ij}$, $v_{iL}$ (see Eq.~(\ref{submatrix}). The 
former submatrix is like the one of the MSSM/NMSSM 
provided that one uses the effective $\mu$ term of Eq.~(\ref{muterm}).
Finally, down- and up-quark mass matrices can also be found in the Appendix.
%Appendices~\ref{downquark} and~\ref{upquark}.

\vspace{0.25cm}

\noindent
{\bf Neutrino physics}

\noindent
%Concerning the neutral fermions, 
We have discussed in the previous section,  
%in Appendix~\ref{minimization}.1
that
effective Majorana masses for right-handed neutrinos of the order of the EWSB scale
are dynamically generated in the $\mn$ (see Eq.~(\ref{majorana})).
In addition, the MSSM neutralinos mix with the left- and righ-handed neutrinos giving rise to the $10\times 10$ neutral fermion (`neutrino') mass matrix shown in 
Eq.~(\ref{neumatrix}), which has the structure of 
%Appendix~\ref{nfms}.
%Notice that the structure of this matrix is that of 
a generalized electroweak-scale seesaw.
%, since it involves not only the right-handed neutrinos but also the neutralinos. 
Because of this structure, data on neutrino physics~\cite{Gonzalez-Garcia:2015qrr,Forero:2014bxa,Capozzi:2013csa} can easily be reproduced at tree level~\cite{LopezFogliani:2005yw,Escudero:2008jg,Ghosh:2008yh,Bartl:2009an,Fidalgo:2009dm,Ghosh:2010zi}, even with diagonal Yukawa couplings~\cite{Ghosh:2008yh,Fidalgo:2009dm}, i.e.
$Y^{\nu}_{ii}=Y^{\nu}_i$ and vanishing otherwise.
Qualitatively, we can understand this in the following way. First of all, the three neutrino masses are going to be very small since the entries of the first three rows (and columns) of the neutrino matrix 
%of Eq.~(\ref{neumatrix}) 
are much smaller than the rest of the entries. 
%Notice in this sense that 
The latter are of the order of the electroweak scale, whereas the former are of the order 
of the Dirac masses for neutrinos (see Eq.~(\ref{dirac})) \cite{LopezFogliani:2005yw,
Escudero:2008jg}. 
Second, from this matrix one can obtain a simplified formula 
%in Eq.~(\ref{Limit no mixing Higgsinos gauginos}) 
for the effective mixing mass matrix of the 
light neutrinos~\cite{Fidalgo:2009dm}:
\begin{eqnarray}
\label{Limit no mixing Higgsinos gauginos}
(m^{\text{eff}}_{\nu})_{ij} 
\simeq \frac{Y^{\nu}_iY^{\nu}_jv_u^2}
{6\sqrt 2 \kappa v_{R}}
                   (1-3 \delta_{ij})-\frac{v_{iL} v_{jL}}{4M^{\text{eff}}}
%&  
-\frac{1}{4M^{\text{eff}}}\left[\frac{v_d\left(Y^{\nu}_iv_{jL}
   +Y^{\nu}_jv_{iL}\right)}{3\lambda}
   +\frac{Y^{\nu}_iY^{\nu}_jv_d^2}{9\lambda^2 }\right]\ ,
   \nonumber\\
  \end{eqnarray}     
with
\begin{eqnarray}
\label{effectivegauginomass}
 M^{\text{eff}}\equiv M -\frac{v^2}{2\sqrt 2 \left(\kappa v_R^2+\lambda v_u v_d\right)
        \ 3 \lambda v_R}\left(2 \kappa v_R^{2} \frac{v_u v_d}{v^2}
        +\frac{\lambda v^2}{2}\right)\ ,
%\nonumber\\
\end{eqnarray} 
where ${M}= \frac{M_1 M_2}{g'^2 M_2 + g^2 M_1}$. 
Here is assumed $\lambda_i = \lambda$, $v_{iR}= v_{R}$, and
$\kappa_{iii}\equiv\kappa_{i}=\kappa$ and vanishing otherwise.
Of the five terms in $(m^{\text{eff}}_{\nu})_{ij}$, the first two 
%in Eq.~(\ref{Limit no mixing Higgsinos gauginos}) is 
are generated through the mixing of left-handed neutrinos $\nu_L$ with right-handed neutrinos $\nu_R$-Higgsinos.
The rest of them also include the gaugino mixing.
Using this approximate formula it is easy
to understand how diagonal Yukawas can give rise to off-diagonal entries in the mass matrix. The key point are clearly the extra contributions with respect to the ordinary seesaw, given by the four terms which are not proportional to
$\delta_{ij}$.
%(all of them except the second one in  Eq.~(\ref{Limit no mixing Higgsinos gauginos}))
%with respect to the ordinary seesaw, where they are absent. 

Under several assumptions, this formula for $(m^{\text{eff}}_{\nu})_{ij}$ can be further simplified.
Notice first that the last two terms
%in brackets in Eq.~(\ref{Limit no mixing Higgsinos gauginos}) 
are proportional to $v_d$, and therefore negligible in the limit of large or even moderate $\tan\beta$ provided that 
$\lambda$ is not too small. Besides,
the second term for $M^{\text{eff}}$
%in brackets in Eq.~(\ref{effectivegauginomass}) 
is also negligible in this limit, and for typical values of the parameters involved in the seesaw also the third one, i.e.
$M^{\text{eff}}\sim M$.
Under this assumption, the third 
term for $(m^{\text{eff}}_{\nu})_{ij}$ 
%in Eq.~(\ref{Limit no mixing Higgsinos gauginos}) 
is generated only through the mixing of left-handed neutrinos with gauginos. Therefore, we arrive to a very simple formula that can be used to understand the
seesaw mechanism in the $\mn$ in a qualitative way, that is
% To continue understanding neutrino physics in a qualitative way in the $\mn$, we can also obtain a simplified formula for the two eigenvalues of the above matrix in the limit $Y^{\nu}_1\rightarrow 0$, $v_{1L}\rightarrow 0$, where the electron neutrino is decoupled from the other two neutrinos, having a negligible mass \cite{Fidalgo:2009dm}:
% \begin{eqnarray}
% m^{eff}_{{\nu}_{2,3}}
% \simeq \left|\frac{{Y^{\nu}_2}^2 v_u^2}
% {6\sqrt 2 \kappa v_{R}}+\frac{v_{2L}^2}{2M}\right| \ \ , \;\;\;
% \frac{{Y^{\nu}_2}^2 v_u^2}
% {2\sqrt 2 \kappa v_{R}}\ , 
% \label{Limit}
% \end{eqnarray}
% where the approximations $Y^{\nu}_2=Y^{\nu}_3$, $v_{2L}=v_{3L}$, have also been used.
\begin{eqnarray}
(m^{\text{eff}}_{\nu})_{ij} 
\simeq \frac{Y^{\nu}_iY^{\nu}_jv_u^2}
{6\sqrt 2 \kappa v_{R}}
                   (1-3 \delta_{ij})-\frac{v_{iL} v_{jL}}{4M}  \ .
\label{Limit no mixing Higgsinos gauginos2}
\end{eqnarray}

% Using this approximate formula it is easy
% to understand how diagonal Yukawas can give rise to off-diagonal entries in the mass matrix. The key point are clearly the extra contributions given by the terms which are not proportional to
% $\delta_{ij}$ (all of them except the second one in  
% Eq.~(\ref{Limit no mixing Higgsinos gauginos}))
% with respect to the ordinary seesaw, where they are absent. Besides, under several assumptions, this formula can be further simplified to become Eq.~(\ref{Limit no mixing Higgsinos gauginos2}).

As we can understand from these equations, neutrino physics in the $\mn$ is
closely related to the parameters and VEVs of the model, 
since the values chosen for them must reproduce current data on neutrino masses
and mixing angles. For example, for the typical values of the parameters and VEVs in Eqs.~(\ref{vevR}),~(\ref{yukawas}) and~(\ref{vevdirac}), neutrino masses $\lsim 0.1$ eV as expected, can easily be reproduced.

Let us finally point out that all these results in the $\mn$ give a kind of answer to the question of why the mixing angles are so different in the quark and lepton sectors. Basically, because no generalized seesaw exists for the quarks.

%%%%%%%%%%%%%%%%%%%%%%%%%%%%%%%%%%%%%%%%%%%%%%%%%%%%%%%%%%%%%%%%
%%%%%%%%%%%%%%%%%%%%%%%%%%%%%%%%%%%%%%%%%%%%%%%%%%%%%%%%%%%%%%%%
\section{The left sneutrino as LSP}
\label{Section:Sneutrino-LSP}
%%%%%%%%%%%%%%%%%%%%%%%%%%%%%%%%%%%%%%%%%%%%%%%%%%%%%%%%%%%%%%%%
%%%%%%%%%%%%%%%%%%%%%%%%%%%%%%%%%%%%%%%%%%%%%%%%%%%%%%%%%%%%%%%%
%

%\subsection{The parameter space}

In this section, we will discuss first the regions of the parameter space of the $\mn$ where the left sneutrino can become the LSP.
Then we will study separately BPs with left sneutrinos co-LSPs of the first two families, and left sneutrino LSP of the third family, since their phenomenology is very different.
% The decay modes of the left sneutrino, depending on
% its nature, scalar or pseudoscalar, are also discussed. 
% In the case of the pseudoscalar state its decay is mainly into neutrinos or 
% leptons, with the later case through tau sneutrinos due to its large Yukawa coupling.
% Unlike the pseudoscalar state, the scalar one can have a sizable decay into photons, in a way not very different from the Higgs.
% Also lepton decays are possible for the scalar tau sneutrino.
For the mass spectrum and decay modes computed in this section, we have used a suitably modified version of
{\tt SARAH} code \cite{Staub:2008uz,Staub:2011dp,Staub:2013tta} 
as well as the {\tt SPheno} 
{{v}}3.3.6 code \cite{Porod:2003um,Porod:2011nf}. These results were linked to 
{\tt MadGraph5\_aMC@NLO} {{v}}2.3.2.2 \cite{Alwall:2014hca} 
and {\tt PYTHIA} 6.428 \cite{Sjostrand:2006za} tools, in order to make the full analysis of detection of signals at the LHC in 
Section~\ref{Section:detectionSneutrino-LSP}.

%%%%%%%%%%%%%%%%%%%%%%%%%%%%%%%%%%%%%%%%%%%%%%%%%%%%%%%%%%%%%%%%
%%%%%%%%%%%%%%%%%%%%%%%%%%%%%%%%%%%%%%%%%%%%%%%%%%%%%%%%%%%%%%%%
%%\subsection{The mass 
%of the left sneutrino and the NLSP
%%}
%%\label{nlsp}
%%%%%%%%%%%%%%%%%%%%%%%%%%%%%%%%%%%%%%%%%%%%%%%%%%%%%%%%%%%%%%%%
%%%%%%%%%%%%%%%%%%%%%%%%%%%%%%%%%%%%%%%%%%%%%%%%%%%%%%%%%%%%%%%%

%%%%%%%%%%%%%%%%%%%%%%%%%%%%%%%%%%%%%%%%%%%%%%%%%%%%%%%%%%%%%%%%
%%%%%%%%%%%%%%%%%%%%%%%%%%%%%%%%%%%%%%%%%%%%%%%%%%%%%%%%%%%%%%%%
% \section{The left-handed sneutrino as the LSP}
% \label{Section:Sneutrino-LSP}
%%%%%%%%%%%%%%%%%%%%%%%%%%%%%%%%%%%%%%%%%%%%%%%%%%%%%%%%%%%%%%%%
%%%%%%%%%%%%%%%%%%%%%%%%%%%%%%%%%%%%%%%%%%%%%%%%%%%%%%%%%%%%%%%%
%

To understand how the left sneutrino can become the LSP in the
$\mn$, we have to discuss first the relevant regions of the parameters space. As pointed out in Section~\ref{masses}, because
of the generalized electroweak-scale seesaw, data on neutrino physics can be reproduced at tree level in the $\mn$, even with diagonal Yukawa couplings $Y^{\nu}_i$. Nevertheless, for this first analysis focused on the detection of the sneutrino LSP at the LHC, it will be operationally simpler to work   
with only one family of right-handed neutrinos and its sneutrino partner. Thus we leave the three-family case for a future work \cite{preparation}, since our LHC analysis
%concerning LHC physics 
is not going to be essentially modified by this simplification.
% and that a complete description of neutrino physics is always possible at three level when the other two families are added.
As a consequence, in addition to $\tan\beta$ we will work with the following non-vanishing parameters of Eq.~(\ref{freeparameters}):
$v_{1R}\equiv v_{R}$, ${\lambda_1} \equiv \lambda $, $A^{\lambda}_1 \equiv A^{\lambda} $, 
${\kappa_{111}} \equiv \kappa$, $A^{\kappa}_{111} \equiv A^{\kappa}$, and
$Y^{\nu}_{i1}$,
$A^{\nu}_{i1}$.
For the last two parameters we will assume universality,
%will work only with non vanishing $\nu^c_1\equiv \nu^c$, and
$Y^{\nu}_{i1}\equiv Y^{\nu}$ and
$A^{\nu}_{i1}\equiv A^{\nu}$, since in this way we will have three
large enough diagonal left sneutrino masses,
mimicking the case of three families of right sneutrinos.
% mimic roughly the three-family case
% with diagonal Yukawa couplings.
% For the rest of the parameters in (\ref{freeparameters}), we will also assume for simplicity that there is no intergenerational mixing
% and that they have
% the same values for the three families:
% ${\lambda_i} \equiv \lambda $, 
% ${\kappa_{iii}} \equiv \kappa$, 
% $A_{\lambda_i} \equiv A_{\lambda} $, 
% $A_{\kappa_{iii}} \equiv A_{\kappa}$.
Summarizing, the free parameters in the neutral scalar sector at the low scale $M_{EWSB}$, are
in our analysis:
\bea
%\lambda, \, \kappa,\, \tan\beta, \, \nu_1, \,  \nu_3, \nu^c, \, A_{\lambda}, \, A_{\kappa}, \, A_\nu\ ,
\lambda, \, \kappa,\, Y^{\nu}, \, \tan\beta, \, v_{iL}, v_{R}, \, A^{\lambda}, \, 
A^{\kappa}, \, A^\nu\ .
\label{freeparameters2}
\eea
%where we have defined ${\lambda} \equiv {\lambda_i}$, $\kappa\equiv {\kappa_{iii}}$, 
%$A_{\lambda} \equiv A_{\lambda_i}$, 
%$A_{\kappa} \equiv A_{\kappa_{iii}}$.
Concerning the soft parameters of Eq.~(\ref{freeparameterssoft}), for simplicity in the computation we will consider that the trilinear ones,
as well as the scalar masses, are universal, i.e. $A^{u,d,e}_{ij}=A^{u,d,e}$ and 
$m_{\tilde Q_{iL},\tilde u_{iR},\tilde d_{iR},\tilde e_{iR}}=
m_{\tilde Q_L,\tilde u_R,\tilde d_R,\tilde e_R}$, respectively.
Altogether, we have the following free parameters:
\bea
M_1, \, M_2,\, M_3, \, m_{\tilde Q_{L}},\, 
m_{\tilde u_{R}}, \, m_{\tilde d_{R}}, \,
m_{\tilde e_{R}}, \,
A^u, \, A^d, \, A^e\ .
\label{freeparameterssoft23}
\eea
Let us remark nevertheless that, in the sake of completeness, the formulas given in the text and 
%Appendix~\ref{Apendix:Sneutrino-masses}, 
Appendices~\ref{Section:munuSSM} and~\ref{masasm}, 
as well as the figures, are for 
the general case of three neutrino generations and without assuming universality of parameters or vanishing 
intergenerational mixing.
The formulas of Appendix~\ref{Section:Coupling} for the interactions are the only ones written for simplicity 
for one family
of right-handed neutrinos and its sneutrino partner.

Once fixed our parameter space,
%In order to understand how the LH sneutrino can become the LSP in some regions of the parameter space of the 
%$\mn$, 
it is necessary to study the neutral scalar and pseudoscalar mass matrices written
in Appendix~\ref{masasm}, 
%\ref{SubAppendix-scalarmasses} and~\ref{SubAppendix:NeutrealScalar}, respectively, 
in order to determine how a left sneutrino can become the LSP.
First of all, 
as discussed in Section~\ref{masses}, 
the $5\times 5$ Higgs-right sneutrino submatrix is almost decoupled from
the $3\times 3$ left sneutrino submatrix, and therefore we can concentrate on the latter\footnote{The LSP is in fact the lightest
mass eigenstate of the whole matrix, but the composition of the sneutrino will dominate over the
others.}.
Second, although 
there is a mass difference between scalar and pseudoscalar sneutrinos, 
it turns out to be negligible because is due to the tiny D-term contribution:
%, and therefore negligible:
%in all entries of the scalar matrix, 
%the pseudoscalar sneutrinos are always lighter, although with a negligible mass difference:
%From Eq.~(\ref{oddLL}), we see:
%
\bea
m_{\widetilde{\nu}^\mathcal{I}_{iL}\widetilde{\nu}^\mathcal{I}_{jL}}^{2} 
=
m_{\widetilde{\nu}^\mathcal{R}_{iL}\widetilde{\nu}^\mathcal{R}_{jL}}^{2} 
-\frac{1}{4}\left(g^{2}+g'^{2}\right) v_{iL} v_{jL}
\approx 
m_{\widetilde{\nu}^\mathcal{R}_{iL}\widetilde{\nu}^\mathcal{R}_{jL}}^{2} 
\ .
\label{evenLRR}
\eea
% Notice nevertheless that this mass difference is very small, basically negligible, given its proportionality to $v^2_{\nu}$, as mentioned before.
% Another thing to notice is that the off-diagonal entries of the mass matrices for both types of LH sneutrinos are negligible compared to the diagonal ones, since the former are suppressed by the above mentioned D-term proportional to $v^2_{\nu}$ and by terms proportional to 
% $Y_{\nu}^2$.
Finally, from Eqs.~(\ref{evenLL}) and~(\ref{oddLL}), we realize that for 
both left sneutrino states the off-diagonal entries of the mass matrices are negligible compared to the diagonal ones,
% Concerning both types of LH sneutrinos, CP-even and CP-odd, we notice from
% Eqs.~(\ref{oddLL}) and~(\ref{evenLL}) that the off-diagonal entries of the mass matrices are negligible compared to the diagonal ones, 
since the former are suppressed by terms proportional to 
$(Y^{\nu})^2$, and (for the scalar sneutrinos) by the D-term contribution proportional to 
$v^2_{L}$.
As a conclusion, both states can be considered co-LSPs with
$m_{\widetilde{\nu}^\mathcal{I}_{iL}\widetilde{\nu}^\mathcal{I}_{iL}}^{2} 
\approx m_{\widetilde{\nu}^\mathcal{R}_{iL}\widetilde{\nu}^\mathcal{R}_{iL}}^{2}$,
if their masses are sufficiently low.

% Now,
% from
% Eqs.~(\ref{evenLL}) and~(\ref{oddLL}), we realize first that for 
% both left sneutrino states, scalar and pseudoscalar, the off-diagonal entries of the mass matrices are negligible compared to the diagonal ones,
% since the former are suppressed by terms proportional to 
% $(Y^{\nu})^2$, and (for the scalar sneutrinos) by the D-term contribution proportional to 
% $v^2_{L}$.
% Second, 
% there is a mass difference between scalar and pseudoscalar sneutrinos due to the tiny D-term contribution

% \bea
% m_{\widetilde{\nu}^\mathcal{I}_{iL}\widetilde{\nu}^\mathcal{I}_{jL}}^{2} 
% =
% m_{\widetilde{\nu}^\mathcal{R}_{iL}\widetilde{\nu}^\mathcal{R}_{jL}}^{2} 
% -\frac{1}{4}\left(g^{2}+g'^{2}\right) v_{iL} v_{jL}
% \approx 
% m_{\widetilde{\nu}^\mathcal{R}_{iL}\widetilde{\nu}^\mathcal{R}_{jL}}^{2} 
% \ ,
% \label{evenLRR}
% \eea
% implying that both states can be considered co-LSPs if one of them has a mass low enough as to be the LSP.

Concerning how low the left sneutrino masses can be, it is worth 
using Eq.~(\ref{evenLL}) for writing the dependence on the soft masses
of their diagonal entries:
% of the above equation:
%the dependence on the soft masses,
%using Eq.~(\ref{evenLL}):
% \cite{Escudero:2008jg}:
%
\bea
m_{\widetilde{\nu}^{\mathcal{I}}_{iL}\widetilde{\nu}^{\mathcal{I}}_{iL}}^{2}
&&\approx  
m_{\widetilde L_{iL}}^2
-\frac{1}{8}\left(g^{2}+g'^{2}\right)\left(v_{u}^{2}-v_{d}^{2}\right)
% +
% \frac{\partial^2V^{(n)}}{\partial v_{\nu_i} \partial v_{\nu_j}}
\ .
\label{even01LUURR}
\eea
From Eq.~(\ref{tadpoles4}) we can write approximately for the tree-level contribution
\bea
m_{\widetilde L_{iL}}^2\approx 
\frac{1}{8}\left(g^{2}+g'^{2}\right)\left(v_{u}^{2}-v_{d}^{2}\right)
+
\frac{Y^{\nu}v_u}{2v_{iL}}v_{R}
\left(
-\sqrt 2 A^{\nu}-\kappa v_{R}
+
\frac{\lambda v_{R}}{\tan\beta}
\right)
% -
% \frac{1}{v_{\nu_i}}
% \frac{\partial V^{(n)}}{\partial v_{\nu_i}}
\ ,
\label{even01LURRR}
\eea
obtaining therefore the expression 
%for (\ref{even01LUURR})
%
\bea
m_{\widetilde{\nu}^{\mathcal{I}}_{iL}\widetilde{\nu}^{\mathcal{I}}_{iL}}^{2}
\approx  
\frac{Y^{\nu}v_u}{2v_{iL}}v_{R}
\left(
-\sqrt 2 A^{\nu}-\kappa v_{R}
+
\frac{\lambda v_{R}}{\tan\beta}
\right)
% +
% \frac{\partial^2V^{(n)}}{\partial v_{\nu_i} \partial v_{\nu_i}}
% -
% \frac{1}{v_{\nu_i}}
% \frac{\partial V^{(n)}}{\partial v_{\nu_i}}
\ ,
\label{evenLLL}
\eea
which coincides as expected with Eqs.~(\ref{oddLL}), and~(\ref{evenLL}), neglecting small terms.

Obviously, it is always possible to tune the parameters in 
Eq.~(\ref{evenLLL}) (or Eq.~(\ref{even01LURRR}))
in such a way that these 
%relevant tree-level 
contributions 
%to the LH sneutrino masses 
% \bea
% m_{\widetilde{\nu}^\mathcal{I}_{i}\widetilde{\nu}^\mathcal{I}_{i}}^{2} 
% \approx
% \frac{Y_{\nu}v_u}{v_{\nu_i}}v_{\nu^c}
% \left(
% -A_{\nu}-\kappa v_{\nu^c}
% +
% \frac{\lambda v_{\nu^c}}{\tan\beta}
% \right)\ ,
% \label{evenLLL}
% \eea
turn out to be sufficiently small. Actually, we will find in the next subsection
% in Section~\ref{Section:Results} 
interesting examples with small soft masses and therefore with small left sneutrino masses. In particular, we will see that for sneutrino 
masses $\lsim$ 310 GeV  
%. It is worth pointing out here
%that this is the adequate range 
is possible to produce and detect the sneutrino LSP at the LHC. 
% As we will discuss in detail in 
% Section~\ref{Section:detectionSneutrino-LSP}, for left sneutrino masses larger than about 300 GeV, the production cross section is not sufficiently large, whereas for masses smaller than 95 GeV the background starts to overshoot the signal. 
%BETTER WRITE HERE THE ANALYSIS OF THE PARAMETER SPACE INCLUDING THE TABLES.
%$m_{\widetilde{\nu}_{i_L}} \lesssim  200 \gev$. 
%There, inspired by the structure of soft SUSY-breaking terms from supergravity, we will 
%assume $a_{\nu_{ij}} = (A_{\nu}Y_{\nu})_{ij}$ in Eq.~(\ref{2:Vsoft}). Thus 
% Notice that in order to get it we can tune for example the quantity
% in brackets (where typically we expect $\kappa v_{\nu^c}, \lambda v_{\nu^c}\sim$ 100 GeV) playing with $A_{\nu}$, to get a value
% of order 10, since 
% $\frac{Y_{\nu}v_u}{v_{\nu_i}}$ is expected to be of order 1 
% ($v_{\nu_i}\sim m_D \sim Y_\nu v_u$) and $v_{\nu^c}\sim 1$ TeV.
% Thus the CP-odd LH sneutrino becomes the LSP with its mass $\sim$ 100 GeV, almost degenerated
% with the one of the CP-even state.
In order to get masses of this order, we can tune for example the quantity
in brackets in Eq.~(\ref{evenLLL}). Since the factors in front of it are $v_{R}\sim 1$ TeV
and
$\frac{Y^{\nu}v_u}{2v_{iL}}\sim 1$ 
(see Eq.~(\ref{vevdirac})),
%(since $v_{\nu_i}\sim m_D \sim Y^\nu v_u$),
we need this quantity of the order of 10 GeV.
%$-\sqrt 2 A^{\nu}-\kappa v_{\nu^c}+\frac{\lambda v_{\nu^c}}{\tan\beta}\sim 10$ GeV.
Given that we expect $\kappa v_{R}, \lambda v_{R}\sim$ 100 GeV, this implies that
$A^\nu\sim -$ 100 GeV is the necessary condition to obtain the pseudoscalar left sneutrino as the LSP with mass $\sim$ 100 GeV.
%, and almost degenerated with the one of the CP-even state.
% Notice that in order to get them we can tune for 
% example $-A_{\nu}$ with the product $\kappa v_{\nu^c}$.
% Thus the CP-odd LH sneutrino becomes the LSP with its mass almost degenerated
% with the one of the CP-even state.
From a theoretical viewpoint, this means that we need a SUSY-breaking mechanism producing low-energy soft parameters of the order of 1 TeV, except for
$m_{\widetilde L_{iL}}$ and $A^{\nu}$ which should be of the order of 100 GeV. Once this is fulfilled, the minimization conditions set the required values for the VEVs, $v_{iL}\sim 10^{-4}$ GeV and $v_{R}\sim 1$ TeV. 

Notice that, in principle, we also could have used a very small value of 
%$Y_{\nu_{i}}$ ($Y_{\nu_i} v_u << \nu_i$) 
$Y^{\nu}$ ($Y^{\nu} v_u \ll v_{iL}$) 
to lower the sneutrino masses. However, this would give rise to a negligible contribution to the mixing 
between right- and left-handed neutrinos (unless a very small effective Majorana mass is assumed), 
making it difficult to reproduce the experimental constraints on neutrino physics.
% coming from neutrino masses and mixing angles.

Summarizing, we have shown that it is viable to obtain in the spectrum of the
$\mn$ left sneutrinos as LSPs. They can in principle belong to any of the three families of the SM. Nevertheless, this can have crucial implications for the signals produced at the LHC, because of the different decay modes of the third family with respect to the first two.
We will study this issue in detail in Section~\ref{decaymodes}. Here we will analyze the strategy to obtain LSPs of different families.

%Concerning to which family of left sneutrinos belongs the LSP, notice that
We can assign different values for the 
input parameters associated to each family. This is the case for example of the left sneutrino VEVs, $v_{iL}$.
Thus, if we choose $v_{1L}=v_{2L} > v_{3L}$, 
%and universal $A_{\nu}$, $Y_{\nu}$,
we obtain from the approximate expression in Eq.~(\ref{evenLLL}) that the electron sneutrino $\widetilde \nu^\mathcal{I}_{eL}$ and the muon sneutrino $\widetilde \nu^\mathcal{I}_{\mu L}$ have masses degenerate and therefore behave as co-LSPs. 
%(the terms neglected in that equation would introduce different contributions for each family, but these are negligible). 
Although this degeneracy is broken by the mixing of the mass matrices and by the loop corrections, the mass difference is going to be negligible.
For example, for the BP in Table~\ref{table:musneu-125}
to be analyzed below, 
$\widetilde \nu^\mathcal{I}_{eL}$ is 0.0002 GeV heavier than
$\widetilde \nu^\mathcal{I}_{\mu L}$.
%this difference is of about 0.0002 GeV. 
Following our discussion in Eq.~(\ref{evenLRR}), where we show that both sneutrino states, scalar and pseudoscalar, are co-LSPs, we conclude that in this case there are four co-LSPs: 
$\widetilde \nu^\mathcal{I}_{\mu L}$, $\widetilde \nu^\mathcal{R}_{\mu L}$,
$\widetilde \nu^\mathcal{I}_{eL}$ and $\widetilde \nu^\mathcal{R}_{eL}$.
%On the other hand, 
Alternatively, if we 
choose $v_{1L}=v_{2L} < v_{3L}$, then we obtain that the tau sneutrinos $\widetilde \nu^\mathcal{I}_{\tau L}$ and $\widetilde \nu^\mathcal{R}_{\tau L}$ 
are co-LSPs. 
% This would also be expected from a high-energy theory with degenerate soft masses for the three families of sneutrinos, due to the negative contributions to the renormalization group equations (RGEs) proportional to the tau Yukawa coupling. 
Obviously, in the case of universal VEVs, 
$v_{iL}=v_{L}$, one obtains that the left sneutrinos of the three families, scalars and pseudoscalars, become co-LSPs.
% with degenerate masses (up to loop corrections). 
%\R{EXPLAIN THAT THE SIGNAL IS GIVING NOTHING NEW WITH RESPECT TO THE OTHER BPs STUDIED.}

Let us finally remark that
another equivalent strategy in order to find sneutrinos of different families as LSPs, is to allow for non-universality of the
parameters $A^{\nu}_{ij}$ or $Y^{\nu}_{ij}$, while keeping $v_{iL}$ universal.
% As we will discus in Section~\ref{decaymodes}, whether
% $\widetilde \nu^\mathcal{I}_{\tau}$ or 
% $\widetilde \nu^\mathcal{I}_{e}$/$\widetilde \nu^\mathcal{I}_{\mu}$
% is the LSP  
% has crucial implications for the signals produced at the LHC.

\vspace{0.25cm}

\noindent
{\bf The NLSP}

\noindent
When a left sneutrino is the LSP, we expect to have a left slepton as the NLSP. We will see in 
Section~\ref{Section:detectionSneutrino-LSP} that this has 
implications for the production of the left sneutrino LSP at the LHC, because the direct production of sleptons and their decays is a relevant source of sneutrinos.
To check that the slepton can be the NLSP, let us point out that 
although sneutrinos and sleptons are in the same $SU(2)$ doublet, sleptons are heavier. 
First of all, 
we discussed in
Section~\ref{masses} that left sleptons
are decoupled from the other charged scalars.
Second, the left slepton submatrix of Eq.~(\ref{chargedmixture}) has
the off-diagonal entries negligible compared to the diagonal ones. Finally, the diagonal entries are always larger than the ones of the left sneutrinos, due to the positive D-term contribution. Altogether, similar to the MSSM one obtains
\bea
m_{\widetilde{e}_{iL} \widetilde{e}_{iL}^{*}}^{2}
\approx 
m_{\widetilde{\nu}^{\mathcal{I}}_{iL}\widetilde{\nu}^{\mathcal{I}}_{iL}}^{2} +
\frac{g^{2}}{4}\left(v_{u}^{2}-v_{d}^{2}\right)
\approx m_{\widetilde{\nu}^{\mathcal{I}}_{iL}\widetilde{\nu}^{\mathcal{I}}_{iL}}^{2}
-m_W^2 \cos 2\beta
% \approx  
% \frac{g_2^{2}}{2}\left(v_{u}^{2}-v_{d}^{2}\right)
% +
% m_{\widetilde{\nu}^{\mathcal{R}}_{i}\widetilde{\nu}^{\mathcal{R}}_{i}}^{2}
\ .
\label{even0LRR}
\eea
%
% \bea
% m_{\widetilde{e}_{i} \widetilde{e}_{i}^{*}}^{2} - \frac{g_2^{2}}{2}\left(v_{u}^{2}-v_{d}^{2}\right)
% \approx 
% m_{\widetilde{\nu}^{\mathcal{I}}_{i}\widetilde{\nu}^{\mathcal{I}}_{i}}^{2}
% \approx  
% m_{\widetilde{\nu}^{\mathcal{R}}_{i}\widetilde{\nu}^{\mathcal{R}}_{i}}^{2}
% \ .
% \label{even0LRR}
% \eea
Since this D-term contribution is small,
%$\sim (10)^2$, 
when a left sneutrino is the LSP, a
left slepton can be naturally the NLSP. 

\vspace{0.25cm}

\noindent
%\subsection 
{\bf Electron and muon sneutrinos co-LSPs}
\label{electronmuon}

\noindent 
Following the discussion above,
% we show here a benchmark point (BP) with the rigth properties to produce   
% an electron/muon sneutrino LSP ($\tilde{\nu}^\mathcal{I}_{e,\mu}$).
% or a tau sneutrino LSP ($\widetilde \nu^\mathcal{I}_{\tau}$).
we show in Table~\ref{table:musneu-125}
a BP with the right properties to produce
$\widetilde\nu^\mathcal{I}_{\mu L}$,
$\widetilde\nu^\mathcal{R}_{\mu L}$ ,
$\widetilde \nu^\mathcal{I}_{e L}$ 
and $\widetilde \nu^\mathcal{R}_{e L}$
co-LSPs, with masses of about 125.4 GeV. 
%an electron/muon left sneutrino LSP,
%$\widetilde \nu^\mathcal{I}_{eL}/\widetilde \nu^\mathcal{I}_{\mu L}$
The input parameters at the low scale $M_{EWSB}$ can be found in the first box of the table. 
Concerning the input soft parameters, 
as discussed below Eq.~(\ref{evenLLL}), the most relevant one for our computation is $A^\nu$ and we have used the value $-A^\nu=386$ GeV 
$\sim \kappa v_{R}/\sqrt 2$.
Other relevant soft parameters are the gaugino masses $M_{1}$ and $M_2$, since
%As will be discussed in detail in Section~\ref{decaymodes}, 
Bino and Wino
mediate the decay channels of the left 
sneutrino (see Fig.~\ref{fig:eff-to-neutrinos}). We take them as
600 and 900 GeV, respectively, and for $M_3$ we choose 1600 GeV.
% Thus small gaugino masses would enhance this kind of decays.
% For the case of the
% $\tilde{\nu}_{e,\mu}$ LSP, the increase in the channel
% $\tilde{\nu}_{e,\mu}^{\mathcal{I}}\to\nu\nu$ would increase the total decay width of the particle, but would not affect crucially the BRs since this channel is already dominant. However, the channels in 
% Fig.~\ref{fig:eff-to-neutrinos} could become dominant 
% also in the case of the $\tilde{\nu}_{\tau}$ LSP when compared with 
% the process of Fig.~\ref{fig:eff-to-leptons-higgsino}.
%  
%  DISCUSS NOW WHICH RANGE IS IMPORTANT (NEUTRINO CONSTRAINTS?) AND HOW WE STUDY IT IN THE TABLES. 
% [and for the gaugino masses only $M_2=1$ TeV will be 
% used as input, whereas the others 
% will be determined by the approximate GUT relations 
% $M_1 = \frac{\alpha_1^2}{\alpha_2^2} M_2$,
% $M_3 =  \frac{\alpha_3^2}{\alpha_2^2} M_2$,
% implying $M_1\approx 0.5 M_2$, $M_3\approx 2.7 M_2$ CHECK THAT WE DON'T DO THIS.]
% [CLARIFY WITH INAKI HOW IMPORTANT IS THIS IN ADDITION TO THE DISCUSSION ABOVE OF THE DECAY CHANNELS:
% The influence of these soft masses in the relevant part of the spectrum takes place in the loop corrections of the sneutrino masses 
% A lower value of $M_1$ or $M_2$ would produce a slightly lower mass of the sneutrinos, smaller than a couple of GeV for M1 or M2 as low as 200 GeV.]
For the rest of trilinear parameters, for simplicity in the computation we assume %consider them universal, $A_{u_{ij},d_{ij},e_{ij}}=A_{u,d,e}$, and 
$A^d=A^e=A^{\lambda}=-A^{\kappa}=1$ TeV
%|$A^{d,e,\lambda,\kappa}$|=1 TeV, 
with the exception of
%a value of their modulus of 1 TeV 
$A^u\sim -3.1$ TeV
%$|A^u|\sim 3.1$ TeV 
in order to reproduce the mass of the Higgs. 
As we will discuss below,
the negative value of $A^{\kappa}$ is necessary to avoid tachyonic pseudoscalar right sneutrinos.
The one of $A^u$ it to avoid tachyonic left sneutrinos due to the loop corrections.
In the same spirit of simplicity, we use $m_{\tilde e_R}=$ 1 TeV and
%$m^2_{\tilde Q_i,\tilde u^c_i,\tilde d^c_i,\tilde e^c_i}=
$m_{\tilde Q_L,\tilde u_R,\tilde d_R}=$ 1.3 TeV.\footnote{
% Because of the large value of $A_{t}$, this point is in a borderline region between a stable and a metastable realistic vacuum. To have a more precise information, one should carry out the analysis at the appropriate renormalization scale to evaluate the existence of charge and color breaking minima~\cite{Casas:1995pd}, and, in that case to compute the lifetime of the vacuum considering the tunneling at zero/finite temperature~\cite{Beuria:2016cdk}, something which is beyond the scope of this work. Nevertheless, this is not crucial for our study of the sneutrino LSP since just slightly increasing the soft masses for stops (from 1.3 TeV to e.g. 1.4 TeV), we will be in the safe stable region, being able at the same time of obtaining the same kind of signals discussed here.
As analyzed in 
Refs.~\cite{Camargo-Molina:2013sta,Blinov:2013fta,Camargo-Molina:2014pwa,
Bobrowski:2014dla,Chattopadhyay:2014gfa,Hollik:2016dcm,Beuria:2016cdk} using a numerical minimization of the potential, 
large values of $A^{t}$ may give rise to an unstable electroweak ground state decaying rapidly to a charge and color breaking (CCB) minima (see e.g. Ref.~\cite{Casas:1995pd}). Using the results of 
Refs.~\cite{Blinov:2013fta,Camargo-Molina:2014pwa}, we have been able to check that  our point around the maximal mixing corresponds to a metastable vacuum. 
The dangerous additional possibility of rapid thermal tunneling~\cite{Camargo-Molina:2014pwa} is dependent on the thermal evolution of the Universe, and to analyze it is beyond the scope of this work. 
Nevertheless, it is worth noticing that CCB minima are not a crucial subject in our study of the sneutrino LSP. We can easily modify $A^{t}$ and/or stop soft masses obtaining the same kind of signals discussed here. For example, keeping 
$A^t\sim -3.1$ TeV but increasing $m_{\tilde t}$ in a few hundred GeV we can enter in a safe stable region~\cite{Camargo-Molina:2013sta,Blinov:2013fta,Camargo-Molina:2014pwa,
Bobrowski:2014dla,Chattopadhyay:2014gfa,Hollik:2016dcm,Beuria:2016cdk}. Also the same situation can be obtained reducing $|A^t|$ and properly changing $m_{\tilde t}$.}
%$A_{u,d,e}=1$ TeV, 
Finally, we have $\tan\beta = 10$, corresponding to the following Higgs VEVs:
$v_u/\sqrt 2=170.84$ GeV, $v_d/\sqrt 2=17.08$ GeV.
%$v_u=241.60$ GeV and $v_d=24.15$ GeV.

%%%%%%%%%%%%%%%%%%%%%%%%%%%%%%%%%%%%%%%%%%%%%%%%%%%%%%%%%%%%%%%%%%%%%%%%%%%%%%%%%

%%   TABLES  125 e mu
\begin{table}[t!]
% \begin{center}
\centering
\caption{Benchmark point producing $\widetilde\nu^\mathcal{I}_{\mu L}$,$\widetilde\nu^\mathcal{R}_{\mu L}$,
$\widetilde \nu^\mathcal{I}_{e L}$ and $\widetilde \nu^\mathcal{R}_{e L}$ co-LSPs, with masses of 125.4 GeV. 
Input parameters, and soft masses obtained from the minimization conditions, are given in the first and second boxes at the low scale
$M_{EWSB}$. Sparticle physical masses are shown in the third box (with their dominant compositions written in brackets).
Sneutrino branching ratios (larger than $10^{-4}$) and decay widths are shown in the fourth and fifth boxes, respectively.
VEVs, soft parameters, sparticle masses and decay widths are given in GeV.}
  \label{table:musneu-125}
{ \scriptsize
 \begin{tabular}{|p{1.95cm}|p{1.95cm}|p{1.95cm}|p{1.95cm}|p{1.95cm}|p{1.95cm}|} \hline
 $\lambda $&$0.2$&$\kappa_{}$&$0.3$&$Y^{\nu}$&$5\times 10^{-7}$\\ \hline
  $v_{1,2 L}/\sqrt 2$&$3\times 10^{-4}$& $v_{3L}/\sqrt 2 $  & $5\times 10^{-6}$ 
& $v_{R}/\sqrt 2$ &$1350$\\ \hline
 tan $\beta$&10&$A^u$&$-3177$&$A^{d,e} $&$1000$\\ \hline 
 $A^{\lambda}$ & $1000$ &$A^{\kappa} $ & $-1000$ &$A^{\nu} $&$-386$\\ \hline
 $M_1$&600&$M_2$&900& $M_3 $&1600\\ \hline 
 $m^2_{\tilde Q_L,\tilde u_R,\tilde d_R} $&$ 1.69\times 10^6$&$m_{\tilde e_R}^2 $&$10^6$
 &--&--\\ \hline 
  \end{tabular}
  \begin{tabular}{|p{1.95cm}|p{1.95cm}|p{1.95cm}|p{1.95cm}|p{1.95cm}|p{1.95cm}|} \hline
 $m_{Hd}^2 $&$3.62\times 10^6$&$m_{Hu}^2 $&$-1.09\times 10^{5}$
 &$m_{\tilde \nu_R}^2 $ & $0.750\times 10^{5}$\\ \hline
 $m_{\tilde L_{e L}}^2 $&$0.968\times 10^{4}$& $m_{\tilde L_{\mu L}}^2 $&$0.968\times 10^{4}$
 & $m_{\tilde L_{\tau L}}^2 $&$0.935\times 10^{6}$\\ \hline
  \end{tabular}
  \begin{tabular}{|p{1.95cm}|p{1.95cm}|p{1.95cm}|p{1.95cm}|p{1.95cm}|p{1.95cm}|} \hline
 $m_{h_1(H^\mathcal{R}_u)} $&$124.2$&$m_{h_2(\widetilde{\nu}^\mathcal{R}_{{\mu L}})}$&$125.4$&$m_{h_3(\widetilde{\nu}^\mathcal{R}_{{eL}})} $&$125.4$  \\ \hline
 $m_{h_4(\widetilde{\nu}^{\mathcal{R}}_R)} $&$501.2$&$m_{h_5(\widetilde{\nu}^\mathcal{R}_{{\tau L}})} $&$972.4$&$m_{h_6(H^\mathcal{R}_d)} $&$1934.4$\\ \hline
--&--& $m_{A^0_2(\widetilde{\nu}^\mathcal{I}_{{\mu L}})} $&$125.4$&$m_{A^0_3(\widetilde{\nu}^\mathcal{I}_{{eL}})}$&$125.4$ \\ \hline
$m_{A^0_4(\widetilde{\nu}^\mathcal{I}_{{\tau L}})} $&$972.4$&
 $m_{A^0_5(\widetilde{\nu}^{\mathcal{I}}_R)}$&$1100$&$m_{A^0_6(H^\mathcal{I}_d)} $&$1933.9$\\ \hline
--&--& $m_{H^-_2(\widetilde{\mu}_L)}$&$145.4$&$m_{H^-_3(\widetilde{e}_L)} $&$145.4$
\\ \hline
 $m_{H^-_4(\widetilde{\tau}_L)} $&$946.9$&
 $m_{H^-_5(\widetilde{e}_R)} $&$1000.9$&$m_{H^-_6(\widetilde{\mu}_R)} $&$1000.9$
\\ \hline     
 $m_{H^-_7(\widetilde{\tau}_R)} $&$1000.9$&
 $m_{H^-_8(H^-_d)} $&$1936.2$&--&--\\ \hline
$m_{\lambda^0_4(\tilde H^0_u/\tilde H^0_d)} $&$264.8$&$m_{\lambda^0_5(\tilde H^0_u/\tilde H^0_d)} $&$279.6$&$m_{\lambda^0_6({\tilde B}^0)} $&$600.3$\\ \hline
$m_{\lambda^0_7(\nu_R)} $&$809.6$&$m_{\lambda^0_8({\tilde W}^0)} $&$919.8$&--&--\\ \hline
$m_{\lambda^{-}_4(\tilde{H}^{-}_d/(\tilde{H}^{+}_u)^c)}$
&$272.1$&$m_{\lambda^{-}_5({\tilde W}^{-})} $&$920$&--&--\\ \hline
$m_{\widetilde{u}_1(\widetilde{t}_L/\widetilde{t}_R)} $&$1112 $&$m_{\widetilde{u}_2(\widetilde{c}_R)} $&$1340 $&$m_{\widetilde{u}_3(\widetilde{u}_R)} $&$1340$\\ \hline
$m_{\widetilde{u}_4(\widetilde{u}_L)} $&$1343$&$m_{\widetilde{u}_5(\widetilde{c}_L)} $&$1343$&$m_{\widetilde{u}_6(\widetilde{t}_L/\widetilde{t}_R)} $&$1465$\\ \hline
$m_{\widetilde{d}_1(\widetilde{b}_L)} $&$1310.6$&$m_{\widetilde{d}_2(\widetilde{b}_R)} $&$1338.8$&$m_{\widetilde{d}_3(\widetilde{s}_R)} $&$1338.8$\\ \hline
$m_{\widetilde{d}_4(\widetilde{d}_R)} $&$1338.8$&$m_{\widetilde{d}_5(\widetilde{s}_L)} $&$1344.9$&$m_{\widetilde{d}_6(\widetilde{d}_L)} $&$1344.9$\\ \hline
$m_{\widetilde{g}} $&$1619.5$&--&--&--&-- \\ \hline 
 \end{tabular}
 \begin{tabular}{|p{4.33cm}|p{1.95cm}|p{4.33cm}|p{1.95cm}|} \hline
 $\text{BR}(A^0_2 \to \nu \nu)$&$0.9744$&$\text{BR}(A^0_3 \to \nu \nu)$&$0.9908$\\ \hline 
$\text{BR}(A^0_2\to {\mu}^\pm e^\mp)$&$0.0058$& $\text{BR}(A^0_2\to {\mu}^+ \mu^-)$ & $0.0055$\\ \hline
$\text{BR}(A^0_2\to {\mu}^\pm \tau^\mp)$&$0.0054$& $\text{BR}(A^0_{2,3}\to {\tau}^+ \tau^-)$ &$0.0003$\\ \hline
$\text{BR}(A^0_{2,3}\to \bar{b}b)$&$0.0017$&$\text{BR}(A^0_{2,3}\to \bar{c} c)$  &$0.0007$ \\ \hline 
$\text{BR}(A^0_{2,3}\to g g)$&$0.0061$&-- &-- \\ \hline
 $\text{BR}(h_{2,3}\to \nu \nu)$&$0.0015$&-- &-- \\ \hline
 $\text{BR}(h_2\to {\tau}^+ \tau^-)$&$0.0863$&$\text{BR}(h_3\to {\tau}^+ \tau^-)$&$0.0828$\\ \hline
$\text{BR}(h_{2,3}\to \bar{b} b)$&$0.468$& $\text{BR}(h_{2,3}\to \bar{c}c)$& $0.033$ \\ \hline
$\text{BR}(h_{2,3}\to g g)$&$0.122$&$\text{BR}(h_{2,3}\to \gamma \gamma)$ & $0.003$\\ \hline
$\text{BR}(h_{2,3}\to W^\pm W^{\mp^*})$&$0.256$& $\text{BR}(h_{2,3}\to Z Z^*)$ & $0.028$\\ \hline \hline
 $\Gamma(h_{2,3})$&$6.7\times 10^{-11}$&$\Gamma(A^0_{2,3})$&$1.0\times 10^{-13}$\\ \hline
\end{tabular}}
\end{table}

%%%%%%%%%%%%%%%%%%%%%%%%%%%%%%%%%%%%%%%%%%%%%%%%%%%%%%%%%%%%%%%%

The soft masses obtained from 
the minimization conditions of Eqs.~(\ref{tadpoles1})--(\ref{tadpoles4}) 
are shown in the second box of the table.
In particular, from Eq.~(\ref{even01LURRR})
%(\ref{tadpoles4}) 
it is easy to check that one can obtain 
%there is a soft mass of the order of the TeV, 
$m_{\widetilde L_{\tau L}}^2\sim$ (1 TeV)$^2$  
corresponding to the VEV 
%$v_{\nu_{3}}= 5 \times 10^{-6}$ GeV,
$v_{3L}/\sqrt 2= 5 \times 10^{-6}$ GeV,
%$\frac{v_{\nu_{3}}}{\sqrt 2}= 5 \times 10^{-6}$ GeV,
and
two smaller soft masses 
$m_{\widetilde L_{e,\mu L}}^2\sim$ (100 GeV)$^2$
%10^{4}$ GeV$^2$ 
corresponding to larger VEVs, $v_{1,2L}/\sqrt 2= 3 \times 10^{-4}$ GeV.

The sparticle physical masses 
%for the physical states 
are shown in the third box of the table, with their dominant compositions written in brackets.
The masses of the neutral `Higgses'
%scalars and pseudoscalars 
can be found in $1^{st}-4^{th}$ rows
of that box. 
There we have followed the notation of 
Appendix~\ref{masasm}.1
%Appendices~\ref{SubAppendix-scalarmasses} and~\ref{SubAppendix:NeutrealScalar} 
for the dominant compositions. The scalar 
%and CP-odd 
mass eigenstates are denoted 
%by $h_{\alpha}$,
by $h_{1,...,6}$,
%with $\alpha=1,...,6$ 
since we are considering only one family of right-handed neutrinos and the scalar partner. 
The pseudoscalars 
%the masses 
are denoted by 
%$A^0_{\alpha}$, 
$A^0_{2,...,6}$, 
%with $\alpha=2,...,6$ 
because 
we associate $A^0_{1}$ to the Goldstone boson eaten by the $Z$.
The composition of the latter is dominated by the $H^\mathcal{I}_u$ (98.9\%), with the second most important composition $H^\mathcal{I}_d$ ($\sim$ 1\%).
% only the masses for $A^0_{2,...,6}$ are shown in the table.
By convention, the masses are labeled
in ascending order, so that for example 
%$m_{h_1}<m_{h_2}<...$ 
$m_{A^0_1}<m_{A^0_2}<...$

% As evident from Eq.~(\ref{even01LUURR}) (or Eq.~(\ref{evenLLL})),
% because of the small values for the soft masses of the first two families of lepton doublets
% $\sim $ 100 GeV, 
% we are able to get a LSP mainly pseudoscalar left sneutrino, 
% $A^0_2(\widetilde\nu^\mathcal{I}_{\mu L})$, with a low mass of 125.4 GeV. Since we use a BP with
% $v_{1L}=v_{2L} > v_{3L}$, we obtain that
% this state is almost degenerate in mass with the
% $A^0_3(\widetilde \nu^\mathcal{I}_{eL})$ 
% and with the scalar partners
% $h_2(\widetilde\nu^\mathcal{R}_{\mu L})$ and $h_3(\widetilde \nu^\mathcal{R}_{eL})$,
% with the loop corrections breaking the tree-level degeneracy.
As evident from Eq.~(\ref{even01LUURR}) (or Eq.~(\ref{evenLLL})),
because of the small low-energy soft masses of the first two families of lepton doublets
$\sim $ 100 GeV (since we assume $v_{1L}=v_{2L} > v_{3L}$), 
we are able to get as LSP with a mass of 125.4 GeV a pseudoscalar state dominated by the muon left sneutrino 
$A^0_2(\widetilde\nu^\mathcal{I}_{\mu L})$, with co-LSPs essentially degenerate in mass
$A^0_3(\widetilde \nu^\mathcal{I}_{eL})$ 
and the scalar partners
$h_2(\widetilde\nu^\mathcal{R}_{\mu L})$, $h_3(\widetilde \nu^\mathcal{R}_{eL})$.
%with the loop corrections breaking the tree-level degeneracy.
On the other hand, because of the large soft mass of the third family of lepton doublets $\sim$ 1 TeV,
we obtain $A^0_4(\widetilde\nu^\mathcal{I}_{\tau L})$ and $h_5(\widetilde\nu^\mathcal{R}_{\tau L})$ with masses of 972.4 GeV.
All these states are very pure left sneutrinos (>99.99\%), confirming
our statement above that the left sneutrino submatrix is almost decoupled from 
the Higgs-right sneutrino submatrix.
Actually, for this BP, because of the not very large value of $\lambda$, the Higgses and right sneutrinos are also almost decoupled.
They are quite pure right sneutrino states (>99.96\%) or Higgs states (>98.86\%).
In particular, we obtain $h_1(H^\mathcal{R}_u)$ as the SM-like Higgs with 
a mass of 
%$\sim 125$ GeV
124.2 GeV, and $h_6(H^\mathcal{R}_d)$ and $A^0_6(H^\mathcal{I}_d)$ as the heavy scalar and pseudoscalar Higgses with masses of about
1934 GeV
% $\sim 1.7$ TeV
(see Eq.~(\ref{prueba2}) vs.~Eqs.~(\ref{prueba1}) and~(\ref{pruebas1}), where the factor $\frac{1}{\tan\beta}$ vs.~$\tan\beta$ is crucial).
The second most important composition for 
these states is ($\sim 1\%$) 
$H^\mathcal{R}_d$, $H^\mathcal{R}_u$ and $H^\mathcal{I}_u$, respectively.
For the right sneutrino states we obtain
%or RH sneutrino states (>99.96\%):
$h_4(\widetilde{\nu}^{\mathcal{R}}_R)$ and 
$A^0_5(\widetilde{\nu}^{\mathcal{I}}_R)$ 
%as CP-even and CP-odd RH sneutrinos 
with masses of 501.2 GeV and 1100 GeV, respectively.
These values can be reproduced using the following approximate formulas
from Eqs.~(\ref{evenrr}) and~(\ref{oddRR}):
\bea
m^2_{\widetilde{{\nu}}^{\mathcal{R}}_R}
\approx   2 \kappa^2 {v^2_{R}} + \frac{1}{\sqrt 2}\kappa A^{\kappa} v_{R}
\ ,\;\;
m^2_{\widetilde{{\nu}}^{\mathcal{I}}_R}
\approx  - \frac{3}{\sqrt 2} \kappa A^{\kappa} v_{R}
\ .
\label{sps-approx2}
\eea

The masses for `charged Higgses'
% charged scalars 
are written in $5^{th}-7^{th}$ rows of the third box of Table~\ref{table:musneu-125}. 
Following the notation of 
Appendix~\ref{masasm}.1, 
%Appendix~\ref{SubAppendix:chargedScalar}, 
they are labeled %as  $H^{\pm}_{\alpha}$, with $\alpha=2,...,8$,
as $H^{-}_{2,...,8}$, 
%The masses are labeled $H^{\pm}_{2,...,8}$ because 
since we associate the first state to the Goldstone bosons eaten by the $W^{\pm}$.
The composition of this first state is dominated by the $H^+_u$ (98.9\%), with the second most important composition ${{H^{-}_d}}^{*}$ ($\sim$ 1\%).
As a consequence of the result in Eq.~(\ref{even0LRR}), there are two light (one heavy) left sleptons associated to the light (heavy) left sneutrinos, 
$H^{-}_2 (\tilde\mu_L)$, $H^{-}_3 (\tilde e_L)$ ($H^{-}_4 (\tilde\tau_L)$), with masses of 145.4 (946.9) GeV. Thus $H^{-}_2 (\tilde\mu_L)$ 
and $H^{-}_3 (\tilde e_L)$ are the co-NLSPs
with masses almost degenerate.
Concerning the right sleptons, they are decoupled from the left ones as 
already discussed in Section~\ref{masses}.
We can also see in 
Eq.~(\ref{pruebass7}) that their masses
are basically determined by the soft masses. As a consequence, they
have negligible off-diagonal entries, and there are three states
$H^{-}_5(\tilde\mu_R)$, $H^{-}_6 (\tilde e_R)$ and $H^{-}_7 (\tilde\tau_R)$ 
with masses of about 1 TeV.
%$m_{\tilde e^c}^2=(950$ GeV)$^2$ BE CAREFUL IF FINALLY WE TAKE THE SAME VALUE AS FOR THE OTHER SCALARS.
Since the slepton and Higgs submatrices are decoupled, all these states are very 
pure sleptons (>99.97\%).
Finally, there is a charged Higgs (>98.9\%), $H^{-}_8 (H^-_d)$, whose second most important composition is ($\sim 1\%$) 
${H^+_u}^*$.
As expected, this state is heavy with a mass of about 1.9 TeV, as those of the heavy neutral scalar $h_6(H^\mathcal{R}_d)$ and pseudoscalar $A^0_6(H^\mathcal{I}_d)$ 
(compare Eq.~(\ref{pruebass1}) with Eqs.~(\ref{prueba1}) and~(\ref{pruebas1})).

The masses for `neutrinos'
%the neutral fermions 
are shown in $8^{th}-9^{th}$ rows of the third box of Table~\ref{table:musneu-125}. 
%We follow the notation of Appendix~\ref{Section:Coupling} and 
The mass eigenstates are denoted as $\lambda^0_{4,...,8}$, 
%with $\alpha=4,...,8$, 
since we associate the first three states to the SM left-handed neutrinos.
% SHOULD WE WRITE SOMEWHERE THE MASS OF THE NEUTRINO WE GET?
% ALSO COMPARING IT WITH THE EQ. OF NEUTRINOS MASSES IN SECTION 2.2
The other five eigenstates arise from the mixing of MSSM-like neutralinos and the right-handed neutrino.
As can be deduced from the matrix in Eq.~(\ref{neumatrix}), we obtain almost pure 
Wino, Bino, Higgsinos and right-handed neutrino states. The $\lambda^0_8({\tilde W}^0)$
and
%$\lambda^0_6({\lambda}_{\tilde B})$ 
$\lambda^0_6({\tilde B}^0)$ 
states with 99.1\% and 99.3\% of Wino and Bino composition, respectively, have masses of 919.8 and 600.3 TeV, respectively, and these are determined approximately by the soft masses 
$M_2$ and $M_1$:
\bea
m_{\tilde W^0}
\approx   M_2\ ,\;\;
%m_{{\lambda}_{\tilde B}}
m_{\tilde B^0}
\approx M_1
\ .
\label{sps-approx22}
\eea 
The Higgsinos have a mixing of order 50\%, and the two states,
 $\lambda^0_4(\tilde H^0_u/\tilde H^0_d)$ and $\lambda^0_5(\tilde H^0_u/\tilde H^0_d)$, have similar masses of 264.8 and 279.6 GeV, respectively, which are determined
approximately by the effective $\mu$ term in Eq.~(\ref{muterm}):
%  \bea
% m_{\tilde H^0_u}\ , m_{\tilde H^0_d}
% \approx   
% \la_i \frac{v_{iR}}{\sqrt 2}=\lambda \frac{v_{R}}{\sqrt 2}
% \ .
% \label{sps-approx223}
% \eea 
 \bea
m_{\tilde H^0_{u,d}}
\approx   
%\la_i \frac{v_{iR}}{\sqrt 2}=
\lambda \frac{v_{R}}{\sqrt 2}
\ .
\label{sps-approx223}
\eea 
Finally, the 
$\lambda^0_7(\nu_R)$ state has a 99.8\% of right-handed neutrino composition. Its mass is 809.6 GeV and can be approximated by the effective Majorana mass of Eq.~(\ref{majorana}) 
 \bea
m_{\nu_R}
\approx 2 \kappa \frac{v_{R}}{\sqrt 2 }
\ .
\label{sps-approx2234}
\eea

Notice that from Eqs.~(\ref{sps-approx2}) and (\ref{sps-approx2234}) we obtain
\bea
m^2_{\widetilde{{\nu}}^{\mathcal{I}}_R}
%\widetilde{{\nu}}^{c\mathcal{I}}}
%^{2}
\approx  - \frac{3}{2} A^{\kappa} m_{\nu_R}
%,\quad
%m_{\widetilde \chi^0_{i+3}} \approx 2\kappa \nu^c
\ ,
\label{sps-approx12}
\eea 
and 
therefore $A^k $ and $m_{\nu_R}$ (i.e. the product $\kappa v_{R}$) 
must have opposite signs in order to avoid tachyonic pseudoscalar right sneutrinos.
In particular, in our BP we choose a negative value for $A^k$.

The masses for `leptons' 
%charged fermions 
are shown in $10^{th}$ row of the third box.
% of the third box of the table.
The mass eigenstates are denoted as $\lambda^{\pm}_{4,5}$, 
%with $\alpha=4,5$, 
since we associate the first three states to the SM leptons.
As discussed in Section~\ref{masses},
%Although the latter mix with the charginos giving rise to the $5\times 5$ charged fermion mass matrix written in Eq.~() GIVE CHARGINO MATRIX?, 
the
$2\times 2$ MSSM-like chargino submatrix is basically decoupled from the
$3\times 3$ lepton submatrix. 
%The former is like the one of the MSSM/NMSSM 
%provided that one uses the effective $\mu$ term in Eq.~(\ref{muterm}).
Thus we obtain almost pure charged Wino, Higgsino.
The 
${\lambda^{-}_5({\tilde W}^{-})}$ mass of 920 GeV can be approximated by the soft mass $M_2$:
\bea
m_{{\tilde W}^{\pm}}
\approx   M_2\ .
\label{sps-approx222}
\eea 
The charged Higgsinos have a mixing of order 50\%,
and the state 
%${\lambda^{\pm}_4(\tilde{H}^{+}_u/\tilde{H}^{-}_d)}$ 
$\lambda^{-}_4(\tilde{H}^{-}_d/(\tilde{H}^{+}_u)^c)$ 
have a mass
of 272.1 GeV, which can be approximated by the value of the 
effective $\mu$ term, as for the neutral Higgsinos 
in Eq.~(\ref{sps-approx223}):
%  \bea
% m_{\tilde H^{\pm}}
% \approx   
% \la_i \frac{v_{iR}}{\sqrt 2}=\lambda \frac{v_{R}}{\sqrt 2}
% \ .
% \label{sps-approx1223}
% \eea 
 \bea
m_{\tilde H^{\pm}}
\approx   
%\la_i \frac{v_{iR}}{\sqrt 2}=
\lambda \frac{v_{R}}{\sqrt 2}
\ .
\label{sps-approx1223}
\eea

The squarks masses are shown in $11^{th}-14^{th}$ rows of the same box. They were
%The squark mass matrices 
discussed in Section~\ref{masses}. 
As a consequence of their structure, 
%are written in Eq.~() GIVE SQUARK MATRICES?. When compared to the MSSM/NMSSM case, their structure is essentially unaffected, provide that one uses the effective $\mu$ term in Eq.~(\ref{muterm}), and neglects
%the terms proportional to $Y_\nu$ and $v_\nu$.
%As a consequence, 
all squark masses are of the order of the corresponding soft masses $\sim$ 1.3 TeV, except the lightest and the heaviest ones which because of the large top Yukawa coupling driven mixing between the left and right stops, obtain masses of the order of 1.1 and 1.4 TeV, respectively. 

The gluinos masses are shown in $15^{th}$ row. They are of the order of 1.6 TeV, determined by the value of $M_3$:
%In our BP these are of order 1.6 TeV
  \bea
m_{\tilde g} \approx M_3
\ .
\label{gluino}
\eea 

Let us finally remark that it is easy
to obtain other masses for the electron and muon sneutrinos co-LSPs, as can be deduced from the discussion below Eq.~(\ref{evenLLL}). A simple way to decrease (increase) the mass of the LSP is to increase (decrease) the left sneutrino VEVs. However, as we will discuss in 
Section~\ref{Section:detectionSneutrino-LSP},
%Subsection~\ref{Subsection:SneutrinosProduction}, 
only the narrow range of masses $118-132$ GeV is relevant for the present work where we focus on prompt decays, since only for that mass range the LSP can be treated as a promptly decaying particle.
On the other hand, for a tau sneutrino LSP the range is broader and a richer collider phenomenology could be obtained.
%, thus we leave the detailed analysis of these other masses for the next subsection below.

\vspace{0.25cm}

\noindent
%\subsection
{\bf Tau sneutrino LSP}
\label{staulsp}

\noindent 
In Table~\ref{table:tau-sneu},
%-\ref{table:tau-sneu-145} 
we have adopted the strategy discussed below Eq.~(\ref{evenLLL})
in order to produce a tau left sneutrino LSP, $\widetilde \nu^\mathcal{I}_{\tau L}$, namely to use
similar input parameters as in Table~\ref{table:musneu-125} but with
$v_{1L}=v_{2L} < v_{3L}$.
% in order to produce a $\widetilde \nu^\mathcal{I}_{\tau}$ LSP. 
In this case, the masses obtained from Eq.~(\ref{evenLLL}) are different
from the ones in Table~\ref{table:musneu-125}, with the mass of the pseudoscalar state essentially degenerate with the mass of its scalar partner $\widetilde \nu^\mathcal{R}_{\tau L}$, but not with the other families of sneutrinos.
In the third box of Table~\ref{table:tau-sneu}, we see that the  $A^0_2(\widetilde\nu^\mathcal{I}_{\tau L})$ is the LSP with a mass of 126.4 GeV basically degenerate with the one of the state $h_2(\widetilde\nu^\mathcal{R}_{\tau L})$, which is the co-LSP.
The next heavier state is now the $h_3(\widetilde{\nu}^{\mathcal{R}}_R)$ with a mass of 501.2 GeV, since the other two families of left sneutrinos have masses of 776.4 GeV. The spectrum for the charged scalars is modified accordingly with respect to Table~\ref{table:musneu-125}, e.g. $H^-_2 (\tilde\tau_L)$ is the NLSP with a mass of 146.9 GeV, and no other state has mass degeneracy with this one.

%%%%%%%%%%%%%%%%%%%%%%%%%%%%%%%%%%%%%%%%%%%%%%%%%%%%%%%%%%%%%%%%%%%%

%125 tau
\begin{table}[t!]
\centering
 \caption{Benchmark point producing $\widetilde \nu^\mathcal{I}_{\tau L}$ 
and $\widetilde \nu^\mathcal{R}_{\tau L}$ co-LSPs, with masses of 126.4 GeV.
Input parameters, and soft masses obtained from the minimization conditions, are given 
in the first and second boxes at the low scale $M_{EWSB}$. Sparticle masses are shown in the third box
(with their dominant compositions written in brackets).
Squark and gluino masses are the same as in Table \ref{table:musneu-125} and not shown.
Sneutrino branching ratios (larger than $10^{-4}$) and decay widths are shown in 
the fourth and fifth boxes, respectively. VEVs, soft parameters, sparticle masses and decay widths
are given in GeV.}
 \label{table:tau-sneu}
{\scriptsize
 \begin{tabular}{|p{1.95cm}|p{1.95cm}|p{1.95cm}|p{1.95cm}|p{1.95cm}|p{1.95cm}|} \hline
 $\lambda $&$0.2$&$\kappa_{}$&$0.3$&$Y^{\nu}$&$5\times 10^{-7}$\\ \hline
  $v_{1,2 L}/\sqrt 2$&$1\times 10^{-5}$& $v_{3L}/\sqrt 2 $  & $4\times 10^{-4}$ 
& $v_{R}/\sqrt 2$ &$1350$\\ \hline
 tan $\beta$&10&$A^u$&$-3177$&$A^{d,e} $&$1000$\\ \hline 
 $A^{\lambda}$ & $1000$ &$A^{\kappa} $ & $-1000$ &$A^{\nu} $&$-400$\\ \hline
 $M_1$&300&$M_2$&500& $M_3 $&1600\\ \hline 
 $m^2_{\tilde Q_L,\tilde u_R,\tilde d_R} $&$1.69\times 10^6
$&$m_{\tilde e_R}^2 $&$10^6$  &--&--\\ \hline
  \end{tabular}
\begin{tabular}{|p{1.95cm}|p{1.95cm}|p{1.95cm}|p{1.95cm}|p{1.95cm}|p{1.95cm}|} \hline
 $m_{Hd}^2 $&$3.62\times 10^6$&$m_{Hu}^2 $&$-1.06\times 10^{5}$
 &$m_{\tilde \nu_R}^2 $ & $0.750\times 10^{5}$\\ \hline
 $m_{\tilde L_{e L}}^2 $&$0.598\times 10^{6}$& $m_{\tilde L_{\mu L}}^2 $&$0.598\times 10^{6}$
 & $m_{\tilde L_{\tau L}}^2 $&$1.35\times 10^{4}$\\ \hline
  \end{tabular}
 \begin{tabular}{|p{1.95cm}|p{1.95cm}|p{1.95cm}|p{1.95cm}|p{1.95cm}|p{1.95cm}|} \hline
 $m_{h_1(H^\mathcal{R}_u)} $&$124.2$&
  $m_{h_2(\widetilde{\nu}^\mathcal{R}_{{\tau L}})} $&$126.4$&$m_{h_3(\widetilde{\nu}^{\mathcal{R}}_R)}$&$501.2$ \\ \hline
  $m_{h_4(\widetilde{\nu}^\mathcal{R}_{{eL}})}$&$776.4$& 
  $m_{h_5(\widetilde{\nu}^\mathcal{R}_{{\mu L}})}$&$776.4 $&
$m_{h_6(H^\mathcal{R}_d)} $&$1934.4$\\ \hline
--&--&
  $m_{A^0_2(\widetilde{\nu}^\mathcal{I}_{\tau L})} $&$126.4$&$m_{A^0_3(\widetilde{\nu}^\mathcal{I}_{{eL}})}$&
$776.4$\\ \hline
$m_{A^0_4(\widetilde{\nu}^\mathcal{I}_{{\mu L}})}$&$776.4$&
  $m_{A^0_5(\widetilde{\nu}^{\mathcal{I}}_R)}$&$1099.9$&
$m_{A^0_6(H^\mathcal{I}_d)} $&$1933.9$\\ \hline
--&--&
  $m_{H^-_2(\widetilde{\tau}_L)}$&$146.9$&$m_{H^-_3(\widetilde{\mu}_L)} $
&$786.8$\\ \hline
$m_{H^-_4(\widetilde{e}_L)} $&$786.8$&
  $m_{H^-_5(\widetilde{e}_R)} $&$1000.4$&$m_{H^-_6(\widetilde{\mu}_R)} $&
$1000.5$\\ \hline
$m_{H^-_7(\widetilde{\tau}_R)} $&$1000.5$&
 $m_{H^-_8(H^-_d)} $&$1936.2$&--&--\\ \hline
  $m_{\lambda^0_4(\tilde H^0_u/\tilde H^0_d)} $&$241.2$&$m_{\lambda^0_5(\tilde H^0_u/\tilde H^0_d)} $&$280.8$&$m_{\lambda^0_6({\tilde B}^0)} $&$317.2$\\ \hline
$m_{\lambda^0_7({\tilde W}^0)} $&$531.4$&$m_{\lambda^0_8(\nu_R)} $&$809.6$&--&--\\ \hline
%$m_{\lambda^{\pm}_4(\tilde{H}^{+}_u/\tilde{H}^{-}_d)} $
$m_{\lambda^{-}_4(\tilde{H}^{-}_d/(\tilde{H}^{+}_u)^c)}$
&$264$&$m_{\lambda^{-}_5({\tilde W}^{-})} $&$531.5$&--&--\\ \hline
  \end{tabular}
 \begin{tabular}{|p{4.33cm}|p{1.95cm}|p{4.33cm}|p{1.95cm}|}\hline
$\text{BR}(A^0_2 \to \nu \nu)$&$0.4430$& $\sum\limits_{l=e,\mu,\tau}\text{BR}(A^0_2\to\tau^\pm l^\mp)$&$0.5548$
\\ \hline 
$\text{BR}(A^0_2 \to  \bar{b}b)$&$0.0008$&$\text{BR}(A^0_2 \to g g)$&$0.0015$\\ \hline 
$\text{BR}(h_2\to \nu \nu)$&$0.0059$&$\sum\limits_{l'=e,\mu}\text{BR}(h_2\to\tau^\pm l'^\mp)$&$0.0048$
\\ \hline 
$\text{BR}(h_2\to {\tau}^+ \tau^-)$&$0.1168$&$\text{BR}(h_2\to {\mu}^+ \mu^-)$&$0.0003$\\ \hline
$\text{BR}(h_2\to \bar{b} b)$&$0.4315$&$\text{BR}(h_2\to \bar{c}c)$&$0.0306$\\ \hline
$\text{BR}(h_2\to g g)$&$0.1143$&$\text{BR}(h_2\to \gamma \gamma)$&$0.003$\\ \hline
$\text{BR}(h_2\to W^\pm W^{\mp^*})$&$0.2624$&$\text{BR}(h_2\to Z Z^*)$&$0.0301$\\ \hline \hline
 $\Gamma(h_2) $ & $6.75 \times 10^{-11}$ & $\Gamma (A^0_2) $ 
 & $9.14\times 10^{-13}$\\ \hline
  \end{tabular}}
\end{table}
%%%%%%%%%%%%%%%%%%%%%%%%%%%%%%%%%%%%%%%%%%%%%%%%%%%%%%%%%%%%%%%%%%%%%

Notice that we have modified in this table the values of the soft masses $M_1$ and $M_2$, with respect to Table~\ref{table:musneu-125}, lowering them to 300 and 500 GeV, respectively. This is because the gaugino masses affect the seesaw mechanism generating neutrinos masses, as discussed in Section~\ref{masses}. 
Therefore, we have to choose the values of $M_1$ and $M_2$ in such a way that the mass of the heavier neutrino is maintained below the upper bound on the sum of neutrino masses $\sim 0.23$ eV \cite{Ade:2015xua}, and above the square root of the mass-squared difference 
%$\Delta m_{21}^2=7.53\times 10^{-3}\mathrm{eV}^2$ \cite{Gando:2013nba} and 
$\Delta m_{atm}^2\sim 2.42\times 10^{-3}\mathrm{eV}^2$ \cite{An:2015rpe}.
%, for each of the three BPs.  

From the discussion below Eq.~(\ref{evenLLL}), one deduces that a simple way to decrease the mass of the LSP is to increase the left sneutrino VEVs. In particular, we show in 
Table~\ref{table:tau-sneu-95} a point similar to the one of Table~\ref{table:tau-sneu} but with 
$v_{3L}/\sqrt 2= 5 \times 10^{-4}$ GeV.
In this way, we obtain 
%$A^0_2(\widetilde\nu^\mathcal{I}_{\mu})$ 
$\widetilde\nu^\mathcal{I}_{\tau L}$ and
$\widetilde\nu^\mathcal{R}_{\tau L}$ co-LSPs
with masses of about 97.8 GeV. 
The mass of the 
%$H^{\pm}_2 (\tilde\mu)$ 
$\tilde\tau_L$ 
NLSP also decreases and becomes 122 GeV.
For this point, the SM-like Higgs is heavier than the LSP and therefore, following our convention, is labeled as $h_2 (H^\mathcal{R}_u)$ in the table.

Following the same strategy, in order to increase the mass of the LSP we can simply decrease the value of the concerned VEV. 
We show in 
Table~\ref{table:tau-sneu-145} the case with $v_{3L}/\sqrt 2= 3\times 10^{-4}$ GeV giving rise to $\widetilde\nu^\mathcal{I}_{\tau L}$ 
and $\widetilde\nu^\mathcal{R}_{\tau L}$  co-LSPs with masses of about 
146 GeV, and a $\tilde\tau_L$ 
NLSP with a mass of 163.6 GeV.
In 
Table~\ref{table:tau-sneu-310} we show another case with a larger sneutrino mass.
For that we take
$v_{3L}/\sqrt 2= 9.48\times 10^{-5}$ GeV obtaining now a $\widetilde\nu^\mathcal{I}_{\tau L}$ and
$\widetilde\nu^\mathcal{R}_{\tau L}$ 
co-LSPs with masses of about 311 GeV.
%, and a $\tilde\tau_L$ NLSP with a mass of 310 GeV, respectively. \R{(same mass for 310?)}.
We have also changed the value of  $M_1$ and $\lambda$ to keep the Bino more massive than the LSP and  to avoid too light mass scale for neutrinos, while having enough multileptonic decays. The value of $A_\lambda$ is chosen in order to minimize the singlet composition of $h_1$ avoiding a decrease in its mass.
%Teixeira debajo eq. 3.2
As we will discuss in 
Section~\ref{Section:detectionSneutrino-LSP}, this range of sneutrino masses of about 95--310 GeV is the appropriate one for our analysis of signal detection.

Similarly, we could have worked with a fix value for $v_{3L}$ but 
varying the value of $A^\nu$. For example, for 
$v_{3L}/\sqrt 2= 5\times 10^{-4}$ GeV as in Table~\ref{table:tau-sneu-95},
with $A^\nu$ in the range between $-385$ and $-435$ GeV one can obtain the 
$\widetilde\nu^\mathcal{I}_{\tau L}$ LSP with a mass in the range of about 95 and 145 GeV.
Needless to say, we could also play around with the other relevant input parameters for our computation, i.e. $\lambda$, $\kappa$, $Y^\nu$,
$\tan\beta$, $v_{R}$, still obtaining this range of masses for the LSP.

%%%%%%%%%%%%%%%%%%%%%%%%%%%%%%%%%%%%%%%%%%%%%%%%%%%%%%%%%%%%%%%%%
%%%%%%%%%%%%%%%%%%%%%%%%%%%%%%%%%%%%%%%%%%%%%%%%%%%%%%%%%%%%%%%%

%%%%%%%%%%%%%%%%%%%%%%%%%%%%%%%%%%%%%%%%%%%%%%%%%%%%%%%%%%%%%%%%%%%%%

%%%%%%%%%%%%%%%%%%%%%%%%%%%%%%%%%%%%%%%%%%%%%%%%%%%%%%%%%%%%%%%%%
%%%%%%%%%%%%%%%%%%%%%%%%%%%%%%%%%%%%%%%%%%%%%%%%%%%%%%%%%%%%%%%%
%95 tau
\begin{table}[t!]
\centering
 \caption{The same as in Table \ref{table:tau-sneu} but for $\widetilde \nu^\mathcal{I}_{\tau L}$ 
and $\widetilde \nu^\mathcal{R}_{\tau L}$ co-LSPs with masses of 97.8 GeV 
{{considering 
$v_{3L}/\sqrt 2 = 5\times 10^{-4}$ GeV and  $A{^\nu} = -385$ GeV.
In the first and second boxes}} we show only the parameters whose values have changed.}
 \label{table:tau-sneu-95}
{\scriptsize
      \begin{tabular}{|p{1.95cm}|p{1.95cm}|p{1.95cm}|p{1.95cm}|p{1.95cm}|p{1.95cm}|} \hline
   $m_{\widetilde L_{eL}}^2 $&$0.454\times 10^{6}$& $m_{\widetilde L_{\mu L}}^2 $&$0.454\times 10^{6}$
 & $m_{\widetilde L_{\tau L}}^2 $&$0.692\times 10^{4}$\\ \hline \hline
$m_{h_1(\widetilde{\nu}^\mathcal{R}_{{\tau L}})}$&$97.8$&$m_{h_2(H^\mathcal{R}_u)}$&$124.7$&$m_{h_3(\widetilde{\nu}^{\mathcal{R}}_R)}$&$501.2$\\ \hline
$m_{h_4(\widetilde{\nu}^\mathcal{R}_{{\mu L}})}$&$676.8 $ &$m_{h_5(\widetilde{\nu}^\mathcal{R}_{eL})}$&$676.8$&--&--\\ \hline
--&--&$m_{A^0_2(\widetilde{\nu}^\mathcal{I}_{{\tau L}})} $&$97.8$&$m_{A^0_3(\widetilde{\nu}^\mathcal{I}_{\mu L})}$&$676.8$\\ \hline
$m_{A^0_4(\widetilde{\nu}^\mathcal{I}_{{eL}})}$&$676.8$&$m_{A^0_5(\widetilde{\nu}^{\mathcal{I}}_R)} $&$1099.9$&--&--\\ \hline
--&--&$m_{H^-_2(\widetilde{\tau}_L)}$&$122$&$m_{H^-_3(\widetilde{\mu}_L)} $&$666.8$\\ \hline
$m_{H^-_4(\widetilde{e}_L)} $&$666.8$&$m_{H^-_5(\widetilde{e}_R)} $&$1000.2$&$m_{H^-_6(\widetilde{\mu}_R)} $&$1000.2$\\ \hline
$m_{H^-_7(\widetilde{\tau}_R)} $&$1000.2$&--&--&--&--\\ \hline
  \end{tabular}
%}
%  \end{table}
  
% \begin{table*}
% \centering
 %{\scriptsize
   \begin{tabular}{|p{4.33cm}|p{1.95cm}|p{4.33cm}|p{1.95cm}|} \hline
  $\text{BR}(A^0_2 \to \nu \nu)$&$0.5515$&$\sum\limits_{l=e,\mu,\tau}\text{BR}(A^0_2\to \tau^\pm l^\mp)$&$0.4483$\\ \hline
$\text{BR}(h_1\to \nu \nu)$&$0.507$&$\sum\limits_{l=e,\mu,\tau}\text{BR}(h_1\to\tau^\pm l^\mp)$&$0.3889$\\ \hline
$\text{BR}(h_1\to \bar{b} b)$&$0.0854$&$\text{BR}(h_1\to \bar{c}c)$&$0.0053$\\ \hline
$\text{BR}(h_1\to g g)$&$0.00112$&$\text{BR}(h_1\to \gamma \gamma)$&$0.0005$\\ \hline
$\text{BR}(h_1\to W^\pm W^{\pm^*})$&$0.0013$&--&--\\ \hline
 $\Gamma(h_1) $ & $9.3\times 10^{-13} $ & $\Gamma (A^0_2) $ 
 & $8.5\times 10^{-13}$\\ \hline
 \end{tabular}}
 \end{table}

%%%%%%%%%%%%%%%%%%%%%%%%%%%%%%%%%%%%%%%%%%%%%%%%%%%%%%%%%%%%

%%%%%%%%%%%%%%%%%%%%%%%%%%%%%%%%%%%%%%%%%

\begin{table}[ht!]
\centering
 \caption{The same as in Table \ref{table:tau-sneu} but for $\widetilde \nu^\mathcal{I}_{\tau L}$ 
and $\widetilde \nu^\mathcal{R}_{\tau L}$ co-LSPs with masses of 146 GeV, 
{{choosing $v_{3L}/\sqrt 2 = 3\times 10^{-4}$ GeV.
In the first and second boxes}} we show only the parameters whose values have been
changed.}
 \label{table:tau-sneu-145}
{\scriptsize
        \begin{tabular}{|p{1.95cm}|p{1.95cm}|p{1.95cm}|p{1.95cm}|p{1.95cm}|p{1.95cm}|} \hline
     $m_{\widetilde L_{eL}}^2 $&$0.590\times 10^{6}$& $m_{\widetilde L_{\mu L}}^2 $&$0.590\times 10^{6}$
 & $m_{\widetilde L_{\tau L}}^2 $&$1.87\times 10^{4}$\\ \hline\hline
 $m_{h_1(H^\mathcal{R}_u)} $&$124.8$&
  $m_{h_2(\widetilde{\nu}^\mathcal{R}_{{\tau L}})} $&$146$&$m_{h_3(\widetilde{\nu}^{\mathcal{R}}_R)}$&$501.2$\\ \hline
$m_{h_4(\widetilde{\nu}^\mathcal{R}_{{eL}})}$&$771.3$&
  $m_{h_5(\widetilde{\nu}^\mathcal{R}_{{\mu L}})}$&$771.13$&--&--\\ \hline
  $m_{A^0_2(\widetilde{\nu}^\mathcal{I}_{{\tau L}})} $&$146$&$m_{A^0_3(\widetilde{\nu}^\mathcal{I}_{{eL}})}$&$771.3$&$m_{A^0_4(\widetilde{\nu}^\mathcal{I}_{{\mu L}})}$&$771.3$\\ \hline
  $m_{A^0_5(\widetilde{\nu}^{\mathcal{I}}_R)} $&$1100$&--&--&--&--\\ \hline
  $m_{H^-_2(\widetilde{\tau}_L)}$&$163.6$&$m_{H^-_3(\widetilde{\mu}_L)} $&$786.8$&$m_{H^-_4(\widetilde{e}_L)} $&$786.8$\\ \hline
  $m_{H^-_5(\widetilde{e}_R)} $&$1000.5$&$m_{H^-_6(\widetilde{\mu}_R)} $&$1000.5$&$m_{H^-_7(\widetilde{\tau}_R)} $&$1000.5$\\ \hline
  \end{tabular}
%}
 % \end{table}
%  
%\begin{table*}
%\centering
%{\scriptsize  
\begin{tabular}{|p{4.33cm}|p{1.95cm}|p{4.33cm}|p{1.95cm}|}\hline
$\text{BR}(A^0_2 \to \nu \nu)$&$0.3250$&$\sum\limits_{l=e,\mu,\tau}\text{BR}(A^0_2\to\tau^\pm l^\mp)$&
$0.6719$\\ \hline 
$\text{BR}(A^0_2 \to \bar{b}b)$&$0.0007$&$\text{BR}(A^0_2\to g g)$&$0.0021$\\ \hline
$\text{BR}(h_2\to \nu \nu)$&$0.1668$&--&--\\ \hline 
$\text{BR}(h_2\to {\tau}^+ \tau^-)$&$0.2492$&
$\sum\limits_{l'=e,\mu}\text{BR}(h_2\to\tau^\pm l'^\mp)$&$0.2284$\\ \hline
$\text{BR}(h_2\to \bar{b} b)$&$0.0716$&$\text{BR}(h_2\to \bar{c}c)$&$0.0056$\\ \hline 
$\text{BR}(h_2\to g g)$&$0.0293$&$\text{BR}(h_2\to \gamma \gamma)$&$0.0007$\\ \hline
$\text{BR}(h_2\to W^\pm W^{\mp^*})$&$0.2198$&$\text{BR}(h_2\to Z Z^*)$&$0.028$\\ \hline
 $\Gamma(h_2) $ & $1.69\times 10^{-12} $ & $\Gamma (A^0_2)$ 
 & $8.7 \times 10^{-13}$\\ \hline
\end{tabular}} 
\end{table}

%%%%%%%%%%%%%%%%%%%%%%%%%%%%%%%%%%%%%%%%%%%%%%%%%%%%%%%%%%%%%%%%%%%%%%%%%%%%%%

\begin{table}[t!]
\centering
 \caption{The same as in Table \ref{table:tau-sneu} but for $\widetilde \nu^\mathcal{I}_{\tau L}$ 
and $\widetilde \nu^\mathcal{R}_{\tau L}$ co-LSPs with masses of 310.9 GeV.
In the first, second and third {{boxes}} we show only the parameters whose values have been changed.}
 \label{table:tau-sneu-310}
{\scriptsize
      \begin{tabular}{|p{1.95cm}|p{1.95cm}|p{1.95cm}|p{1.95cm}|p{1.95cm}|p{1.95cm}|} \hline
    $\lambda$&0.35&$A^{\lambda}$&$3714$&$v_{3L}/\sqrt 2 $ & $9.48\times 10^{-5}$\\ \hline
    $M_1$&500&--&--&--&--\\ \hline \hline
        $m_{Hd}^2 $&$1.90\times 10^{7}$ & $m_{Hu}^2 $&$-1.42\times 10^{5}$ 
&$m_{\tilde \nu_R}^2 $ & $-4.345\times 10^{4}$\\ \hline
     $m_{\widetilde L_{eL}}^2 $&$0.590\times 10^{6}$& $m_{\widetilde L_{\mu L}}^2 $&$0.590\times 10^{6}$
 & $m_{\widetilde L_{\tau L}}^2 $&$7.55\times 10^{4}$\\ \hline \hline
 $m_{h_1(H^\mathcal{R}_u)} $&$125.3$&
  $m_{h_2(\widetilde{\nu}^\mathcal{R}_{{\tau L}})} $&$310.9$&$m_{h_3(\widetilde{\nu}^{\mathcal{R}}_R)}$&$523.3$\\ \hline
$m_{h_4(\widetilde{\nu}^\mathcal{R}_{{eL}})}$&$778.1$&
  $m_{h_5(\widetilde{\nu}^\mathcal{R}_{{\mu L}})}$&$778.1$&
$m_{h_6(H^\mathcal{R}_d)} $&$4423$\\ \hline
  $m_{A^0_2(\widetilde{\nu}^\mathcal{I}_{{\tau L}})} $&$310.9$&$m_{A^0_3(\widetilde{\nu}^\mathcal{I}_{{eL}})}$&$778.1$&$m_{A^0_4(\widetilde{\nu}^\mathcal{I}_{{\mu L}})}$&$778.1$\\ \hline
  $m_{A^0_5(\widetilde{\nu}^{\mathcal{I}}_R)} $&$1079.4$&
$m_{A^0_6(H^\mathcal{I}_d)} $&$4420.7$&--&--\\ \hline
  $m_{H^-_2(\widetilde{\tau}_L)}$&$311$&$m_{H^-_3(\widetilde{\mu}_L)} $&$780.4$&$m_{H^-_4(\widetilde{e}_L)} $&$780.4$\\ \hline
  $m_{H^-_5(\widetilde{e}_R)} $&$985.2$&$m_{H^-_6(\widetilde{\mu}_R)} $&$985.2$&$m_{H^-_7(\widetilde{\tau}_R)} $&$988.3$\\ \hline
  $m_{H^-_8(H^-_d)} $&$4421.4$&--&--&--&--\\ \hline
    $m_{\lambda^0_4(\tilde H^0_u/\tilde H^0_d)} $&$421.6$&$m_{\lambda^0_5(\tilde H^0_u/\tilde H^0_d)} $&$484.8$&$m_{\lambda^0_6({\tilde B}^0)} $&$501.4$\\ \hline
$m_{\lambda^0_7({\tilde W}^0)} $&$567.1$&$m_{\lambda^0_8(\nu_R)} $&$814.5$&--&--\\ \hline
$m_{\lambda^{-}_4(\tilde{H}^{-}_d/(\tilde{H}^{+}_u)^c)}$
&$436.7$&$m_{\lambda^{-}_5({\tilde W}^{-})} $&$563.4$&--&--\\ \hline
  \end{tabular}
%}
%  \end{table}
  
%\begin{table*}
% \centering
%{\scriptsize
\begin{tabular}{|p{4.33cm}|p{1.95cm}|p{4.33cm}|p{1.95cm}|}\hline
$\text{BR}(A^0_2 \to \nu \nu)$&$0.0569$&$\sum_{l=e,\mu,\tau} \text{BR}(A^0_2\to  {\tau}^\pm l^\mp )$&
$0.7565$\\ \hline 
$\text{BR}(A^0_2 \to \bar{b}b)$&$0.0002$&$\text{BR}(A^0_2\to g g)$&$0.0070$\\ \hline
$\text{BR}(h_2\to \nu \nu)$&$0.0374$&$\text{BR}(h_2\to h_1 h_1)$&$0.2877$
\\ \hline 
$\text{BR}(h_2\to {\tau}^+ \tau^-)$&$0.1846$&
$\sum_{l'=e,\mu}\text{BR}(h_2\to\tau^\pm l'^\mp)$&$0.3308$\\ \hline
$\text{BR}(h_2\to \bar{b} b)$&$0.0005$&
$\text{BR}(h_2\to g g)$&$0.005$\\ \hline
$\text{BR}(h_2\to W^\pm W^{\mp^*})$&$0.1017$&$\text{BR}(h_2\to Z Z^*)$&$0.0475$\\ \hline\hline
 $\Gamma(h_2) $ & $8.13\times 10^{-13} $ & $\Gamma (A^0_2)$ 
 & $5.3 \times 10^{-13}$\\ \hline
\end{tabular}}
\end{table}

%%%%%%%%%%%%%%%%%%%%%%%%%%%%%%%%%%%%%%%%%%%%%%%%%%%%%%%%%%%%%%%%%
%%%%%%%%%%%%%%%%%%%%%%%%%%%%%%%%%%%%%%%%%%%%%%%%%%%%%%%%%%%%%%%%

%%%%%%%%%%%%%%%%%%%%%%%%%%%%%%%%%%%%%%%%%%%%%%%%%%%%%%%%%%%%%%%%%
%%%%%%%%%%%%%%%%%%%%%%%%%%%%%%%%%%%%%%%%%%%%%%%%%%%%%%%%%%%%%%%%

%%%%%%%%%%%%%%%%%%%%%%%%%%%%%%%%%%%%%%%%%%%%%%%%%%%%%%%%%%%%%%%%
%%%%%%%%%%%%%%%%%%%%%%%%%%%%%%%%%%%%%%%%%%%%%%%%%%%%%%%%%%%%%%%%
\section{Decay modes}

\label{decaymodes}
%%%%%%%%%%%%%%%%%%%%%%%%%%%%%%%%%%%%%%%%%%%%%%%%%%%%%%%%%%%%%%%%
%%%%%%%%%%%%%%%%%%%%%%%%%%%%%%%%%%%%%%%%%%%%%%%%%%%%%%%%%%%%%%%%

\begin{figure}[t]
   \begin{center}
%    \begin{tabular}{cc}
%       \epsfig{file=figures/snI-u_u.eps,height=3.6cm}
%      \hspace*{-1mm} & \hspace*{-8mm}
      \epsfig{file=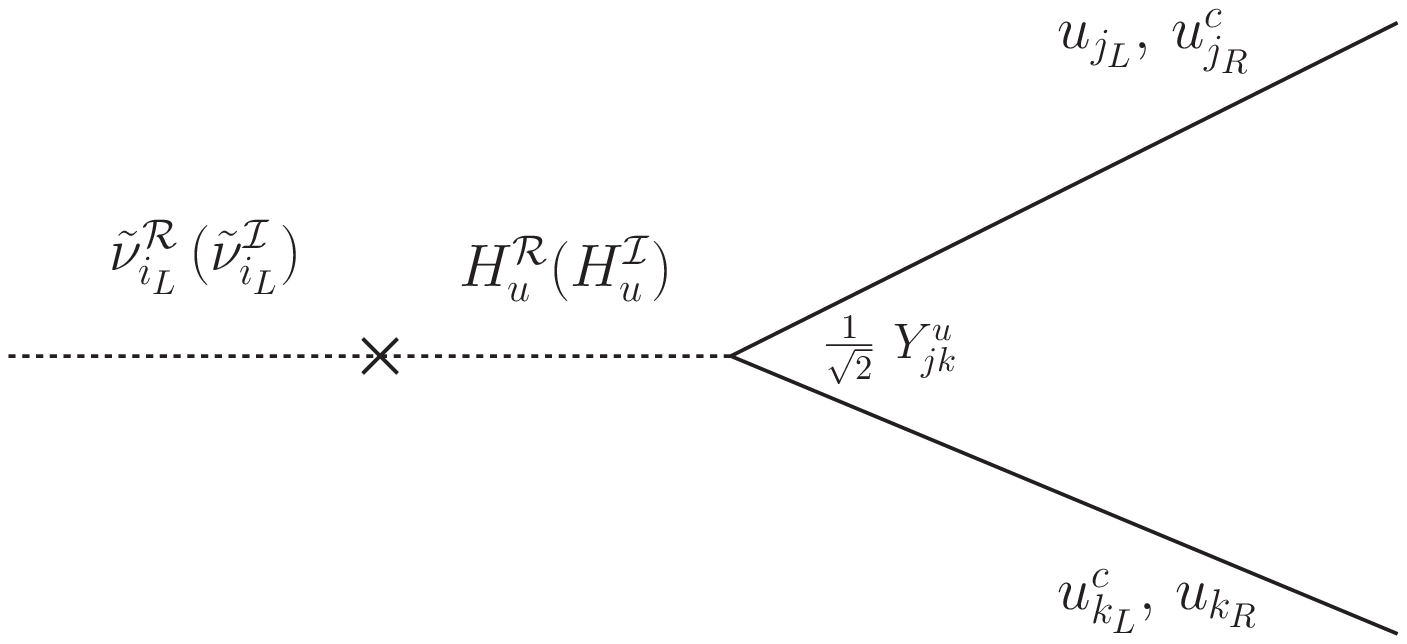,height=3.6cm}
%      \\ & \\
%      (a)\hspace*{2mm}&\hspace*{1mm} (b)
%    \end{tabular}
     \captions{
% Feynman diagram contributing to the decay of the CP-odd LH sneutrino LSP into up quarks, through its mixing with $H^{\mathcal{I}}_u$. 
Feynman diagram contributing to the decay of the scalar (pseudoscalar) left sneutrino  into up-type quarks, through its mixing with $H^{\mathcal{R}}_u$ ($H^{\mathcal{I}}_u$). 
}
%      (a) quarks (b) leptons.}
%}
     \label{fig:efflambda-to-quarksu}
   \end{center}
\end{figure}

\begin{figure}[t]
   \begin{center}
%     \begin{tabular}{cc}
%       \epsfig{file=figures/snI-d_d.eps,height=3.6cm}
%       \hspace*{-1mm}&\hspace*{-8mm}
      \epsfig{file=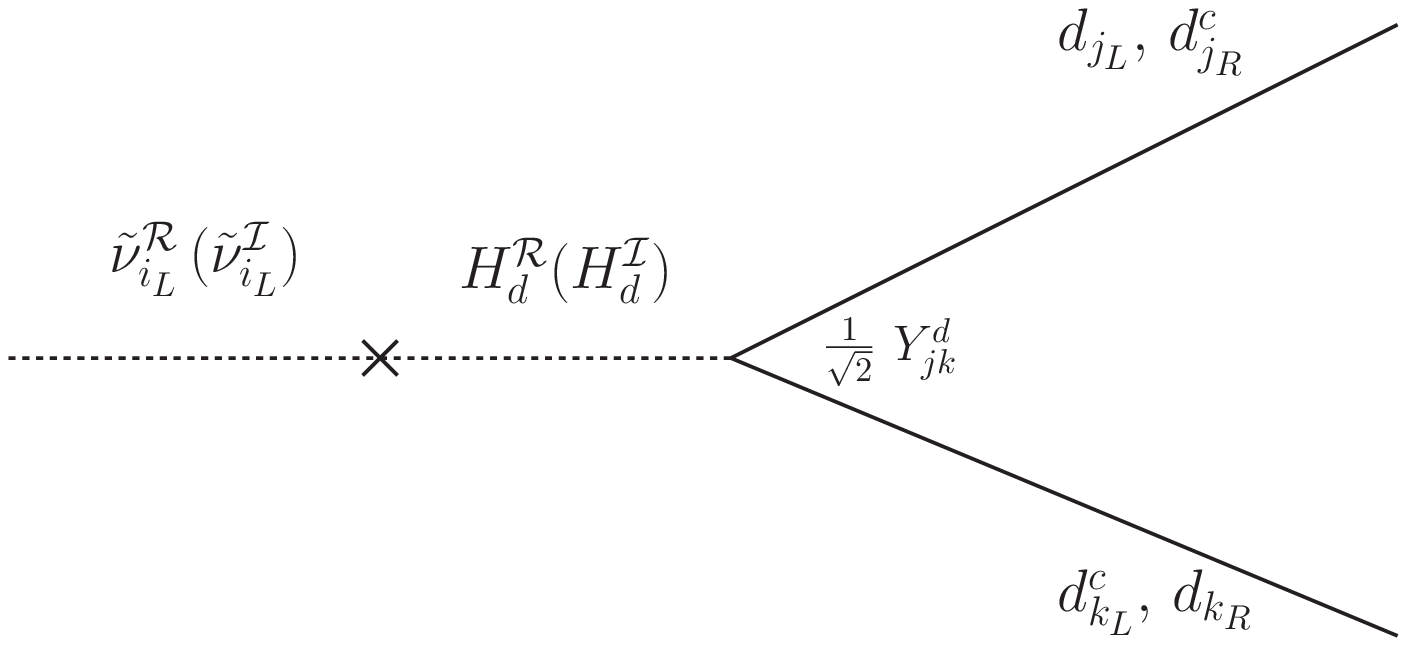,height=3.6cm}
%       \\ & \\
%       (a)\hspace*{2mm}&\hspace*{1mm} (b)
%     \end{tabular}
     \captions{
% Feynman diagram contributing to the decay of the CP-odd
%  LH sneutrino LSP into down quarks, through its mixing with $H^{\mathcal{I}}_d $.
Feynman diagram contributing to the decay of the scalar (pseudoscalar)
 left sneutrino into down-type quarks, through its mixing with
$H^{\mathcal{R}}_d $ ($H^{\mathcal{I}}_d)$.
       }
%}
     \label{fig:efflambda-to-quarksd-Hd}
   \end{center}
\end{figure}

\begin{figure}[t]
   \begin{center}
%     \begin{tabular}{cc}
%       \epsfig{file=figures/snI-eL_eR.eps,height=3.6cm}
%       \hspace*{-1mm}&\hspace*{-8mm}
       \epsfig{file=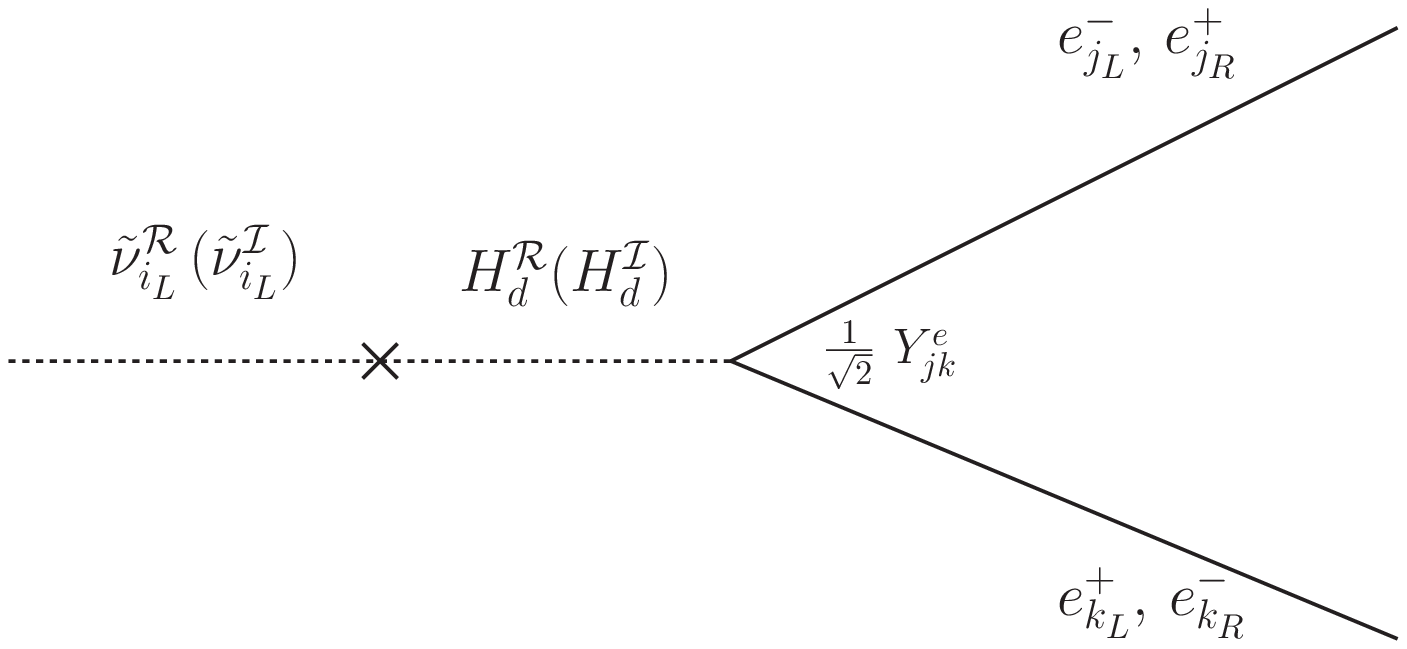,height=3.6cm}
%      \\ & \\
%      (a)\hspace*{2mm}&\hspace*{1mm} (b)
%     \end{tabular}
     \captions{
% Feynman diagram contributing to the decay of the CP-odd LH sneutrino LSP into charged leptons, through its mixing with $H^{\mathcal{I}}_d$.
Feynman diagram contributing to the decay of the scalar (pseudoscalar) left sneutrino into leptons, through its mixing with $H^{\mathcal{R}}_d$ ($H^{\mathcal{I}}_d$).
}
     \label{fig:efflambda-to-leptons-HuHd}
   \end{center}
\end{figure}

\begin{figure}[t]
   \begin{center}
%    \begin{tabular}{cc}
%      \epsfig{file=figures/snL-eL_eR.eps,height=3.6cm}
%      \hspace*{-1mm}&\hspace*{-8mm}
       \epsfig{file=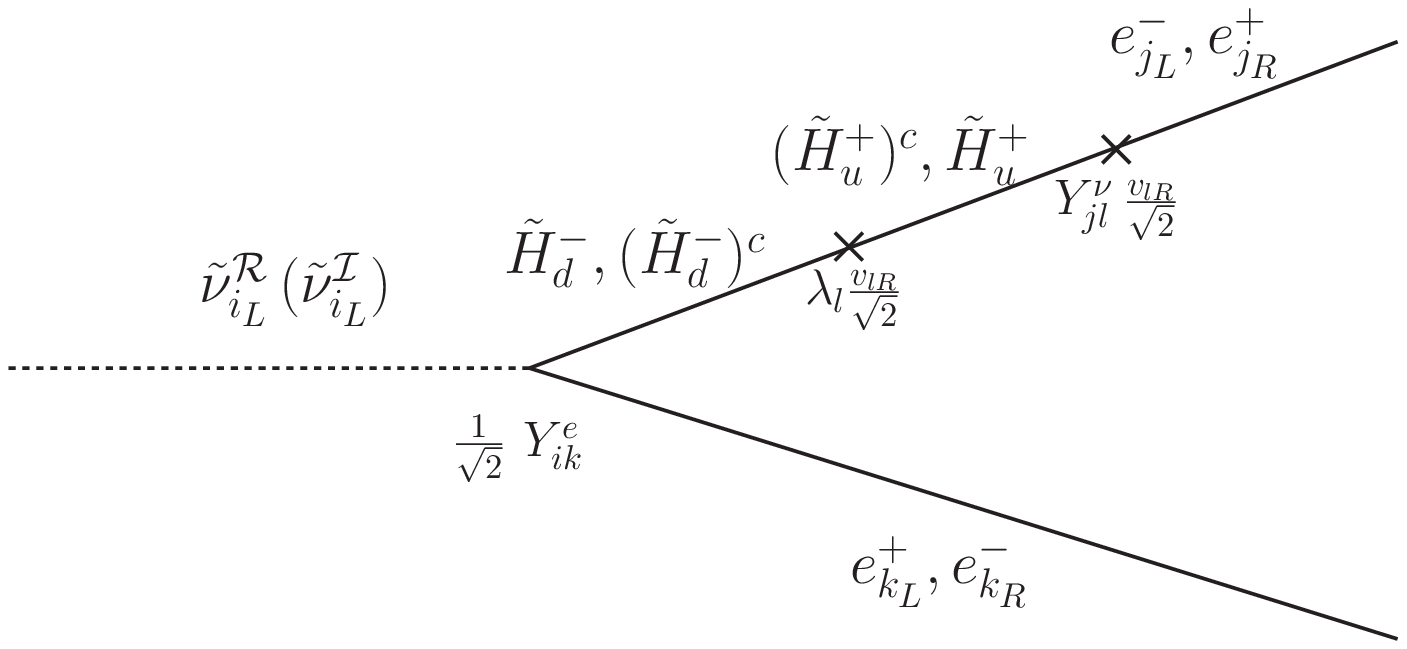,height=3.6cm}
%      \\ & \\
%      (a)\hspace*{2mm}&\hspace*{1mm} (b)
%    \end{tabular}
     \captions{Feynman diagram contributing to the decay of 
the scalar and pseudoscalar left sneutrino into leptons, through 
%a double mass insertion
the mixing between 
charged Higgsinos
%$\tilde %H_d^-$ 
and leptons.
%      (a) quarks (b) leptons.}
}
     \label{fig:eff-to-leptons-higgsino}
   \end{center}
\end{figure}

\begin{figure}[t]
   \begin{center}
     \begin{tabular}{cc}
       \epsfig{file=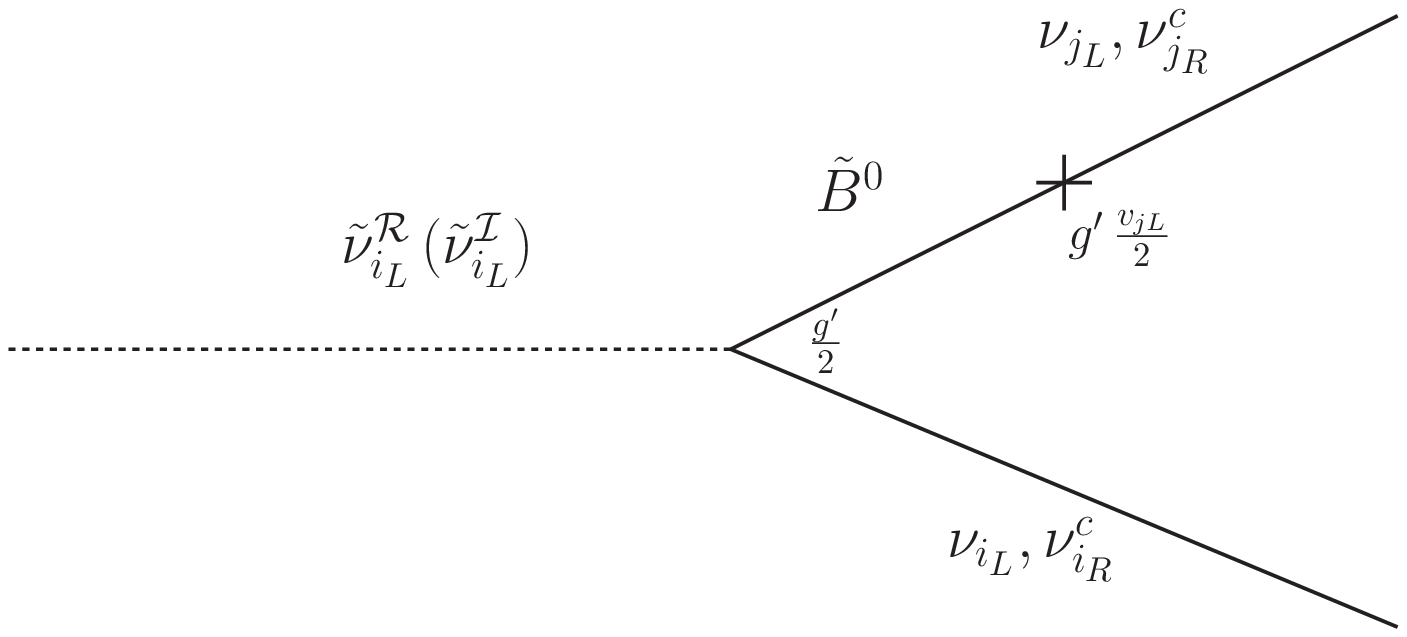,height=3.6cm}
       \hspace*{-1mm}&\hspace*{-8mm}
       \epsfig{file=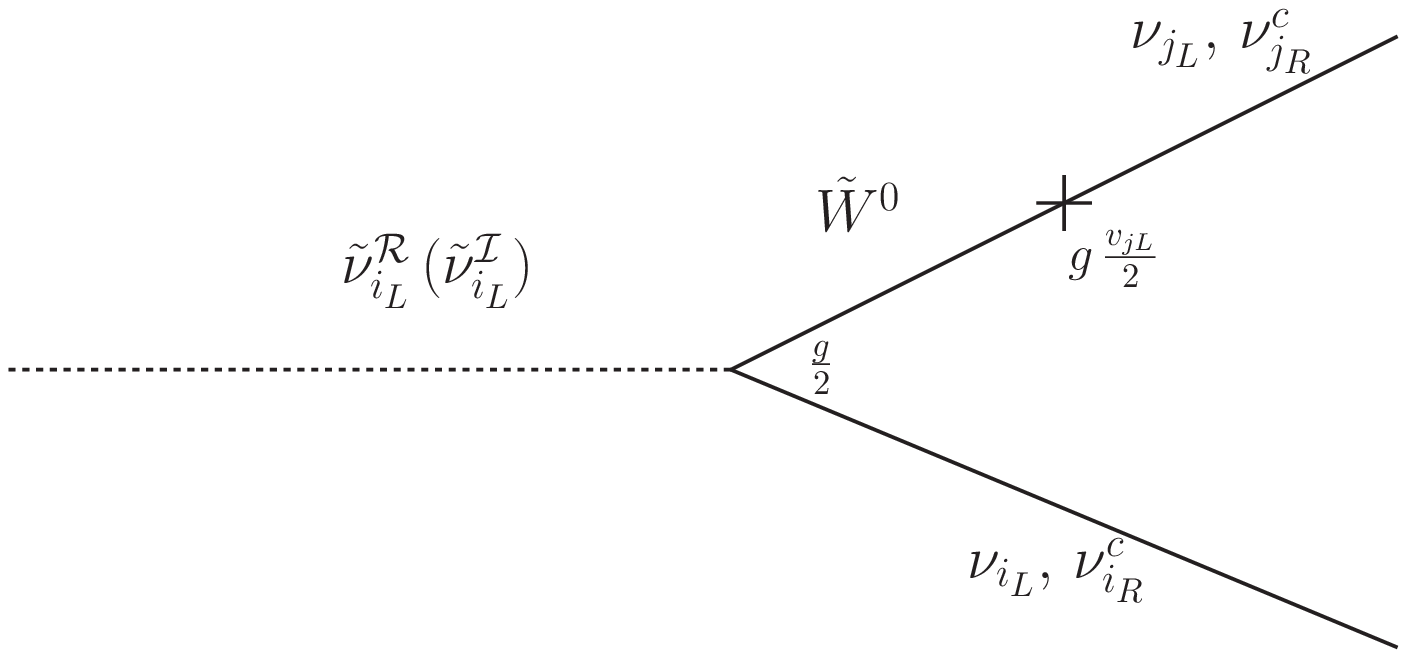,height=3.6cm}
       \\ & \\
       (a)\hspace*{2mm}&\hspace*{1mm} (b)
     \end{tabular}
     \captions{Feynman diagrams contributing to the decay of 
the scalar (pseudoscalar) left sneutrino into neutrinos, through the mixing between (a)
$\tilde B$ and neutrinos, (b) $\tilde W^0$ and neutrinos.}
     \label{fig:eff-to-neutrinos}
   \end{center}
\end{figure}

The interactions relevant for our analysis of the detectable decay of a left sneutrino LSP
into SM particles, are given in Appendix \ref{Section:Coupling}.
Since scalar and pseudoscalar states have essentially degenerate masses and therefore 
are co-LSPs,
%play the role of LSPs indistinctly, 
both are studied in the Apppendix.
% Both, scalar and pseudoscalar Higgses-sneutrinos (`Higgses') are considered. In this sense,
% let us emphasize again that both types of sneutrino states have almost degenerate masses, and therefore they can play the role of LSPs indistinctly. 
Although the expressions are quite involved, specially for the vertices with charged fermions (`leptons') and neutral fermions (`neutrinos'), one can 
straightforwardly 
identify the most important contributions for the decays as we will discuss below. These are schematically shown in 
Figs.~\ref{fig:efflambda-to-quarksu}--\ref{fig:eff-to-neutrinos}
using the gauge basis
(signs are neglected in the couplings since they are not relevant for the discussion).
%Notice nevertheless that we label the states as if they were in the mass or physical basis, since
%the mixing with the other states is negligible. 
%in Section~\ref{Section:detectionSneutrino-LSP}.
As explained in Section~\ref{masses}, the mixing between left sneutrinos and Higgses/right sneutrinos is small,
implying that the left sneutrino dominates the LSP composition. Nevertheless, although small, this mixing could still be sufficient as to produce a detectable decay into quarks and leptons, as shown in 
Figs.~\ref{fig:efflambda-to-quarksu}--\ref{fig:efflambda-to-quarksd-Hd} and 
Fig.~\ref{fig:efflambda-to-leptons-HuHd}, respectively. 
Similarly, the mixing between leptons and charginos (and between neutrinos and neutralinos)
%-right-handed neutrinos) 
is also small, but could still be sufficient as to produce a detectable decay of the left sneutrino into leptons (and neutrinos), as shown in Fig.~\ref{fig:eff-to-leptons-higgsino} (and Fig.~\ref{fig:eff-to-neutrinos}).
% {\it This means that the physical state giving the LSP is a very pure LH sneutrino, however the mixture with the other chargless scalars (pseudoscalars) it is not zero. The figures are describing the relevant admixture between the LH sneutrinos and the others neutral states.}

For a qualitative discussion of the left sneutrino decay in the $\mn$, we will use wherever is possible the mass insertion approximation. 
Note in this respect that only for
Figs.~\ref{fig:eff-to-leptons-higgsino} and~\ref{fig:eff-to-neutrinos},
the leading couplings in this approximation ($\equiv$ x) are written explicitly.
%, but they are not shown in Figs.~\ref{fig:efflambda-to-quarksu}-\ref{fig:efflambda-to-leptons-HuHd}.
The reason being that the analysis of 
Figs.~\ref{fig:efflambda-to-quarksu}--\ref{fig:efflambda-to-leptons-HuHd} is more subtle because scalar and pseudoscalar states behave in a different way, and the mass insertion technique cannot always be used.
Actually, this fact is directly related to the critical issue we face in this discussion:  
although both states are basically degenerate in mass, relevant decay modes (specially into quarks)
%in (\ref{decaymodes}) 
are different depending on the state analyzed.
This can have important implications for the detection of the sneutrino LSP at the LHC, as we discuss below.

%MENTION HERE THE DISPLACED VERTICES SINCE WE ARE REVIEWING THE SITUATION?
\vspace{0.25cm}

\noindent
%%%%%%%%%%%%%%%%%%%%%%%%%%%%%%%%%%%%%%%%%%%%%%%%%%%%%%%%%%%%%%%%
%%%%%%%%%%%%%%%%%%%%%%%%%%%%%%%%%%%%%%%%%%%%%%%%%%%%%%%%%%%%%%%%
{\bf Decays into quarks}
%%%%%%%%%%%%%%%%%%%%%%%%%%%%%%%%%%%%%%%%%%%%%%%%%%%%%%%%%%%%%%%%
%%%%%%%%%%%%%%%%%%%%%%%%%%%%%%%%%%%%%%%%%%%%%%%%%%%%%%%%%%%%%%%%

\noindent
% In the mass insertion approximation,
% denoting by $\Delta$ ($\equiv$ x in Figs.~\ref{fig:efflambda-to-quarksu}-\ref{fig:efflambda-to-leptons-HuHd}) the appropriate off-diagonal terms in the scalar mass matrices of
% Appendix~\ref{Apendix:Sneutrino-masses}, the
% sneutrino propagator
% can be expanded as a series in terms 
% of $\Delta/\tilde m^2$, where $\tilde m$ is a kind of average
% scalar mass to be discussed for each particular case below.
% If $\Delta$ is sufficiently small compared to $\tilde m^2$, it is enough to consider only the first term in the expansion.
The expressions for the interactions given in Appendix \ref{Section:Coupling} seem to be similar for both scalar and pseudoscalar states, however there is a crucial difference arising from the rotations 
which switch the flavor basis to the mass basis. This is due to the presence of the would be Goldstone boson in the pseudoscalar state, which is dominated by the composition of the pseudoscalar $H^{\mathcal{I}}_u$.
%As a consequence, in a crude approximation the CP-odd LH sneutrino cannot have decay channels as those shown in
%Figs.~\ref{fig:efflambda-to-quarksu}, \ref{fig:efflambda-to-quarksd-Hd}b and \ref{fig:efflambda-to-leptons-HuHd}b exchanging $H^{\mathcal{R}}_u$ by $H^{\mathcal{I}}_u$.
%This generates an asymmetry between CP-odd and CP-even state decays.
This  fact gives rise to a suppression of the couplings 
of pseudoscalar sneutrinos of the three families $\widetilde{\nu}^{\mathcal{I}}_{iL}$ to quarks.
% The composition 
% of the would be Goldstone boson is mainly $H^{\mathcal{I}}_u$.
%This implies that 
Although an estimation of 
%$\Delta$
the interaction in Fig.~\ref{fig:efflambda-to-quarksu} for the pseudoscalar sneutrino $\widetilde{\nu}^{\mathcal{I}}_{iL}$
decay into up-type quarks, using the mass insertion approximation, 
is not possible 
because of its mixing with $H^{\mathcal{I}}_u$,
in a crude approximation one would say that $\widetilde{\nu}^{\mathcal{I}}_{iL}$ cannot decay. In practice,
the Goldstone boson is not a pure $H^{\mathcal{I}}_u$, and therefore, although very small, a mixing with the sneutrino is present.
Thus the pseudoscalar sneutrino is able to decay into up-type quarks, but with a very small 
branching ratio (BR). We can see for example in (the fourth box of) Table~\ref{table:musneu-125} that, for the BP analyzed there,
the result of the numerical computation for the decays of the light left sneutrinos into charm quarks, using the interaction in Eq.~(\ref{eq:coupling_odd_sn->uu}),
%Eq.~(\ref{eq:coupling_odd_sn->uu}), 
is
BR($A_{2,3}^0\to \bar c c) =0.0007$.
%1.7\times 10^{-3}$. 
Of course, decays into top quarks are kinematically forbidden, given the sneutrino mass considered there of 125.4 GeV.


The scalar sneutrino $\widetilde{\nu}^{\mathcal{R}}_{iL}$ can
%in Subsection~\ref{decayscpeven}, 
have nevertheless a sizable mixing with $H^{\mathcal{R}}_u$.
Although the mixing of the latter with $H^{\mathcal{R}}_d$ makes
also the mass insertion approximation unreliable
for this case shown in Fig.~\ref{fig:efflambda-to-quarksu},
the numerical computation using the interaction in Eq.~(\ref{eq:coupling_even_sn->uu})
 gives BR($h_{2,3}\to \bar c c) =0.033$
%\times 10^{-2}$ for the BP of
for the BP of Table~\ref{table:musneu-125}.
This can be compared with the much smaller result above of the pseudoscalar sneutrino.
Because of the same reason as before, the mass insertion approximation cannot be used to estimate the interaction
in Fig.~\ref{fig:efflambda-to-quarksd-Hd} for the decay of the 
$\widetilde{\nu}^{\mathcal{R}}_{iL}$ into down-type quarks,
through its mixing with $H^{\mathcal{R}}_d$.
The numerical result of Table~\ref{table:musneu-125} 
using the interaction in 
Eq.~(\ref{eq:coupling_even_sn->dd}),
shows that the BR into bottom quarks is the largest one with a value
BR($h_{2,3}\to \bar b b) =0.468$.
% This
% can produce a detectable signal in combination with neutrinos or leptons from the decay of the pseudoscalar or scalar state, as we will discuss in 
% Section~\ref{Section:detectionSneutrino-LSP}.

% Notice that the second largest BR with a value of 0.256 corresponds to the decay into a real and a 
% virtual $W^{\pm}$. 
% This decay could proceed directly through the diagram coupling the sneutrino and the gauge bosons, which is proportional to $g^2v_L$, or through the mixing mainly with the Higgs
% $H_u$, which couples to the gauge bosons as $g^2 v_u $.

The mass insertion approximation can be used nevertheless for 
% Figs.~\ref{fig:efflambda-to-quarksd-Hd} and~\ref{fig:efflambda-to-leptons-HuHd}.
the 
%pseudoscalar sneutrino 
$\widetilde{\nu}^{\mathcal{I}}_{iL}$
decay into down-type quarks shown in 
Fig.~\ref{fig:efflambda-to-quarksd-Hd}, because,
in a crude approximation, one could say that $H^{\mathcal{I}}_d$
cannot mix with $H^{\mathcal{I}}_u$. Then, from Eq.~(\ref{pruebas6}) one can deduce that the dominant value of the %$\Delta$ 
coupling for the mixing between the sneutrino $\widetilde{\nu}^{\mathcal{I}}_{i}$ and the Higgs $H^{\mathcal{I}}_d$ is given by 
$\frac{1}{2}Y^\nu_{il}\lambda_{m}v_{lR}v_{mR}$. This is suppressed by the neutrino Yukawa coupling with respect to the squared mass of Eq.~(\ref{pruebas1}), 
thus, we can take advantage of the mass insertion technique and,
their ratio determines the value of  
the sneutrino propagator. Armed with that, one can straightforwardly obtain the following expression for the largest effective interaction of the pseudoscalar sneutrino decay into down-type quarks:
\bea
\frac{Y^b \, Y^\nu}
{\lambda } 
\frac{\lambda v_{R}}{(2A^\lambda + \sqrt 2\kappa v_{R})\tan\beta}\ .
\label{cp1}
\eea
The complete interaction that we use for the numerical computation can be found in 
Eq.~(\ref{eq:coupling_odd_sn->dd}). For the BP of 
Table~\ref{table:musneu-125}, we obtain
BR($A_{2,3}^0\to \bar b b) =0.0017$.

Let us finally remark that 
for a tau sneutrino LSP of a similar mass of about 126 GeV as the electron and muon sneutrinos in Table~\ref{table:musneu-125}, we can see in 
Table~\ref{table:tau-sneu} that the results for these BRs are very similar.

%%%%%%%%%%%%%%%%%%%%%%%%%%%%%%%%%%%%%%%%%%%%%%%%
% The diagrams for the CP-even state are similar (exchanging ${\mathcal{I}}$
% by ${\mathcal{R}}$) but with different effective couplings in some of the figures, as we will discuss below.

%%%%%%%%%%%%%%%%%%%%%%%%%%%%%%%%%%%%%%%%%%%%%%%%%%%%%%%%%%%%%%%%
%%%%%%%%%%%%%%%%%%%%%%%%%%%%%%%%%%%%%%%%%%%%%%%%%%%%%%%%%%%%%%%%
% \begin{figure}[t]
%   \begin{center}
%       \epsfig{file=figures/snL-nu_nu.eps,height=3.4cm}
%       \epsfig{file=figures/snL-nu_nu-2.eps,height=3.4cm}
%     \captions{Relevant Feynman diagrams contributing to the decay of the LH CP-even (CP-odd) 
%     sneutrino  LSP into neutrinos through \bl{(left)} the mixing between $\widetilde B$ and neutrinos, 
%     \bl{(right)} the mixing between $\widetilde W^0$ and neutrinos. 
%     \bl{[PG Note: Just draw these figures in latex twice/thrice
%     to make them bold looking].}}
%     \label{fig:eff-to-neutrinos}
%   \end{center}
% \end{figure}
%%%%%%%%%%%%%%%%%%%%%%%%%%%%%%%%%%%%%%%%%%%%%%%%%%%%%%%%%%%%%%%%
%%%%%%%%%%%%%%%%%%%%%%%%%%%%%%%%%%%%%%%%%%%%%%%%%%%%%%%%%%%%%%%%
%%%%%%%%%%%%%%%%%%%%%%%%%%%%%%%%%%%%%%%%%%%%%%%%%%%%%

\vspace{0.25cm}

\noindent
%%%%%%%%%%%%%%%%%%%%%%%%%%%%%%%%%%%%%%%%%%%%%%%%%%%%%%%%%%%%%%%%
%%%%%%%%%%%%%%%%%%%%%%%%%%%%%%%%%%%%%%%%%%%%%%%%%%%%%%%%%%%%%%%%
{\bf Decay into photons}
%%%%%%%%%%%%%%%%%%%%%%%%%%%%%%%%%%%%%%%%%%%%%%%%%%%%%%%%%%%%%%%%
%%%%%%%%%%%%%%%%%%%%%%%%%%%%%%%%%%%%%%%%%%%%%%%%%%%%%%%%%%%%%%%%

\noindent
As a consequence of the mixing of the scalar sneutrino with $H^{\mathcal{R}}_{u,d}$ discussed above, a sizeable decay channel  
into photons is generated
\bea
\widetilde{\nu}^\mathcal{R}_{iL} \rightarrow \gamma \gamma
\ ,
\label{decaymodesphotons}
\eea
mainly through $W^{\pm}$ and top-quark loops, 
with 
BR($h_{2,3}\to \gamma\gamma) =0.003$ as shown in 
Tables~\ref{table:musneu-125} and~\ref{table:tau-sneu}.
For an early work analyzing $\tilde\nu_L\rightarrow \gamma\gamma$ 
in the context of trilinear $\rpv$, see Ref.~\cite{BarShalom:1998xz}, 
where a negligible BR$\sim 10^{-6}$ was obtained.
This decay of the scalar sneutrino into two photons in a way not very different from the Higgs,
can be very interesting for our purposes. Let us recall in this sense that the Higgs 
was discovered thanks to this kind of decay.
% its decay into two photons, 
Although the associated BR 
is far from being the dominant one, the diphoton signal is very clear 
and easy to disentangle from the SM backgrounds. 
%Inspired by this fact, we will search for a signal given  by diphoton plus MET.

Notice however, that for other masses of the sneutrino as in Tables~\ref{table:tau-sneu-95}--\ref{table:tau-sneu-310}, the BR to photons is decreased. 
For example, in Table~\ref{table:tau-sneu-95} one obtains
BR($h_{2,3}\to \gamma\gamma) =0.0005$.
This is because the mixing of the scalar left sneutrino with the SM Higgs is reduced when the separation between their masses is increased.
%In the mass insertion language, and in a very rough approximation, we might think of this as if the average scalar mass $\tilde m^2$ 
%is better expressed as the difference between the squared masses of the sneutrino and the SM Higgs. 
In addition, the BR of the $H^\mathcal{R}_u$ to diphoton is maximal in the vicinity of the mass of the SM Higgs.
% Nevertheless, as we will discuss below, the BR to leptons which is determined by the diagrams in Figs.~\ref{fig:efflambda-to-leptons-HuHd} and~\ref{fig:eff-to-leptons-higgsino}
% is increased.
% The latter channel could produce a detectable multilepton signal in combination 
% with leptons from the decay of the pseudoscalar state or another scalar state.

\vspace{0.25cm}

\noindent
%%%%%%%%%%%%%%%%%%%%%%%%%%%%%%%%%%%%%%%%%%%%%%%%%%%%%%%%%%%%%%%%
%%%%%%%%%%%%%%%%%%%%%%%%%%%%%%%%%%%%%%%%%%%%%%%%%%%%%%%%%%%%%%%%
{\bf Decays into leptons}
%%%%%%%%%%%%%%%%%%%%%%%%%%%%%%%%%%%%%%%%%%%%%%%%%%%%%%%%%%%%%%%%
%%%%%%%%%%%%%%%%%%%%%%%%%%%%%%%%%%%%%%%%%%%%%%%%%%%%%%%%%%%%%%%%

\noindent
The mass insertion approximation can be used as before for the
$\widetilde{\nu}^{\mathcal{I}}_{iL}$
decay into down-type quarks, but now for its decay
%the $\widetilde{\nu}^{\mathcal{I}}_{iL}$ decay 
into leptons shown in
Fig.~\ref{fig:efflambda-to-leptons-HuHd}.
Therefore, the largest effective interaction is given by 
Eq.~(\ref{cp1}) exchanging the bottom Yukawa coupling $Y^b$ by
the tau Yukawa coupling $Y^{\tau}$:
\bea
\frac{Y^{\tau} \, Y^\nu}
{\lambda } 
\frac{\lambda v_{R}}{(2A^\lambda + \sqrt 2\kappa v_{R})\tan\beta}
%\frac{1}{\tan\beta}
\ .
\label{cp2}
\eea
For the interaction shown in Fig.~\ref{fig:eff-to-leptons-higgsino}, which also contributes to the decay of the left sneutrino into leptons, the discussion is more subtle.
First of all, it is necessary to proceed through a double mass insertion in the fermion propagator.
%, each one contributing with $\Delta/\tilde m$, with $\tilde m$ determined in this case by the Higgsino masses. 
Second,
the Yukawa coupling in the vertex is directly related to the
family of the LSP, 
% i.e. whether it
%is a $\widetilde{\nu}_{\tau_L}$, $\widetilde{\nu}_{\mu_L}$ or $\widetilde{\nu}_{e_L}$ (CHECK THE NOTATION).
thus for an electron or muon sneutrino,
% For the case of a $\widetilde{\nu}_{\tau_L}$, 
the effective interaction for the pseudoscalar (and the scalar) sneutrino
decay
%in Fig.~\ref{fig:eff-to-leptons-higgsino} 
is given respectively by
\bea
\frac{Y^{e,\mu} \, Y^{\nu}}{\sqrt 2 \lambda}\ ,
\label{cplepton}
\eea
whereas for a pseudoscalar (and scalar) sneutrino of the third family one obtains
\bea
\frac{Y^{\tau} \, Y^{\nu}}{\sqrt 2 \lambda}\ .
\label{cptau}
\eea
The complete interactions for the decays 
%of the pseudoscalar left sneutrino into leptons 
can be found in 
Appendix~\ref{Section:Coupling}.
%and~\ref{SubSection:Coupling-All}.
In particular, the relevant diagrams shown in
Figs.~\ref{fig:eff-to-leptons-higgsino} 
and~\ref{fig:efflambda-to-leptons-HuHd} correspond
to the first and third
%$1^{\rm st}$ and $7^{\rm th}$ 
term, respectively, multiplying 
the projectors $P_{L,R}$ in Eqs.~(\ref{eq:coupling_even_sn->ll}) and~(\ref{eq:coupling_odd_sn->ll}).

%, which is proportional to $Y_{\tau}$.
For an electron sneutrino LSP the contribution in Eq.~(\ref{cplepton}) 
is proportional to $Y^{e}$, and therefore suppressed with respect to the one in 
Eq.~(\ref{cp2}), where typically $\frac{\lambda v_{R}}{(2A^\lambda + \sqrt 2\kappa v_{R})\tan\beta}\approx 10^{-2}$. For a muon sneutrino LSP, however, $Y^{\mu}$ is larger
than $Y^e$ and we obtain for the BP in Table~\ref{table:musneu-125} that the contribution in Eq.~(\ref{cplepton})
is a factor of order 3 larger than that of Eq.~(\ref{cp2}).
See e.g. in Table~\ref{table:musneu-125} that
BR($A_{2}^0\to {\mu}^\pm \mu^\mp)=0.0055$ whereas
BR($A_{2}^0\to {\tau}^\pm \tau^\mp) =0.0003$.
Notice also that BR($A_{2}^0\to \bar b b) =0.0017$ is larger than the later mainly because of the factor 3 of color that has to be included in the computation. 

On the other hand, for a tau sneutrino LSP, the contribution in Eq.~(\ref{cptau}) is larger than those in Eqs.~(\ref{cp1}) and (\ref{cp2}). 
As a consequence, in this case one dominant decay channel for the pseudoscalar (and scalar) left sneutrino is the one 
shown in 
Fig.~\ref{fig:eff-to-leptons-higgsino} into leptons 
\bea
\widetilde{\nu}^{\mathcal{I},\mathcal{R}}_{\tau L} \rightarrow \tau^{+}_{L} l^{-}_{L}\ ,
\tau^{-}_{R} l^{+}_{R}\ ,
\label{decaymodesleptons}
\eea
where $l=e,\mu,\tau$. 
In Table~\ref{table:tau-sneu}, we see for example 
that 
$\sum_{l=e,\mu,\tau} BR(A^0_2\to {\tau}^{\pm} l^\mp) = 0.55$.
Similar results, 0.44, 0.67, and 0.75, are obtained in 
Tables~\ref{table:tau-sneu-95}--\ref{table:tau-sneu-310}, respectively, where other tau sneutrino masses are analyzed.

\vspace{0.25cm}

\noindent
%%%%%%%%%%%%%%%%%%%%%%%%%%%%%%%%%%%%%%%%%%%%%%%%%%%%%%%%%%%%%%%%
%%%%%%%%%%%%%%%%%%%%%%%%%%%%%%%%%%%%%%%%%%%%%%%%%%%%%%%%%%%%%%%%
{\bf Decays into neutrinos}
%%%%%%%%%%%%%%%%%%%%%%%%%%%%%%%%%%%%%%%%%%%%%%%%%%%%%%%%%%%%%%%%
%%%%%%%%%%%%%%%%%%%%%%%%%%%%%%%%%%%%%%%%%%%%%%%%%%%%%%%%%%%%%%%%

\noindent
The Feynman diagrams shown in Fig.~\ref{fig:eff-to-neutrinos} describe 
%the effective coupling that makes possible 
the decay of the scalar and pseudoscalar left 
sneutrino into neutrinos. The approximate expressions for the effective couplings can easily be deduced, as:
\bea
%g_{1,2}\frac{\nu}{M_{1,2}} \left(g_2\frac{\nu}{M_2}\right)\ .
g'^2\frac{v_{L}}{4M_{1}}\ , \;\;\;\;  g^2\frac{v_{L}}{4M_2}\ .
\label{cpneutrinos}
\eea
The complete interactions are given in 
Appendix~\ref{Section:Coupling}.
%and~\ref{SubSection:Coupling-Ann}. 
The
above values are a rough approximation of the first and second
terms, respectively, multiplying the projectors $P_{L,R}$ in  
Eqs.~(\ref{eq:coupling_even_sn->nn}) and~(\ref{eq:coupling_odd_sn->nn}).

For an electron or muon pseudoscalar sneutrino LSP, the contributions in 
Eq.~(\ref{cpneutrinos})
are of the order of 10 larger than the largest one in Eq.~(\ref{cplepton}) which is proportional to 
$Y^{\mu}$. This is the reason why in Table~\ref{table:musneu-125} we obtain
BR($A_{2}^0\to \nu \nu) = 0.97$ and BR($A_{3}^0\to \nu \nu) = 0.99$.
As a consequence, the dominant decay channel is the one shown in 
Fig.~\ref{fig:eff-to-neutrinos} into neutrinos
%\ref{fig:eff-to-leptons-higgsino}.
%DISCUSS BRIEFLY THE OTHER TWO TABLES.
%For a scalar sneutrino
%
% \bea
% \widetilde{\nu}^\mathcal{I}_{e,\mu L} \rightarrow \nu_{e,\mu_L} \nu_{i_L}\ ,
% \nu_{e,\mu_R}^c \nu_{i_R}^c
% \ .
% \label{decaymodesneutrinos}
% \eea
%
%
\bea
\widetilde{\nu}^\mathcal{I}_{i_L} \rightarrow \nu_{i_L} \nu_{j_L}\ ,
\nu_{i_R}^c \nu_{j_R}^c
\ .
\label{decaymodesneutrinos}
\eea
For a tau sneutrino LSP, the contributions in Eq.~(\ref{cpneutrinos})
are of the same order as that in Eq.~(\ref{cptau}), and therefore
%are smaller than that in (\ref{cplepton}) for a tau sneutrino LSP.
there are two dominant decay channels, those in Eqs.~(\ref{decaymodesneutrinos}) and~(\ref{decaymodesleptons}).
% , and those shown in Fig.~\ref{fig:eff-to-neutrinos} into neutrinos 
% \bea
% \widetilde{\nu}^\mathcal{I}_{\tau L} \rightarrow \nu_{\tau_L} \nu_{i_L}\ ,
% \nu_{\tau_R}^c \nu_{i_R}^c
% \ .
% \label{decaymodesneutrinos2}
% \eea
The relative size between them depends on the values of the gaugino masses $M_1$ and $M_2$ necessary to reproduce the correct neutrino physics as discussed in 
Section~\ref{Section:Sneutrino-LSP}, and the left sneutrino VEVs $v_{iL}$. 
In Table~\ref{table:tau-sneu}, we see for example
that 
%the one for the decays into charged leptons is
$\sum_{l=e,\mu,\tau} BR(A^0_2\to {\tau}^{\pm} l^\mp) = 0.55$
vs.
BR($A_{2}^0\to \nu \nu) = 0.44$.
%, where we denote $l=e,\mu$.
%for the decays into charged leptons, neutrinos, bottom quarks and gluons, 
%computed numerically using the couplings in Appendix~\ref{Section:Coupling}.
%DISCUSS BRIEFLY THE OTHER TWO TABLES.
% Similar results are obtained in Tables~\ref{table:tau-sneu-95} and~\ref{table:tau-sneu-145}, where other tau sneutrino masses are analyzed.
% \\

% The diagram in Fig.~\ref{fig:eff-to-neutrinos}(a) gives a effective 
% coupling of order $\sim g_1 \frac{\nu}{M_1}$, and the one in Fig.~\ref{fig:eff-to-neutrinos}(b) 
% goes as $\sim g_2 \frac{\nu}{M_2}$. Where $\nu$ represents a typical value for the LH sneutrinos 
% VEV $\nu \sim 10^{-4} \gev$. 

\vspace{0.5cm}

As we will discuss in detail in the next section,
%~\ref{Section:detectionSneutrino-LSP},
scalar and pseudoscalar sneutrinos can be produced in pairs at the 
LHC, and as
%(see Figs.~\ref{fig:production}-\ref{fig:production3}), and 
a consequence, some of
the above decay modes can give rise to detectable signals. In particular, this is the case of diphoton plus missing transverse energy (MET) from sneutrinos of any family combining the channel in 
Eq.~(\ref{decaymodesphotons}) with that of Eq.~(\ref{decaymodesneutrinos}),
%and~(\ref{decaymodesneutrinos2}),
and
diphoton plus leptons
% for the case of $\widetilde\nu_{eL}$/$\widetilde\nu_{\mu L}$ LSP.
from tau sneutrinos 
%$\widetilde\nu_{\tau L}$ 
combining the channels
in Eqs.~(\ref{decaymodesphotons}) and~(\ref{decaymodesleptons}).
An interesting multilepton signal can also be produced combining the decay channels for 
scalar and pseudoscalar tau sneutrinos in Eq.~(\ref{decaymodesleptons}).
% As for the pseudoscalar state of the third family, for the scalar state
% $\widetilde{\nu}^{\mathcal{R}}_{\tau L}$
% the decay channel into leptons in 
% Fig.~\ref{fig:eff-to-leptons-higgsino}
% can also be relevant, producing as well
% \bea
% \widetilde{\nu}^\mathcal{R}_{\tau L} \rightarrow 
% \tau^{+}_{L} l^{-}_{L}\ ,
% \tau^{-}_{R} l^{+}_{R}\ ,
% \label{decaymodesleptonss}
% \eea
% and giving rise to a multilepton signal in combination with the decay channel of Eq.~(\ref{decaymodesleptons}).
% Actually, as we will discuss in 
% Section~\ref{Section:detectionSneutrino-LSP},
% when the production of sneutrino LSPs occurs through slepton NLSPs, scalar or pseudoscalar states can be produced in pairs as well
% giving rise to non-negligible 
% contributions to the signals with leptons 
% (see Figs.~\ref{fig:production2}c-d and~\ref{fig:production3}b-c).
Although
%Let us finally mention that,
%as can be easily deduced from the above discussion, 
a signal with leptons plus MET is in principle possible, it suffers from a significant background and is unlikely to be observed.
To have signals of diphoton plus jets or multijets is also possible,
but disfavored with respect to the signals discussed here.

\begin{figure}[t!]
\centering
\includegraphics[scale=0.45]{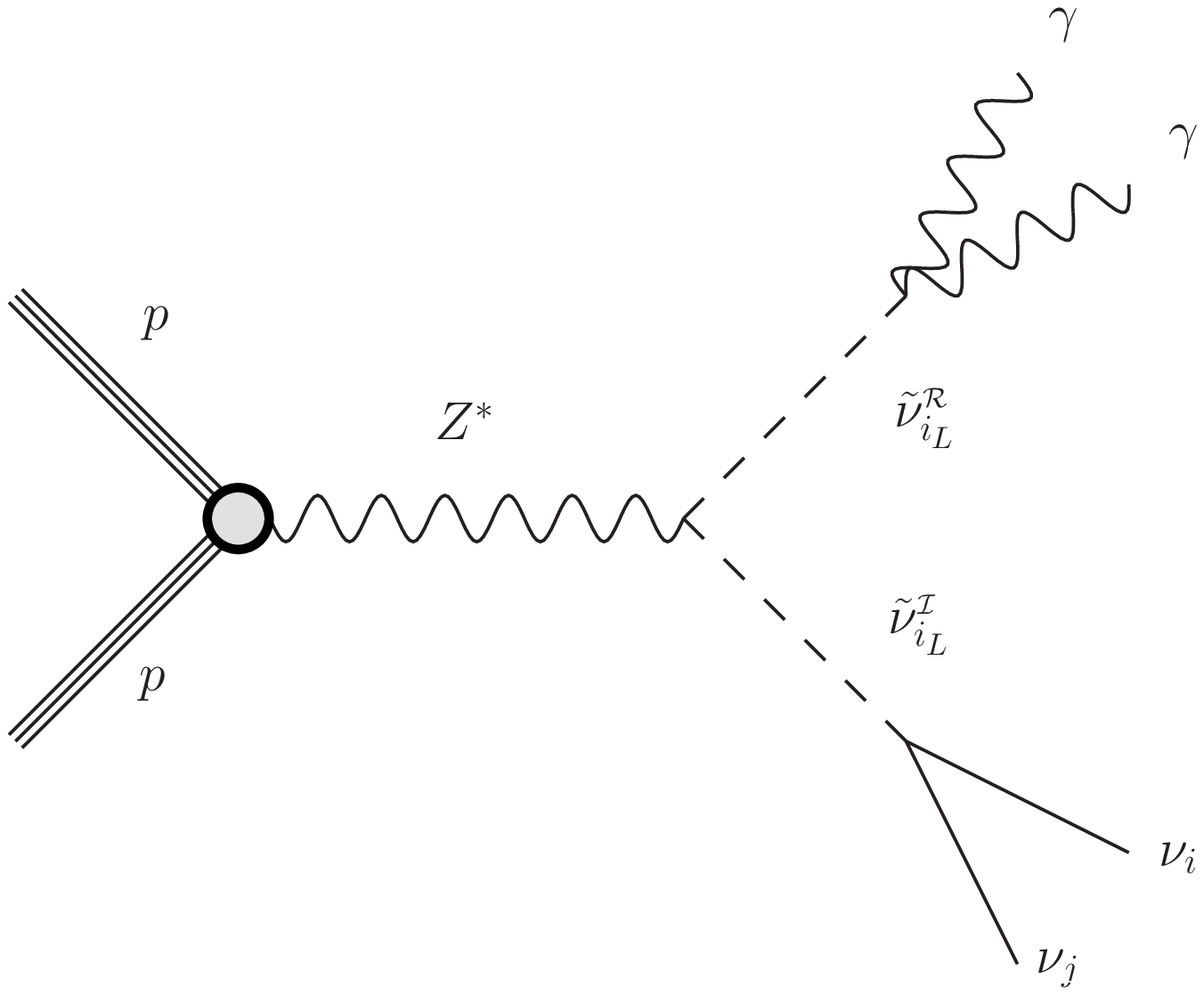}a)  
\includegraphics[scale=0.45]{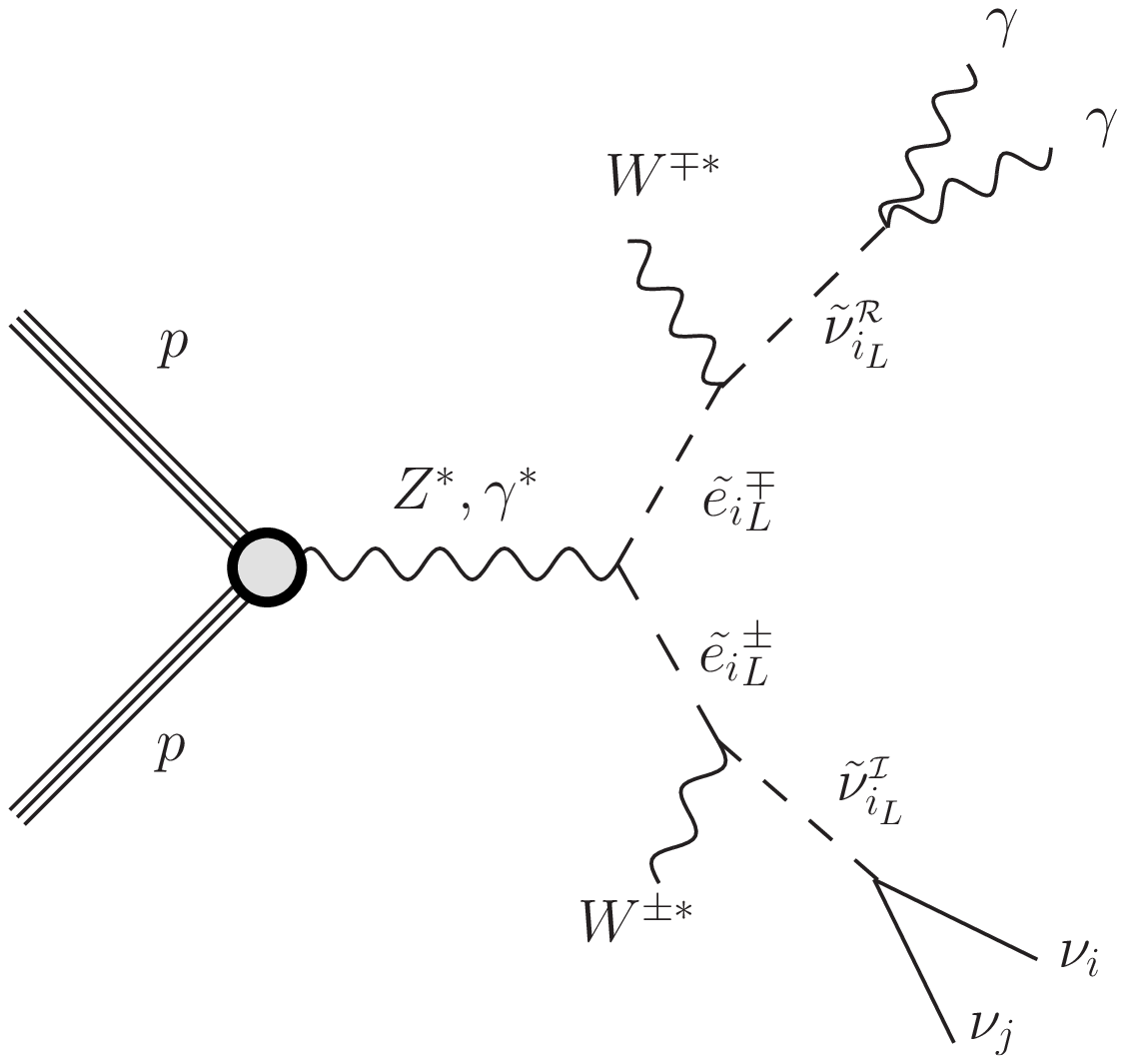}b)  
\includegraphics[scale=0.45]{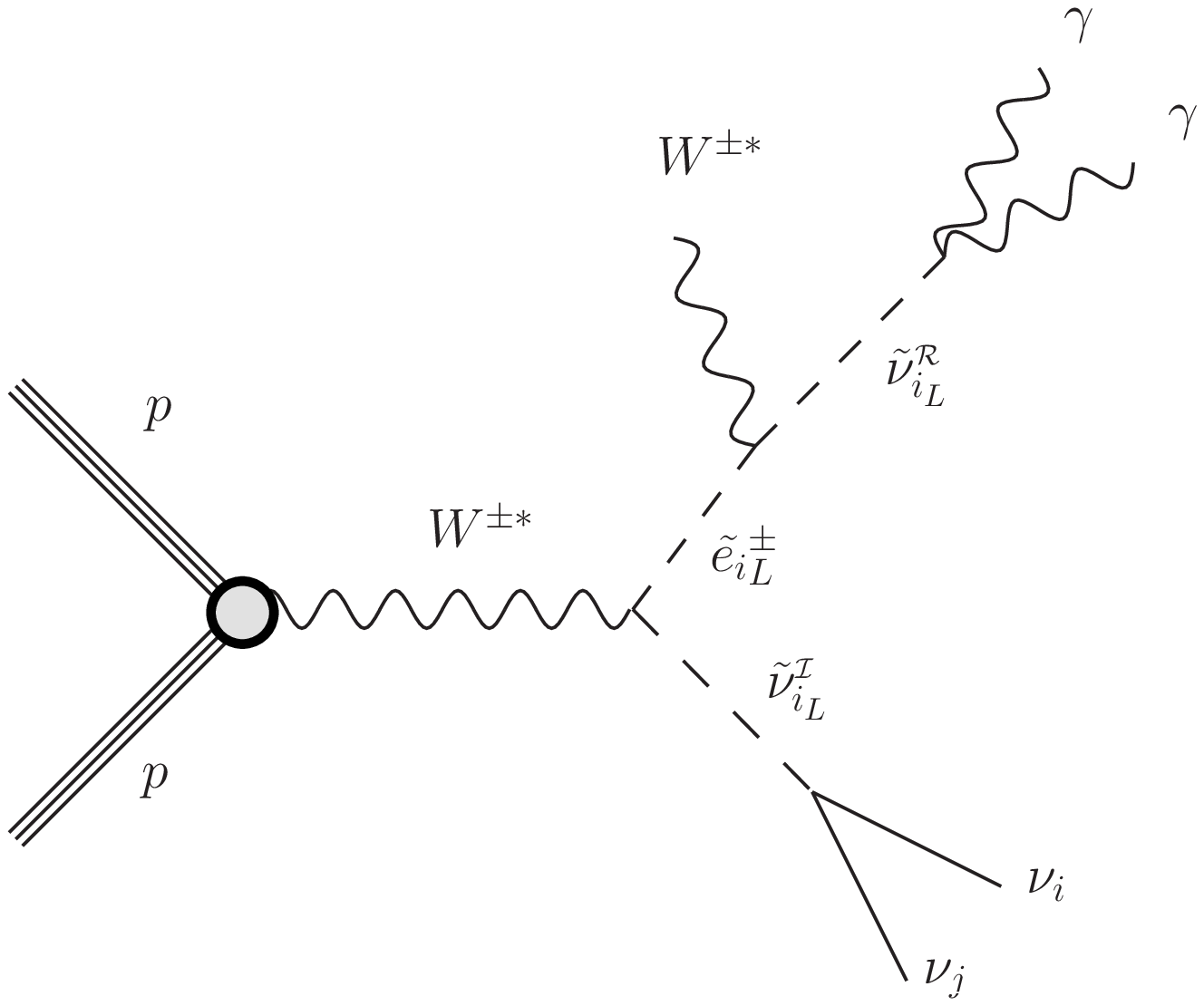}c)
\includegraphics[scale=0.45]{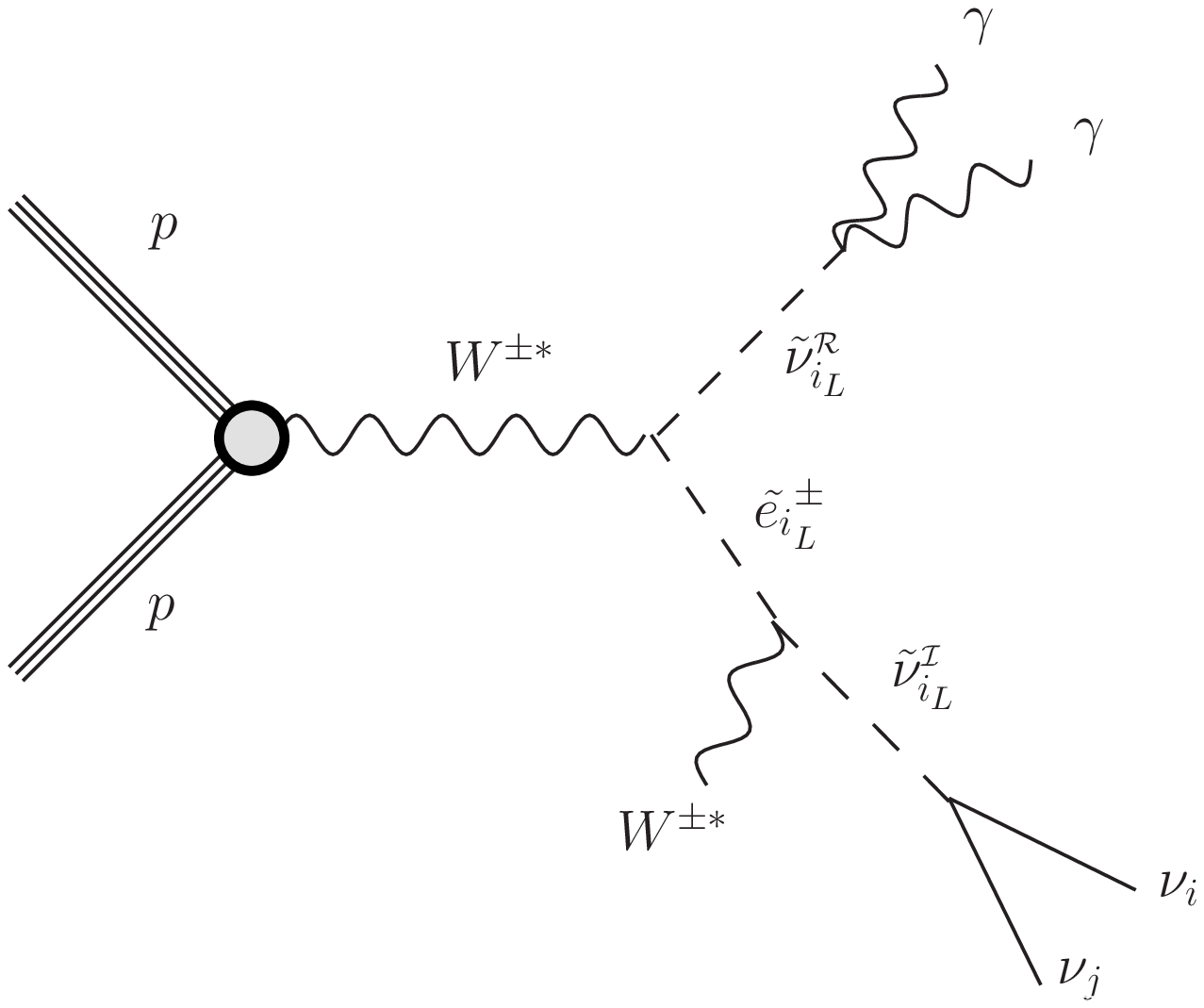}d) 
\caption{Dominant decay channels into diphoton plus neutrinos from a pair production at the LHC
of scalar and pseudoscalar sneutrinos LSP of any of the three families, $\widetilde\nu^\mathcal{I}_{iL}$, 
with $i=e,\mu,\tau$.  
Filled circles indicate effective
interactions.
%type  $\widetilde\nu_{eL}$ or $\widetilde\nu_{\mu L}$.
% Dominant production channels at the LHC for a pair of scalar and pseudoscalar 
% sneutrino LSPs, 
% $\widetilde\nu_{eL}/\widetilde\nu_{\mu L}$, including the relevant decay signal
% into diphoton state and neutrinos. 
%\bl{[PG Note: Just draw these figures in latex twice/thrice
 %   to make them bold looking].}
}
\label{fig:production}
\end{figure} 

\begin{figure}[t!]
\centering
\includegraphics[scale=0.45]{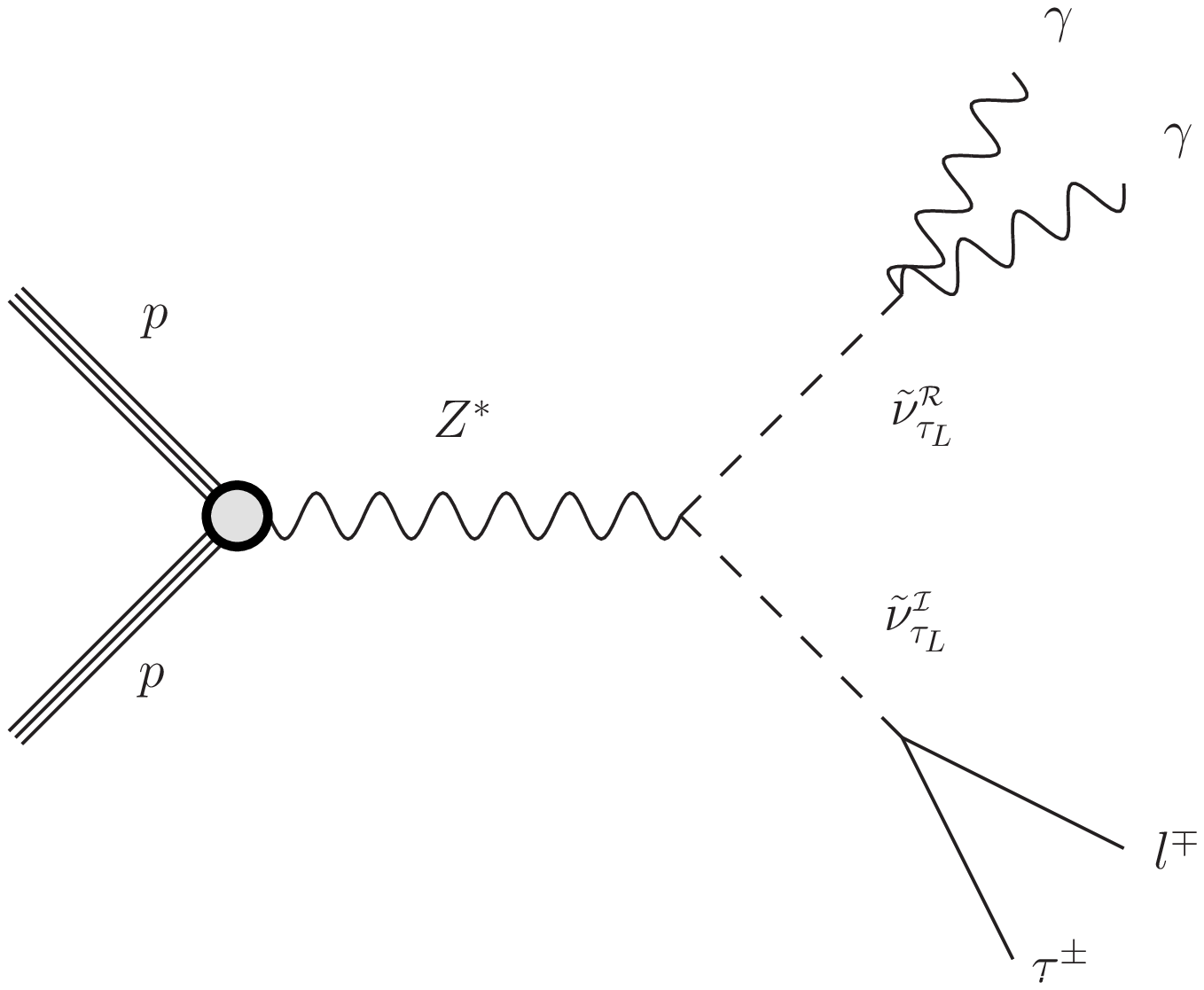}a)  
\includegraphics[scale=0.45]{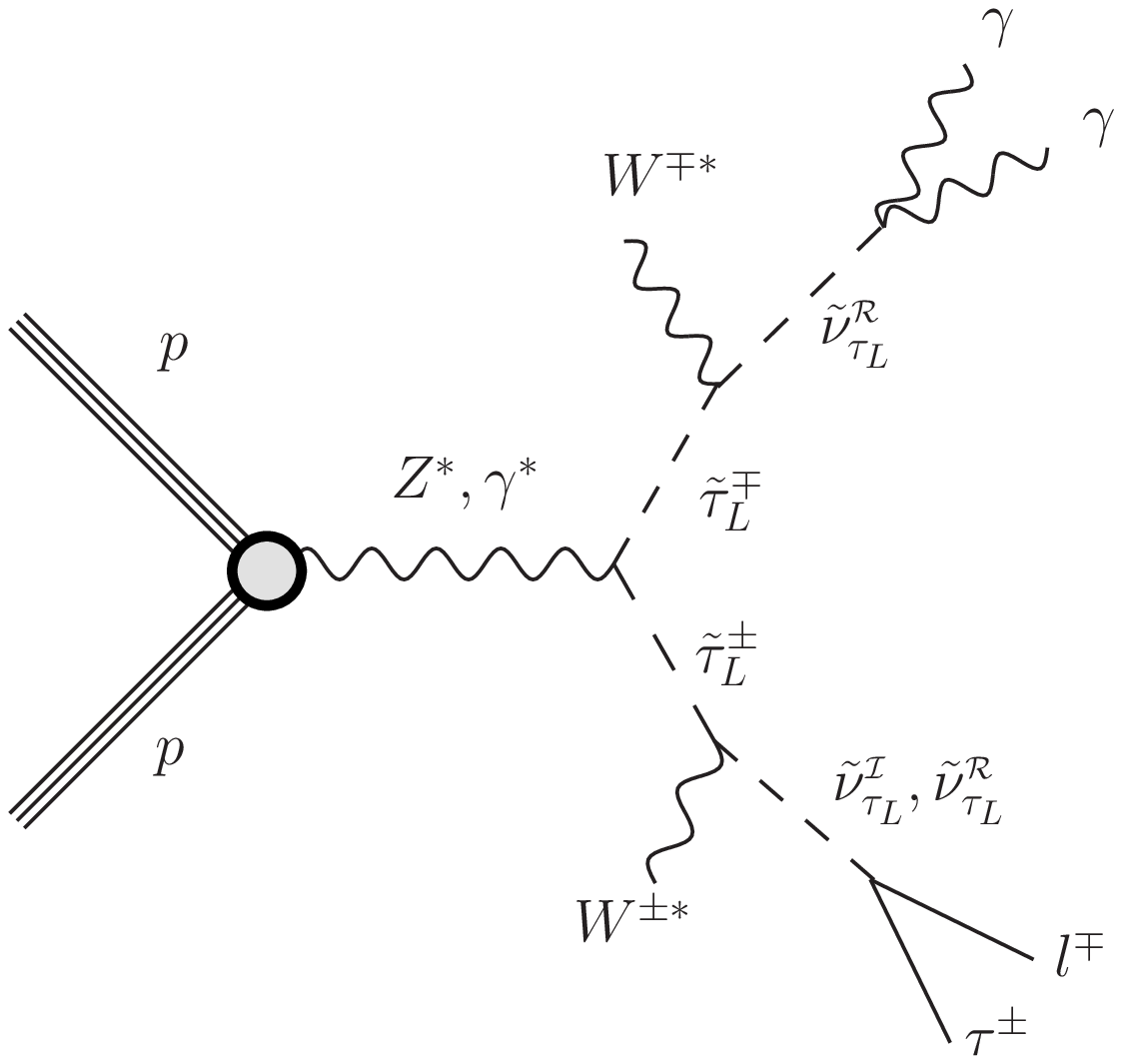}b)  
\includegraphics[scale=0.45]{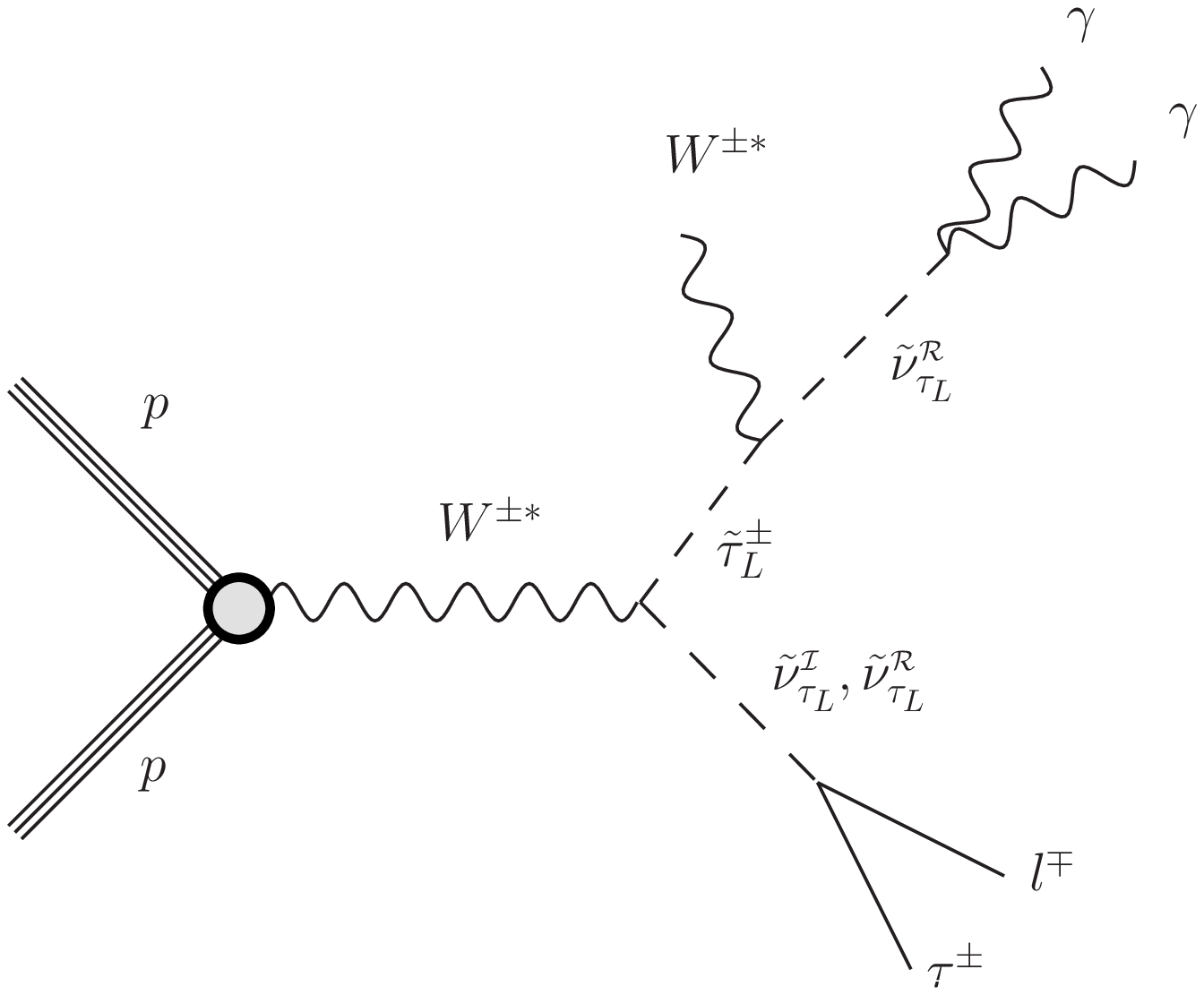}c)
\includegraphics[scale=0.45]{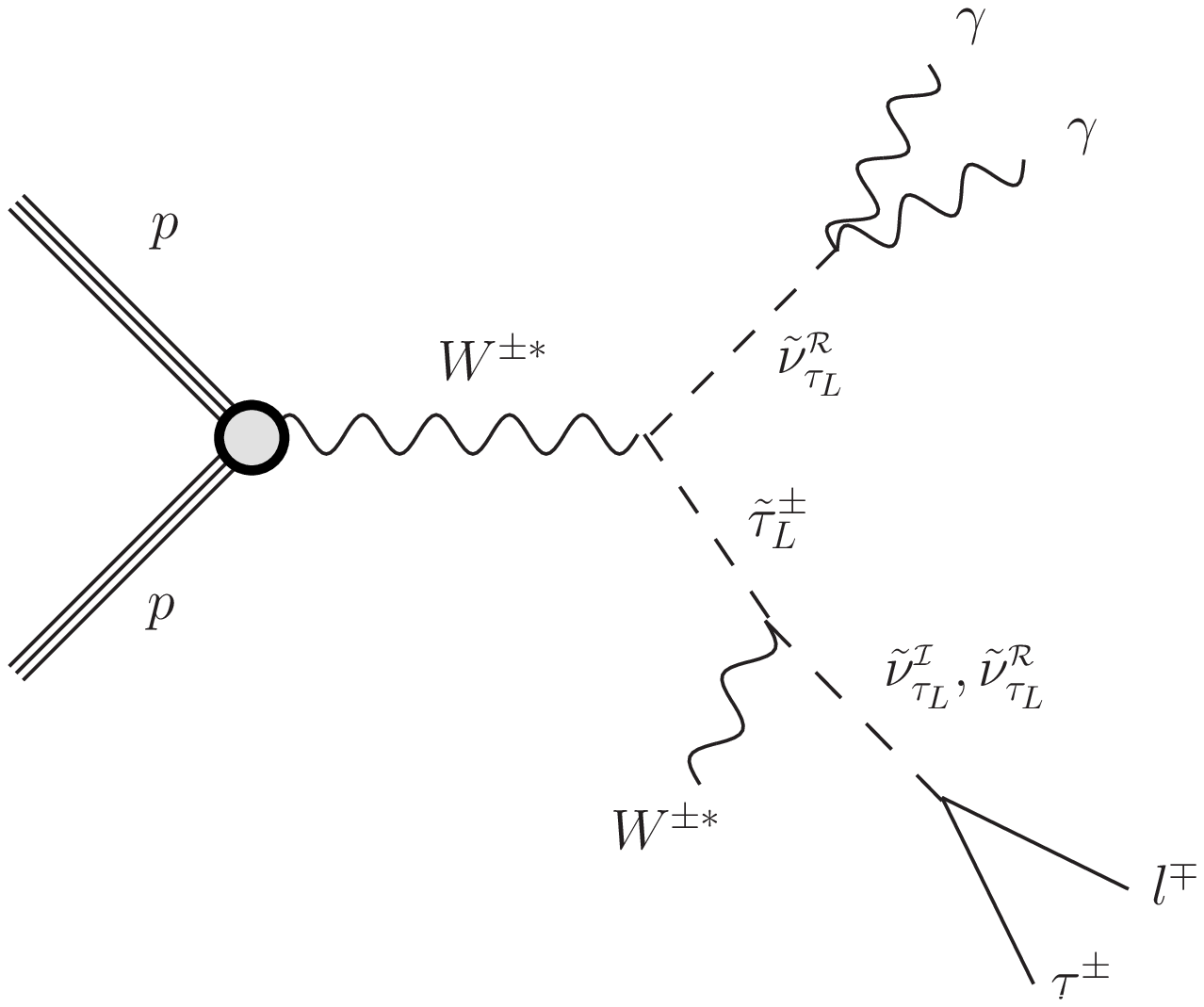}d) 
\caption{Dominant decay channels into diphoton plus 
leptons ($l=e,\mu,\tau$) 
% and diphoton state and neutrinos (a-d), and into multileptons (e-g),
from a pair production at the LHC
of scalar and pseudoscalar sneutrinos LSP of the third family $\widetilde\nu_{\tau L}$. Filled circles indicate effective
interactions.
% \R{I think that in figures b,c,d, in the channel to leptons, a scalar sneutrino could be present in addition to the pseudoscalar one.
% Similar to figures 8b,c}.
%In the figure, $l=e,\mu,\tau$. 
%Dominant production channels at the LHC for a pair of scalar and pseudoscalar 
% sneutrino LSPs, $\widetilde\nu_{\tau L}$, including the relevant decay signals
% into diphoton state and leptons, diphoton state and neutrinos \R{ERASE THE ELECTRON/MUON}, and multileptons, where $l=e,\mu,\tau$. 
%\bl{[PG Note: Just draw these figures in latex twice/thrice
 %   to make them bold looking].}
}
\label{fig:production2}
\end{figure} 
\section{Detection at the LHC}
\label{Section:detectionSneutrino-LSP}
%%%%%%%%%%%%%%%%%%%%%%%%%%%%%%%%%%%%%%%%%%%%%%%%%%%%%%%%%%%%%%%%
%%%%%%%%%%%%%%%%%%%%%%%%%%%%%%%%%%%%%%%%%%%%%%%%%%%%%%%%%%%%%%%%

%%%%%%%%%%%%%%%%%%%%%%%%%%%%%%%%%%%%%%%%%%%%%%%%%%%%%%%%%%%%%%%%
%%%%%%%%%%%%%%%%%%%%%%%%%%%%%%%%%%%%%%%%%%%%%%%%%%%%%%%%%%%%%%%%
% \subsection{Production}
% \label{Subsection:SneutrinosProduction}
%%%%%%%%%%%%%%%%%%%%%%%%%%%%%%%%%%%%%%%%%%%%%%%%%%%%%%%%%%%%%%%%
%%%%%%%%%%%%%%%%%%%%%%%%%%%%%%%%%%%%%%%%%%%%%%%%%%%%%%%%%%%%%%%%

%%%%%%%%%%%%%%%%%%%%%%%%%%%%%%%%%%%%%%%%%%%%%%%%%%%%%%%%%%%%%%%%
%%%%%%%%%%%%%%%%%%%%%%%%%%%%%%%%%%%%%%%%%%%%%%%%%%%%%%%%%%%%%%%%

\begin{figure}[t!]
\centering
\includegraphics[scale=0.45]{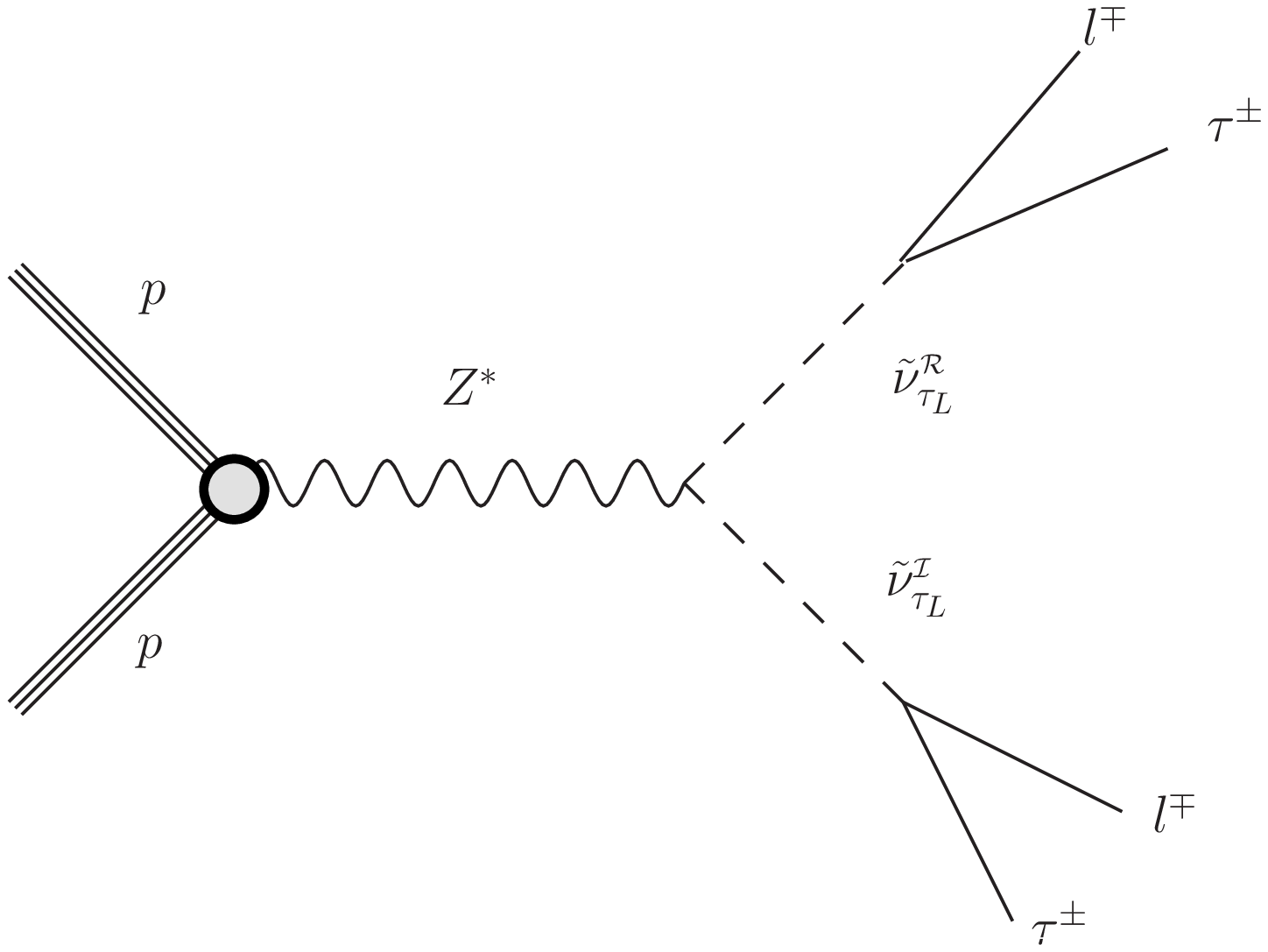}a) 
\includegraphics[scale=0.45]{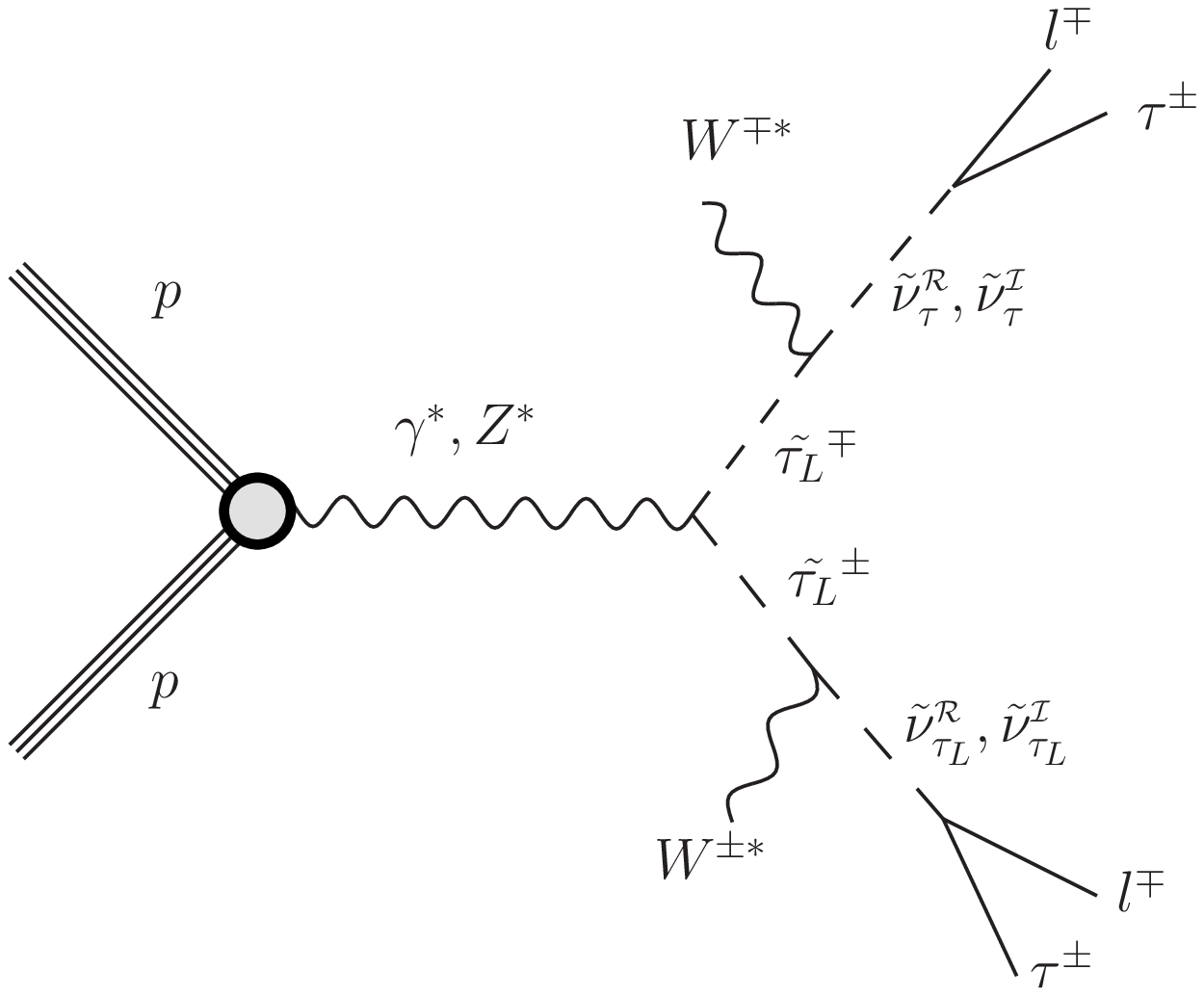}b) 
\includegraphics[scale=0.45]{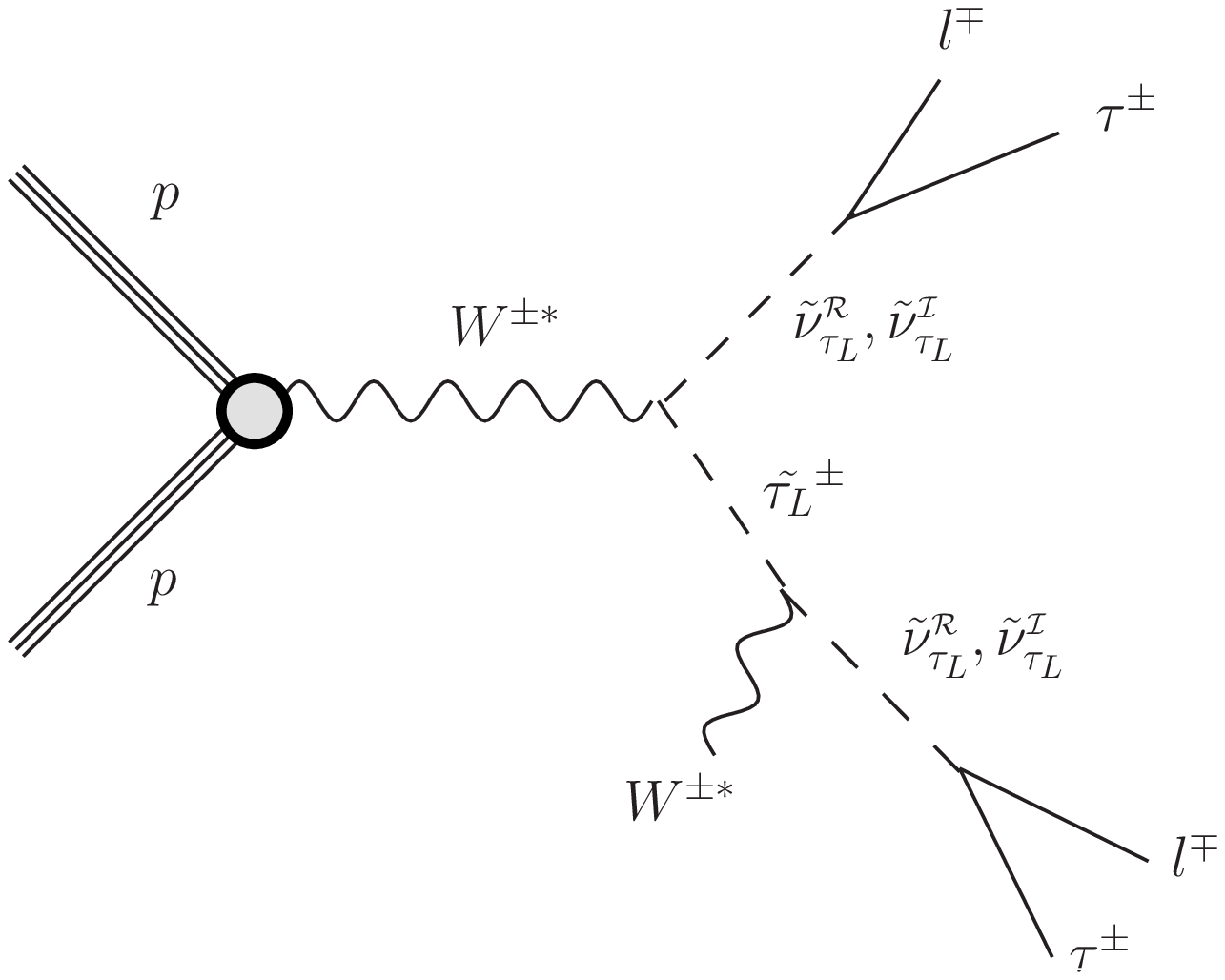}c)
\caption{Decay channels into multileptons ($l=e,\mu,\tau$)
from a pair production at the LHC
of scalar and pseudoscalar sneutrinos LSP of the third family $\widetilde\nu_{\tau L}$.
Filled circles indicate effective
interactions.
%, with $l=e,\mu,\tau$,.
%In the figure, $l=e,\mu,\tau$. 
%Dominant production channels at the LHC for a pair of scalar and pseudoscalar 
% sneutrino LSPs, $\widetilde\nu_{\tau L}$, including the relevant decay signals
% into diphoton state and leptons, diphoton state and neutrinos \R{ERASE THE ELECTRON/MUON}, and multileptons, where $l=e,\mu,\tau$. 
%\bl{[PG Note: Just draw these figures in latex twice/thrice
 %   to make them bold looking].}
}
\label{fig:production3}
\end{figure}

{The dominant pair production 
channels of sleptons at large hadron colliders were studied in Refs.~\cite{Dawson:1983fw,Eichten:1984eu,delAguila:1990yw,Baer:1993ew,Baer:1997nh,Bozzi:2004qq}.
In Figs.~\ref{fig:production}--\ref{fig:production3}, we show the detectable signals discussed above from a pair production at the LHC of sneutrinos LSP.
%the production 
%channels at the LHC for the case of the sneutrino LSP, which is the one interesting for our analysis.
The sparticles are denoted in the figures by their dominant composition.

Concerning the sneutrino production,
the direct one of e.g. Fig.~\ref{fig:production}a 
% (or Fig.~\ref{fig:production2}). 
%They 
occurs via a $Z$ channel giving rise to a pair of scalar and pseudoscalar left sneutrinos. 
As discussed in Section~\ref{Section:Sneutrino-LSP},  
these states
%scalar and 
%pseudoscalar left sneutrinos 
have 
essentially degenerate masses and therefore are co-LSPs.
%, thus when talking about left sneutrino LSP we refer to both indistinctly.
On the other hand, since the left slepton
in the same $SU(2)$ doublet as the left sneutrino, it becomes the NLSP, and its direct production and decay is
%as shown in Subsection~\ref{nlsp}, 
another important source
of the sneutrino LSP.
%is the direct production of the NLSP and its decays. 
In particular, pair production can be obtained  
through a $\gamma$ or a $Z$ decaying into 
%$\widetilde{l}^+ \widetilde{l}^-$
$\tilde{e}^+_{iL} \tilde{e}^-_{iL}$
(Fig.~\ref{fig:production}b),
with the sleptons dominantly decaying into 
a (scalar or pseudoscalar) sneutrino plus an off-shell $W^\pm$ producing a soft meson or a pair
of a lepton and a neutrino
($\tilde e^{\pm}_{iL}\to e^{\pm}_{j}\ \nu_{k}\ 
\widetilde{\nu}^{\mathcal{R},\mathcal{I}}_{lL}$), which are usually undetectable.
Besides, sneutrinos can be pair produced 
%and~\ref{fig:production2}b) 
through a $W^{\pm}$ decaying into %$\widetilde{l}^{\pm}  \widetilde{\nu}$ 
$\tilde{e}^{\pm}_{iL}  \widetilde{\nu}_{jL}$
(Figs.~\ref{fig:production}c-d), with the slepton decaying as before.}
%\ref{fig:production}d 
%and~\ref{fig:production2}c-d),
%,\ref{fig:production2}d
% with the slepton dominantly decaying into 
% a (scalar or pseudoscalar) sneutrino plus an off-shell $W^\pm$ (producing an undetectable lepton and a neutrino as follows:
% $\tilde e^{\pm}_{iL}\to e^{\pm}_{j}\ \nu_{k}\ 
% \widetilde{\nu}^{\mathcal{R},\mathcal{I}}_{lL}$).

Concerning the signals, we will study first diphoton plus MET arising from the production and decay of a pair of sneutrinos $\widetilde\nu^\mathcal{I}_{i_L}$ $\widetilde\nu^\mathcal{R}_{i_L}$ of any family, $i=e,\mu,\tau$, as
shown in Fig.~\ref{fig:production}.
Second, we will focus on other channels that can be produced via the $\widetilde{\nu}_{\tau_L}$ LSP, given the large value of the tau Yukawa coupling.
This is the case of diphoton plus leptons, and multileptons, as shown in Figs.~\ref{fig:production2} and~\ref{fig:production3}, respectively.
% where again sparticles are denoted by their dominant composition.
% Notice that in Figs.~\ref{fig:production2}c-d
% (Figs.~\ref{fig:production3}b-c),
% where left sleptons are involved, a pair of scalar sneutrinos (or pseudoscalar sneutrinos) has been included in the production, since they have a non-negligible contribution to the number of events.
% This is different from the case of the direct pair production in 
% Fig.~\ref{fig:production2}a
% (Fig.~\ref{fig:production3}a), where always a pair of scalar and pseudoscalar sneutrinos is produced.

These signatures for a sneutrino LSP are similar to the final states presented in several analysis of ATLAS and CMS. In particular, those including photons plus MET/leptons 
(see for example Refs.~\cite{Aaboud2016,Aad:2016tuk,Khachatryan:2017qgo,Khachatryan:2016iqn,Khachatryan:2016ojf,ATLAS-CONF-2016-096}). However, these searches are designed typically towards the production of colored sparticles in the context of $R_p$ conservation. Therefore, the analysis normally requires a large amount of MET, several energetic jets or a large effective mass. Thus, these searches are inefficient looking for events of direct pair production of the sneutrino in our scenario.

We have also confronted {all our BPs with LHC searches \cite{Aad:2016tuk,Aaboud:2016uro,Aaboud:2016tnv,Aaboud:2016zdn,Aad:2016qqk,Aad:2016eki,Aaboud:2016lwz,ATLAS-CONF-2015-082,ATLAS-CONF-2016-013,ATLAS-CONF-2016-050,ATLAS-CONF-2016-076,ATLAS-CONF-2016-096} using {\tt{CheckMATE}} {{2}} \cite{Dercks:2016npn,deFavereau:2013fsa,Cacciari:2011ma,Cacciari:2005hq,Cacciari:2008gp,Read:2002hq}, and LEP searches using 
{\tt{HiggsBounds-4.3.1}} \cite{Bechtle:2008jh,Bechtle:2011sb,Bechtle:2013gu,Bechtle:2013wla,Bechtle:2015pma}.}
In the case of the multilepton signal, there exist generic searches for production of three or more leptons, which include also signal regions with a low missing transverse momentum and total transverse energy (see Refs.~\cite{Chatrchyan:2012mea,Chatrchyan:2014aea}). 
In these works, by lepton is meant $e$, $\mu$ or hadronically 
decaying $\tau$ ($\tau_{\mathrm{h}}$) candidate.
These searches are close to be sensitive to our signal, and an updated analysis with current data could put constraints on the sneutrino LSP scenario. 
Let us finally remark that past collider searches in the context of trilinear $\rpv$ 
couplings \cite{Aaltonen:2010fv,Abazov:2010km,
Achard:2001ek,Heister:2002kq,Heister:2002jc,
Abbiendi:2003rn,Abdallah:2003xc,CMS:2015neg,Aad:2014aqa, Khachatryan:2015dcf,ATLAS:2015nsi,
Aad:2015pfa,Khachatryan:2016ovq} 
are ineffectual for our scenario.

%%%%%%%%%%%%%%%%%%%%%%%%%%%%%%%%%%%%%%%%%%%%%%%%%%%%%%%%%%%%%%%%%
%%%%%%%%%%%%%%%%%%%%%%%%%%%%%%%%%%%%%%%%%%%%%%%%%%%%%%%%%%%%%%%%
\begin{table}[t!]
\centering
\caption{Madgraph cuts. $P_T$ is given in GeV.}
\label{table:1.6}
 {\scriptsize
\begin{tabular}{|p{1.95cm}|p{1.95cm}|p{1.95cm}|p{1.95cm}|p{1.95cm}|p{1.95cm}|}
\hline $P_T$ for $jets$ & $|\eta|$  for $jets$ &  $P_T$ for $e$, $\mu$
& $|\eta|$ for $e$, $\mu$ & $P_T$ for $\gamma$& $|\eta|$ for $\gamma$\\ \hline
$>20$& $<5$ & $>10$  & $<2.5$ & $>10$ & $<2.5$ \\
 \hline
 \end{tabular}} 
\end{table} 
%%%%%%%%%%%%%%%%%%%%%%%%%%%%%%%%%%%%%%%%%%%%%%%%%%%%%%%%%%%%%%%%%
%%%%%%%%%%%%%%%%%%%%%%%%%%%%%%%%%%%%%%%%%%%%%%%%%%%%%%%%%%%%%%%%

%%%%%%%%%%%%%%%%%%%%%%%%%%%%%%%%%%%%%%%%%%%%%%%%%%%%%%%%%%%%%%%%%
%%%%%%%%%%%%%%%%%%%%%%%%%%%%%%%%%%%%%%%%%%%%%%%%%%%%%%%%%%%%%%%%
\begin{table}[t!]
\centering
\caption{{\tt PGS} configuration. {{ECAL and HCAL stand for 
Electromagnetic Calorimeter and Hadronic Calorimeter, respectively.}}}
\label{table:1.7}
{\scriptsize
 \begin{tabular}{|p{4.73cm}|p{1.6cm}|p{4.73cm}|p{1.6cm}|}
 \hline $\eta$ cells in calorimeter & 81 &$\phi$ cells in calorimeter&63\\
 \hline $\eta$ width of calorimeter cells&$0.1$&$\phi$ width of calorimeter cells& {0.09973}\\
 \hline{{ECAL}} resolution&0.01&{{ECAL}} resolution $\times \sqrt E$ (GeV$^{1/2}$)&0.1\\
 \hline {{HCAL}} resolution $\times \sqrt E$ (GeV$^{1/2}$)& 0.8 &MET resolution&0.2\\
 \hline Calorimeter cell edge crack fraction& 0.00 &Jet finding algorithm& anti-$k_t$ \cite{Cacciari:2008gp}\\
 \hline Calorimeter trigger cluster&3.0 &Calorimeter trigger cluster &0.5\\
  finding seed threshold  &&finding shoulder threshold  &\\
 \hline Calorimeter $k_t$ cluster finder&0.7&Outer radius of tracker (m)&1.0\\
  one size ($\Delta$R)&&&\\
 \hline Magnetic field (T) & 2.0 &Sagitta resolution (m)& {\bf{$5\times 10^{-6}$}}\\
 \hline Track finding efficiency & 0.98 &Minimum track $P_T$ (GeV/c)&0.30\\
 \hline Tracking $\eta$ coverage & 2.5 &e/gamma $\eta$ coverage&3.0\\
 \hline Muon $\eta$ coverage & 2.4 &Tau $\eta$ coverage&2.0\\
 \hline 
 \end{tabular}}
\end{table} 
%%%%%%%%%%%%%%%%%%%%%%%%%%%%%%%%%%%%%%%%%%%%%%%%%%%%%%%%%%%%%%%%%
%%%%%%%%%%%%%%%%%%%%%%%%%%%%%%%%%%%%%%%%%%%%%%%%%%%%%%%%%%%%%%%%

%Let us finally discuss 
The strategy that we will follow for the analyses of the sneutrino signals in the $\mn$ is the following. Ten thousand  
%Now, in order to carry out the analyses of these signals, 10000 
events are generated for each case 
with {\tt MadGraph5\_aMC@NLO}~\cite{Alwall:2014hca} 
at leading order (LO) of perturbative QCD simulating 
the production of the described process. 
{We include the next-to-leading order (NLO)~\cite{Baer:1997nh} and next-to-leading logarithmic accuracy (NLL) ~\cite{Fuks:2013lya} results using a $K$-factor of about 1.2.}
The hard process simulation is then passed for decay and 
hadronization to {\tt PYTHIA}~\cite{Sjostrand:2006za}. The output is passed through a naive and fast detector simulation 
({\tt PGS})~\cite{pgs}. The standard card for {\tt MadGraph5\_aMC@NLO} is used, which includes the cuts presented 
in Table~\ref{table:1.6}. {\tt PYTHIA}
is executed with initial state radiation (ISR), final state radiation (FSR) 
and multiple interactions switched on. Besides, {\tt PYTHIA} will consider the $\tau$ 
lepton as stable to make it decay with the {\tt TAUOLA} \cite{Jadach:1990mz,Jadach:1993hs} 
routine within 
{\tt PGS}. The package {\tt PGS} is finally 
executed using a card designed for ATLAS, as shown in Table~\ref{table:1.7}.
The output of {\tt PGS} is passed through some selection criteria to avoid overlapping and to
discard the events outside the detector coverage according to Ref.~\cite{Aad:2014iza}. That is, first 
candidate events should pass the requirements of Table~\ref{table:1.8}.
After the previous process, overlapping objects are removed applying the following requirements in this precise order:  
First, if two electrons as candidates are identified within $\Delta R =0.05$ of each other, 
the one with lower transverse momentum ($P_T$) is discarded.
Here $\Delta R$ is defined as $\sqrt{(\Delta \Phi)^2+(\Delta \eta)^2}$, where
$\Delta \Phi$ is the difference in involved azimuthal angles while
$\Delta \eta$ is the difference of concerned pseudo-rapidities. 
Then if an electron and a jet candidates are within $\Delta R=0.2$ 
of each other, the jet is discarded. All remaining leptons are required to be separated by more 
than $\Delta R=0.4$ from the closest remaining jet. Whenever an electron and a muon candidates 
overlap within $\Delta R=0.01$, both are discarded. Also, if two muons are separated by less than 
$\Delta R=0.05$, both are removed.
$\tau$'s as candidates are required to be separated by more than $\Delta R=0.2$ from the closest 
$e$ or $\mu$; otherwise they are discarded. Finally, photons are required to be separated by $\Delta R=0.4$ 
from any reconstructed jet and $\Delta R=0.01$ from any $e$ \cite{Aad:2015hea}.
A similar process, with a higher number of events when required by precision, is implemented  
to generate background samples at {NLO}.

%%%%%%%%%%%%%%%%%%%%%%%%%%%%%%%%%%%%%%%%%%%%%%%%%%%%%%%%%%%%%%%%%
%%%%%%%%%%%%%%%%%%%%%%%%%%%%%%%%%%%%%%%%%%%%%%%%%%%%%%%%%%%%%%%%
%%%%%%%%%%%%%%%%%%%%%%%%%%%%%%%%%%%%%%%%%%%%%%%%%%%%%%%%%%%%%%%%
\begin{table}%[t!]
\centering
\caption{Event filtering. $P_T$ is given in GeV.}
\label{table:1.8}
{\scriptsize
 \begin{tabular}{|c|c|c|c|c|c|c|c|c|c|} \hline
 $P_T$ & $|\eta|$ & $P_T$ & $|\eta|$
 & $P_T$ & $|\eta|$  & $P_T$  & $|\eta|$ 
 & $P_T$  & $|\eta|$  \\ 
 for $jets$ & for $jets$ & for $e$ & for $e$
 & for $\mu$ & for $\mu$ & for $\tau_{\mathrm{h}}$  & for $\tau_{\mathrm{h}}$
 & for $\gamma$ & for $\gamma$ \\ \hline 
 $> 30$ & $<4.5$ & $>15$ & $<2.47$   & $>10$ 
 & $<2.5$ & $> 10$  & $<2.5$ & $> 25$  & $2.37$ \\
  &&& \& outside &&& &&& \& outside\\
 &&&$1.37-1.52$&&&&&& $1.37-1.52$\\
 \hline
 \end{tabular}}
\end{table}

%%%%%%%%%%%%%%%%%%%%%%%%%%%%%%%%%%%%%%%%%%%%%%%%%%%%%%%%%%%%%%%%%%%%%%%%%%%%%%%%%%5
%%%%%%%%%%%%%%%%%%%%%%%%%%%%%%%%%%%%%%%%%%%%%%%%%%%%%%%%%%%%%%%%%
%%%%%%%%%%%%%%%%%%%%%%%%%%%%%%%%%%%%%%%%%%%%%%%%%%%%%%%%%%%%%%%%
\vspace{0.25cm}

\noindent
%\subsection
{\bf Diphoton plus MET}
%%%%%%%%%%%%%%%%%%%%%%%%%%%%%%%%%%%%%%%%%%%%%%%%%%%%%%%%%%%%%%%%%
%%%%%%%%%%%%%%%%%%%%%%%%%%%%%%%%%%%%%%%%%%%%%%%%%%%%%%%%%%%%%%%%

\noindent
The pair production of left sneutrinos can generate one scalar and one pseudoscalar, 
as shown in Fig.~\ref{fig:production}.
This opens the 
possibility of the pseudoscalar sneutrino 
%of any generation
%($\widetilde{\nu}^\mathcal{I}_{eL}$ or $\widetilde{\nu}^\mathcal{I}_{\mu L}$) 
decaying  into neutrinos, i.e., producing
MET, 
and the scalar sneutrino decaying into two photons in a way not very different from the Higgs.
% For an early work analyzing $\tilde\nu_L\rightarrow \gamma\gamma$ 
% in the context of trilinear $\rpv$, see 
% Ref.~\cite{BarShalom:1998xz} where a negligible BR$\sim 10^{-6}$ was obtained.

%%%%%%%%%%%%%%%%%%%%%%%%%%%%%%%%%%%%%%%%%%%%%%%%%%%%%%%%%%%%%%%%%

%%   TABLES  125 e mu
\begin{table}[t!]
 \centering
 \caption{Analysis of the signal with diphoton plus MET from production and decay of a pair of 
sneutrino co-LSPs of the type $\widetilde\nu_{eL}$ or $\widetilde\nu_{\mu L}$, corresponding to the
BP in Table \ref{table:musneu-125}. Production cross sections are shown in fb in 
the first box for a $\widetilde\nu_{eL}/\widetilde\nu_{\mu L}$ mass of 125.4 GeV. 
The number of events  of the signal and background is shown in 
the second box, together with the effect of a set of cuts, assuming 13 TeV center-of-mass energy with $\mathcal{L}=300$ fb$^{-1}$. 
Energies, momenta and invariant mass are given in GeV.}
 \label{table2:musneu-125}
 {\scriptsize
%\begin{tabular}{|p{9.5cm}|p{2.5cm}|} \hline
  \begin{tabularx}{0.982\linewidth}{|X|C|}\hline
$\sigma(pp\to Z^* \to h_2A^0_2)$ &\qquad $107.08$ 
 \\ \hline
$\sigma(pp \to \gamma^*,Z^* \to H_2^+ H_2^- \to h_2 A_2^0 + W^+_{\text{soft}} W^-_{\text{soft}})$ &\qquad $21.89$ 
 \\ \hline
$\sigma(pp \to W^{\pm^*} \to H_2^{\pm}\ h_2/A_2^0 \to h_2 A_2^0 + W^{\pm}_{\text{soft}})$ & 
\qquad$142.8$ \\ \hline
$\sigma(pp\to Z^* \to h_3A^0_3)$ & \qquad$106.536$ \\ \hline
$\sigma(pp \to \gamma^*,Z^* \to H_3^+ H_3^- \to h_3 A_3^0 + W^+_{\text{soft}}  W^-_{\text{soft}})$ &
\qquad$20.12$  \\ \hline
$\sigma(pp \to W^{\pm^*} \to H_3^{\pm}\ h_3/A_3^0 \to h_3 A_3^0 + W^{\pm}_{\text{soft}})$&\qquad142.4 \\ \hline
\end{tabularx}
\begin{tabular}{|c|c|c|c|c|c|c|c|c|}\hline
 Dataset&$\mathrm{N_{ev}}$&$
{E}_{T}^{\mathrm{miss}}$&$P_{T1}^{\gamma_1}$&$P_{T2}^{\gamma_2}$&$N_\gamma$=2&$N_l=0$&$\Delta R$
 &$M_{\gamma\gamma}\in$ \\ 
 &&  $> 200$& $> 100$& $> 50$&&&$<1.5$&$[115,135]$\\ \hline
 Signal&449.45$\pm 0.02$&103.6$\pm 0.8$&80.3$\pm 0.7$&41.0$\pm 0.5$&41.0$\pm 0.5$&36.4$\pm 0.5$&35.9$\pm 0.5$
 &34.1$\pm 0.5$\\ \hline
 2$jets$&$10^{7}$&0&0&0&0&0&0&0\\ %\hline
+I/FSR&&&&&&&&\\ \hline
 $jet$&$10^{7}$&0&0&0&0&0&0&0\\ %\hline
 +I/FSR&&&&&&&&\\ \hline
 H (ggF)&5424&0&0&0&0&0&0&0\\ \hline
 $Z$+H&120.8$\pm$0.4&6.9$\pm$0.3&5.9$\pm$0.3&3.3$\pm$0.2&3.3$\pm$0.2&3.2$\pm$0.2&3.1$\pm$0.2&2.9$\pm$0.2\\ \hline
 $Z$+ISR&11310$\pm$40&104$\pm$11&97$\pm$10&33$\pm$6&33$\pm$6&33$\pm$6&8$\pm$3&$1\pm$1\\ \hline
 $W$+FSR& $2.14\times 10^5$&60$\pm$9&57$\pm$9&13$\pm$4&13$\pm$4&6$\pm$3&1.4$\pm$1.4&0\\ 
 %\hline
  &$\pm 76$&&&&&&&\\ \hline
 $\frac{S}{\sqrt{B}}$&---&7.9$\pm$0.5&6.3$\pm$0.4&5.9$\pm$0.7&5.9$\pm$0.7&5.6$\pm$0.7&10$\pm$2
 &17$\pm3$\\
 \hline
 \end{tabular}}
\end{table}

%%%%%%%%%%%%%%%%%%%%%%%%%%%%%%%%%%%%%%%%%%%%%%%%%%%%%%%%%%%%%

In what follows, we will discuss first the case of sneutrinos co-LSPs of the first two families
($\widetilde{\nu}_{eL}$, $\widetilde{\nu}_{\mu L}$) with masses of about 125 GeV
%$ m_{\widetilde{\nu}} \approx 125 \gev$
as representative in order to search for a signal. This is because the sensible range of masses turns out to be 
%about $118-132$ GeV, as discussed above. 
%  arising from the production and decay of a pair of sneutrinos $\widetilde\nu^\mathcal{I}_{iL}$ $\widetilde\nu^\mathcal{R}_{iL}$ of any family, $i=e,\mu,\tau$, as
% shown in Fig.~\ref{fig:production}. 
%The sensible range of masses for this search of {\it diphoton plus MET} is
\bea
118 \lesssim m_{\widetilde{\nu}_{iL}} \lesssim 132\ \gev \ ,
\label{rangesneutrinomuon}
\eea 
%As discussed at the end of Subsection~\ref{decaymodes},
%for the first two families of sneutrinos, 
%$\widetilde{\nu}_{eL}$ or $\widetilde{\nu}_{\mu L}$,
%this mass range is necessary 
in order to
treat the sneutrinos as promptly decaying particles with a decay length
$\lsim$ 0.1 mm. 
For the case of the tau sneutrino, $\widetilde{\nu}_{\tau_L}$, where decay lengths of this order can be obtained for masses $\gsim$ 95 GeV, the above mass range is still valid because outside it the number of events turns out to be too small, as we will discuss below.

The case of $\widetilde{\nu}_{e_L}$ and $\widetilde{\nu}_{\mu_L}$ co-LSPs is 
shown in Table~\ref{table2:musneu-125}.
% A left sneutrino of the first 
% two families ($\widetilde{\nu}_{eL}$ or $\widetilde{\nu}_{\mu L}$) generates a signal consisting of two photons plus MET 
% at the LHC.
% As discussed in Section~\ref{decaymodes}, this is because the effective interaction of the pseudoscalars proportional to the value of the Yukawa couplings in Eq.~(\ref{cplepton})
% (see also Fig.~\ref{fig:eff-to-leptons-higgsino})
% is clearly sub-dominant with respect to the one proportional to the value of the gauge couplings in Eq.~(\ref{cpneutrinos}) (see also Fig.~\ref{fig:eff-to-neutrinos}).
%In what follows, we will discuss the case of a sneutrino with mass of about 125 GeV
%$ m_{\widetilde{\nu}} \approx 125 \gev$
%as representative in order to search for a signal, since the sensible range of masses is about $118-132$ GeV, as discussed above. 
%This case is shown in Table~\ref{table2:musneu-125}.
% with the $\widetilde{\nu}_{e}$ and the  $\widetilde{\nu}_{\mu}$ almost degenerated in mass.
The cross sections for the pair production of sneutrinos calculated by {\tt MadGraph5\_aMC@NLO} 2.3.2.2 at LO for 13 TeV center-of-mass energy, {including a $K$-factor of 1.2 for the NLO results}, are shown in the first box of that Table.
%~\ref{table:musneu-125}.
The first, second and third rows of that box correspond to the diagrams in Figs.~\ref{fig:production}a,~\ref{fig:production}b and~\ref{fig:production}c-d, respectively. 
%As can be seen, the production of sneutrinos through a $Z$ decaying into a pair of
%left sleptons is very small, with cross sections of the order of 3 fb.
%However, for the direct production via a $Z$ channel, the values increase to about 88 fb,
%and even larger values of about 120 fb 
%are obtained when the process is mediated by a $W^\pm$ decaying into a slepton and a sneutrino.
Taking into account these values for the cross sections, the BRs
of the corresponding Table~\ref{table:musneu-125}, and using an integrated luminosity of $\mathcal{L}=300$ fb$^{-1}$, we obtain a signal with about 449 events.
%374.54 $\pm$ 0.02.
Although this BP suffers from a significant SM background mainly due to the Z+$H$ channel which decays in a similar way,
%and the fact that the scalar left sneutrino is decaying effectively in a prompt way (WE WANT TO SAY THAT IF IT WERE DISPLACED WE WOULD HAVE LESS BACKGROUND?), 
we found that
the number 
of expected events for the signal is still sufficient to give a significant evidence. 
% The final results are properly 
% rescaled to the appropriate number of events, corresponding to 
% $\mathcal{L}=300$ fb$^{-1}$. 
The effect of a set of cuts on 
missing transverse energy
${E}_{T}^{\mathrm{miss}}$,
%${E}_{T} \hspace{-0.38cm}\slash\hspace{0.2cm}$,
% ${E}_T$, 
$P_T$ for the leading and sub-leading photons, a lepton veto, a maximum angular separation of photons, 
%of $\Delta R$, 
and a selection cut
on the invariant mass of the diphoton system, is summarized 
%for 13 TeV 
in the second box of Table~\ref{table2:musneu-125}.
%~\ref{table:musneu-125}.
As a final result of the analysis, we obtain {34.1 $\pm 0.5$ 
events with a significant evidence of
$\frac{S}{\sqrt B}= 17 \pm 3$}.
For $\mathcal{L}=100$ fb$^{-1}$ to be reached in Run 2 we just have to rescale the number of events by a factor $1/3$ and correspondingly  the significance by $1/\sqrt 3$.
For completeness, we show in Table~\ref{table:musneu2-125} the results for this BP with 14 TeV center-of-mass energy.

%%%%%%%%%%%%%%%%%%%%%%%%%%%%%%%%%%%%%%%%%%%%%

%%%14 TeV
\begin{table}[t!]
\centering
\caption{The same as in Table \ref{table2:musneu-125} but showing the production cross sections and the event 
sample generated with 14 TeV center-of-mass energy.}
 \label{table:musneu2-125}
 {\scriptsize
 %\begin{tabular}{|p{9.5cm}|p{2.5cm}|} \hline
  \begin{tabularx}{0.98\linewidth}{|X|C|} \hline
$\sigma(pp\to Z^* \to h_2A^0_2)$ & \qquad$119.95$ 
 \\ \hline
$\sigma(pp \to \gamma^*,Z^* \to H_2^+ H_2^- \to h_2 A_2^0 + W^+_{\text{soft}} W^-_{\text{soft}})$ & \qquad$25.43$ 
\\ \hline
$\sigma(pp \to W^{\pm^*} \to H_2^{\pm}\ h_2/A_2^0 \to h_2 A_2^0 + W^{\pm}_{\text{soft}})$ &\qquad $160.7$  
\\ \hline
 $\sigma(pp\to Z^* \to h_3A^0_3)$ & \qquad $119.83$ \\ \hline
 $\sigma(pp \to \gamma^*,Z^* \to H_3^+ H_3^- \to h_3 A_3^0 + W^+_{\text{soft}}  W^-_{\text{soft}})$ &\qquad
$23.35$ \\ \hline
$\sigma(pp \to W^{\pm^*} \to H_3^{\pm}\ h_3/A_3^0 \to h_3 A_3^0 + W^{\pm}_{\text{soft}})$&\qquad 158.9 \\ \hline
\end{tabularx} 
 \begin{tabular}{|c|c|c|c|c|c|c|c|c|}\hline
 Dataset&$\mathrm{N_{ev}}$&$
{E}_{T}^{\mathrm{miss}}$&$P_{T1}^{\gamma_1}$&$P_{T2}^{\gamma_2}$&$N_\gamma$=2&$N_l=0$&$\Delta R$
 &$M_{\gamma\gamma}\in$ \\ 
 && $>200$&$>100$ & $>50$&&&$<1.5$&$[115,135]$\\ \hline
  Signal&503.40$\pm$0.02&116.3$\pm$0.9&95.0$\pm$0.8&50.8$\pm$0.6&50.5$\pm$0.6&43.8$\pm$0.6&43.3$\pm$0.6
 &38.8$\pm$0.6\\  \hline
 2$jets$&$10^{7}$&0&0&0&0&0&0&0\\ %\hline
 +I/FSR&&&&&&&&\\ \hline
 $jet$&$10^{7}$&0&0&0&0&0&0&0\\ %\hline
 +I/FSR&&&&&&&&\\ \hline
 H (ggF)&6104&0&0&0&0&0&0&0\\ \hline
 $Z$+H&133.76&7.9$\pm$0.3&6.7$\pm$0.3&3.5$\pm$0.2&3.5$\pm$0.2&3.4$\pm$0.2&3.3$\pm$0.2&3.0$\pm$0.2\\ %\hline
 &$\pm0.01$&&&&&&&\\ \hline
 $Z$+ISR&9284.91&90$\pm$3&82$\pm$3&26$\pm$2&26$\pm$2&25$\pm$2&8$\pm$1&$1.2\pm$0.3\\ %\hline
 &$\pm0.09$&&&&&&&\\ \hline
 $W$+FSR&23708$\pm2$&57$\pm$9&54$\pm$9&5$\pm$3&5$\pm$3&1.6$\pm1.61$&0&0\\  \hline
 $\frac{S}{\sqrt{B}}$&---&9.3$\pm$0.5&8.0$\pm$0.4&8.7$\pm$0.7&8.7$\pm$0.7&8.1$\pm$0.6&13.0$\pm$0.8
 &19$\pm1$\\
 \hline
 \end{tabular}}
\end{table}

%%%%%%%%%%%%%%%%%%%%%%%%%%%%%%%%%%%%%%%%%%%%%%%%%%%%%%%%%%%%%%%%%%%%

Concerning the case of the $\widetilde{\nu}_{\tau_L}$ LSP
of a similar mass,
we can see in the fourth box of
Table~\ref{table:tau-sneu} that it has a significant BR to neutrinos.
After a straightforward computation, we obtain a number of events {of $7.5 \pm 0.2$}. The background is the same as in Table~\ref{table2:musneu-125}, and therefore {we obtain $\frac{S}{\sqrt B}\sim 3.8 \pm 0.6$}. This BP can also give rise to a signal with diphoton plus leptons, to be analyzed subsequently, implying that a tau left sneutrino LSP could be distinguished from electron and muon left sneutrinos co-LSPs.
For the other $\widetilde{\nu}_{\tau L}$ masses studied in Tables~\ref{table:tau-sneu-95} 
and~\ref{table:tau-sneu-145}, although the BRs to neutrinos are still significant, the number of events of the signal diphoton plus MET turns out to be too small to be detected.

The case with all sneutrinos degenerate in mass would give rise to a superposition of the signals discussed so far. For instance, if the three families of sneutrinos have a mass of 126 GeV, the number of events expected for the signal diphoton plus MET will be the sum of both contributions discussed above, that {is $41.6 \pm 0.5$ events with a significance 
of $\frac{S}{\sqrt B}=21 \pm 1$}. In addition, the signal with diphoton plus leptons, specific for the $\widetilde{\nu}_{\tau L}$, would also be present.

\vspace{0.25cm}

\noindent
{\bf Diphoton plus leptons}
%\label{Subsection:photon-plus-leptons}
%%%%%%%%%%%%%%%%%%%%%%%%%%%%%%%%%%%%%%%%%%%%%%%%%%%%%%%%%%%%%%%%%
%%%%%%%%%%%%%%%%%%%%%%%%%%%%%%%%%%%%%%%%%%%%%%%%%%%%%%%%%%%%%%%%

% On the other hand, for a $\widetilde{\nu}_{\tau L}$ LSP, 
% since the tau Yukawa coupling is large, 
% significant BRs of scalar and pseudoscalar sneutrinos to leptons are feasible.
% Thus in this case we will search, in addition to the above signal with diphoton plus MET, for signals given by diphoton plus leptons, and by multileptons, as shown in Figs.~\ref{fig:production2} and~\ref{fig:production3}, where again sparticles are denoted by their dominant composition.
% Notice that in Figs.~\ref{fig:production2}c-d
% (Figs.~\ref{fig:production3}b-c),
% where left sleptons are involved, a pair of scalar sneutrinos (or pseudoscalar sneutrinos) has been included in the production, since they have a non-negligible contribution to the number of events.
% This is different from the case of the direct pair production in 
% Fig.~\ref{fig:production2}a
% (Fig.~\ref{fig:production3}a), where always a pair of scalar and pseudoscalar sneutrinos is produced.

\noindent
For the case of the left sneutrino LSP dominated by the tau composition,
$\widetilde{\nu}_{\tau L}$, 
%the BP 
%given in Table~\ref{table:tau-sneu}, where 
another expected signal is diphoton plus leptons,
%$2\gamma+\tau^\pm+l^\mp$, with $l=e,\mu,\tau$, 
as shown in 
Fig~\ref{fig:production2}.
%Which, if not ruled out, could be an interesting signal. 
For this signal 
%with {\it diphoton plus leptons}
the adequate range of masses turns out ot be
\bea
95 \lesssim m_{\widetilde{\nu}_{\tau_L}} \lesssim 145\ \gev \ .
\label{rangesneutrinotau}
\eea 
For the lower bound, notice that the selection cuts used to discriminate the decay of the sneutrino from the background require energetic photons and a large amount of missing energy. Therefore, a sneutrino with a small mass would lead to a small boost of the final photons and neutrinos. Thus reducing the mass of the sneutrino reduces the number of events in the signal region, although the cross section increases.
Moreover, 
%as discussed in Subsection~\ref{decaymodes}, 
when
the separation between the masses of the scalar left sneutrino and the SM Higgs is increased, the BR to diphoton is decreased. Altogether, the number of events drops fast when the mass of the left sneutrino is below 95 GeV. 
Actually, we already mentioned that about this mass is also the limit where the LSP cannot be treated as a promptly decaying particle.
On the other hand,
the decrease of the cross section for large sneutrino masses, and therefore of the number of events, gives rise to the upper bound of 145 GeV.

%%%%%%%%%%%%%%%%%%%%%%%%%%%%%%%%%%%%%%%%%%%%%%%%%%%%%%%%%%%%%%%
%aqui van las nuevas
%125 tau
\begin{table}[t!]
\centering
 \caption{ Analysis of the signal with two photons plus leptons from production and decay of a pair of 
$\widetilde \nu_{\tau L}$ LSPs, corresponding to the BP in Table \ref{table:tau-sneu}.
Production cross sections are shown in fb in the first box for a $\widetilde\nu_{\tau L}$ mass of 126.4 GeV.
The number of events of the signal and background is shown in the second box,
together with the effect of a set of cuts, assuming 13 TeV of energy with $\mathcal{L}=300$ fb$^{-1}$.
Momenta and invariant mass are given in GeV.}
 \label{table2:tau-sneu}
{\scriptsize
  \begin{tabularx}{0.99\linewidth}{|X|C|}\hline
 $\sigma(pp\to Z^* \to h_2A^0_2)$&103.58 \\ \hline
 $\sigma(pp \to \gamma^*,Z^* \to H_2^+H_2^- \to h_2A^0_2\ + W^+_{\text{soft}} W^-_{\text{soft}})$ &21.91 \\ \hline 
  $\sigma(pp \to W^{\pm^*} \to H_2^{\pm}h_2/A^0_2 \to h_2A^0_2\ + W^{\pm}_{\text{soft}})$ & $138.36$  \\ \hline
  $\sigma(pp \to W^{\pm^*} \to H_2^{\pm}h_2 \to h_2h_2\ + W^{\pm}_{\text{soft}})$ & $69.18$  \\ \hline
\end{tabularx}
}{\scriptsize
 \begin{tabular}{|c|c|c|c|c|c|c|c|}\hline
 Dataset&$\mathrm{N_{ev}}$&$P_{T1}^{\gamma_1}$&$P_{T2}^{\gamma_2}$&$N_\gamma$=2&$N_{\tau_{had}}= 1\, \&$
 &$\Delta R<1.5$&$M_{\gamma\gamma}\in$\\ 
  && $>100$ & $>50$&& $N_{e,\mu,\tau_{had}}>1$&&$[115,135]$\\ \hline  
 Signal&128.136$\pm$0.007&67.8$\pm$0.4&25.4$\pm0.3$&25.4$\pm$0.3&5.9$\pm$0.2&4.9$\pm0.1$&4.7$\pm$0.1\\ \hline
 $Z$+H&73.26$\pm$0.06&35.4$\pm$0.3&10.0$\pm$0.3&10.0$\pm$0.3&0.54$\pm$0.06&0.21$\pm$0.04&0.21$\pm$0.04\\ \hline
 $W$+H&151.2$\pm$0.5&71.3$\pm$0.3&19.9$\pm$0.1&19.9$\pm$0.1&0.28$\pm$0.03&0.14$\pm$0.01&0.13$\pm$0.01\\ \hline
 $Z$+ISR&53949$\pm$40&1394$\pm$42&210$\pm$17&210$\pm$17&7$\pm$3&0&0\\ \hline
 $W$+FSR&71414$\pm$204&8776$\pm$116&1922$\pm$58&1922$\pm$58&17$\pm$5&0$\pm$0&$0\pm$0\\ 
 \hline
 $\frac{S}{\sqrt{B}}$&---&0.67$\pm$0.01&$0.55\pm0.02$&0.55$\pm$0.02&1.2$\pm$0.2&8.4$\pm$0.9&8.1$\pm0.9$\\
 \hline
 \end{tabular}}
\end{table}
%%%%%%%%%%%%%%%%%%%%%%%%%%%%%%%%%%%%%%%%%%%%%%%%%%%%%%%%%%%%%%

%95 tau
\begin{table}[t!]
\centering
  \caption{
The same as in Table~11 but for the BP in Table \ref{table:tau-sneu-95} corresponding to a
$\widetilde \nu_{\tau L}$ LSP with a mass of 97.8 GeV.}
 \label{table2:tau-sneu-95}
{\scriptsize
  \begin{tabularx}{0.97\linewidth}{|X|C|} \hline
 $\sigma(pp\to Z^* \to h_1A^0_2)$&265.92  \\ \hline
 $\sigma(pp \to \gamma^*,Z^* \to H_2^+ H_2^- \to h_1A^0_2\ + W^+_{\text{soft}} W^-_{\text{soft}})$&42.67 \\ \hline
 $\sigma(pp \to W^{\pm^*} \to h_1^{\pm}h_1/A^0_2 \to h_1A^0_2\ + W^{\pm}_{\text{soft}})$ & $325.2$ \\ \hline 
 $\sigma(pp \to W^{\pm^*} \to h_1^{\pm}h_1 \to h_1 h_1\ + W^{\pm}_{\text{soft}})$ & $162.6$ \\
 \hline
 \end{tabularx}
 \begin{tabular}{|c|c|c|c|c|c|c|c|}\hline
 Dataset&$\mathrm{N_{ev}}$&$P_{T1}^{\gamma_1}$&$P_{T2}^{\gamma_2}$&$N_\gamma$=2&$N_{\tau_{had}}=1\,\&$
 &$\Delta R$&$M_{\gamma\gamma}\in$\\ 
  && $>100$ & $>50$& & $N_{e,\mu,\tau_{had}}>1$&$<1.5$&$[85,105]$\\ \hline  
 Signal&44.438$\pm$0.002&14.4$\pm$0.1&3.96$\pm0.06$&3.96$\pm$0.06&0.82$\pm$0.03&0.81$\pm0.03$&0.78$\pm$0.03\\ \hline
 $Z$+H&73.26$\pm$0.06&35.4$\pm$0.3&10.0$\pm$0.3&10.0$\pm$0.3&0.54$\pm$0.06&0.21$\pm$0.04&0.03$\pm$0.01\\ \hline
 $W$+H&151.2$\pm$0.5&71.28$\pm$0.3&19.9$\pm$0.1&19.9$\pm$0.1&0.28$\pm$0.03&0.14$\pm$0.01&0$\pm$0\\ \hline
 $Z$+ISR&53949$\pm$40&1394$\pm$42&210$\pm$17&210$\pm$17&7$\pm$3&0&0\\ \hline
 $W$+FSR&71415$\pm$204&8776$\pm$116&1922$\pm$58&1922$\pm$58&17$\pm$5&0$\pm$0&$0\pm$0\\ 
 \hline
 ${S}/{\sqrt{B}}$&---&0.14&$0.085$&0.0085&0.17$\pm$0.03&1.4$\pm$0.2&5$\pm1$ \\
 &&$\pm 0.002$&$\pm 0.003$&$\pm 0.003$&&& \\  \hline
 \end{tabular}} 
\end{table}

%%%%%%%%%%%%%%%%%%%%%%%%%%%%%%%%%%%%%%%%%%%%%%%%%%%%%%%%%%%%%%%%%%%%%%%%
%145 tau
%
\begin{table}[t!]
\centering
   \caption{The same as in Table \ref{table2:tau-sneu} but for the BP in Table \ref{table:tau-sneu-145} corresponding to a $\widetilde \nu_{\tau L}$ LSP
with a mass of 146 GeV.}
 \label{table2:tau-sneu-145}
{\scriptsize
\begin{tabularx}{0.96\linewidth}{|X|C|}\hline
$\sigma(pp\to Z^* \to h_2A^0_2)$&60.48\\ \hline
$\sigma(pp \to \gamma^*,Z^* \to H_2^+H_2^- \to h_2A^0_2\ + W^+_{\text{soft}} W^-_{\text{soft}})$&14.69 \\ \hline 
  $\sigma(pp \to W^{\pm^*} \to H_2^{\pm}h_2/A^0_2 \to h_2A^0_2\ + W^{\pm}_{\text{soft}})$ & $87.24$  \\ \hline 
    $\sigma(pp \to W^{\pm^*} \to H_2^{\pm}h_2 \to h_2h_2\ + W^{\pm}_{\text{soft}})$ & $43.62$ \\
 \hline
\end{tabularx}
 \begin{tabular}{|c|c|c|c|c|c|c|c|}\hline
 Dataset&$\mathrm{N_{ev}}$&$P_{T1}^{\gamma_1}$&$P_{T2}^{\gamma_2}$&$N_\gamma$=2&$N_{\tau_{had}}=1\, \&$
 &$\Delta R$&$M_{\gamma\gamma}\in$\\ 
  &&$>100$ & $>50$&& $ N_{e,\mu,\tau_{had}}>1$&$<1.5$&$[135,155]$\\ \hline  
 Signal&24.47$\pm$0.01&15.51$\pm$0.06&6.72$\pm0.06$&6.72$\pm$0.06&1.68$\pm$0.03&1.09$\pm0.03$
 &1.01$\pm$0.03\\ \hline
 $Z$+H&73.26$\pm$0.06&35.4$\pm$0.3&10.0$\pm$0.3&10.0$\pm$0.3&0.54$\pm$0.06&0.21$\pm$0.04&0.03$\pm$0.01\\ \hline
 $W$+H&151.2$\pm$0.5&71.3$\pm$0.3&19.9$\pm$0.1&19.9$\pm$0.1&0.28$\pm$0.03&0.14$\pm$0.01&0$\pm$0\\ \hline
 $Z$+ISR&53949$\pm$40&1394$\pm$42&210$\pm$17&210$\pm$17&7$\pm$3&0&0\\ \hline
 $W$+FSR&71414$\pm$204&8776$\pm$116&1922$\pm$58&1922$\pm$58&17$\pm$5&0$\pm$0&$0\pm$0\\ 
 \hline
 ${S}/{\sqrt{B}}$&---&0.153&$0.145$&0.145&0.34$\pm$0.06
 &1.8$\pm$0.2&6$\pm2$\\
  &&$\pm 0.002$&$\pm 0.004$&$\pm 0.004$&&& \\  \hline
 \end{tabular}} 
\end{table}

The results for a sneutrino mass of about 126 GeV, similar to the one studied above, 
are shown in Table~\ref{table2:tau-sneu}.
%for the different BPs with sneutrino masses $m_{\widetilde{\nu}}=125, 95$, and 145 GeV, respectively.
The discussion is similar to that above,
%in the previous subsection, 
although in this case we do not have two families of sneutrinos with degenerate masses, and therefore the different production mechanisms will only give rise to
%$h_2A_2^0$, i.e.
$\widetilde{\nu}_{{\tau}}\widetilde{\nu}_{{\tau}}$, thus reducing the number of events.
These are further suppressed by the BR($A^0_2\to {\tau}^\pm l^\mp$) compared to BR($A^0_2 \to \nu\nu$) in the case of $\widetilde{\nu}_{e,\mu}$ LSP.
Nevertheless, this signal with photons plus leptons is very attractive and worth to be searched at the LHC.

Now, a different set of cuts is taken into account for convenience, as shown in the second box of Table~\ref{table2:tau-sneu}. 
To distinguish the signal from the background in this case, instead of using the missing energy coming from neutrinos, we require two leptons in the final state of which one of them must be an hadronically decaying tau. Since every leptonic decay of the tau sneutrino includes at least one tau, we expect to reduce significantly more the background than the signal itself.
Using an integrated luminosity of $\mathcal{L}=300$ fb$^{-1}$,
%for $m_{\widetilde{\nu}}=125$, 95 and 145 GeV, 
we obtain $4.7 \pm 0.1$ events with a significant {evidence of 
$\frac{S}{\sqrt B}= 8.1 \pm 0.9$}.

In order to confirm the range of sneutrino masses of about 95--145 GeV
%(\ref{rangesneutrino}) 
adequate to observe this kind of signal, we have also analyzed in Tables~\ref{table2:tau-sneu-95} and~\ref{table2:tau-sneu-145} the two extreme cases of about 98 and 146 GeV, respectively.
Note that for both cases the BR of the scalar sneutrino decaying into photons is supressed with respect to the previous case of 126 GeV. 
Although
for the case of 98 GeV, the cross sections are increased with respect to the case of 126 GeV in 
Table~\ref{table2:tau-sneu},
% from
% 86.32 to
% 221.6 fb for the direct production of the sneutrino pair, and from
% 115.3 to 271 and 57.65 to 135.5
% fb for the 
% production through a slepton. 
%Even though the cross sections are highly increased for this point, 
the final products would have less $P_T$, and $E_T$, thus the efficiency of the selection cuts would 
be smaller. 
% Also, for $m_{\widetilde{\nu}} < 98$ GeV, the presence of new particles with a mass below $M_Z$ could increase the well 
% measured width of the $Z$ boson. Altogether we consider that left sneutrinos with masses below 
% 98 GeV are disfavored for detection at the LHC. 
We apply the same set of selection cuts 
to the signal calculated with this new point as in the previous case, but selecting now a new invariant mass window for the 
diphoton sistem of $\pm10$ around 98 GeV. 
The rest of the analysis is completely analogous, and the results are presented in the second box of 
Table~\ref{table2:tau-sneu-95}. For this extreme case we still {obtain 0.78 $\pm$ 0.03 events with a significant evidence of
$\frac{S}{\sqrt B}= 5 \pm 1$}.

Finally, to explore the largest possible value of the sneutrino mass,
%$m_{\widetilde{\nu}}$, 
we have considered the case of 146 GeV. We  
show the final results in Table~\ref{table2:tau-sneu-145}. As can be seen, the production cross sections are reduced 
with the increase of the mass.
%to $\sigma\approx 50.4$ fb for the direct production of sneutrino 
%pairs, and $\sigma\approx 72.7$ fb for the production through a slepton. 
We are not considering 
points with sneutrino masses larger than $146 \gev$ because the possible signal gets likely lost behind 
the SM backgrounds. The results of the different selection cuts for this extreme case are presented in
the second box of Table~\ref{table2:tau-sneu-145}. As a final {result, 1.01 $\pm$ 0.03 events with a
significant evidence of
$\frac{S}{\sqrt B}= 6 \pm 2$ are obtained}. 

% \R{A similar analysis for the previous case of two photons plus MET,
% varying the sneutrino mass, confirms that the range in Eq.~(\ref{rangesneutrino}) is the right one in order to have a measurable signal. THIS HAS TO BE MODIFIED DEPENDING ON OUR ANALYSES OF DISPLACED VERSUS PROMPT AND THE FIGURE OF IQAKI.}

% --

% [ DLF: {\bf   The  scalar and pseudoscalar} left sneutrinos LSP can be prompt or displaced. We need to 
% be careful with neutrino physics in order to make a estimation of the displaced vertices. 
% In  principle is similar to the case discussed above but now also the pseudoscalar gives a 
% visible signal and the decay width must be estimated taken into account neutrino physics 

% WE REALLY NEED TO THINK ON THIS PERHAPS THERE ARE MORE INTERESTING SIGNALS OR ARE REALLY EXCLUDED ....]

%%%%%%%%%%%%%%%%%%%%%%%%%%%%%%%%%%%%%%%%%%%%%%%%%%%%%%%%%5

\begin{table}[t!]
 \centering
  \caption{Analysis of the signal with multileptons from production and decay of a pair of 
$\widetilde \nu_{\tau L}$ LSPs. The number of events of the signal and background is shown,
together with the effect of a set of cuts, assuming 13 TeV center-of-mass energy 
with $\mathcal{L}=20$ fb$^{-1}$. The subindex $l$ in the dataset denotes leptonically decaying tops and gauge bosons.
Three possible masses of $\widetilde \nu_{\tau L}$, 132, 146 and 311 GeV are analyzed, with the
last two obtained using the BPs of Tables \ref{table:tau-sneu-145} and \ref{table:tau-sneu-310}.}
 \label{table2:tau-sneu-310}
{\scriptsize
\begin{tabular}{|c|c|c|c|c|c|c|c|}\hline
Dataset&$N_l \geq4\ \&\ N_{\tau_h}\geq2$&$N_b=0$&THT$\leq 20$ GeV&W-veto&Z-veto&$\frac{S}{\sqrt{B}}$\\ \hline
${\bar{t}t}_{l}$&$306\pm66$&$174\pm50$&14$\pm14$&$0\pm0$&$0\pm0$&--\\\hline
${\bar{t}th}$&$3\pm2$&0$\pm0$&$0\pm0$&$0\pm0$&$0\pm0$&--\\\hline
${\bar{t}t\bar{t}t}$&$0.8\pm0.5$&$0\pm0$&$0\pm0$&$0\pm0$&$0\pm0$&--\\\hline
${\bar{t}tV}_{l}$&$1.2\pm0.3$&$0.6\pm0.2$&$0.12\pm0.09$&$0.12\pm0.09$&$0.12\pm0.09$&--\\\hline
${VV}_{l}$&$6\pm4$&$6\pm4$&$3\pm3$&$3\pm3$&$3\pm3$&--\\\hline
$VVV$&$2\pm1$&$0.8\pm0.8$&$0\pm0$&$0\pm0$&$0\pm0$&--\\\hline
${tV}_{l}$&$15\pm5$&$14\pm5$&$8\pm4$&$8\pm4$&$8\pm4$&--\\\hline
${tVV}_{l}$&$1.0\pm0.3$&$0.6\pm0.2$&$0.10\pm0.08$&$0.10\pm0.08$&$0.10\pm0.08$&--\\\hline
Total&$334\pm66$&$196\pm64$&$25\pm15$&$11\pm5$&$11\pm5$&--\\\hline\hline
Signal 132 GeV&$36\pm2$&$36\pm2$&$20\pm2$&$17\pm1$&$16\pm1$&$4.8\pm2.2$\\\hline
Signal 146 GeV&$68\pm3$&$66\pm3$&$37\pm2$&$31\pm2$&$29\pm2$&$8.8\pm4.0$\\\hline
Signal 311 GeV&$18.2\pm0.5$&$17.9\pm0.5$&$8.9\pm0.4$&$7.6\pm0.4$&$7.5\pm0.2$&$2.2\pm1.0$\\\hline
\end{tabular}}
\end{table}

%%%%%%%%%%%%%%%%%%%%%%%%%%%%%%%%%%%%%%%%%%%%%%%%%%%%%%%%%%%5

\vspace{0.25cm}

\noindent
{\bf Multileptons}

\noindent 
For the tau left sneutrino,
we can see in Tables~\ref{table:tau-sneu-145} and~\ref{table:tau-sneu-310} that the BRs for the
decay of the scalar state $\widetilde{\nu}^\mathcal{R}_{\tau_L}$ into leptons are significant. This gives rise to a non negligible number of events with both sneutrinos decaying into leptons, as shown in 
Fig~\ref{fig:production3}. With the appropriate analysis, these events could constitute a possible signal to be detected at the LHC.
Moreover, these 
%dominant 
decay channels 
of the LSP
%into leptons 
include always at least one $\tau$, a feature that can be exploited to unravel the signal. 
% \R{This also happens in diphoton plus leptons and we do not mention it. Actually this is related with my comment above about explaining a little bit more the cuts}

The main backgrounds for this type of signature would be the production of top quarks through the channels $\bar{t}t$ and $\bar{t}t\bar{t}t$; the production of gauge bosons ZZ, WW and ZW; the associated production of both $\bar{t}t V$, $tV$ and $tVV$; and the top associated Higgs production $\bar{t}th$.  
Since the proposed hard process would not produce quarks, we expect a hadronic activity in the events significantly smaller than the one associated with background events including a leptonically decaying top $t_l$. We will show that it is possible to separate the multilepton signal 
%from the decay of tau sneutrinos 
from the SM backgrounds. 
This is particularly true for sneutrinos with large masses, since the produced leptons are then expected to be more energetic than the ones produced in the decay of gauge bosons.

The Monte Carlo events generated and processed as in the previous signals, but in this case with an integrated luminosity of 
$\mathcal{L}=20$ fb$^{-1}$, are analyzed and summarized in 
Table~\ref{table2:tau-sneu-310} for three different sneutrino masses of 132, 146 and 311 GeV. 
Production cross sections for the case of 146 GeV are already shown in Table~\ref{table2:tau-sneu-145}. For 310 GeV these are much lower.
%, with the first one via a $Z$ channel of 2.64 fb, and the third one via a $W$ channel of 4.46 fb.
At first we select events with at least 4 leptons with $P_T$
%transverse momentum 
$\geq$ 100, 80, 40 and 40 GeV, respectively, requiring also at least two of them to be $\tau_{\mathrm{h}}$'s. The second selection rejects events with $b$-tagged jets in order to reduce backgrounds coming from top decays. In the next step we reject events with a total transverse hadronic energy (THT) greater than 20 GeV. Finally we apply a veto to the transverse mass and invariant mass of the light leptons, compatible with the mass of the W and Z respectively. 
% The result of the analysis for the SM backgrounds and signal events is shown in Table~\ref{table2:tau-sneu-310}.
% for three different sneutrino masses. 
% Summarizing the results shown in Table~\ref{table2:tau-sneu-310}, it is possible to detect 
% $\widetilde{\nu}_{\tau_L}$ decaying leptonically with a significance $\frac{S}{\sqrt{B}}$ greater than 3, in the mass range 
% \bea
% 130 \lesssim m_{\widetilde{\nu}_{\tau L}} \lesssim 310\ \gev \ .
% \label{rangesneutrinotau2}
% \eea 
Summarizing the results shown in Table~\ref{table2:tau-sneu-310}, it is possible to detect 
$\widetilde{\nu}_{\tau_L}$ 
in the mass range 
\bea
130 \lesssim m_{\widetilde{\nu}_{\tau L}} \lesssim 310\ \gev \ ,
\label{rangesneutrinotau2}
\eea 
decaying leptonically with a significance $\frac{S}{\sqrt{B}}$ greater than 3.

% Let us finally discuss briefly a scenario with muon and electron left sneutrinos slightly heavier than the tau left sneutrino, in order to be the NLSPs. In this case,
% for a mass difference of the order of 10 GeV, the three-body decay of the heavier sneutrinos into the lighter one plus neutrinos through a neutralino could entail a significant BR, enhancing therefore the total number of tau sneutrino decays. However, the different kinematic of the event make it inefficient to pass the selection cuts of the analysis presented here, leaving finally about two extra events at most. 
% Thus this scenario is unlikely to improve significantly the results we have discussed before.

\vspace{0.5cm}

Equations (\ref{rangesneutrinomuon}), (\ref{rangesneutrinotau}) and (\ref{rangesneutrinotau2}) 
establish the adequate range of left sneutrino masses for our analysis of the BPs introduced in 
Section~\ref{Section:Sneutrino-LSP}, and
Tables~\ref{table:musneu-125}--\ref{table:tau-sneu-310}.
As we can see, the masses overlap in some ranges, and in these cases the corresponding BP can give rise to different detectable signals.

\section{Conclusions and outlook}
\label{Section:Summary-Conclusion}
%%%%%%%%%%%%%%%%%%%%%%%%%%%%%%%%%%%%%%%%%%%%%%%%%%%%%%%%%%%%%%%%%
%%%%%%%%%%%%%%%%%%%%%%%%%%%%%%%%%%%%%%%%%%%%%%%%%%%%%%%%%%%%%%%%

We have carried out an analysis of the LHC phenomenology associated to the left sneutrino LSP in the $\mn$.
% After reviewing 
% in Section~\ref{Section:munuSSM}
% the structure of the model, in order to clarify our assumptions and notation, 
%we have focused in this work on a left sneutrino as the LSP, 
We have studied 
%in Sections~\ref{Section:Sneutrino-LSP} and~\ref{Section:detectionSneutrino-LSP} its 
the dominant pair production channels, prompt decays, and the detection
of the new signals.

As a result of the different behaviors of scalar and pseudoscalar sneutrino states, a diphoton signal in combination with neutrinos (producing missing transverse energy), or a diphoton with leptons, can appear at the 
LHC.
% (see Figs.~\ref{fig:production}-\ref{fig:production2}).
The former can be detected with a center-of-mass energy of 13 TeV and the integrated luminosity of 100 fb$^{-1}$, for a sneutrino LSP of any family in the mass range 118--132 GeV.
The diphoton plus leptons signal can be probed for the case of a 
tau sneutrino LSP with a mass in the range 95--145 GeV. 
%Heavier sneutrinos would imply that the production cross sections are not sufficiently large.
%In Tables~\ref{table:musneu-125}-\ref{table:tau-sneu-145}, 
We have discussed
%show 
several benchmark points producing these signals, which 
undoubtedly deserve proper experimental attention. 
%Actually, the one in Table~\ref{table:tau-sneu} produces both signals at the same time.
%The corresponding Tables~\ref{table2:musneu-125}-\ref{table2:tau-sneu-145} show 
We have also shown that the number of expected events are capable of giving a significant evidence.
%, even at the current Run 2 when the luminosity reaches 100 fb$^{-1}$.
%,which we think deserves experimental attention. 
% Let us finally remark that long decay chains in this scenario of LH sneutrino LSP might give rise to other potentially interesting signals. 
% This analysis is beyond the scope of this paper, but we plan to cover it in a forthcoming publication [].

A multilepton signal 
%(see Fig.~\ref{fig:production3})
from a tau sneutrino LSP can also appear
detectable at the LHC with a center-of-mass energy of 13 TeV, even with the integrated luminosity of 20 fb$^{-1}$.
%The benchmark point of Table~\ref{table:tau-sneu-145} produces this signal together with the one above of diphoton plus dilepton. 
%Another benchmark point is shown in Table~\ref{table:tau-sneu-310}.
%The results for detection at the LHC are shown in Table \ref{table2:tau-sneu-310}.
%Summarizing, 
It is possible to detect it
in the mass range of 130--310 GeV.
% , it is possible to detect the sneutrino decaying leptonically with a significance $\frac{S}{\sqrt{B}}$ greater than 3. 
We have discussed that existing generic searches at the LHC are close to be sensitive to this lepton signal, suggesting that they deserve experimental attention. An updated analysis with current data could constrain the sneutrino LSP scenario. 

Displaced vertices of the order of the millimeter can appear for sneutrino masses $\lsim 100$ GeV.
Imposing in addition that the sneutrino mass is larger than 45 GeV, not to disturb 
the experimentally well measured decay width of the $Z$, 
we have found that the number of events can be large.
% in that range. 
%Thus in this case, for the range of masses $45-100$ GeV, where the lower limit is due to the mass of the $Z$,
%These cases have the clear advantage that backgrounds are negligible, 
%In this cases 
%the number of events can be large. 
For example, {more than 1000 multilepton events at the parton level} from the production and decay of a %$\widetilde\nu_{\tau L}$ 
tau sneutrino pair
can emerge for an integrated luminosity of 20 fb$^{-1}$ and 13 TeV center-of-mass energy.
These events have the clear advantage that the SM backgrounds are negligible and hence the 
signal significance is high.
However, the analysis of displaced vertices turns out to be quite complicated, and dedicated studies are necessary. 
The efficiency identifying events characterized by the presence of a displaced vertex has a nontrivial dependence on the position of the vertex, as well as the number of tracks and the mass associated to them, among others. Therefore, a reliable analysis requires a precise simulation of the decay length, the boost of the long-lived particle, and the particles produced in the secondary vertex. This analysis, in our model, is expected to depend on the parameters correlated with neutrino physics and is clearly beyond the scope of the present work, although we plan to cover it in a forthcoming publication \cite{prepa}.

%%%%%%%%%%%%%%%%%%%%%%%%%%%%%%%%%%%%%%%%%%%%%%%%%%%%%%%%%%%%%%%%%%%%%%%%%%
%%%%%%%%%%%%         Acknowledgments              %%%%%%%%%%%%%%%%%%%%%%%%
%%%%%%%%%%%%%%%%%%%%%%%%%%%%%%%%%%%%%%%%%%%%%%%%%%%%%%%%%%%%%%%%%%%%%%%%%%

\begin{acknowledgments}
PG acknowledges the support received from P2IO Excellence Laboratory (LABEX) during
the development of this project.
The work of IL and CM was supported in part by the State Research Agency through the grants 
FPA2015-65929-P (MINECO/FEDER, UE) and IFT Centro de Excelencia Severo Ochoa SEV-2016-0597. The work 
of DL was supported by the Argentinian CONICET,and he also acknowledges the support of the Spanish grant FPA2015-65929-P (MINECO/FEDER, UE). The work of RR was supported by the 
Ram\'on y Cajal 
program of the Spanish MINECO, and also thanks the support of the grant
%s 
FPA2014-57816-P,
%and FPA2013-44773, 
and the Program 
SEV-2014-0398 `Centro de Excelencia  Severo Ochoa'. 
The authors also acknowledge the support of the MINECO's Consolider-Ingenio  2010 Programme under 
grant MultiDark CSD2009-00064.
CM gratefully acknowledges the hospitality and support of LPT Orsay during whose stay in August 2017 the last stages of this work were carried out.
 
\end{acknowledgments}

%%%%%%%%%%%%%%%%%%%%%%%%%%%%%%%%%%%%%%%%%%%%%%%%%%%%%%%%%%%%%%%%%%%%%%%%%%%%
%%%%%%%%%%%%%%%%          APPENDIX              %%%%%%%%%%%%%%%%%%%%%%%%%%%%
%%%%%%%%%%%%%%%%%%%%%%%%%%%%%%%%%%%%%%%%%%%%%%%%%%%%%%%%%%%%%%%%%%%%%%%%%%%%
\appendix

%%%%%%%%%%%%%%%%%%%%%%%%%%%%%%%%%%%%%%%%%%%%%%%%%%%%%%%%%%%%%%
%%%%%%%%%%%%%%%%%%%%%%%%%%%%%%%%%%%%%%%%%%%%%%%%%%%%%%%%%%%%%%%%
\section{The Superpotential and Soft Terms} 
\label{Section:munuSSM}
%%%%%%%%%%%%%%%%%%%%%%%%%%%%%%%%%%%%%%%%%%%%%%%%%%%%%%%%%%%%%%
%%%%%%%%%%%%%%%%%%%%%%%%%%%%%%%%%%%%%%%%%%%%%%%%%%%%%%%%%%%%%%%%

We review in this Appendix the superpotential of the model and the associated soft terms,
following the works of Refs.~\cite{LopezFogliani:2005yw,Escudero:2008jg,Lopez-Fogliani:2017qzj}. 
% We discuss first 
% the superpotential of the $\mn$ and the associated soft terms,
% and second
% the scalar potential.

% \vspace{0.25cm}

% \noindent 
% {\bf The Superpotential} 
% \label{superpotentialsoft}

% Following the work of Ref.~\cite{LopezFogliani:2005yw,Escudero:2008jg,Lopez-Fogliani:2017qzj}, we introduce in this section the main characteristics of the $\mn$.
Given the gauge symmetry group of the SM, $SU(3)_C\times SU(2)_L\times U(1)_Y$,
with subscripts $C$, $L$ and $Y$ referring to color, left chirality and weak hypercharge, respectively,
the superpotential of the $\mn$ can be written as~\cite{Lopez-Fogliani:2017qzj}
%of the $\mn$
\bea\label{superpotentialc}
W = && 
%&
\epsilon_{ab} \left(Y^e_{IJk} \, \hat L_I^a \, \hat L_J^b \, \hat e_k^c +
Y^d_{Ijk}\, \delta_{\alpha\beta}\, \hat L_I^a \, \hat Q_{j\alpha}^b \, \hat d_{k\beta}^c 
+
Y^u_{ij} \, \delta_{\alpha\beta}\, \hat L^{^{cb}}_4
%{\overbar L_4} 
\, \hat Q_{i\alpha}^a  \, \hat u_{j\beta}^c\right) 
\nonumber\\
&+&
\epsilon_{ab}\, Y^\nu_{Ij} \, \hat L^{^{cb}}_4
%{\overbar L_4} 
\, \hat L_I^a \, \hat \nu^c_j
%\nonumber\\
%& 
+
\frac{1}{3}
\kappa_{ijk}  \, \hat \nu^c_i  \, \hat \nu^c_j  \, \hat \nu^c_k
\ ,
\eea
%
% \bea\label{superpotentialc}
% W = && 
%%&
% \epsilon_{ab} \left(Y^e_{IJk} \, \hat L_I^a \, \hat L_J^b \, \hat e_k^c +
% Y^d_{Ijk}\, \delta_{\alpha\beta}\, \hat L_I^a \, \hat Q_{j\alpha}^b \, \hat d_{k\beta}^c 
% +
% Y^u_{ij} \, \delta_{\alpha\beta}\, \hat L^{^{cb}}_4
%%{\overbar L_4} 
% \, \hat Q_{i\alpha}^a  \, \hat u_{j\beta}^c \right.
% \nonumber\\
% &+&
%%\epsilon_{ab}\, 
% \left. Y^\nu_{Ij} \, \hat L^{^{cb}}_4
%%{\overbar L_4} 
% \, \hat L_I^a \, \hat \nu^c_j\right)
%%\nonumber\\
%%& 
% +
% \frac{1}{3}
% \kappa_{ijk}  \, \hat \nu^c_i  \, \hat \nu^c_j  \, \hat \nu^c_k
% \ ,
% \eea
% \bea\label{superpotentialc}
% W = && 
%%&
% \epsilon_{ab} \left(Y^e_{IJk} \, \hat L_I^a \, \hat L_J^b \, \hat e_k^c +
% Y^d_{Ijk}\, \delta_{\alpha\beta}\, \hat L_I^a \, \hat Q_{j\alpha}^b \, \hat d_{k\beta}^c 
% +
% Y^u_{ij} \, \delta_{\alpha\beta}\, \hat L^{^{cb}}_4
%%{\overbar L_4} 
% \, \hat Q_{i\alpha}^a  \, \hat u_{j\beta}^c +
% Y^\nu_{Ij} \, \hat L^{^{cb}}_4
% \, \hat L_I^a \, \hat \nu^c_j\right) 
% \nonumber\\
% &+&
%%\epsilon_{ab}\, 
% \frac{1}{3}
% \kappa_{ijk}  \, \hat \nu^c_i  \, \hat \nu^c_j  \, \hat \nu^c_k
% \ ,
% \eea
%
where the summation convention is implied on repeated indexes, with 
$\alpha, \beta = 1,2,3$ $SU(3)_C$ indexes, 
$a,b=1,2$ $SU(2)_L$ indexes with $\epsilon_{ab}$ the totally antisymmetric tensor $\epsilon_{12}=1$, and $I=i,4$ ($J=j,4$)
%, $J=j,4$ 
with $i,j,k=1,2,3$ the usual family indexes of the SM and with the vector-like Higgs doublet superfields interpreted as a fourth family of
vector-like lepton superfields\footnote{An extension of the $\mn$ by adding to the spectrum of this fourth family a vector-like quark doublet representation has also been discussed, together with its new signals at the LHC, in 
Refs.~\cite{Lopez-Fogliani:2017qzj,Aguilar-Saavedra:2017giu}.}
$\hat L_4=(\hat \nu_4, \hat e_4)=(\hat H_d^0, \hat H_d^-)=\hat H_d$ and 
$\hat L_4^c=(\hat e^c_4, \hat \nu^c_4)=(\hat H_u^+, \hat H_u^0)=\hat H_u$.
This interpretation is possible in the $\mn$ because right-handed neutrinos are 
present producing the violation of $R_p$, and as a consequence all fields in the spectrum with the same color, electric charge and spin mix together. 
In particular, Higgses mix with sleptons and Higgsinos with leptons. From the theoretical viewpoint, this seems to be more satisfactory than the situation in usual SUSY models, where the Higgses are `disconnected' from the rest of the matter and do not have a three-fold replication\footnote{For alternative constructions with three superymmetric families of Higgses, see works~\cite{Escudero:2005hk,Escudero:2005ku,Escudero:2007db} and references therein.}.
As pointed out in Ref.~\cite{Lopez-Fogliani:2017qzj}, in this SUSY framework the first scalar particle
discovered at the LHC is mainly a sneutrino belonging to a fourth-family vector-like doublet representation.

 %and the notation for the Yukawa couplings is self-explanatory.

In order to make contact with the usual (three-family) notation 
of the $\mn$~\cite{LopezFogliani:2005yw,Escudero:2008jg}, we can decompose the terms given by the couplings $Y^e_{IJk}$, $Y^d_{Ijk}$ and $Y^{\nu}_{Ij}$ in two type of terms: Yukawa couplings generating fermion masses, and lepton-number violating couplings.
This is possible because, as discussed above, the superfields $L_i$ and $H_d$ have the same gauge quantum numbers, and therefore $\hat L_I = \hat L_i, \hat H_d$.
Thus, we can write superpotential (\ref{superpotentialc}) as 
follows~\cite{LopezFogliani:2005yw,Escudero:2008jg}:
% \cite{LopezFogliani:2005yw, Escudero:2008jg}:
%%%%%%%%%%%%%%%%%%%%%%%%%%%%%%%%%%%%%%%%%%%%%%%%%%%%%%%%%%%%%%
%%%%%%%%%%%%%%%%%%%%%%%%%%%%%%%%%%%%%%%%%%%%%%%%%%%%%%%%%%%%%%%%
% \bea\label{superpotential}
% W  = 
% \ \epsilon_{ab} & \left(
% Y_{u_{ij}} \, \hat H_u^b\, \hat Q^a_i \, \hat u_j^c +
% Y_{d_{ij}} \, \hat H_d^a\, \hat Q^b_i \, \hat d_j^c +
% Y_{e_{ij}} \, \hat H_d^a\, \hat L^b_i \, \hat e_j^c 
% \right. \nonumber \\
% &
% \left.
% + Y_{\nu_{ij}} \, \hat H_u^b\, \hat L^a_i \, \hat \nu^c_j -   
% \lambda_{i} \, \hat \nu^c_i\,\hat H_d^a \hat H_u^b\right)+
% \frac{1}{3}
% \kappa{_{ijk}} 
% \hat \nu^c_i\hat \nu^c_j\hat \nu^c_k\ ,
% \eea
% \bea
% W &=& 
%%&&
%%\epsilon_{ab} \left(
% \ - Y^d \, \hat d^c \, \hat Q\cdot \hat H_1 
% - Y^e \, \hat e^c \, \hat L\cdot \hat H_1
% + Y^u \, \hat u^c\, \hat Q\cdot \hat H_2 
%%\right)
% \nonumber\\
% &&+
% \ Y^{\nu} \, \hat \nu^c \,  \hat L\cdot \hat H_2
% +   
% \lambda \, \hat \nu^c\,\hat H_2\cdot \hat H_1
% +
% \frac{1}{3}
% \kappa \,
% \hat \nu^c \, \hat \nu^c \, \hat \nu^c\ ,
% \label{superpotential}
% \eea
% \bea
% W =
%%&&
%%\epsilon_{ab} \left(
% - Y^d \, \hat d^c \, \hat Q\cdot \hat H_1 
% - Y^e \, \hat e^c \, \hat L\cdot \hat H_1
% + Y^u \, \hat u^c\, \hat Q\cdot \hat H_2 
%%\right)
% + Y^{\nu} \, \hat \nu^c \,  \hat L\cdot \hat H_2
% +   
% \lambda \, \hat \nu^c\,\hat H_2\cdot \hat H_1
% +
% \frac{1}{3}
% \kappa \,
% \hat \nu^c \, \hat \nu^c \, \hat \nu^c\ ,
% \nonumber\\
% \label{superpotential}
% \eea
\bea
W = &&
\epsilon_{ab} \left(
Y^e_{ij} \, \hat H_d^a\, \hat L^b_i \, \hat e_j^c +
Y^d_{ij} \, \delta_{\alpha\beta}\, \hat H_d^a\, \hat Q^{b}_{i \alpha} \, \hat d_{j \beta}^{c} 
+
Y^u_{ij} \, \delta_{\alpha\beta}\, \hat H_u^b\, \hat Q^{a}_{i \alpha} \, \hat u_{j \beta}^{c}
\right)
\nonumber\\
% &+&
% \epsilon_{ab} Y^{\nu}_{ij} \, \hat H_u^b\, \hat L^a_i \, \hat \nu^c_j -
&+&   
\epsilon_{ab} \left(\lambda_{ijk} \hat L_i^a \hat L_j^b \hat e^c_k 
+ 
 \lambda'_{ijk} \delta_{\alpha\beta}\, \hat L_i^a \hat Q_{j \alpha}^{b} \hat d^{c}_{k \beta} \right)
\nonumber\\
&+&   
\epsilon_{ab} \left(
Y^{\nu}_{ij} \, \hat H_u^b\, \hat L^a_i \, \hat \nu^c_j 
-
%\epsilon_{ab}
\lambda_{i} \, \hat \nu^c_i\, \hat H_u^b \hat H_d^a
\right)
+
\frac{1}{3}
\kappa{_{ijk}} 
\hat \nu^c_i\hat \nu^c_j\hat \nu^c_k
\ ,
\label{superpotentiallb}
\eea
% \bea
% W =
%%&&
%%\epsilon_{ab} \left(
% - Y^d  \hat d^c  \hat Q \hat H_1 
% - Y^e  \hat e^c  \hat L \hat H_1
% + Y^u  \hat u^c \hat Q \hat H_2 
%%\right)
% + Y^{\nu} \hat \nu^c   \hat L \hat H_2
% +   
% \lambda \hat \nu^c \hat H_2 \hat H_1
% +
% \frac{1}{3}
% \kappa 
% \hat \nu^c  \hat \nu^c  \hat \nu^c\, ,
%%\nonumber\\
% \label{superpotential}
% \eea
% \bea
% W = &&
% \epsilon_{ab} \left(
% Y^u_{ij} \, \hat H_2^b\, \hat Q^a_i \, \hat u_j^c +
% Y^d_{ij} \, \hat H_1^a\, \hat Q^b_i \, \hat d_j^c +
% Y^e_{ij} \, \hat H_1^a\, \hat L^b_i \, \hat e_j^c \right)
% \nonumber\\
% &+&
% \epsilon_{ab} Y^{\nu}_{ij} \, \hat H_2^b\, \hat L^a_i \, \hat \nu^c_j -   
% \epsilon_{ab}
% \lambda_{i} \, \hat \nu^c_i\,\hat H_1^a \hat H_2^b
% +
% \frac{1}{3}
% \kappa{_{ijk}} 
% \hat \nu^c_i\hat \nu^c_j\hat \nu^c_k\ ,
% \label{superpotential}
% \eea
% where
% the dimensionless complex trilinear couplings form the Yukawa matrices 
% $Y^u$, $Y^d$, $Y^e$ and $Y^{\nu}$,
% a vector $\lambda$ and a totally symmetric 
% tensor $\kappa$. In our convention, 
% for example the term $Y^u \, \hat u^c\, \hat Q \hat H_2$ 
% is interpreted as
% $Y^u_{ij} \epsilon_{ab} \delta_{\alpha\beta}\, \hat u_i^c{^{\alpha}}\, 
% \hat Q_j^{\beta a}\, \hat H_2^b$, with $\alpha, \beta$ color indexes and 
% $a,b=1,2$ $SU(2)_L$ indexes 
% with $\epsilon_{ab}$ the totally antisymmetric tensor, $\epsilon_{12}=1$. 
% The last two terms are interpreted as 
% $\epsilon_{ab}\lambda_{i} \, \hat \nu^c_i\,\hat H_2^a \hat H_1^b$
% and 
% $\frac{1}{3}
% \kappa{_{ijk}} 
% \hat \nu^c_i\hat \nu^c_j\hat \nu^c_k$.
% The summation convention is implied on repeated indexes.
where we have decomposed (in a self-explanatory notation)
$Y^e_{IJk}\rightarrow \lambda_{ijk}, Y^e_{ij}$;
$Y^d_{Ijk}\rightarrow \lambda'_{ijk}, Y^d_{ij}$;  
and 
$Y^{\nu}_{Ij}\rightarrow Y^{\nu}_{ij}, -\lambda_{i}$.
The dimensionless complex trilinear couplings form a
vector $\lambda_i$, the Yukawa matrices $Y^{\nu}_{ij}$, $Y^e_{ij}$, $Y^d_{ij}$, $Y^u_{ij}$, and
the tensors $\lambda_{ijk}$, $\lambda'_{ijk}$, $\kappa_{ijk}$ with $\kappa$ totally symmetric and $\lambda_{ijk}$ antisymmetric with respect to their first two indexes.
% In (\ref{superpotentiallb}), the summation convention is implied on repeated indexes, with $i,j,k=1,2,3$ family indexes, $\alpha, \beta = 1,2,3$ 
% $SU(3)_C$ 
% indexes, and 
% $a,b=1,2$ $SU(2)_L$ indexes where $\epsilon_{ab}$ is the totally antisymmetric tensor, $\epsilon_{12}=1$.
%In addition,

In Eq.~(\ref{superpotentiallb}) (and (\ref{superpotentialc})),
we have defined $\hat u_i$, $\hat d_i$, $\hat \nu_i$, $\hat e_i$, 
and $\hat u^c_i$, $\hat d^c_i$, $\hat e^c_i$, $\hat \nu^c_i$,
% with    
% $i=1,2,3$ a family index, 
as the 
left-chiral superfields whose fermionic components are the left-handed fields
of the corresponding quarks, leptons, and antiquarks, antileptons, respectively. 
For example, the superfield $\hat d_2$ contains the 2-component complex spinor field $s_L$ (and the complex scalar field $\tilde s_L$), 
whereas $\hat d^c_2$ contains the spinor
${s^c}_L=(s_R)^c=i\sigma^2 s_R^*$ (and the scalar 
${\tilde s}_R^* = ({\tilde s_R})^c$),
where
the superscripts $c$ and $*$ indicate charge conjugate and complex conjugate, respectively, with $\sigma^2$ the Pauli matrix.
%${\widetilde{s}_R}^{^*}$,
Needless to say, the subscripts $L$ and $R$ on the scalar fields refer to the chirality of the corresponding fermion fields.
The superfields $\hat u_i$, $\hat d_i$, and $\hat \nu_i$, $\hat e_i$ form the 
$SU(2)_L$ doublets
$\hat Q_i=(\hat u_i, \hat d_i)$ and $\hat L_i=(\hat \nu_i, \hat e_i)$, 
respectively, and the others are $SU(2)_L$ singlets.
% The Higgs doublet superfields are defined as
% $\hat H_d=(\hat H_d^0, \hat H_d^-)$ and $\hat H_u=(\hat H_u^+, \hat H_u^0)$.

In the $\mn$ superpotential, the $\mu$ term is absent, as well as Majorana masses for neutrinos.
This can be obtained invoking a $Z_3$ symmetry as in the case of the NMSSM, which implies that only trilinear terms are allowed.
Actually, this is what one would expect from a high-energy theory where
the low-energy modes should be massless and the massive modes of the order of the high-energy scale.
As pointed out in Ref.~\cite{Lopez-Fogliani:2017qzj}, this is precisely the situation in 
%the low-energy limit of 
string constructions,
%Given the relevance of string theory as a possible underlying unified theory,
%for the unification of interactions including gravity,
% a robust argument pointed out in Ref.~\cite{propuvSSM} in the above direction, is that the unique presence of trilinear terms is actually what happens in the low-energy limit of string constructions, 
where the massive modes have huge masses of the order of the string scale and the massless ones have only trilinear terms at the renormalizable level. Thus one ends up with an accidental $Z_3$ symmetry in the low-energy theory.

The three terms in the first line of the superpotential in Eq.~(\ref{superpotentiallb}) are the usual Dirac Yukawa couplings for quarks and 
%charged 
leptons
of the MSSM.
The two terms in the second line are
the conventional trilinear $\rpv$ couplings
%$\lambda_{ijk}$ and $\lambda'_{ijk}$ 
(see Ref.~\cite{Barbier:2004ez} 
for a review).
As is well known, if the lepton-number violating term, 
$ \lambda'_{ijk} \delta_{\alpha\beta} \epsilon_{ab} \hat L_i^a \hat Q_{j \alpha}^{b} \hat d^{c}_{k \beta}$,
%$\lambda'_{ijk}$, 
appears together with 
the baryon-number violating term, $\lambda''_{ijk} \epsilon^{\alpha\beta\gamma}\hat d^c_{i\alpha} \hat d^c_{j\beta} \hat u^c_{k\gamma}$
where $\epsilon^{\alpha\beta\gamma}$ is the totally antisymmetric 
tensor $\epsilon^{123}=1$, they could give rise to experimentally excluded fast proton decay.
Nevertheless, as discussed in detail in Ref.~\cite{Lopez-Fogliani:2017qzj}, $\lambda''_{ijk}$
%the latter term 
can be naturally forbidden, for example
through $Z_3$ Baryon-parity or stringy selection rules.
% whereas the former is allowed. 
 %in the second line,
% are naturally present associated to Yukawa couplings $Y^{e}_{ij}$  
% and $Y^{d}_{ij}$, respectively, 
% since the superfields $L_i$ and $H_d$ have the same gauge quantum numbers.
%These $\rpv$ terms also violate lepton number, and 
Finally, 
the three terms 
%, and the $\,4^{\rm th}-6^{\rm th}$ terms 
in the third line 
%of superpotential~(\ref{superpotentiallb}) 
are characteristic of the $\mn$.
In particular, the first one 
%$\,4^{\rm th}$ term 
contains the Dirac Yukawa couplings for neutrinos, 
and the last two 
%$\,5^{\rm th}$ and $6^{\rm th}$ terms 
generate dynamically
the $\mu$ term and Majorana masses for neutrinos, respectively. 
Since sparticles do not appear in pairs in these two terms,
they generate $\rpv$ couplings.\footnote{Notice that the couplings
% in the context of the MSSM, 
$\lambda_{ijk}$ and
$\lambda'_{ijk}$
%$\lambda_{ijk}$ and $\lambda'_{ijk}$, 
cannot generate decays into up-type quarks and neutrinos as those shown in 
%unlike the
%$\mn$ where this is possible through the diagrams shown in 
Figs.~\ref{fig:efflambda-to-quarksu} and~\ref{fig:eff-to-neutrinos}. 
Also decays of the sneutrino into leptons with $i=j$, i.e. the sneutrino and one of the leptons belonging to the same family as in Figs.~\ref{fig:efflambda-to-leptons-HuHd} and~\ref{fig:eff-to-leptons-higgsino}, are not possible because of the antisymmetry of $\lambda_{ijk}$ with respect to their first two indexes. 
These decays present through the terms characteristic of the $\mn$ are however crucial for generating the signals at the LHC analyzed in this work.}
%$R_p$ turns out to be explicitly broken.
Nevertheless, $\lambda_i$ and $\kappa_{ijk}$ are obviously harmless with respect to proton decay.
% by the $\,5^{\rm th}$ and $6^{\rm th}$ terms. 
% Following the same argument as above for the natural presence of lepton-number violating terms, given the same quantum numbers of $L_i$ and $H_d$,
% the presence of the $\lambda_i$ term is also natural once the Yukawa couplings for neutrinos $Y^{\nu}_{ij}$ are allowed in the superpotential.

Unlike
$\lambda_i$ and $\kappa_{ijk}$, the 
couplings ${\lambda}_{ijk}$ and ${\lambda}'_{ijk}$ 
%in the second line of superpotential (\ref{superpotentiallb}) 
are not useful 
to solve neither the $\mu$ problem nor to generate neutrino masses and mixing (which is the only confirmed source of new physics).
In addition, they
%The  
%couplings $\lambda_{ijk}$ and $\lambda'_{ijk}$ 
are constrained by existing bounds on quadratic coupling constant products 
$\lambda_{ijk} \lambda_{lmn}$, $\lambda_{ijk}\lambda'_{lmn} $ 
and $\lambda'_{ijk}\lambda'_{lmn}$ (see Refs.~\cite{Chemtob:2004xr,Barbier:2004ez,Dreiner:2012mx} for reviews).
Thus, one can neglect them for simplicity in the superpotential,\footnote{
Although $\lambda_{ijk}$ and $\lambda'_{ijk}$ will appear through loop processes even if they are not present at tree level, as shown in Ref.~\cite{Escudero:2008jg}, their contributions are
smaller than order $10^{-9}$. 
Obviously, 
all existing bounds on quadratic coupling constant products 
are satisfied, but these contributions are anyway negligible for studying physical processes.
Let us remark that $\lambda'_{ijk}$ are generated at one loop through the equation~\cite{Escudero:2008jg}
$\frac{d}{dt} \lambda'_{ijk}=\frac{1}{16 \pi^2 } \, Y_{d_{jk}} \, \gamma^{H_d}_{L_i}$, with
$\gamma^{H_d}_{L_i} = - Y_{\nu_{il}} \lambda_l$. 
However, for $\lambda_{ijk}$ higher order contributions are necessary.
The antisymmetric character under 
$i \leftrightarrow j$ of $\lambda_{ijk}$, makes 
the one-loop contribution identically zero, as can be seen
from the fact that the one-loop equation
$\frac{d}{dt} \lambda_{ijk}=\frac{1}{16 \pi^2 }\left(Y_{e_{jk}} \, \gamma^{H_d}_{L_i} + Y_{e_{ik}} 
\,  \gamma^{H_d}_{L_j}\right)$ cannot generate antisymmetric contributions.}, 
and use:
\bea
W = &&
\epsilon_{ab} \left(
Y^e_{ij} \, \hat H_d^a\, \hat L^b_i \, \hat e_j^c +
Y^d_{ij} \, \delta_{\alpha\beta}\, \hat H_d^a\, \hat Q^{b}_{i\alpha} \, \hat d_{j\beta}^{c} 
+
Y^u_{ij} \, \delta_{\alpha\beta}\, \hat H_u^b\, \hat Q^{a}_{i\alpha} \, \hat u_{j\beta}^{c}
\right)
\nonumber\\
% &+&
% \epsilon_{ab} Y^{\nu}_{ij} \, \hat H_u^b\, \hat L^a_i \, \hat \nu^c_j -
&+&   
\epsilon_{ab} \left(
Y^{\nu}_{ij} \, \hat H_u^b\, \hat L^a_i \, \hat \nu^c_j 
-
%\epsilon_{ab}
\lambda_{i} \, \hat \nu^c_i\, \hat H_u^b \hat H_d^a
\right)
+
\frac{1}{3}
\kappa{_{ijk}} 
\hat \nu^c_i\hat \nu^c_j\hat \nu^c_k
% \nonumber\\
% &+&   
% \epsilon_{ab} \left(\lambda_{ijk} \hat L_i^a \hat L_j^b \hat e^c_k 
% + 
% \delta_{\alpha\beta} \lambda'_{ijk} \hat L_i^a \hat Q_j^{b \alpha} \hat d^{c \beta}_k \right)
\ .
\label{superpotential}
\eea
By the same token, the soft trilinear parameters in Eq.~(\ref{trilinear}) 
%and~(\ref{tmunu}) 
below can also be neglected.
% Nevertheless, the formulas given in the text and 
% Appendix~\ref{masasm} include for completeness the contributions from ${\lambda}_{ijk}$, ${\lambda'}_{ijk}$ and  their corresponding soft trilinear parameters.

% In the limit $Y^\nu_{ij} \to 0$, $\hat \nu^c_i$ can be identified in the superpotential above as pure singlet superfields without lepton number, similar to the case of the NMSSM where one extra singlet is added to the spectrum of the MSSM and $R_p$ is not broken.Thus $Y^\nu_{ij}$ are the parameters which determine the $\rpv$ in the superpotential,
% and this violation is therefore small.

\vspace{0.25cm}

\noindent 
{\bf Soft terms}

\noindent
Working in the framework of a typical low-energy SUSY, the Lagrangian 
containing the soft SUSY-breaking terms related to the 
superpotential in Eq.~(\ref{superpotentiallb}) is given by:
\bea
-\mathcal{L}_{\text{soft}}  =&&
\epsilon_{ab} \left(
T^e_{ij} \, H_d^a  \, \widetilde L^b_{iL}  \, \widetilde e_{jR}^* +
T^d_{ij} \, H_d^a\,   \widetilde Q^b_{iL} \, \widetilde d_{jR}^{*} 
+
T^u_{ij} \,  H_u^b \widetilde Q^a_{iL} \widetilde u_{jR}^*
%^{^*} 
%\widetilde d_{jR}^{^{^*}} 
%^{^*}  
+ \text{h.c.}
\right)
\nonumber \\
&+&
\epsilon_{ab} \left(
T^{\nu}_{ij} \, H_u^b \, \widetilde L^a_{iL} \widetilde \nu_{jR}^*
%^{^*} 
- 
T^{\lambda}_{i} \, \widetilde \nu_{iR}^*
%^{^*}
\, H_d^a  H_u^b
+ \frac{1}{3} T^{\kappa}_{ijk} \, \widetilde \nu_{iR}^*
%^{^*} 
\widetilde \nu_{jR}^*
%^{^*} 
\widetilde \nu_{kR}^*
%^{^*}
\
+ \text{h.c.}\right)
\nonumber\\
&+&   
\epsilon_{ab} \left(T^{\lambda}_{ijk}\, \widetilde L_{iL}^a \, \widetilde L_{jL}^b \, \widetilde e_{kR}^*
%\hat e^c_k 
+ 
T^{\lambda'}_{ijk} \, \widetilde L_{iL}^a \, \widetilde Q_{jL}^{b} \,  \widetilde d_{kR}^{*} 
\
+ \text{h.c.}
%\hat d^{c}_{k} 
\right)
\nonumber \\
&+& 
\left(m_{\widetilde{Q}_L}^2\right)_{ij} 
%\widetilde{Q}_{iL}^{a^{^*}}  
\widetilde{Q}_{iL}^{a*}
\widetilde{Q}^a_{jL}
+\left(m_{\widetilde{u}_R}^{2}\right)_{ij} \widetilde{u}_{iR}^*
%^{^*}  
\widetilde u_{jR}
+ \left(m_{\widetilde{d}_R}^2\right)_{ij}  \widetilde{d}_{iR}^*
%^{^{^*}}  
\widetilde d_{jR}
+
\left(m_{\widetilde{L}_L}^2\right)_{ij}  
%\widetilde{L}_{iL}^{a^{^*}}  
\widetilde{L}_{iL}^{a*}  
\widetilde{L}^a_{jL}
\nonumber\\
&+&
\left(m_{\widetilde{\nu}_R}^2\right)_{ij} \widetilde{\nu}_{iR}^*
%^{^*} 
\widetilde\nu_{jR} 
+
\left(m_{\widetilde{e}_R}^2\right)_{ij}  \widetilde{e}_{iR}^*
%^{^*}  
\widetilde e_{jR}
+ 
m_{H_d}^2 {H^a_d}^*
%{^*} 
H^a_d + m_{H_u}^2 {H^a_u}^*
%^{^*} 
H^a_u
\nonumber \\
&+&  \frac{1}{2}\, \left(M_3\, {\widetilde g}\, {\widetilde g}
+
M_2\, {\widetilde{W}}\, {\widetilde{W}}
+M_1\, {\widetilde B}^0 \, {\widetilde B}^0 + \text{h.c.} \right)\ ,
% &+&  \frac{1}{2}\, \left(M_3\, \lambda_{\tilde g}\, \lambda_{\tilde g}
% +
% M_2\, \lambda_{\tilde{W}}\, \lambda_{\tilde{W}}
% +M_1\, \lambda_{\tilde B} \, \lambda_{\tilde B} + \text{c.c.} \right)\ ,
%&-&  \frac{1}{2}\, \left(M_3\, \widetilde\lambda_3\, \widetilde\lambda_3+M_2\,
%  \widetilde\lambda_2\, \widetilde
%\lambda_2
%+M_1\, \widetilde\lambda_1 \, \widetilde\lambda_1 + \text{c.c.} \right) \,.
\label{2:Vsoft}
\eea
where an implicit sum over the (undisplayed) color indexes is assumed in the terms involving squarks and gluinos.
The complex trilinear parameters $T^{\lambda}_i$, 
$T^{d,e,u,\nu}_{ij}$ and $T^{\kappa,\lambda,\lambda'}_{ijk}$
are 
%$3\times 3$ matrices, a vector and a symmetric tensor, respectively, 
in correspondence with
the trilinear couplings of the superpotential.
The squared sfermion
%squark and slepton 
masses are required to be $3\times 3$ hermitian matrices in family space, whereas $m_{H_{u,d}}$ are the real Higgs mass parameters.
% , and
% $m_{\footnotesize{\widetilde{\psi}}}^2$ with
% $\widetilde{\psi}$ the scalar components of the superfields,
% $\hat{Q}, \hat{L}, \hat{u}^c, \hat{d}^c, \hat{e}^c, \hat{\nu}^c$,
% are $3\times 3$ hermitian matrices in family space (SHOW IT).
The parameters
$M_{3,2,1}$ are the (generally complex) Majorana masses of the 2-component gluino, Wino and Bino fields, and
an implicit sum over the (undisplayed) adjoint representation gauge indexes on the gluino and Wino fields is assumed.

Soft masses of the type
$m^{2}_{H_d \tilde L_{iL}} {H_d^a}^*  \tilde L_{iL}^a + \text{h.c.}$,
could have been included in Eq.~(\ref{2:Vsoft}). 
However, they would contribute to the minimization equations of the left sneutrinos with terms $m^{2}_{H_d \tilde L_{iL}}\langle H_d^0 \rangle$ (in the right-hand side of Eq.~(\ref{tadpoles4})), generating VEVs $\sim$ TeV for them.
This would spoil the generalized electroweak-scale seesaw present
in the $\mn$, where correct neutrino masses require the VEVs of the left sneutrinos to be small,
$\langle \tilde\nu_L \rangle\lsim 10^{-4}$ GeV,
 driven dynamically by the Yukawa couplings.
%as in (\ref{vevdirac}).
As discussed in Eq.~(\ref{vevdirac}),
%Eq.~\ref{Limit no mixing Higgsinos gauginos} of Section~\ref{masses}, 
these small VEVs are necessary because neutrino masses acquire a term of the 
order of ${\langle \tilde\nu_L \rangle}^2/M$, with $M\sim$ gaugino masses.
Thus we will assume that the above soft masses are not present in our Lagrangian or that they are negligible.\footnote{
Although they will appear through loop processes even if they are not present at tree level (see e.g. Ref.~\cite{deCarlos:1996ecd}), their contributions are negligible.}
Notice that a similar destabilization of the left sneutrino VEVs would arise
with trilinear parameters $T^{\nu}\sim$ TeV.
This can be avoided for example if the 
$T^{\nu}$ are proportional to the small $Y^{\nu}$, i.e. $T^{\nu}=A^{\nu}Y^{\nu}$ where $A_{\nu}$ can be $\sim$ TeV. 

Both assumptions above about the parameters $T^{\nu}$ and
$m_{H_d \tilde L}^2$ 
are reliable in the framework of the current studies of SUSY.
%below Eq.~(\ref{2:Vsoft}).
Let us recall in this sense that
strong upper bounds upon the intergenerational scalar mixing 
exist (see e.g. Ref.~\cite{Gabbiani:1996hi}), implying that one has to assume that 
such mixings are negligible, and therefore that the squared sfermion mass 
matrices in Eq.~(\ref{2:Vsoft}) are diagonal in the flavor space.
%,
%$m_{\widetilde{\psi}_{ij}}^2=\delta_{ij} m_{\widetilde{\psi}_{j}}^2$.
% We will use this assumption in order to simplify the analysis when discussing the neutral scalar mass matrices
% in Subsection~\ref{masses}, and the parameter space of the model
% in Subsection~\ref{parametre}. 
Actually, diagonal squared mass matrices occur in general in supergravity models when the observable matter fields have a diagonal 
K$\ddot{a}$hler metric, such as in several string compactifications
%or when the dilaton field is the source of SUSY breaking 
(for a review see Ref.~\cite{Brignole:1997dp}).
%This occurs for example in several string compactifications as a consequence of 
%having diagonal
%Kahler metrics, or when the dilaton is the source of SUSY breaking \cite{bim}. 
Also in this case of a diagonal metric, the soft trilinear parameters turn out to be directly proportional to the couplings present in the superpotential. Even with a 
general K$\ddot{a}$hler metric, these parameters are already functions of the couplings and their derivatives with respect to the hidden sector fields.
Inspired by this structure of supergravity, 
and also by the interpretation of the Higgs $H_d$ as a fourth-family slepton $\tilde L_4$, one can consider that
soft masses of the type $m^{2}_{H_d \tilde L_i}$ are not present in the Lagrangian, and also assume
the values for soft trilinear parameters given in 
Eqs.~(\ref{tyukawa}),~(\ref{tmunu}), and
\bea
% T^{e}_{ij} &=& A^{e}_{ij} Y^{e}_{ij}\ , \;\;
% T^{d}_{ij} = A^{d}_{ij} Y^{d}_{ij}\ , \;\;
% T^{u}_{ij} = A^{u}_{ij} Y^{u}_{ij}\ ,
% \label{tyukawa}
% \\
% T^{\nu}_{ij} &=& A^{\nu}_{ij} Y^{\nu}_{ij}\ , \;\;
% T^{\lambda}_i= A^{\lambda}_i\lambda_i\ , \;\;
% T^{\kappa}_{ijk}= A^{\kappa}_{ijk} \kappa_{ijk}\ ,
% \label{tmunu}
%\\
T^{\lambda}_{ijk} &=& A^{\lambda}_{ijk} \lambda_{ijk}\ , \;\;
T^{\lambda'}_{ijk}= A^{\lambda'}_{ijk} \lambda'_{ijk}\ .
\label{trilinear}
\eea
%where the summation convention on repeated indexes does not apply for this case.

% We have reviewed at this point the Lagrangian of the $\mn$.
% Let us then discuss in the next subsection the minimization of the scalar potential.

%%%%%%%%%%%%%%%%%%%%%%%%%%%%%%%%%%%%%%%%%%%%%%%%%%%%%%%%%%%%%%
%%%%%%%%%%%%%%%%%%%%%%%%%%%%%%%%%%%%%%%%%%%%%%%%%%%%%%%%%%%%%%%%

% \vspace{0.25cm}

% \noindent 
% {\bf The Scalar Potential}
% \label{minimization}

\section{Mass Matrices}

\label{masasm}

We write below the tree-level mass matrices generated in the $\mn$.
Upon EWSB, fields with the same color, electric charge and spin mix. To name them 
%when mixing occurs 
we follow the 
%SARAH 
convention of using for the eigenstates the names of detected particles: Higgs, neutrinos, leptons.
In what follows we use $i,j,k,l,m,n$ as family indexes,
and $a,b$ as the indices for the physical states (mass eigenstates), not to be confused with $a,b=1,2$ used in Appendix~\ref{Section:munuSSM} as $SU(2)_L$ index.
We include in the formulas for completeness the contribution due to lepton-number violating couplings $\lambda_{ijk}$ and $\lambda'_{ijk}$ 
in the superpotential of Eq.~(\ref{superpotentiallb}) and
soft Lagrangian of Eq.~(\ref{2:Vsoft}).

%%%%%%%%%%%%%%%%%%%%%%%%%%%%%%%%%%%%%%%%%%%%%%%%%%%%%%%%%%%%%%%%%
%%%%%%%%%%%%%%%%%%%%%%%%%%%%%%%%%%%%%%%%%%%%%%%%%%%%%%%%%%%%%%%%

\vspace{0.5cm}

\noindent 
%\subsection
{\bf B.1 Scalar Mass Matrices}
%\label{SubAppendix:ScalarMasses}
\label{Apendix:Sneutrino-masses}
%%%%%%%%%%%%%%%%%%%%%%%%%%%%%%%%%%%%%%%%%%%%%%%%%%%%%%%%%%%%%%%%%
%%%%%%%%%%%%%%%%%%%%%%%%%%%%%%%%%%%%%%%%%%%%%%%%%%%%%%%%%%%%%%%%
% \section{Mass matrices \label{appx:mass}}
% \label{Apendix:Sneutrino-masses}
%%%%%%%%%%%%%%%%%%%%%%%%%%%%%%%%%%%%%%%%%%%%%%%%%%%%%%%%%%%%%%%%

The scalar mass matrices generated in the $\mu\nu$SSM were computed 
in Appendix A.1 of Ref.~\cite{Escudero:2008jg} with the assumption of CP conservation for simplicity.
% the scalar mass matrices generated in the $\mu\nu$SSM are written. 
In this Appendix, we write those equations and
%for simplicity of the computation, 
%using those matrices (SHOULD WE ALSO COPY THEM?) we write the results 
replace the values of the soft masses obtained through the minimization conditions in Eqs.~(\ref{tadpoles1})-(\ref{tadpoles4}), assuming that
slepton soft mass matrices are diagonal in flavor space.

\vspace{0.25cm}
\noindent
%%%%%%%%%%%%%%%%%%%%%%%%%%%%%%%%%%%%%%%%%%%%%%%%%%%%%%%%%%%%%%%%%
%%%%%%%%%%%%%%%%%%%%%%%%%%%%%%%%%%%%%%%%%%%%%%%%%%%%%%%%%%%%%%%%
%\subsubsection
{\bf Mass Matrix for Higgses
%CP-even neutral scalars
}
\label{SubAppendix-scalarmasses}
%%%%%%%%%%%%%%%%%%%%%%%%%%%%%%%%%%%%%%%%%%%%%%%%%%%%%%%%%%%%%%%%%
%%%%%%%%%%%%%%%%%%%%%%%%%%%%%%%%%%%%%%%%%%%%%%%%%%%%%%%%%%%%%%%%

\noindent
Higgses mix with left and right sneutrinos. In the basis  
$S^T=(H_{d}^\mathcal{R},H_{u}^\mathcal{R},\tilde{\nu}^{\mathcal{R}}_{iR},
\tilde{\nu}_{jL}^\mathcal{R})$, one obtains the following mass terms for scalar Higgses in the
Lagrangian:
%The scalar potential includes the following quadratic term for the CP-even
%neutral scalars:
%%%%%%%%%%%%%%%%%%%%%%%%%%%%%%%%%%%%%%%%%%%%%%%%%%%%%%%%%%%%%%%%%
%%%%%%%%%%%%%%%%%%%%%%%%%%%%%%%%%%%%%%%%%%%%%%%%%%%%%%%%%%%%%%%%
% \beq
% \mathbf{S}^\mathcal{R}_{\alpha} m^2_{S^\mathcal{R}_{\alpha} S^\mathcal{R}_{\beta}} \mathbf{S}^\mathcal{R}_{\beta}\ ,
% \label{matrix1}
% \eeq
\begin{equation}
-\frac{1}{2} S^T 
%\mathcal{M}_{\mathrm{n}}
%\mathcal{M}_{\chi^0} 
%\mathcal{M}_{\nu} 
{m}^2_{h} S\, ,
\label{matrix1}
\end{equation}
%%%%%%%%%%%%%%%%%%%%%%%%%%%%%%%%%%%%%%%%%%%%%%%%%%%%%%%%%%%%%%%%%
%%%%%%%%%%%%%%%%%%%%%%%%%%%%%%%%%%%%%%%%%%%%%%%%%%%%%%%%%%%%%%%%
where ${m}^2_{h}$ is the 
$8\times 8$ (symmetric) matrix obtained computing the second derivative of the scalar potential of Eq.~(\ref{finalpotential}) with respect to the fields
\begin{align}
m_{h}^2= 
  \left( \begin{array}{cccc} 
  m_{H_{d}^\mathcal{R}H_{d}^\mathcal{R}}^{2} & 
m_{H_{d}^\mathcal{R}H_{u}^\mathcal{R}}^{2} & 
m_{H_{d}^\mathcal{R}\widetilde{\nu}_{jR}^{\mathcal{R}}}^{2} & 
m_{H_{d}^\mathcal{R}\widetilde{\nu}_{jL}^\mathcal{R}}^{2}
\\
m_{H_{u}^\mathcal{R}H_{d}^\mathcal{R}}^{2} & 
m_{H_{u}^\mathcal{R}H_{u}^\mathcal{R}}^{2}& 
m_{H_{u}^\mathcal{R}\widetilde{\nu}_{jR}^\mathcal{R}}^{2}  & 
m_{H_{u}^\mathcal{R}\widetilde{\nu}_{jL}^\mathcal{R}}^{2}
\\
 m_{\widetilde{{\nu}}^{\mathcal{R}}_{iR} H_{d}^\mathcal{R}}& 
m_{\widetilde{{\nu}}^{\mathcal{R}}_{iR} H_{u}^\mathcal{R}}&   
m_{\widetilde{{\nu}}^{\mathcal{R}}_{iR} \widetilde{{\nu}}^{\mathcal{R}}_{jR}}^{2} &   
m_{\widetilde{{\nu}}^{\mathcal{R}}_{iR} \widetilde{{\nu}}^{\mathcal{R}}_{jL}}^{2} 
\\
m_{\widetilde{\nu}_{iL}^\mathcal{R} H_{d}^\mathcal{R}}^{2}  & 
 m_{\widetilde{\nu}_{iL}^\mathcal{R} H_{u}^\mathcal{R}}^{2} & 
m_{\widetilde{\nu}_{iL}^\mathcal{R} \widetilde{\nu}^{\mathcal{R}}_{jR}}^{2} &   
m_{\widetilde{\nu}_{iL}^\mathcal{R} \widetilde{\nu}_{jL}^\mathcal{R}}^{2}
         \end{array} \right)\ ,  
\end{align}
{\footnotesize
\bea
m_{H_{d}^\mathcal{R}H_{d}^\mathcal{R}}^{2} &=&  
m_{H_d}^{2}+\frac{1}{8}(g^2 + g'^2)(3v_{d}^{2}-v_{u}^{2}+v_{iL}v_{iL})+
\frac{1}{2}\lambda_{i}\lambda_{j}v_{iR} v_{jR} 
+\frac{1}{2}\lambda_{i}\lambda_{i}v_{u}^{2}
%+ V^{(n)}_{v_dv_d}
\nonumber\\
%&&
&=&
\frac{1}{4}(g^2 + g'^2) v_{d}^{2}
+
v_{iR} \mathrm{tan} 
\beta\left(\frac{1}{\sqrt 2}T^{\lambda}_i+\frac{1}{2}\lambda_{j}\kappa_{ijk}v_{kR}\right)
+Y^{\nu}_{ij}\frac{v_{iL}}{2v_d}\left(\lambda_k v_{jR}v_{kR}+\lambda_j v_u^2\right)
%  \nonumber\\
% &&
% + V^{(n)}_{v_dv_d}-\frac{\sqrt 2}{v_d}V^{(n)}_{v_d}
\ ,
   \label{prueba1}
\\ 
\nonumber
\\
m_{H_{u}^\mathcal{R}H_{u}^\mathcal{R}}^{2}&=&
m_{H_u}^{2}+\frac{1}{8}(g^2 + g'^2)(-v^2_{d}+3v_{u}^{2}-v_{iL}v_{iL})+ 
\frac{1}{2}\lambda_{i}\lambda_{j} v_{iR} v_{jR} +\frac{1}{2}\lambda_{i}\lambda_{i}v_{d}^2 
\nonumber \\ 
&&
-Y^{\nu}_{ij} \lambda_{j}v_{d}v_{iL}+ \frac{1}{2}Y^{\nu}_{ik} Y^{\nu}_{ij}v_{jR} 
v_{kR}+\frac{1}{2}Y^{\nu}_{ik}Y^{\nu}_{jk} v_{iL}v_{jL}
%+ V^{(n)}_{v_uv_u}
\nonumber\\
&=&
\frac{1}{4}(g^2 + g'^2) v_{u}^{2}
+
v_{iR} \frac{1}{\mathrm{tan} \beta}
\left(\frac{1}{\sqrt 2}T^{\lambda}_i
+\frac{1}{2}\lambda_{j}\kappa_{ijk} v_{kR}\right)
\nonumber\\
&&
-\frac{v_{iL}}{v_u}\left(\frac{1}{\sqrt 2}T^{\nu}_{ij} v_{jR}+
\frac{1}{2}Y^{\nu}_{ij}\kappa_{ljk}v_{lR}v_{kR}\right)
%  \nonumber\\
% &&
% + V^{(n)}_{v_uv_u}-\frac{\sqrt 2}{v_u}V^{(n)}_{v_u}
\ ,
   \label{prueba2}
\\ 
\nonumber
\\
m_{H_{u}^\mathcal{R} H_{d}^\mathcal{R}}^{2}&=&-
\frac{1}{4}(g^2 + g'^2) v_{d}v_{u}
-\frac{1}{\sqrt 2}T^{\lambda}_{i} v_{iR} - \frac{1}{2}\lambda_{k}\kappa_{ijk} v_{iR} v_{jR}
+v_{d}v_{u}\lambda_{i}\lambda_{i}
-Y^{\nu}_{ij}\lambda_{{j}}v_{u}v_{iL}
% +V^{(n)}_{v_dv_u}
\ , 
   \label{prueba3}
\\ 
\nonumber
\\
m_{\widetilde{\nu}_{iR}^{\mathcal{R}} H_{d}^\mathcal{R}}^{2}&=&-\frac{1}{\sqrt 2}T^{\lambda}_{i}v_{u}-\lambda_{k} \kappa_{ijk}v_{u}v_{jR}+\lambda_{i}\lambda_{j}v_{d}v_{jR} 
-\frac{1}{2}Y^{\nu}_{ji}\lambda_{k}v_{jL}v_{kR}
-\frac{1}{2}Y^{\nu}_{jk}\lambda_{i}v_{jL}v_{kR}
%+ V^{(n)}_{v_d v_{iR}}
\ ,
   \label{Adr}
\\ 
\nonumber
\\
m_{\widetilde{\nu}_{iR}^{\mathcal{R}} H_{u}^\mathcal{R}}^{2}&=&
-\frac{1}{\sqrt 2}T^{\lambda}_{i}v_{d}+ \frac{1}{\sqrt 2}T^{\nu}_{ji} v_{jL}-\lambda_{k}\kappa_{ilk}v_{d}v_{lR}
+\lambda_{i}\lambda_{j}v_{u}v_{jR}
+ Y^{\nu}_{jk} \kappa_{ilk} v_{jL} v_{lR}  
+ Y^{\nu}_{jk} Y^{\nu}_{ji} v_{u} v_{kR}
%  \nonumber\\
% && 
% + V^{(n)}_{v_u v_{iR}}
\ ,
\label{Aur}
\\ 
\nonumber
\\
m_{\widetilde{{\nu}}^{\mathcal{R}}_{iR} \widetilde{{\nu}}^{\mathcal{R}}_{jR}}^{2}&=&
\left(m^2_{\widetilde{\nu}_R}\right)_{ij}
+{\sqrt 2}T^{\kappa}_{ijk}v_{kR}-\lambda_{k}\kappa_{ijk}v_{d}v_{u}+ \kappa_{ijk}\kappa_{lmk}v_{lR}v_{mR}+2\kappa_{ilk}\kappa_{jmk}v_{lR}v_{mR}
\nonumber \\
&&
+
\frac{1}{2}\lambda_{i}\lambda_{j}(v_{d}^{2}+v_{u}^{2})
+Y^{\nu}_{lk}\kappa_{ijk}v_{u}v_{lL}
-\frac{1}{2}(Y^{\nu}_{kj}\lambda_{i}+Y^{\nu}_{ki}\lambda_{j})v_{d}v_{kL}
+\frac{1}{2}Y^{\nu}_{ki}Y^{\nu}_{kj}v_{u}^{2}
+\frac{1}{2}Y^{\nu}_{ki}Y^{\nu}_{lj}v_{kL}v_{lL}
%+V^{(n)}_{v_{iR}v_{jR}}
\nonumber\\
&=&
{\sqrt 2}T^{\kappa}_{ijk} v_{kR}-\lambda_{k}\kappa_{ijk}v_{d}v_{u}+\kappa_{ijk}\kappa_{lmk}v_{lR}v_{mR}+2\kappa_{ilk}\kappa_{jmk}v_{lR}v_{mR}
+\frac{1}{2}\lambda_{i}\lambda_{j}(v_{d}^{2}+v_{u}^{2})
\nonumber\\
&&
-Y^{\nu}_{lk}\kappa_{ijk}v_{u}v_{lL}
-\frac{1}{2}
\left(Y^{\nu}_{kj}\lambda_{i}+Y^{\nu}_{ki}\lambda_{j}\right) v_{d}v_{kL}
+\frac{1}{2}Y^{\nu}_{ki}Y^{\nu}_{kj}v_{u}^{2}+\frac{1}{2}Y^{\nu}_{li}Y^{\nu}_{kj}v_{kL}v_{lL}
\nonumber\\
&&
+\frac{\delta_{ij}}{v_{jR}}
\left[-\frac{1}{\sqrt 2}T^{\nu}_{ki}v_{kL}v_u+\frac{1}{\sqrt 2}T^{\lambda}_iv_uv_d
-\frac{1}{\sqrt 2}T^{\kappa}_{ilk}v_{lR}v_{kR}
+\lambda_{l}\kappa_{ilk}v_{d}v_{u}v_{kR}
\right.\nonumber\\
&&
-\kappa_{lim}\kappa_{lnk}v_{mR}v_{nR}v_{kR}
-\frac{1}{2}\lambda_{i}\lambda_{l}(v_{d}^{2}+v_{u}^{2})v_{lR}
-Y^{\nu}_{lk}\kappa_{ikm}v_{u}v_{lL}v_{mR}
\nonumber\\
&&\left.
+\frac{1}{2}\left(Y^{\nu}_{kl}\lambda_{i}
+Y^{\nu}_{ki}\lambda_{l}\right)v_{d}v_{kL}v_{lR}
-\frac{1}{2}Y^{\nu}_{ki}Y^{\nu}_{kl}v_{u}^{2}v_{lR}
-\frac{1}{2}Y^{\nu}_{ki}Y^{\nu}_{lm}v_{kL}v_{lL}v_{mR}\right]
%+V^{(n)}_{v_{iR}v_{jR}}-\frac{{\sqrt 2}\delta_{ij}}{v_{iR}}V^{(n)}_{v_{iR}}
\ ,
\label{evenrr}
\\  
\nonumber
\\
m_{\widetilde{\nu}_{iL}^\mathcal{R} H_{d}^\mathcal{R}}^{2}&=&
\frac{1}{4}(g^2 + g'^2) v_{d}v_{iL}-\frac{1}{2}Y^{\nu}_{ij}\lambda_{j}v_{u}^{2}
-\frac{1}{2}Y^{\nu}_{ij}\lambda_{k}v_{kR}v_{jR}
%+V^{(n)}_{v_dv_{iL}}
\ ,
   \label{prueba5}
\\ 
\nonumber
\\
m_{\widetilde{\nu}_{iL}^\mathcal{R} H_{u}^\mathcal{R}}^{2}&=&-
\frac{1}{4}(g^2 + g'^2) v_{u}v_{iL} 
+ \frac{1}{\sqrt 2}T^{\nu}_{ij}v_{jR}
+\frac{1}{2}Y^{\nu}_{ik}\kappa_{ljk}v_{lR}v_{jR}
-Y^{\nu}_{ij}\lambda_{j}v_{d}v_{u}
+Y^{\nu}_{ij}Y^{\nu}_{kj}v_{u}v_{kL}
%+V^{(n)}_{v_u v_{iL}}
\ ,
\label{upleading}
\\ 
\nonumber
\\
% m_{\widetilde{\nu}_{jR}^\mathcal{R} \widetilde{\nu}^{\mathcal{R}}_{iL}}^{2}
% &=&
m_{\widetilde{\nu}_{iL}^\mathcal{R} \widetilde{\nu}^{\mathcal{R}}_{jR}}^{2} &=&
\frac{1}{\sqrt 2}T^{\nu}_{ij}v_{u}
-\frac{1}{2}Y^{\nu}_{ij}\lambda_{k} v_{d}v_{kR}-\frac{1}{2}Y^{\nu}_{ik}\lambda_{j}v_{d}v_{kR}
+Y^{\nu}_{ik}\kappa_{jlk}v_{u}v_{lR}
+\frac{1}{2}Y^{\nu}_{ij}Y_{\nu_{kl}}v_{kL}v_{lR}
\nonumber\\
&&
+\frac{1}{2}Y^{\nu}_{il}Y^{\nu}_{kj}v_{kL}v_{lR}
%+V^{(n)}_{v_{iL} v_{jR}}
\ , 
   \label{prueba6}
\\ 
\nonumber
\\
m_{\widetilde{\nu}_{iL}^\mathcal{R} \widetilde{\nu}_{jL}^\mathcal{R}}^{2} &=&
\left(m^2_{\widetilde{L}_L}\right)_{ij}
+
\frac{1}{4}(g^2 + g'^2)v_{iL}v_{jL}+\frac{1}{8}(g^2 + g'^2)
(v_{kL}v_{kL}+v_{d}^{2}-v_{u}^{2})\delta_{ij}
\nonumber\\
&&
+
\frac{1}{2}Y^{\nu}_{ik}Y^{\nu}_{jk}v^2_{u}+
\frac{1}{2}Y^{\nu}_{ik}Y^{\nu}_{jl}v_{kR}v_{lR}
%+V^{(n)}_{v_{iL} v_{jL}}
\nonumber\\
&=&
\frac{1}{4}(g^2 + g'^2) v_{iL}v_{jL}
+
\frac{1}{2}Y^{\nu}_{ik}Y^{\nu}_{jk}v^2_{u}
+
\frac{1}{2}Y^{\nu}_{ik}Y^{\nu}_{jl}v_{kR}v_{lR}
+\frac{\delta_{ij}}{v_{jL}}
\left[-\frac{1}{\sqrt 2}T^{\nu}_{ik}v_u v_{kR}
\right.
\nonumber\\
&&
\left.
+\frac{1}{2}Y^{\nu}_{ik}
\left(\lambda_lv_d v_{kR} v_{lR}
+\lambda_k v_d v_u^2
-\kappa_{klm} v_u v_{lR}v_{mR}
-Y^{\nu}_{mk}v_{mL} v^2_{u}-Y^{\nu}_{ml} v_{mL}v_{lR}v_{kR}\right)
\right]
%+V^{(n)}_{v_{iL} v_{jL}}-\frac{{\sqrt 2}\delta_{ij}}{v_{iL}}V^{(n)}_{v_{iL}}
\ .
\label{evenLL}
\eea
}
%%%%%%%%%%%%%%%%%%%%%%%%%%%%%%%%%%%%%%%%%%%%%%%%%%%%%%%%%%%%%%%%%
%%%%%%%%%%%%%%%%%%%%%%%%%%%%%%%%%%%%%%%%%%%%%%%%%%%%%%%%%%%%%%%%
% Here $V^{(n)}_{xy} \equiv \frac{\partial^2V^{(n)}}{\partial x \partial y}$, 
% $V^{(n)}_{y} \equiv \frac{\partial V^{(n)}}{\partial y}$ with $x,\,y=v_d,\,v_u,\,v_{iL},\,v_{iR}$, and 
% $V^{(n)}$ represents the $n$--loop
% radiative correction to the potential $V=V^{(0)} + V^{(n)}$. 
This matrix is diagonalized by an orthogonal matrix
$Z^H$:
\bea
%\left(m^2_{S_{\alpha} S_{\beta}}\right)^{\text{dia}}
Z^H
m^2_{h}\ {Z^H}^{^T}=
\left(m^2_{h}\right)^{\text{dia}}
\ ,
\label{scalarhiggs}
\eea
with 
\begin{equation}
S = {Z^H}^{^T} h\, ,
\label{physscalarhiggses}
\end{equation}
where the 8 entries of the matrix $h$ are the `Higgs' mass eigenstate fields.
% In particular,
% \begin{equation}
% H_{d}^\mathcal{R} = Z^H_{k1} h_k\, ,\;\;\;
% H_{u}^\mathcal{R} = Z^H_{k2} h_k\, ,\;\;\;
% \tilde{\nu}_{iR}^\mathcal{R} = Z^H_{ki} h_k\, ,\;\;\;
% \tilde{\nu}_{jL}^\mathcal{R} = Z^H_{kj} h_k\, .
% \label{physscalarhiggsesw}
% \end{equation}
% In the case of considering only one generation of right-handed neutrinos as we do in 
% Appendix~\ref{Section:Coupling}, these formulas can be written as
% \begin{equation}
% H_{d}^\mathcal{R} = Z^H_{j1} h_j\, ,\;\;\;
% H_{u}^\mathcal{R} = Z^H_{j2} h_j\, ,\;\;\;
% \tilde{\nu}_{R}^\mathcal{R} = Z^H_{j3} h_j\, ,\;\;\;
% \tilde{\nu}_{3+aL}^\mathcal{R} = Z^H_{j3+a} h_j\, ,
% \label{physscalarhiggsesww}
% \end{equation}
% where we have defined $a=1,2,3$.
%%, not to be confused with $a=1,2$ used in Subsection~\ref{superpotentialsoft} as $SU(2)_L$ index.
In particular,
\begin{equation}
H_{d}^\mathcal{R} = Z^H_{b1} h_b\, ,\;\;\;
H_{u}^\mathcal{R} = Z^H_{b2} h_b\, ,\;\;\;
\tilde{\nu}_{iR}^\mathcal{R} = Z^H_{bi} h_b\, ,\;\;\;
\tilde{\nu}_{jL}^\mathcal{R} = Z^H_{bj} h_b\, .
\label{physscalarhiggsesws}
\end{equation}
% where we are defining $a,b=1,...,8$, not to be confused with $a,b=1,2$ used in Subsection~\ref{superpotentialsoft} as $SU(2)_L$ index.
In the case of considering only one family of right-handed neutrinos as we do in 
Appendix~\ref{Section:Coupling}, the last two equalities can be written as
\begin{equation}
% H_{d}^\mathcal{R} = Z^H_{b1} h_b\, ,\;\;\;
% H_{u}^\mathcal{R} = Z^H_{b2} h_b\, ,\;\;\;
\tilde{\nu}_{R}^\mathcal{R} = Z^H_{b3} h_b\, ,\;\;\;
\tilde{\nu}_{3+iL}^\mathcal{R} = Z^H_{b3+i} h_b\, .
\label{physscalarhiggseswwss}
\end{equation}

% ---

% $G^2\equiv g^2 + g'^2$, and the functions $V^{(n)}_{xy}$ and $V^{(n)}_{y}$ are the same
% as defined in Appendix \ref{SubAppendix:NeutrealScalar}.

% --

%Then the mass eigenvectors are
%%%%%%%%%%%%%%%%%%%%%%%%%%%%%%%%%%%%%%%%%%%%%%%%%%%%%%%%%%%%%%%%%
%%%%%%%%%%%%%%%%%%%%%%%%%%%%%%%%%%%%%%%%%%%%%%%%%%%%%%%%%%%%%%%%
% \bea
%  \mathbf{h}_{\alpha}=Z^H_{\alpha\beta} \mathbf{S}^\mathcal{R}_\beta\ ,
% \eea
%%%%%%%%%%%%%%%%%%%%%%%%%%%%%%%%%%%%%%%%%%%%%%%%%%%%%%%%%%%%%%%%%
%%%%%%%%%%%%%%%%%%%%%%%%%%%%%%%%%%%%%%%%%%%%%%%%%%%%%%%%%%%%%%%%
%with the diagonal mass matrix 
%with the eigenvalues $m^2_{h_{\alpha}}$ obtained from the diagonal mass matrix 
%%%%%%%%%%%%%%%%%%%%%%%%%%%%%%%%%%%%%%%%%%%%%%%%%%%%%%%%%%%%%%%%%
%%%%%%%%%%%%%%%%%%%%%%%%%%%%%%%%%%%%%%%%%%%%%%%%%%%%%%%%%%%%%%%%
%\bea
%\left(m^2_{S_{\alpha} S_{\beta}}\right)^{\text{dia}}
% \left(m^2_{h_{\alpha\beta}}\right)^{\text{dia}}
% =Z^H_{\alpha \gamma} 
% m^2_{S^\mathcal{R}_{\gamma} S^\mathcal{R}_{\delta}} Z^H_{\beta \delta}\ .
% \eea
%%%%%%%%%%%%%%%%%%%%%%%%%%%%%%%%%%%%%%%%%%%%%%%%%%%%%%%%%%%%%%%%%
%%%%%%%%%%%%%%%%%%%%%%%%%%%%%%%%%%%%%%%%%%%%%%%%%%%%%%%%%%%%%%%%

\vspace{0.25cm}
\noindent
%%%%%%%%%%%%%%%%%%%%%%%%%%%%%%%%%%%%%%%%%%%%%%%%%%%%%%%%%%%%%%%%%
%%%%%%%%%%%%%%%%%%%%%%%%%%%%%%%%%%%%%%%%%%%%%%%%%%%%%%%%%%%%%%%%
%\subsubsection
{\bf Mass Matrix for Pseudoscalar Higgses
%CP-odd neutral scalars
}
\label{SubAppendix:NeutrealScalar}
%%%%%%%%%%%%%%%%%%%%%%%%%%%%%%%%%%%%%%%%%%%%%%%%%%%%%%%%%%%%%%%%%
%%%%%%%%%%%%%%%%%%%%%%%%%%%%%%%%%%%%%%%%%%%%%%%%%%%%%%%%%%%%%%%%

\noindent
Following similar arguments as above,
%in Appendix~\ref{SubAppendix-scalarmasses},
in the basis  
$P^T=(H_{d}^\mathcal{I},H_{u}^\mathcal{I},\tilde{\nu}^{\mathcal{I}}_{iR},
\tilde{\nu}_{jL}^\mathcal{I})$, one obtains the following mass terms for pseudoscalar Higgses in the
Lagrangian:
\begin{equation}
-\frac{1}{2} P^T 
%\mathcal{M}_{\mathrm{n}}
%\mathcal{M}_{\chi^0} 
%\mathcal{M}_{\nu} 
{m}^2_{A^0} P\, ,
\label{matrix1p}
\end{equation}
%%%%%%%%%%%%%%%%%%%%%%%%%%%%%%%%%%%%%%%%%%%%%%%%%%%%%%%%%%%%%%%%%
%%%%%%%%%%%%%%%%%%%%%%%%%%%%%%%%%%%%%%%%%%%%%%%%%%%%%%%%%%%%%%%%
where ${m}^2_{A^0}$ is the $8\times 8$ (symmetric) matrix
\begin{align}
m^2_{A^0}= 
  \left( \begin{array}{cccc} 
  m_{H_{d}^\mathcal{I}H_{d}^\mathcal{I}}^{2} & 
m_{H_{d}^\mathcal{I}H_{u}^\mathcal{I}}^{2} & 
m_{H_{d}^\mathcal{I}\widetilde{\nu}_{jR}^{\mathcal{I}}}^{2} & 
m_{H_{d}^\mathcal{I}\widetilde{\nu}_{jL}^\mathcal{I}}^{2}
\\
m_{H_{u}^\mathcal{I}H_{d}^\mathcal{I}}^{2} & 
m_{H_{u}^\mathcal{I}H_{u}^\mathcal{I}}^{2}& 
m_{H_{u}^\mathcal{I}\widetilde{\nu}_{jR}^\mathcal{I}}^{2}  & 
m_{H_{u}^\mathcal{I}\widetilde{\nu}_{jL}^\mathcal{I}}^{2}
\\
 m_{\widetilde{{\nu}}^{\mathcal{I}}_{iR} H_{d}^\mathcal{I}}& 
m_{\widetilde{{\nu}}^{\mathcal{I}}_{iR} H_{u}^\mathcal{I}}&   
m_{\widetilde{{\nu}}^{\mathcal{I}}_{iR} \widetilde{{\nu}}^{\mathcal{I}}_{jR}}^{2} &   
m_{\widetilde{{\nu}}^{\mathcal{I}}_{iR} \widetilde{{\nu}}^{\mathcal{I}}_{jL}}^{2} 
\\
m_{\widetilde{\nu}_{iL}^\mathcal{I} H_{d}^\mathcal{I}}^{2}  & 
 m_{\widetilde{\nu}_{iL}^\mathcal{I} H_{u}^\mathcal{I}}^{2} & 
m_{\widetilde{\nu}_{iL}^\mathcal{I} \widetilde{\nu}^{\mathcal{I}}_{jR}}^{2} &   
m_{\widetilde{\nu}_{iL}^\mathcal{I} \widetilde{\nu}_{jL}^\mathcal{I}}^{2}
         \end{array} \right)\ , 
\label{matrixscalar2} 
\end{align}

{\footnotesize
\bea
m_{H_{d}^\mathcal{I}H_{d}^\mathcal{I}}^{2} &=& 
m_{H_{d}^\mathcal{R}H_{d}^\mathcal{R}}^{2}
- \frac{1}{4}(g^2 + g'^2) v_{d}^{2}
   \label{pruebas1}
\\ 
\nonumber
\\
m_{H_{u}^\mathcal{I}H_{u}^\mathcal{I}}^{2} &=& 
m_{H_{u}^\mathcal{R}H_{u}^\mathcal{R}}^{2}
- \frac{1}{4}(g^2 + g'^2) v_{u}^{2}
\ ,
   \label{pruebas2}
\\ 
\nonumber
\\
m_{H_{u}^\mathcal{I}H_{d}^\mathcal{I}}^{2}&=&
\frac{1}{\sqrt 2}T^{\lambda}_{i} v_{iR}+\frac{1}{2}\lambda_{k}\kappa_{ijk}v_{iR}v_{jR}
%+ V^{(n)}_{v_dv_u}
\ ,
   \label{pruebas3}
\\ 
\nonumber
\\
m_{\widetilde{\nu}^{\mathcal{I}}_{iR}H_{d}^\mathcal{I}}^{2}&=&
\frac{1}{\sqrt 2}T^{\lambda}_{i} v_{u}-\lambda_{k}\kappa_{ijk}v_{u}v_{jR}
-\frac{1}{2}Y^{\nu}_{ji}\lambda_{k} v_{jL} v_{kR}+
\frac{1}{2}Y^{\nu}_{jk}\lambda_{i}v_{jL} v_{kR}
%+ V^{(n)}_{v_d v_{iR}}
\ ,
   \label{pruebas4}
\\ 
\nonumber
\\
m_{\widetilde{\nu}^{\mathcal{I}}_{iR}H_{u}^\mathcal{I}}^{2}&=&
\frac{1}{\sqrt 2}T^{\lambda}_{i} v_{d}-\frac{1}{\sqrt 2}T^{\nu}_{ji} v_{jR}
-\lambda_{k}\kappa_{ilk}v_{d}v_{lR}+Y^{\nu}_{jk}\kappa_{ilk} v_{jL}v_{lR}
%+V^{(n)}_{v_uv_{iR}}
\ ,
   \label{pruebas5}
\\ 
\nonumber
\\
m_{\widetilde{{\nu}}^{\mathcal{I}}_{iR} \widetilde{{\nu}}^{\mathcal{I}}_{jR}}^{2}&=&
m_{\widetilde{{\nu}}^{\mathcal{R}}_{iR} \widetilde{{\nu}}^{\mathcal{R}}_{jR}}^{2}
-
2\left({\sqrt 2}T^{\kappa}_{ijk}v_{kR}-\lambda_{k}\kappa_{ijk}v_{d}v_{u}+\kappa_{ijk}\kappa_{lmk}v_{lR}v_{mR}\right)
\ ,
\label{oddRR}
\\ 
\nonumber
\\
m_{\widetilde{\nu}_{iL}^\mathcal{I}H_{d}^\mathcal{I}}^{2}&=&
-\frac{1}{2}Y^{\nu}_{ij}\lambda_{j} v_{u}^{2}-\frac{1}{2}Y^{\nu}_{ij}\lambda_{k}v_{kR}v_{jR}
%+V^{(n)}_{v_dv_{iL}}
\ ,
   \label{pruebas6}
\\ 
\nonumber
\\
m_{\widetilde{\nu}_{iL}^\mathcal{I}H_{u}^\mathcal{I}}^{2}&=&-\frac{1}{\sqrt 2}T^{\nu}_{ij}v_{jR} - \frac{1}{2}
Y^{\nu}_{ik}\kappa_{ljk}v_{lR}v_{jR}
%+V^{(n)}_{v_uv_{iL}}
\ ,
\label{figure2down}
   \label{pruebas7}
\\ 
\nonumber
\\
% m_{\widetilde{\nu}_{jR}^\mathcal{I} \widetilde{\nu}^{\mathcal{I}}_{iL}}^{2}
% &=&
m_{\widetilde{\nu}_{iL}^\mathcal{I} \widetilde{\nu}^{\mathcal{I}}_{jR}}^{2} &=&
-\frac{1}{\sqrt 2}T^{\nu}_{ij}v_{u}
+\frac{1}{2}Y^{\nu}_{ij}\lambda_{k}v_{d}v_{kR}
-\frac{1}{2}Y^{\nu}_{ik}\lambda_{j}v_{d}v_{kR}
+Y^{\nu}_{il}\kappa_{jlk}v_{u}v_{kR}
\nonumber\\
&&
-\frac{1}{2}Y^{\nu}_{ij}Y^{\nu}_{lk}v_{lL}v_{kR}
+\frac{1}{2}Y^{\nu}_{ik}Y^{\nu}_{lj}v_{lL}v_{kR}
%+V^{(n)}_{v_{iL}v_{jR}}
\ ,
   \label{pruebas8}
\\ 
\nonumber
\\
m_{\widetilde{\nu}_{iL}^\mathcal{I} \widetilde{\nu}_{jL}^\mathcal{I}}^{2} &=&
m_{\widetilde{\nu}_{iL}^\mathcal{R}\widetilde{\nu}_{jL}^\mathcal{R}}^{2}
-\frac{1}{4}(g^2 + g'^2) v_{iL}v_{jL}
\ ,
\label{oddLL}
\eea
}
and, in order to simplify some of these formulas, the entries of the mass matrix for
Higgses 
%of Appendix~\ref{SubAppendix-scalarmasses} 
are used when appropriate. 
%%%%%%%%%%%%%%%%%%%%%%%%%%%%%%%%%%%%%%%%%%%%%%%%%%%%%%%%%%%%%%%%%
%%%%%%%%%%%%%%%%%%%%%%%%%%%%%%%%%%%%%%%%%%%%%%%%%%%%%%%%%%%%%%%%%
%%%%%%%%%%%%%%%%%%%%%%%%%%%%%%%%%%%%%%%%%%%%%%%%%%%%%%%%%%%%%%%%
% where $V^{(n)}_{xy} \equiv \frac{\partial^2V^{(n)}}{\partial x \partial y}$, 
% $V^{(n)}_{y} \equiv \frac{\partial V^{(n)}}{\partial y}$ with $x,\,y=v_d,\,v_u,\,v_{\nu_i},\,v_{\nu^c_i}$. 
% Here $V^{(n)}$ represents the $n$--loop
% radiative correction to the potential, $V=V^{(0)} + V^{(n)}$. 
The matrix of Eq.~(\ref{matrixscalar2}) is diagonalized by an orthogonal matrix
$Z^A$:
\bea
%\left(m^2_{S_{\alpha} S_{\beta}}\right)^{\text{dia}}
Z^A
m^2_{A^0}\ {Z^A}^{^T}
=
\left(m^2_{A^0}\right)^{\text{dia}}
\ ,
\label{scalarhiggss}
\eea
with 
\begin{equation}
P = {Z^A}^{^T} A^0\, ,
\label{physscalarhiggsess}
\end{equation}
where the 8 entries of the matrix $A^0$ are the `pseudoscalar Higgs' mass eigenstate fields.
In particular,
\begin{equation}
H_{d}^\mathcal{I} = Z^A_{b1} h_b\, ,\;\;\;
H_{u}^\mathcal{I} = Z^A_{b2} h_b\, ,\;\;\;
\tilde{\nu}_{iR}^\mathcal{I} = Z^A_{bi} h_b\, ,\;\;\;
\tilde{\nu}_{jL}^\mathcal{I} = Z^A_{bj} h_b\, .
\label{physscalarhiggseswp}
\end{equation}
% where we are defining $a,b=1,...,8$, not to be confused with $a,b=1,2$ used in Subsection~\ref{superpotentialsoft} as $SU(2)_L$ index.
In the case of considering only one family of right-handed neutrinos as we do in 
Appendix~\ref{Section:Coupling}, the last two equalities
can be written as
\begin{equation}
% H_{d}^\mathcal{I} = Z^A_{b1} h_b\, ,\;\;\;
% H_{u}^\mathcal{I} = Z^A_{b2} h_b\, ,\;\;\;
\tilde{\nu}_{R}^\mathcal{I} = Z^A_{b3} h_b\, ,\;\;\;
\tilde{\nu}_{3+iL}^\mathcal{I} = Z^A_{b3+i} h_b\, .
\label{physscalarhiggseswwpp}
\end{equation}

\vspace{0.25cm}
\noindent
%%%%%%%%%%%%%%%%%%%%%%%%%%%%%%%%%%%%%%%%%%%%%%%%%%%%%%%%%%%%%%%%%
%%%%%%%%%%%%%%%%%%%%%%%%%%%%%%%%%%%%%%%%%%%%%%%%%%%%%%%%%%%%%%%%
%\subsubsection
{\bf Mass Matrix for Charged Higgses}
\label{SubAppendix:chargedScalar}
%%%%%%%%%%%%%%%%%%%%%%%%%%%%%%%%%%%%%%%%%%%%%%%%%%%%%%%%%%%%%%%%%
%%%%%%%%%%%%%%%%%%%%%%%%%%%%%%%%%%%%%%%%%%%%%%%%%%%%%%%%%%%%%%%%

\noindent
Charged Higgses mix with left and right sleptons. 
In the basis  
$C^T=
({H^-_d}^*,{H^+_u},\widetilde{e}_{iL}^*,\widetilde{e}_{jR}^*)$, one obtains the following mass terms in the
Lagrangian:
\begin{equation}
-{C^*}^T 
{m}^2_{H^+} C\, ,
\label{matrix122}
\end{equation}
%%%%%%%%%%%%%%%%%%%%%%%%%%%%%%%%%%%%%%%%%%%%%%%%%%%%%%%%%%%%%%%%%
%%%%%%%%%%%%%%%%%%%%%%%%%%%%%%%%%%%%%%%%%%%%%%%%%%%%%%%%%%%%%%%%
% where the expressions for the entries of the mass matrix ${m}^2_{H^+}$ 
% (using the entries of
% Appendix~\ref{SubAppendix-scalarmasses} when appropriate) are
% are obtained computing the second derivative of the scalar potential 
% (\ref{finalpotential}) with respect to the fields, with the result:
where $m_{H^+}^2$ is the $8\times 8$ (symmetric) matrix
\begin{align}
m_{H^+}^2= 
  \left( \begin{array}{cccc} 
  m_{H_{d}^-{H^{-}_d}^{*}}^{2} & m_{H_{d}^- H_{u}^+}^{2} & 
m_{{H_d^-} \widetilde{e}^*_{jL}}^2 & m_{{H_d^-} \widetilde{e}^*_{jR}}^2  \\
 m_{{H_{u}^+}^* {H_d^-}^*}^{2}  & m_{{H^{+}_u}^{*} H_{u}^{+}}^{2} & 
m_{{H_u^+}^*\widetilde{e}^*_{jL}}^{2} & m_{{H_u^+}^*\widetilde{e}^*_{jR}}^{2} \\
 m_{\widetilde{e}_{iL} {{H^{-}_d}}^{*}}^2 & m_{\widetilde{e}_{iL} H_u^+}^2 &   
m_{\widetilde{e}_{iL} \widetilde{e}_{jL}^{*}}^{2} &   
m_{\widetilde{e}_{iL} \widetilde{e}_{jR}^{*}}^{2}\\
 m_{\widetilde{e}_{iR} {{H^{-}_d}}^{*}}^2 & m_{\widetilde{e}_{iR} H_u^+}^2 & 
  m_{\widetilde{e}_{iR} \widetilde{e}_{jL}^{*}}^{2} &   
m_{\widetilde{e}_{iR} \widetilde{e}_{jR}^{*}}^{2}
         \end{array} \right)\ ,  
         \label{matrixcharged2}
\end{align}

% The scalar potential includes the following quadratic term for the charged scalars:
% \beq
% {\mathbf{C^{*}_{\alpha }}} m^2_{C_{\alpha}C_{\beta}} \mathbf{C_{\beta}}\ ,
% \label{matrixchargedscalars}
% \eeq
% where
% \beq
% \mathbf{C}_{\alpha}=({{H^{-}_d}}^{*},H^+_u,\widetilde{e}^{*}_i,\widetilde{e}^{c^*}_{i})\ ,
% \eeq 
% is in the unrotated basis. 
% Then the mass eigenvectors are
% \bea
% \mathbf{H}_{\alpha}^+=Z^{+}_{\alpha \beta} \mathbf{C}_{\beta}\ , 
% \eea
% with the eigenvalues $m^2_{H^{\pm}_{\alpha}}$ obtained from the diagonal mass matrix 
% \bea
% \left(m^2_{H^{\pm}_{\alpha\beta}}\right)^{\text{diag}}
% =Z^{+}_{\alpha \gamma} 
% m^2_{C_{\gamma}C_{\delta}}  
% Z^{+}_{\beta \delta}\ .
% \eea
% The expressions for the 
% entries of $m^2_{C_{\alpha}C_{\beta}}$ (using the entries of
% Appendix~\ref{SubAppendix-scalarmasses} when appropriate) are:
%%%%%%%%%%%%%%%%%%%%%%%%%%%%%%%%%%%%%%%%%%%%%%%%%%%%%%%%%%%%%%%%%
%%%%%%%%%%%%%%%%%%%%%%%%%%%%%%%%%%%%%%%%%%%%%%%%%%%%%%%%%%%%%%%%
{\footnotesize
\bea
m_{H_{d}^-{{H^{-}_d}}^{*}}^{2}&=&
%m_{H_{d}^\mathcal{I}H_{d}^\mathcal{I}}^{2}
m_{H_{d}^\mathcal{R}H_{d}^\mathcal{R}}^{2}
- \frac{1}{4}(g^2 + g'^2) v_{d}^{2}
+
\frac{g^2}{4}(v_u^2-v_{iL}v_{iL})
%+v_{\nu^c_i} \mathrm{tan} \beta(a_{\lambda_i}+\lambda_{j}\kappa_{ijk}v_{\nu^c_k})
-\frac{1}{2}\lambda_i\lambda_j v_u^2
%\nonumber\\
%&&
%+Y_{\nu_{ij}}\frac{v_{\nu_i}}{v_d}(\lambda_kv_{\nu^c_j}v_{\nu^c_k}+\lambda_j v_u^2)
+\frac{1}{2}Y^{e}_{ik}Y^{e}_{jk}v_{iL}v_{jL}
%+V^{(n)}_{v_d v_d}-\frac{1}{v_d}V^{(n)}_{v_d}
\ ,
   \label{pruebass1}
\\ 
\nonumber
\\
m_{{{H^{+}_u}}^{*} H_{u}^{+}}^{2}&=&
%m_{H_{u}^\mathcal{I}H_{u}^\mathcal{I}}^{2}
m_{H_{u}^\mathcal{R}H_{u}^\mathcal{R}}^{2}
- \frac{1}{4}(g^2 + g'^2) v_{u}^{2}
+
\frac{g^2}{4}(v_d^2+v_{iL}v_{iL})
-\frac{1}{2}\lambda_{i}\lambda_{i}v_{d}^2 +Y^{\nu}_{ij}\lambda_{j}v_{d}v_{iL}
\nonumber\\
&&
-\frac{1}{2}Y^{\nu}_{ik}Y^{\nu}_{jk}v_{iL}v_{jL}
%+V^{(n)}_{v_uv_u}-\frac{1}{v_u}V^{(n)}_{v_u}
\ ,
   \label{pruebass2}
\\ 
\nonumber
\\
%m_{H_{d}^- H_{u}^+}^{2}&=&
m_{{H_{u}^+}^* {H_{d}^-}^*}^{2}
&=&
\frac{g^{2}}{4}v_{d}v_{u}+\frac{1}{\sqrt 2}
T^{\lambda}_iv_{iR}
+ \frac{1}{2}\lambda_{k} \kappa_{ijk}v_{iR} v_{jR}
-\frac{1}{2}\lambda_i \lambda_i  v_d v_u
+\frac{1}{2}Y^{\nu}_{ij}\lambda_jv_u v_{iL}
%+ V^{(n)}_{v_dv_u}
\ ,
   \label{pruebass3}
   \\ 
\nonumber
\\
m_{\widetilde{e}_{iL} {{H^{-}_d}}^{*}}^2&=&
%m_{{H^{-}_d} \widetilde{e}^*_{iL}}^2=
\frac{g^{2}}{4}v_{d}v_{iL} 
-\frac{1}{2}Y^{\nu}_{ij}\lambda_{k}v_{kR}v_{jR}
-\frac{1}{2}Y^{e}_{ij}Y^{e}_{kj}v_{d}v_{kL}
%+V^{(n)}_{v_{\nu_i}v_d}
\ ,
   \label{pruebass11}
\\ 
\nonumber
\\
m_{\widetilde{e}_{iL} H_u^+}^2&=&
%m_{{H_u^+}^* \widetilde{e}^*_{iL}}^2=
\frac{g^{2}}{4}v_{u}v_{iL}
-\frac{1}{\sqrt 2}T^{\nu}_{ij}v_{jR}
-\frac{1}{2}Y^{\nu}_{ij} \kappa_{ljk} v_{lR} v_{kR}+\frac{1}{2}Y^{\nu}_{ij} \lambda_j v_d v_u
-\frac{1}{2}Y^{\nu}_{ik} Y^{\nu}_{kj}v_u v_{jL}
%+V^{(n)}_{v_{\nu_i}v_u}
\ ,
   \label{pruebass12}
\\ 
\nonumber
\\
m_{\widetilde{e}_{iR} {{H^{-}_d}}^{*}}^2&=&
%m_{H^{-}_d \widetilde{e}^*_{iR}}^2=
- \frac{1}{\sqrt 2}T^{e}_{ji}v_{jL}-\frac{1}{2}Y^{e}_{ki}Y^{\nu}_{kj}v_{u}v_{jR}
%+V^{(n)}_{\widetilde{e}^{c^*}_{i}v_d}
\ ,
   \label{pruebass13}
\\ 
\nonumber
\\
m_{\widetilde{e}_{iR} H_u^+}^2&=&
%m_{{H_u^+}^* \widetilde{e}^*_{iR}}^2=
-\frac{1}{2}Y^{e}_{ki}(\lambda_{j}v_{kL}v_{jR}
+Y^{\nu}_{kj}v_{d}v_{jR})
+\lambda_{lni} Y^{\nu}_{lk} v_{nL} v_{kR}
%+V^{(n)}_{\widetilde{e}^{c^*}_{i}v_u}
\ ,
\label{olvidado}
\\ 
\nonumber
\\
%m_{\widetilde{e}_{jR} \widetilde{e}_{iL}^*}^{2} &=&
m_{\widetilde{e}_{iL} \widetilde{e}^{*}_{jR}}^{2} &=&
\frac{1}{\sqrt 2}T^{e}_{ij}v_{d}
-\frac{1}{2}Y^{e}_{ij}\lambda_{k}v_{u}v_{kR}
+\frac{2}{\sqrt 2}T^{\lambda}_{kij}v_{kL}
%+ V^{(n)}_{v_{v_i}\widetilde{e}^{c^*}_{j}}
\ , 
   \label{pruebass5}
\\ 
\nonumber
\\
% m_{\widetilde{e}_{jR} \widetilde{e}_{iL}^*}^{2}&=&
% m_{\widetilde{e}_{iL} \widetilde{e}^{*}_{jR}}^{2}
% +V^{(n)}_{\widetilde{e}^{c^*}_{i}v_{v_j}}\ ,
%    \label{pruebass6}
% \\ 
% \nonumber
% \\
m_{\widetilde{e}_{iR} \widetilde{e}^{*}_{jR}}^2&=&
%m_{\widetilde{e}^c_{ij}}^{2}
\left(m_{\widetilde{e}_R}^2\right)_{ij} 
+
\frac{g'^{2}}{4}(v_{u}^{2}-v_{d}^{2}-v_{kL} v_{kL})\delta_{ij}+\frac{1}{2}Y^{e}_{ki}Y^{e}_{kj}v_{d}^{2}
+\frac{1}{2}Y^{e}_{li}Y^{e}_{kj}v_{kL}v_{lL}
\nonumber\\
&&
+2\lambda_{mlj}\lambda_{nli}v_{mL}v_{nL}
%+V^{(n)}_{\widetilde{e}^{c^*}_{i}\widetilde{e}^{c^*}_{j}}
\ ,
   \label{pruebass7}
   \\ 
\nonumber
\\   
   m_{\widetilde{e}_{iL} \widetilde{e}_{jL}^{*}}^{2}&=&
   m_{\widetilde{\nu}_{iL}^\mathcal{R}\widetilde{\nu}_{jL}^\mathcal{R}}^{2}
-\frac{1}{4}(g^2 + g'^2) v_{iL}v_{jL}  
%   m_{\widetilde{\nu}_{i}^\mathcal{I}\widetilde{\nu}_{j}^\mathcal{I}}^{2}
+
\frac{g^{2}}{4}(v_{u}^{2}-v_{d}^{2} 
-v_{kL} v_{kL})
\delta_{ij}
+\frac{g^{2}}{4}v_{iL}v_{jL}
%+Y_{\nu_{il}}Y_{\nu_{jk}}v_{\nu^c_l}v_{\nu^c_k}
\nonumber\\
&&
-\frac{1}{2}Y^{\nu}_{ik}Y^{\nu}_{jk}v^2_{u}
+\frac{1}{2}Y^{e}_{il}Y^{e}_{jl}v_{d}^{2}
+2\lambda_{iml}\lambda_{jnl}v_{mL}v_{nL}
%\nonumber\\
% &&
% +
% \frac{\delta_{ij}}{v_{\nu_j}}\left[
% -a_{\nu_{ik}}{v_uv_{\nu^c_k}}+
% {Y_{\nu_{ik}}}
% \left(\lambda_lv_dv_{\nu^c_k}v_{\nu^c_l}
% +\lambda_kv_dv_u^2
% -\kappa_{klm}v_uv_{\nu^c_l}v_{\nu^c_m}
% -Y_{\nu_{mk}}v_{\nu_m} v^2_{u}-Y_{\nu_{ml}}v_{\nu_m}v_{\nu^c_l}v_{\nu^c_k}
% \right)
% \right]
% \nonumber\\
% &&
% +V^{(n)}_{v_{\nu_i}v_{\nu_j}}-\frac{\delta_{ij}}{v_{\nu_i}}V^{(n)}_{v_{\nu_i}}
\ ,
\label{chargedmixture}
\eea
}
and, in order to simplify some of these formulas, the entries 
of the mass matrix for
Higgses 
%matrix of Appendix~\ref{SubAppendix-scalarmasses} 
are used when appropriate. 
Matrix of Eq.~(\ref{matrixcharged2}) is diagonalized by an orthogonal matrix
$Z^+$:
\bea
%\left(m^2_{S_{\alpha} S_{\beta}}\right)^{\text{dia}}
Z^+
m^2_{H^+}\ {Z^+}^{^T}
=
\left(m^2_{H^+}\right)^{\text{dia}}
\ ,
\label{scalarhiggs22}
\eea
with 
\begin{equation}
C = {Z^+}^{^T} H^+\, ,
\label{physscalarhiggsesss}
\end{equation}
where the 8 entries of the matrix $H^+$ are the `charged Higgs' mass eigenstate fields.
In particular,
\begin{equation}
H_{d}^- = Z^+_{b1} H_b^{-}\, ,\;\;\;
H_{u}^+ = Z^+_{b2} H^+_b\, ,\;\;\;
\widetilde{e}_{iL} = Z^+_{bi} H_b^-\, ,\;\;\;
\widetilde{e}_{jR} = Z^+_{bj} H_b^-\, .
\label{physscalarhiggseswc}
\end{equation}
\vspace{0.25cm}
\noindent
%\subsubsection
{\bf Mass Matrix for Down-Squarks}
\label{squarks}

\noindent
Left and right down-squarks are mixed. In the basis  
${\tilde d}^{^T}=\left(\widetilde{d}_{iL},\widetilde{d}_{jR}\right)$, one obtains the following mass terms in the
Lagrangian:
%The scalar potential includes the following quadratic term for the CP-even
%neutral scalars:
%%%%%%%%%%%%%%%%%%%%%%%%%%%%%%%%%%%%%%%%%%%%%%%%%%%%%%%%%%%%%%%%%
%%%%%%%%%%%%%%%%%%%%%%%%%%%%%%%%%%%%%%%%%%%%%%%%%%%%%%%%%%%%%%%%
% \beq
% \mathbf{S}^\mathcal{R}_{\alpha} m^2_{S^\mathcal{R}_{\alpha} S^\mathcal{R}_{\beta}} \mathbf{S}^\mathcal{R}_{\beta}\ ,
% \label{matrix1}
% \eeq
\begin{equation}
- {\tilde d}^{^T} 
%\mathcal{M}_{\mathrm{n}}
%\mathcal{M}_{\chi^0} 
%\mathcal{M}_{\nu} 
{m}^2_{\widetilde d}\ {\widetilde d}^*\, ,
\label{matrix1d}
\end{equation}
%%%%%%%%%%%%%%%%%%%%%%%%%%%%%%%%%%%%%%%%%%%%%%%%%%%%%%%%%%%%%%%%%
%%%%%%%%%%%%%%%%%%%%%%%%%%%%%%%%%%%%%%%%%%%%%%%%%%%%%%%%%%%%%%%%
where ${m}^2_{\widetilde d}$ is the $6\times 6$ (symmetric) matrix 
\begin{align}
m_{\widetilde{d}}^2= 
  \left( \begin{array}{cc} 
m^2_{\widetilde{d}_{iL}\widetilde{d}_{jL}^*} & m^2_{\widetilde{d}_{iL} \widetilde{d}_{jR}^*}\\
       m^2_{\widetilde{d}_{iR} \widetilde{d}_{jL}^*}
 & m^2_{\widetilde{d}_{iR} \widetilde{d}_{jR}^*}
         \end{array} \right)\ ,  
         \label{matrixsq1}
\end{align}
%
% expressions for the entries of the (symmetric) mass matrix ${m}^2_{\tilde d}$ 
% are given by:
%{\footnotesize
\bea
m^2_{\widetilde{d}_{iL}\widetilde{d}_{jL}^*}&=&
\left(m_{\widetilde{Q}_L}^2\right)_{ij} 
%    m^2_{\widetilde{Q}_{ij}}
-\frac{1}{24} \left(3g^2
    + g'^2\right)\left(v_d^2-v_u^2+v_{kL}v_{kL}\right) 
    + \frac{1}{2}Y^{d}_{ik} Y^{d}_{jk}  v_d^2
    \nonumber\\
    &&
+\frac{1}{2}\lambda'_{nil} \lambda'_{mjl} v_{nL} v_{mL}
%+\frac{1}{2}\lambda'_{nil} Y^d_{jl} v_{nL} v_d
%+\frac{1}{2}\lambda'_{njl} Y^d_{il} v_{nL} v_d
+\frac{1}{2}\left(\lambda'_{nil} Y^d_{jl}+\lambda'_{njl} Y^d_{il}\right) v_{nL} v_d
\ ,
   \label{pruebass12n}
\\ 
\nonumber
\\
% \ ,
% \nonumber\\ 
m^2_{\widetilde{d}_{iR} \widetilde{d}_{jR}^*}&=& 
\left(m_{\widetilde{d}_R}^2\right)_{ij} 
%    m^2_{\widetilde{d}_{ij}}
- \frac{g'^2}{12}\left(v_d^2-v_u^2+v_{kL}v_{kL}\right)
    + \frac{1}{2}Y^{d}_{ki} Y^{d}_{kj}v_d^2
\nonumber\\
    &&
+\frac{1}{2} \lambda'_{mli} \lambda'_{nlj} v_{mL} v_{nL}
%+\frac{1}{2}\lambda'_{nlj} Y^d_{li} v_{nL} v_d
+\frac{1}{2}\left(\lambda'_{nli} Y^d_{lj}+\lambda'_{nlj} Y^d_{li}\right) v_{nL} v_d
\ ,
   \label{pruebass13n}
\\ 
\nonumber
\\
% \nonumber\\
m^2_{\widetilde{d}_{iL} \widetilde{d}_{jR}^*} &=&
m^2_{\widetilde{d}_{jR} \widetilde{d}_{iL}^*} =
    \frac{1}{\sqrt 2}T^{d}_{ij} v_d -\frac{1}{2}Y^{d}_{ij}\lambda_k v_u v_{kR}
+\frac{1}{\sqrt 2}T^{\lambda'}_{kij} v_{kL} 
% \ ,
%    \label{pruebass14}
% \\ 
% \nonumber
% \\
% m^2_{\widetilde{d}_{jR} \widetilde{d}_{iL}^*} &=& m^2_{\widetilde{d}_{iL} \widetilde{d}_{jR}^*}
\ .
\eea
%} 
%
%where $a_{u_{ij}}\equiv (A_uY_u)_{ij}$ and $a_{d_{ij}}\equiv (A_dY_d)_{ij}$. 
%This $6\times 6$ (symmetric) matrix 
Matrix of Eq.~(\ref{matrixsq1}) is diagonalized by an orthogonal matrix $Z^D$:
\bea
%\left(m^2_{S_{\alpha} S_{\beta}}\right)^{\text{dia}}
Z^D
m^2_{\tilde d}\ {Z^D}^{^T}
=
\left(m^2_{\tilde d}\right)^{\text{dia}}
\ ,
\label{scalarhiggsss}
\eea
with 
\begin{equation}
\tilde d = {Z^D}^{^T} \tilde D\, ,
\label{physscalarhiggsesn}
\end{equation}
where the 6 entries of the matrix $\tilde D$ are the down-squark mass eigenstate fields.
In particular,
\begin{equation}
\tilde d_{iL} = Z^D_{bi} \tilde D_b\, ,\;\;\;
\tilde d_{jR} = Z^D_{bj} \tilde D_b\, .
\label{physscalarhiggseswy}
\end{equation}
% where we are defining $a,b=1,...,8$, not to be confused with $a,b=1,2$ used in Subsection~\ref{superpotentialsoft} as $SU(2)_L$ index.

% For the mass state $\widetilde{\mathbf{q}}_i$ we have
% \begin{align}
% \widetilde{\mathbf{q}}_i = R^{\widetilde{q}}_{ij} \widetilde{q}_j\ , \end{align}
% with the diagonal mass matrix
% \begin{align}
%  (M^{\text{diag}}_{\widetilde{q}})^2_{ij} 
%    = R^{\widetilde{q}}_{il}  M^2_{\widetilde{q}_{lk}} R^{\widetilde{q}}_{jk}\ .\end{align}
%

% It is worth noticing here that if we allow
% the presence of the baryon number violating terms in the superpotential
% discussed in the Introduction, 
% $\lambda'_{ijk}\hat{L}_{i}\hat{Q}_{j}\hat{d}^c_{k}$,
% they would contribute to the above squark masses.
% Actually, even if they are set to zero,
% one-loop corrections will generate them, as discussed in Appendix
% \ref{appx:rges}. However, these contributions are negligible.

% We must comment that a term 
%$\frac{\lambda_{ijk}}{2}\hat{L}_{i} \hat{L}_{j} \hat{e}_{k}^{c} %+\frac{\lambda'_{ijk}}{2}\hat{L}_{i}\hat{Q}_{j}\hat{d}^c_{k}$ in 
%the superpotential will appear at the loop level and by the RGE affected 
%the charged masses, but we neglected this contributions since are 
%really very small.]

\vspace{0.25cm}
\noindent
%\subsubsection
{\bf Mass Matrix for Up-Squarks}

\label{squarks2}

\noindent
Left and right up-squarks are mixed.
%Left and right down-squarks are mixed. 
In the basis  
${\widetilde u}^{^T}=\left(\widetilde{u}_{iL},\widetilde{u}_{jR}\right)$, one obtains the following mass terms in the
Lagrangian:
%The scalar potential includes the following quadratic term for the CP-even
%neutral scalars:
%%%%%%%%%%%%%%%%%%%%%%%%%%%%%%%%%%%%%%%%%%%%%%%%%%%%%%%%%%%%%%%%%
%%%%%%%%%%%%%%%%%%%%%%%%%%%%%%%%%%%%%%%%%%%%%%%%%%%%%%%%%%%%%%%%
% \beq
% \mathbf{S}^\mathcal{R}_{\alpha} m^2_{S^\mathcal{R}_{\alpha} S^\mathcal{R}_{\beta}} \mathbf{S}^\mathcal{R}_{\beta}\ ,
% \label{matrix1}
% \eeq
\begin{equation}
- {\widetilde u}^{^T} 
%\mathcal{M}_{\mathrm{n}}
%\mathcal{M}_{\chi^0} 
%\mathcal{M}_{\nu} 
{m}^2_{\widetilde u}\ {\widetilde u}^*\, ,
\label{matrix1s}
\end{equation}
%%%%%%%%%%%%%%%%%%%%%%%%%%%%%%%%%%%%%%%%%%%%%%%%%%%%%%%%%%%%%%%%%
%%%%%%%%%%%%%%%%%%%%%%%%%%%%%%%%%%%%%%%%%%%%%%%%%%%%%%%%%%%%%%%%
where ${m}^2_{\widetilde u}$ is the $6\times 6$ (symmetric) matrix 
\begin{align}
m_{\widetilde{u}}^2= 
  \left( \begin{array}{cc} 
m^2_{\widetilde{u}_{iL}\widetilde{u}_{jL}^*} & m^2_{\widetilde{u}_{iL} \widetilde{u}_{jR}^*}\\
       m^2_{\widetilde{u}_{iR} \widetilde{u}_{jL}^*}
 & m^2_{\widetilde{u}_{iR} \widetilde{u}_{jR}^*}
         \end{array} \right)\ , 
\label{matrixsq2} 
\end{align}
%where the expressions for the entries of the (symmetric) mass matrix ${m}^2_{\tilde u}$ 
% i.e. $m_{\phi_d\phi_d}$,
% $m_{\phi_d\phi_u}$, $m_{\phi_d\phi_{iR}}$, etc., 
%are obtained computing the second derivative of the scalar potential 
%(\ref{finalpotential}) with respect to the fields, with the result:
%are given by:
\begin{eqnarray}
m^2_{\widetilde{u}_{iL}\widetilde{u}_{jL}^*}&=&
\left(m_{\widetilde{Q}_L}^2\right)_{ij}  
+ \frac{1}{24}\left(3g^2-g'^2\right)\left(v_d^2-v_u^2 + v_{kL}v_{kL}\right) + 
    \frac{1}{2}Y^{u}_{ik} Y^{u}_{jk} v_u^2
    \ ,
   \label{pruebass1222}
\\ 
\nonumber
\\ 
    m^2_{\widetilde{u}_{iR} \widetilde{u}_{jR}^*}&=& 
\left(m_{\widetilde{u}_R}^2\right)_{ij} 
+ \frac{g'^2}{6}\left(v_d^2-v_u^2+v_{kL}v_{kL}\right)
    + \frac{1}{2}Y^{u}_{ki} Y^{u}_{kj}v_u^2
  \ ,
   \label{pruebass12222}
\\ 
\nonumber
\\
    m^2_{\widetilde{u}_{iL} \widetilde{u}_{jR}^*} &=&
    m^2_{\widetilde{u}_{jR} \widetilde{u}_{iL}^*} =
    \frac{1}{\sqrt 2}T^{u}_{ij}v_u -\frac{1}{2}Y^{u}_{ij}\lambda_k v_dv_{kR}  
    + \frac{1}{2}Y^{u}_{ij}Y^{\nu}_{lk}v_{lL}v_{kR}
% \ ,
%    \label{pruebass1444}
% \\ 
% \nonumber
% \\
% m^2_{\widetilde{u}_{jR} \widetilde{u}_{iL}^*} &=& m^2_{\widetilde{u}_{iL} \widetilde{u}_{jR}^*}
\ .
\end{eqnarray} 
Matrix of Eq.~(\ref{matrixsq2}) is diagonalized by an orthogonal matrix
$Z^U$:
\bea
%\left(m^2_{S_{\alpha} S_{\beta}}\right)^{\text{dia}}
Z^U
m^2_{\tilde u}\ {Z^U}^{^T}
=
\left(m^2_{\tilde u}\right)^{\text{dia}}
\ ,
\label{scalarhiggs2}
\eea
with 
\begin{equation}
\tilde u = {Z^U}^{^T} \tilde U\, ,
\label{physscalarhiggsesnn}
\end{equation}
where the 6 entries of the matrix $\tilde U$ are the up-squark mass eigenstate fields.
In particular,
\begin{equation}
\tilde u_{iL} = Z^U_{bi} \tilde U_b\, ,\;\;\;
\tilde u_{jR} = Z^U_{bj} \tilde U_b\, .
\label{physscalarhiggseswyy}
\end{equation}

\vspace{0.5cm}

\noindent 
%\subsection
{\bf B.2 Fermion Mass Matrices
%Neutralino mass matrix
}
\label{fms}

The neutrino and lepton mass matrices were computed in Appendix A.2 of 
Ref.~\cite{Escudero:2008jg} with the assumption of CP conservation.
In this Appendix we write the general fermion mass matrices, including the quarks matrices, without assuming CP conservation. 
%The contribution due to lepton-number violating couplings $\lambda_{ijk}$ and $\lambda'_{ijk}$ are also included for completeness.
To obtain the results, we apply the standard rotation in the gauge sector:
% \beq
% \lambda_{\tilde{W}_{\alpha}} = Z^{\tilde W}_{\alpha\beta}
% \ ,
% \nonumber\\
% \eeq
\begin{equation}
% \left( \begin{array}{c} 
%                            \lambda_{\tilde{W}_{1}}  \\
%                            \lambda_{\tilde{W}_{2}} \\
%                            \lambda_{\tilde{W}_{3}} 
%                          \end{array}\right)            
\left( \begin{array}{c} 
                           {\widetilde{W}_{1}}  \\
                           {\widetilde{W}_{2}} \\
                           {\widetilde{W}_{3}} 
                         \end{array}\right)              
 = Z^{\widetilde W}
      \left( \begin{array}{c} 
                           {\widetilde W}^-   \\
                         {\widetilde W}^+   \\
                         {\widetilde W}^0 
                         \end{array}
                  \right)\ ,         
\label{rotations}
\nonumber\\
\end{equation} 
where the mixing matrix $Z^{\widetilde W}$ is parametrized by
\begin{equation}
Z^{\widetilde W} = 
%\frac{1}{\sqrt 2}
\begin{pmatrix}
\frac{1}{\sqrt 2} & \frac{1}{\sqrt 2} & 0 \\
\frac{-i}{\sqrt 2} & \frac{i}{\sqrt 2} & 0 \\
0 & 0 & 1             
\label{rotations2}
\end{pmatrix} 
% \begin{pmatrix}
% 1 & 1 & 0 \\
% {-i} & {i}& 0 \\
% 0 & 0 & \sqrt 2            
% \label{rotations2}
% \end{pmatrix} 
\, ,
\nonumber\\
\end{equation}
and 
%$\lambda_{\tilde{W}}$ 
${\widetilde W}_{1,2,3}$ 
are the 2--component wino fields in the soft 
Lagrangian of Eq.~(\ref{2:Vsoft}).

\vspace{0.25cm}
\noindent
%\subsubsection
{\bf
Mass Matrix for Neutrinos
%Neutralino mass matrix
}
\label{nfms}

% $\chi^{0}=(\tilde{B^{0}},\tilde{W^{0}},\tilde{H_{d}},
%\tilde{H_{u}},\tilde{\nu_{R_{1}}},
%\tilde{\nu_{R_{2}}},\tilde{\nu_{R_{3}}},
%\tilde{\nu_{1}},\tilde{\nu_{2}},\tilde{\nu_{3}}$)
\noindent 
The usual left-handed neutrinos of the SM mix with the
right-handed neutrinos and the neutral gauginos and higgsinos. Working in the basis of
2--component spinors\footnote{Since both helicities are present for
neutrinos, it is convenient to introduce here the notation where  
$\varphi_{\alpha}$ is a left-handed spinor and $\bar\eta^{\dot\alpha}$ a 
right-handed spinor. Thus we are using in $({\chi^{0}})^T$,
$\varphi^{\alpha}_{\nu_i}\equiv ({\nu_{iL})^c}^{^*}$ and
$\eta^{\alpha}_{\nu_j}\equiv \nu_{jR}^*$,
and in $\chi^{0}$,
${\varphi_{\nu_i}}_{\alpha}\equiv \nu_{iL}$ and
${\eta_{\nu_j}}_{\alpha}\equiv (\nu_{jR})^c$.},
% $({\chi^{0}})^T=\left(({\nu_{iL})^c}^{^*},\widetilde B^0,
% \widetilde W^{0},\widetilde H_{d}^0,\widetilde H_{u}^0,\nu_{iR}^*\right)$,
%
%$({\chi^{0}})^T=\left(\nu_{iL},\widetilde B^0,
%\widetilde W^{0},\widetilde H_{d}^0,\widetilde H_{u}^0,\nu_{iR}\right)$,
%
$({\chi^{0}})^T=\left(\varphi_{\nu_i},\widetilde B^0,
\widetilde W^{0},\widetilde H_{d}^0,\widetilde H_{u}^0,\eta_{\nu_j}\right)$,
one obtains the following neutral fermion mass terms in the Lagrangian:
%$\mathcal{L}_{\mathrm{neutral}}^{\mathrm{mass}}
%^{\tilde \chi^0} 
%=
\begin{equation}
-\frac{1}{2} ({\chi^{0}})^T 
%\mathcal{M}_{\mathrm{n}}
%\mathcal{M}_{\chi^0} 
%\mathcal{M}_{\nu} 
{m}_{\nu} 
\chi^0 + \mathrm{h.c.}\, ,
\label{matrixneutralinos}
\end{equation}
where ${m}_{\nu}$ is the $10\times 10$ (symmetric) matrix 
{\footnotesize
\begin{align}
%M_{\mathcal{M}}
m_\nu
=
\left(
\begin{array}{cccccc}
0_{3\times 3} & -\frac{1}{\sqrt 2}g'\langle \widetilde \nu_{iL}\rangle^* & \frac{1}{\sqrt 2}g \langle \widetilde \nu_{iL}\rangle^* & 0_{3\times 1} & 
Y^{\nu}_{ik}\langle \widetilde \nu_{kR}\rangle^* & 
\langle H_u^0 \rangle Y^{\nu}_{ij}
\\
 -\frac{1}{\sqrt 2}g'\langle \widetilde \nu_{jL}\rangle^*  & M_{1} & 0 & -\frac{1}{\sqrt 2}g'\langle H_d^0 \rangle^* & \frac{1}{\sqrt 2}g'\langle H_u^0\rangle^* & 0_{1\times 3}\\
\frac{1}{\sqrt 2}g\langle \widetilde \nu_{jL}\rangle^*  & 0 & M_{2} & \frac{1}{\sqrt 2}g\langle H_d^0\rangle^* & -\frac{1}{\sqrt 2}g \langle H_u^0 \rangle^* & 0_{1\times 3}\\
 0_{1\times 3}   & -\frac{1}{\sqrt 2}g'\langle H_d^0 \rangle^* & \frac{1}{\sqrt 2}g \langle H_d^0 \rangle^* & 0 & -\lambda_{k}\langle \widetilde \nu_{kR}\rangle^* & -\lambda_{j}\langle H_u^0\rangle\\
Y^{\nu}_{jk}\langle \widetilde \nu_{kR}\rangle^*  & \frac{1}{\sqrt 2}g'\langle H_u^0\rangle^* & -\frac{1}{\sqrt 2}g\langle H_u^0\rangle^* &
 -\lambda_{k}\langle \widetilde \nu_{kR}\rangle^* & 0 & -\lambda_{j}\langle H_d^0\rangle +Y^{\nu}_{kj}\langle \widetilde 
\nu_{kL}\rangle \\
%\frac{1}{\sqrt 2}\langle H_u^0 \rangle {Y^{\nu}}^{^T}_{3\times 3}  
\langle H_u^0 \rangle (Y^{\nu}_{ij}  )^T
& 0_{3\times 1} & 0_{3\times 1} &  -\lambda_{i}\langle H_u^0\rangle & -\lambda_{i}
\langle H_d^0\rangle + Y^{\nu}_{ki} \langle \widetilde \nu_{kL}\rangle & 2\kappa_{ijk}\langle \widetilde \nu_{kR}\rangle^*
\end{array}
\right).
\label{neumatrix}
\end{align}
}
%This $10\times 10$ (symmetric) matrix 
%${m}_{\nu}$ 
This is diagonalized by an unitary matrix $U^V$:
\begin{equation}
{U^V}^{^*} m_{\nu}\ {U^V}^{^\dagger} = m_{\nu}^{\text{dia}}\, ,
\label{diagmatrixneutralinos}
\end{equation}
with
\begin{equation}
\chi^{0} = {U^V}^{^\dagger} \lambda^0\, ,
\label{physneutralinos}
\end{equation}
where the 10 entries of the matrix $ \lambda^0$ are
the 2-component `neutrino' mass eigenstate fields.
In particular,
\bea
\nu_{iL} &=& {U_{bi}^V}^{^*} \lambda^0_b\, ,\;\;\;
\widetilde B^0 = {U_{b4}^V}^{^*} \lambda^0_b\, ,\;\;\;
 \widetilde W^{0} = {U_{b5}^V}^{^*} \lambda^0_b\, ,
 \nonumber\\
\widetilde H_{d}^0 &=& {U_{b6}^V}^{^*} \lambda^0_b\, ,\;\;\;
\widetilde H_{u}^0 = {U_{b7}^V}^{^*}   \lambda^0_b\, ,\;\;\;
({\nu_{jR})^c}^{^*} = U^{V}_{bj} {\lambda_b^0}^{^*}\ .
\label{physscalarhiggseswpn}
\eea
% where we are defining $a,b=1,...,8$, not to be confused with $a,b=1,2$ used in Subsection~\ref{superpotentialsoft} as $SU(2)_L$ index.
In the case of considering only one family of right-handed neutrinos as we do in 
Appendix~\ref{Section:Coupling}, the last equality can be written as
\begin{equation}
% \nu_{iL} = U^{V,*}_{bi} \lambda^0_b\, ,\;\;\;
% \widetilde B^0 = U^{V,*}_{b4} \lambda^0_b\, ,\;\;\;
%  \widetilde W^{0} = U^{V,*}_{b5} \lambda^0_b\, ,\;\;\;
% \widetilde H_{d}^0 = U^{V,*}_{b6} \lambda^0_b\, ,\;\;\;
% \widetilde H_{u}^0 = U^{V,*}_{b7} \lambda^0_b\, ,\;\;\;
({\nu_{R})^c}^{^*} = U^{V}_{b8} {\lambda_b^0}^{^*}\ .
\label{physscalarhiggseswpnn}
\end{equation}

\vspace{0.25cm}
\noindent
%\subsubsection
{\bf
Mass Matrix for Leptons
%Chargino mass matrix
}
\label{cfms}

\noindent
The usual leptons of the SM mix with charged gauginos and higgsinos. In the basis of 2--component spinors\footnote{Following the convention of the previous footnote, we have in this case
$\varphi^{\alpha}_{e_i}\equiv ({e_{iL})^c}^{^*}$ and
${\eta_{e_j}}_{\alpha}\equiv (e_{jR})^c$.},
% $(\mathbf{\chi}^-)^T = \left({(e_{iL})^c}^{^*}, \widetilde{W}^-, \widetilde{H}^-_1\right)$ 
%
%$(\mathbf{\chi}^-)^T = \left(e_{iL}, \widetilde{W}^-, \widetilde{H}^-_1\right)$
%
$(\mathbf{\chi}^-)^T = \left(\varphi_{e_i}, \widetilde{W}^-, \widetilde{H}^-_d\right)$  
and 
$(\mathbf{\chi}^+)^T = \left(\eta_{e_j}, 
\widetilde{W}^+, \widetilde{H}^+_u\right)$, 
%
%$(\mathbf{\chi}^+)^T = \left(e_{jR}, 
%\widetilde{W}^+, \widetilde{H}^+_2\right)$, 
%and
%${\Psi^+}^T =(-i \tilde \lambda^{+}, \tilde H_u^+ ,  e_R^{+} , \mu_R^{+}, \tau_R^{+})$
%and 
%${\Psi^-}^T = (-i \tilde \lambda^-, \tilde H_d^- ,  e_L^-,  \mu_L^-,  \tau_L^-)$,
one obtains the following charged fermion mass terms in the Lagrangian: 
\begin{equation}
-({\chi^{-}})^T 
%\mathcal{M}_{\mathrm{n}}
%\mathcal{M}_{\chi^0} 
%\mathcal{M}_{\nu} 
{m}_{e}
\chi^+ + \mathrm{h.c.}\, ,
\label{matrixcharginos0}
\end{equation}
% \begin{align}
% -\frac{1}{2} ({\psi^+}^T,{\psi^-}^T) \left( \begin{array}{cc} 0&M^T_C\\M_C & 0  
% \end{array} \right) \left( \begin{array}{cc} {\psi^+}^T\\{\psi^-}^T\\  
% \end{array} \right)\ ,  
% \label{matrixcharginos}
% \end{align}
where ${m}_{e}$ is the $5\times 5$ matrix 
% \begin{align}
% m_e=
% \left(\begin{array}{ccccc}
% M_{2} & g v_{u} & 0 & 0 & 0\\
% g v_{d} & \lambda_{i}\nu^c_{i} & -Y_{e_{i1}}\nu_{i} & -Y_{e_{i2}}\nu_i & -Y_{e_{i3}}\nu_{i}\\
% g\nu_{1} & -Y_{\nu_{1i}}\nu^c_{i} & Y_{e_{11}}v_{d} &  Y_{e_{12}}v_{d} &  Y_{e_{13}}v_{d}\\
% g\nu_{2} & -Y_{\nu_{2i}}\nu^c_{i} &  Y_{e_{21}}v_{d}& Y_{e_{22}}v_{d} &  Y_{e_{23}}v_{d}\\
% g\nu_{3} & -Y_{\nu_{3i}}\nu^c_{i} &  Y_{e_{31}}v_{d} &  Y_{e_{32}}v_{d} & Y_{e_{33}}v_{d}
% \end{array}\right)\ .
% \label{submatrix}
% \end{align}
% \begin{align}
% m_e=
% \left(\begin{array}{ccccc}
% \frac{1}{\sqrt 2}Y^{e}_{11}v_{1} &  \frac{1}{\sqrt 2}Y^{e}_{12}v_{1} &  \frac{1}{\sqrt 2}Y^{e}_{13}v_{1} & \frac{1}{\sqrt 2}gv_{1L} & -\frac{1}{\sqrt 2}Y^{\nu}_{1i}v_{iR}\\
% \frac{1}{\sqrt 2}Y^{e}_{21}v_{1}& \frac{1}{\sqrt 2}Y^{e}_{22}v_{1} &  \frac{1}{\sqrt 2}Y^{e}_{23}v_{1} & \frac{1}{\sqrt 2}gv_{2L} & -\frac{1}{\sqrt 2}Y^{\nu}_{2i}v_{iR}\\
% \frac{1}{\sqrt 2}Y^{e}_{31}v_{1} &  \frac{1}{\sqrt 2}Y^{e}_{32}v_{1} & \frac{1}{\sqrt 2}Y^{e}_{33}v_{1} & \frac{1}{\sqrt 2}gv_{3L} & -\frac{1}{\sqrt 2}Y^{\nu}_{3i}v_{iR}\\
%  0 & 0 & 0 & M_{2} & \frac{1}{\sqrt 2}g v_{2}\\
% -\frac{1}{\sqrt 2}Y^{e}_{i1}v_{iL} & -\frac{1}{\sqrt 2}Y^{e}_{i2}v_{iL} & -\frac{1}{\sqrt 2}Y^{e}_{i3}v_{iL} & \frac{1}{\sqrt 2}g v_{1} & \frac{1}{\sqrt 2} \lambda_{i}v_{iR} 
% \end{array}\right)\, .
% \label{submatrix}
% \end{align}
\begin{align}
m_e=
\left(\begin{array}{ccc}
\langle H_d^0\rangle Y^e_{ij} + 2\langle \widetilde \nu_{lL}\rangle \lambda_{lij}  & g\langle \widetilde \nu_{iL}\rangle^* & -Y^{\nu}_{ik}\langle \widetilde \nu_{kR}\rangle^*\\
0_{1\times 3} & M_{2} & g \langle H_u^0\rangle^*\\
-Y^{e}_{kj}\langle \widetilde \nu_{kL}\rangle  & g \langle H_d^0\rangle^* & \lambda_{k}\langle \widetilde \nu_{kR}\rangle^* 
\end{array}\right)\, .
\label{submatrix}
\end{align}
%This $5\times 5$ matrix 
This is diagonalized by two unitary matrices $U_L^e$ and $U_R^e$:
\begin{equation}
{U_R^e}^{^*} m_{e} {U_L^e}^{^\dagger} = m_{e}^{\text{dia}}
\, ,
\label{diagmatrixneutralinosn}
\end{equation}
%where $m_{e}^{\text{dia}}$ has real nonnegative entries, and
with
\begin{eqnarray}
\chi^{+} = {U_L^e}^{^\dagger} \lambda^+\, ,
\\
\chi^{-} = {U_R^e}^{^\dagger} \lambda^-\, ,
\label{physcharginos}
\end{eqnarray}
%with the 5 entries of the matrices $\lambda^+$, $\lambda^-$, 
%the 2-component `lepton' mass eigenstate fields.
where the 5 entries of the matrices $\lambda^+$, $\lambda^-$, 
are the 2-component `lepton' mass eigenstate fields.
In particular,
\bea
{(e_{jR})^c}^{^*} &=& {U_L^e}_{b4}  {\lambda^+_b}^*\, ,\;\;\;
\widetilde{W}^+ = {U_L^e}^{*}_{b4}  \lambda^+_b\, ,\;\;\;
\widetilde{H}^+_u = {U_L^e}^{*}_{b5}  \lambda^+_b\, ,
 \nonumber\\
 e_{iL}  &=& {U_R^e}^{*}_{bi} \lambda_b^-\ ,\;\;\;
\widetilde{W}^- = {U_R^e}^{*}_{b4}  \lambda^-_b\, ,\;\;\;
\widetilde{H}^-_d= {U_R^e}^{*}_{b5}    \lambda^-_b\, .
\label{physscalarhiggseswpnl}
\eea

\vspace{0.25cm}
\noindent
%\subsubsection
{\bf Mass Matrix for Down-Quarks}

\label{downquark}

\noindent
In the basis of 2--components spinors $(d_L^*)^T=\left(d_{iL}^*\right)$, $(d_R)^T=\left(d_{jR}\right)$, one obtains the following 
down-quark mass terms in the Lagrangian:
\begin{equation}
-(d_{L}^*)^T
%\mathcal{M}_{\mathrm{n}}
%\mathcal{M}_{\chi^0} 
%\mathcal{M}_{\nu} 
{m}_{d} d_R + \mathrm{h.c.}\, ,
\label{matrixdownquarks}
\end{equation}
where ${m}_{d}$ is the $3\times 3$ matrix
\begin{equation}
m_d=\left(\langle H_d^0\rangle^{^*}  {Y^{d^{*}}_{ij}}
+
\langle \widetilde \nu_{lL}\rangle^* 
\lambda'^*_{lij}
\right)
\, .
\label{massmatrixdownquarks}
\end{equation}
This is diagonalized by two unitary matrices $U_L^d$ and $U_R^d$:
\begin{equation}
{U_L^d}^{^{\dagger}} m_{d} {U_R^d} = m_{d}^{\text{dia}}
\, ,
\label{diagmatrixdownquarks}
\end{equation}
with
\begin{eqnarray}
d_R &=& {U_R^d} D_R\, ,
\\
d_L &=& {U_L^d} D_L\, .
\label{physdownquarks}
\end{eqnarray}
where the 3 entries of the matrices $D_L$, $D_R$ are the 2-component down-quark mass eigenstate fields.
In particular,
\bea
d_{jR} &=& {U_R^d}_{jb} D_{bR}\, ,
\nonumber\\
d_{iL} &=& {U_R^d}_{ib} D_{bL}\, .
\label{physscalarhiggseswydq}
\eea

% unrotated bases for down-quarks and up-quarks:
%%%%%%%%%%%%%%%%%%%%%%%%%%%%%%%%%%%%%%%%%%%%%%%%%%%%%%%%%%%%%%%%
%%%%%%%%%%%%%%%%%%%%%%%%%%%%%%%%%%%%%%%%%%%%%%%%%%%%%%%%%%%%%%%
% \beq
% {\mathcal{D}_i}_{L,\alpha} = \left(d_{L}\right)\ ,\;\;
% {\mathcal{D}_i}^*_{R,\beta} = \left({d^*_a}_{R,\beta}\right)\ ,
% \nonumber\\
% \eeq 

% \beq
% {\mathcal{U}_i}_{L,\alpha} = \left({u_a}_{L,\alpha}\right)\ ,\;\; 
% {\mathcal{U}_i}^*_{R,\beta} = \left({u^*_a}_{R,\beta}\right)\ ,
% \nonumber\\
% \eeq

% with the corresponding mass matrices diagonalized by 
% $U_L^d$ and $U_R^d$,
% $U_L^u$ and $U_R^u$, respectively.

\vspace{0.25cm}
\noindent
%\subsubsection
{\bf Mass Matrix for Up-Quarks}

\label{upquark}

\noindent
In the basis of 2--components spinors $(u_L^*)^T=\left(u_{iL}^*\right)$, $(u_R)^T=\left(u_{jR}\right)$, one obtains the following 
up-quark mass terms in the Lagrangian:
\begin{equation}
-(u_{L}^*)^T
%\mathcal{M}_{\mathrm{n}}
%\mathcal{M}_{\chi^0} 
%\mathcal{M}_{\nu} 
{m}_{u} u_R + \mathrm{h.c.}\, ,
\label{matrixupquarks}
\end{equation}
where ${m}_{u}$ is the $3\times 3$ matrix
\begin{equation}
m_u=\left(\langle H_u^0\rangle^{^*}  {Y^{u^{*}}_{ij}}\right)
\, .
\label{massmatrixdownquarkss}
\end{equation}
This is diagonalized by two unitary matrices $U_L^u$ and $U_R^u$:
\begin{equation}
{U_L^u}^{^{\dagger}} m_{u} {U_R^u} = m_{u}^{\text{dia}}
\, ,
\label{diagmatrixdownquarkss}
\end{equation}
with
\begin{eqnarray}
u_R = {U_R^u} U_R\, ,
\\
u_L = {U_L^u} U_L\, .
\label{physdownquarkss}
\end{eqnarray}
where the 3 entries of the matrices $U_L$, $U_R$ are the 2-component up-quark mass eigenstate fields.
In particular,
\bea
u_{jR} &=& {U_R^u}_{jb} U_{bR}\, ,
\nonumber\\
u_{iL} &=& {U_R^u}_{ib} U_{bL}\, .
\label{physscalarhiggseswydqq}
\eea

% \begin{equation}
% m_u=\left(\frac{1}{\sqrt 2} {Y^u}^{^*}H_u^0\right)
% \, ,
% \label{massmatrixdownquarks}
% \end{equation}
% is diagonalized by the two unitary matrices $U_L^u$ and $U_R^u$:
% \begin{equation}
% {U_L^u}^{^*} m_{u} {U_R^u}^{^\dagger} = m_{u}^{\text{dia}}
% \, ,
% \label{diagmatrixdownquarks}
% \end{equation}
% with
% \begin{eqnarray}
% u_L = {U_L^u}^{^\dagger} U_L\, ,
% \\
% u_R = {U_R^u}^{^\dagger} U_R^{^*}\, ,
% \label{physdownquarks}
% \end{eqnarray}
% where the 3 entries of the matrices $U_L$, $U_R$ are the 2-component up-quark mass eigenstate fields.

%%%%%%%%%%%%%%%%%%%%%%%%%%%%%%%%%%%%%%%%%%%%%%%%%%%%%%%%%%%%%%%%%
%%%%%%%%%%%%%%%%%%%%%%%%%%%%%%%%%%%%%%%%%%%%%%%%%%%%%%%%%%%%%%%%
\section{One Scalar/Pseudoscalar Higgs-Two Fermion--Interactions}
\label{Section:Coupling}
%%%%%%%%%%%%%%%%%%%%%%%%%%%%%%%%%%%%%%%%%%%%%%%%%%%%%%%%%%%%%%%%%
%%%%%%%%%%%%%%%%%%%%%%%%%%%%%%%%%%%%%%%%%%%%%%%%%%%%%%%%%%%%%%%%

% In this Appendix we write the relevant interactions for the decays of the left sneutrino.
% We follow the same notation as in Appendix~\ref{masasm}, where
% $\alpha, \beta$ are color indexes, $i,j$ are family indexes, and $a,b,c$ are the indexes for the physical states (mass eigenstates). 
% Only one family of right-handed neutrinos $\nu_R$, and the corresponding scalar and pseudoscalar sneutrino states
% $\tilde{\nu}^{\mathcal{R}}_R$, $\tilde{\nu}^{\mathcal{I}}_R$,
% are considered for the computation of the interactions below. 
% The latter can be obtained with the usual procedure of computing first the couplings in the Lagrangian and then rotate the gauge eigenstates to the mass eigenstates.
% Taking all this into account, in the basis of 4-component spinors with the projectors 
% $P_{L,R}=(1\mp\gamma_5)/2$, the interactions relevant for our computation are as follows.

In this Appendix we write the relevant interactions for our computation of the decays of the left sneutrino. For consistency with the computation of 
Section~\ref{Section:Sneutrino-LSP}
where the {\tt SARAH} code was used,
%we give the couplings generated in the $\mu\nu$SSM
% between the neutral scalars (Higgses-sneutrinos)
% and two quarks, two charged fermions (charginos-leptons), two neutral fermions (neutralinos-neutrinos). 
we follow its notation 
\cite{Staub:2008uz,Staub:2011dp,Staub:2013tta}.
% \cite{Staub:2008uz}, 
In particular, 
%$\alpha, \beta=1,2,3$ are color indexes, 
opposite to our convention in Appendix~\ref{masasm},
now
$a,b=1,2,3$ are family indexes, and $i,j,k$ are the indexes for the physical states.
% (mass eigenstates). 
Only one family of right-handed neutrinos $\nu_R$, and the corresponding scalar and pseudoscalar sneutrino states
$\tilde{\nu}^{\mathcal{R}}_R$, $\tilde{\nu}^{\mathcal{I}}_R$,
are considered for the computation of the interactions below.  
% Thus in this case $U^V$ is a $8\times 8$ neutrino matrix with the elements $U^V_{aj}$, $U^V_{4j}$, $U^V_{5j}$, $U^V_{6j}$, $U^V_{7j}$, $U^V_{8j}$ related to the gauge eigenstates of the three left-handed neutrinos, the four neutralinos and the right-handed neutrino, respectively.
% $U_{L}^e$ and $U_{R}^e$ are
% $5\times 5$ lepton matrices, where e.g. the elements
% $U_{R,aj}^e$, $U_{R,4j}^e$, $U_{R,5j}^e$ are related to the
% gauge eigenstates of the three left-handed leptons and the two charginos, respectively.
% $Z^H$, $Z^A$ are $6\times 6$ scalar and pseudoscalar Higgs matrices, respectively, where e.g. the elements
% $Z^H_{1j}$, $Z^H_{2j}$, $Z^H_{3j}$ and $Z^H_{3+a\;j}$
% are related to the gauge eigenstates of the two scalar Higgses, 
% the right scalar sneutrino and the three left scalar sneutrinos.
% As usual, the elements of the $3\times 3$ quark
% matrices are $U_{R,aj}^d$, $U_{L,aj}^d$, $U_{R,aj}^u$ 
% and $U_{L,aj}^u$.
Notice that the definitions of {\tt SARAH} used in this Appendix for Yukawa, lepton and quark matrices are not the same as those in Appendix~\ref{fms}.
Taking all this into account, in the basis of 4--component spinors with the projectors 
$P_{L,R}=(1\mp\gamma_5)/2$, the interactions for the mass eigenstates are as follows.

\vspace{0.25cm}

\noindent
%%%%%%%%%%%%%%%%%%%%%%%%%%%%%%%%%%%%%%%%%%%%%%%%%%%%%%%%%%%%%%%%%
%%%%%%%%%%%%%%%%%%%%%%%%%%%%%%%%%%%%%%%%%%%%%%%%%%%%%%%%%%%%%%%%
%\subsection
{\bf One Higgs-Two Up Quark--Interaction}
\label{SubSection:Coupling-Huu}
%%%%%%%%%%%%%%%%%%%%%%%%%%%%%%%%%%%%%%%%%%%%%%%%%%%%%%%%%%%%%%%%%
%%%%%%%%%%%%%%%%%%%%%%%%%%%%%%%%%%%%%%%%%%%%%%%%%%%%%%%%%%%%%%%%

%%%%%%%%%%%%%%%%%%%%%%%%%%%%%%%%%%%%%%%%%%%%%%%%%%%%%%%%%%%%%%%%%
%%%%%%%%%%%%%%%%%%%%%%%%%%%%%%%%%%%%%%%%%%%%%%%%%%%%%%%%%%%%%%%%
{\footnotesize
\beq 
-i \frac{1}{\sqrt{2}} \delta_{\alpha \beta} \sum_{a,b=1}^{3}Y^*_{u,{a b}} 
  U_{R,{j a}}^{u}  U_{L,{i b}}^{u}  Z_{{k 2}}^{H} P_R
   \,
   -i \frac{1}{\sqrt{2}} \delta_{\alpha \beta} \sum_{b=1}^{3}U^{u,*}_{L,{j b}} 
 \sum_{a=1}^{3}U^{u,*}_{R,{i a}} Y_{u,{a b}}   Z_{{k 2}}^{H} P_L 
\ .
\label{eq:coupling_even_sn->uu}
  \eeq 
  }
%%%%%%%%%%%%%%%%%%%%%%%%%%%%%%%%%%%%%%%%%%%%%%%%%%%%%%%%%%%%%%%%%
%%%%%%%%%%%%%%%%%%%%%%%%%%%%%%%%%%%%%%%%%%%%%%%%%%%%%%%%%%%%%%%%

\vspace{0.25cm}

\noindent
%%%%%%%%%%%%%%%%%%%%%%%%%%%%%%%%%%%%%%%%%%%%%%%%%%%%%%%%%%%%%%%%%
%%%%%%%%%%%%%%%%%%%%%%%%%%%%%%%%%%%%%%%%%%%%%%%%%%%%%%%%%%%%%%%%
%\subsection
{\bf One Pseudoscalar Higgs-Two Up Quark--Interaction}
\label{SubSection:Coupling-Auu}
%%%%%%%%%%%%%%%%%%%%%%%%%%%%%%%%%%%%%%%%%%%%%%%%%%%%%%%%%%%%%%%%%
%%%%%%%%%%%%%%%%%%%%%%%%%%%%%%%%%%%%%%%%%%%%%%%%%%%%%%%%%%%%%%%%

%%%%%%%%%%%%%%%%%%%%%%%%%%%%%%%%%%%%%%%%%%%%%%%%%%%%%%%%%%%%%%%%%
%%%%%%%%%%%%%%%%%%%%%%%%%%%%%%%%%%%%%%%%%%%%%%%%%%%%%%%%%%%%%%%%
{\footnotesize
\beq
- \frac{1}{\sqrt{2}} \delta_{\alpha \beta} \sum_{a,b=1}^{3}
  Y^*_{u,{a b}} U_{R,{j a}}^{u}  U_{L,{i b}}^{u}  Z_{{k 2}}^{A} P_R
 \,+\frac{1}{\sqrt{2}} \delta_{\alpha \beta} \sum_{b=1}^{3}U^{u,*}_{L,{j b}} 
 \sum_{a=1}^{3}U^{u,*}_{R,{i a}} Y_{u,{a b}}   Z_{{k 2}}^{A} P_L 
\ .
\label{eq:coupling_odd_sn->uu}
\eeq }
%%%%%%%%%%%%%%%%%%%%%%%%%%%%%%%%%%%%%%%%%%%%%%%%%%%%%%%%%%%%%%%%%
%%%%%%%%%%%%%%%%%%%%%%%%%%%%%%%%%%%%%%%%%%%%%%%%%%%%%%%%%%%%%%%%

\vspace{0.25cm}

\noindent
%%%%%%%%%%%%%%%%%%%%%%%%%%%%%%%%%%%%%%%%%%%%%%%%%%%%%%%%%%%%%%%%%
%%%%%%%%%%%%%%%%%%%%%%%%%%%%%%%%%%%%%%%%%%%%%%%%%%%%%%%%%%%%%%%%
%\subsection
{\bf One Higgs-Two Down Quark--Interaction}
\label{SubSection:Coupling-Hdd}
%%%%%%%%%%%%%%%%%%%%%%%%%%%%%%%%%%%%%%%%%%%%%%%%%%%%%%%%%%%%%%%%%
%%%%%%%%%%%%%%%%%%%%%%%%%%%%%%%%%%%%%%%%%%%%%%%%%%%%%%%%%%%%%%%%

%%%%%%%%%%%%%%%%%%%%%%%%%%%%%%%%%%%%%%%%%%%%%%%%%%%%%%%%%%%%%%%%%
%%%%%%%%%%%%%%%%%%%%%%%%%%%%%%%%%%%%%%%%%%%%%%%%%%%%%%%%%%%%%%%%
{\footnotesize
\beq 
-i \frac{1}{\sqrt{2}} \delta_{\alpha \beta} \sum_{a,b=1}^{3}
  Y^*_{d,{a b}} U_{R,{j a}}^{d}  U_{L,{i b}}^{d}  Z_{{k 1}}^{H} P_R
   \,
 -i \frac{1}{\sqrt{2}} \delta_{\alpha \beta} \sum_{b=1}^{3}U^{d,*}_{L,{j b}} 
 \sum_{a=1}^{3}U^{d,*}_{R,{i a}} Y_{d,{a b}}   Z_{{k 1}}^{H} P_L
\ .
\label{eq:coupling_even_sn->dd}  
  \eeq
  }
%%%%%%%%%%%%%%%%%%%%%%%%%%%%%%%%%%%%%%%%%%%%%%%%%%%%%%%%%%%%%%%%%
%%%%%%%%%%%%%%%%%%%%%%%%%%%%%%%%%%%%%%%%%%%%%%%%%%%%%%%%%%%%%%%%  

\vspace{0.25cm}

\noindent
%%%%%%%%%%%%%%%%%%%%%%%%%%%%%%%%%%%%%%%%%%%%%%%%%%%%%%%%%%%%%%%%%
%%%%%%%%%%%%%%%%%%%%%%%%%%%%%%%%%%%%%%%%%%%%%%%%%%%%%%%%%%%%%%%%
%\subsection
{\bf One Pseudoscalar Higgs-Two Down Quark--Interaction}
\label{SubSection:Coupling-Add}
%%%%%%%%%%%%%%%%%%%%%%%%%%%%%%%%%%%%%%%%%%%%%%%%%%%%%%%%%%%%%%%%%
%%%%%%%%%%%%%%%%%%%%%%%%%%%%%%%%%%%%%%%%%%%%%%%%%%%%%%%%%%%%%%%%

%%%%%%%%%%%%%%%%%%%%%%%%%%%%%%%%%%%%%%%%%%%%%%%%%%%%%%%%%%%%%%%%%
%%%%%%%%%%%%%%%%%%%%%%%%%%%%%%%%%%%%%%%%%%%%%%%%%%%%%%%%%%%%%%%%
{\footnotesize
\beq
- \frac{1}{\sqrt{2}} \delta_{\alpha \beta} \sum_{a,b=1}^{3}
  Y^*_{d,{a b}} U_{R,{j a}}^{d}  U_{L,{i b}}^{d}  Z_{{k 1}}^{A} P_R
  \,
 +\frac{1}{\sqrt{2}} \delta_{\alpha \beta} \sum_{b=1}^{3}U^{d,*}_{L,{j b}} 
 \sum_{a=1}^{3}U^{d,*}_{R,{i a}} Y_{d,{a b}}   Z_{{k 1}}^{A} P_L
   \ .
\label{eq:coupling_odd_sn->dd}
\eeq }
%%%%%%%%%%%%%%%%%%%%%%%%%%%%%%%%%%%%%%%%%%%%%%%%%%%%%%%%%%%%%%%%%
%%%%%%%%%%%%%%%%%%%%%%%%%%%%%%%%%%%%%%%%%%%%%%%%%%%%%%%%%%%%%%%%

\vspace{0.25cm}

\noindent
%%%%%%%%%%%%%%%%%%%%%%%%%%%%%%%%%%%%%%%%%%%%%%%%%%%%%%%%%%%%%%%%%
%%%%%%%%%%%%%%%%%%%%%%%%%%%%%%%%%%%%%%%%%%%%%%%%%%%%%%%%%%%%%%%%
%\subsection
{\bf One Higgs-Two Lepton--Interaction}
\label{SubSection:Coupling-Hll}
%%%%%%%%%%%%%%%%%%%%%%%%%%%%%%%%%%%%%%%%%%%%%%%%%%%%%%%%%%%%%%%%%
%%%%%%%%%%%%%%%%%%%%%%%%%%%%%%%%%%%%%%%%%%%%%%%%%%%%%%%%%%%%%%%%

%%%%%%%%%%%%%%%%%%%%%%%%%%%%%%%%%%%%%%%%%%%%%%%%%%%%%%%%%%%%%%%%%
%%%%%%%%%%%%%%%%%%%%%%%%%%%%%%%%%%%%%%%%%%%%%%%%%%%%%%%%%%%%%%%%
{\footnotesize
\bea 
 &&-i \frac{1}{\sqrt{2}} \Big\{
 - U^{e,*}_{R,{j 5}} \sum_{a,b=1}^{3}U^{e,*}_{L,{i a}} Y_{e,{a b}} Z_{{k 3 + b}}^{H}
  + g_2 U^{e,*}_{L,{i 4}} \sum_{a=1}^{3}U^{e,*}_{R,{j a}} Z_{{k 3 + a}}^{H}
  \nonumber\\
  &&
+ \sum_{a,b=1}^{3}U^{e,*}_{R,{j b}} U^{e,*}_{L,{i a}} Y_{e,{a b}} Z_{{k 1}}^{H}
+g_2 U^{e,*}_{L,{i 4}} U^{e,*}_{R,{j 5}} Z_{{k 1}}^{H}  
 +g_2 U^{e,*}_{R,{j 4}} U^{e,*}_{L,{i 5}} Z_{{k 2}}^{H}
  \nonumber \\ 
 && 
 + \lambda U^{e,*}_{R,{j 5}}  U^{e,*}_{L,{i 5}} Z_{{k 3}}^{H}  
 - U^{e,*}_{L,{i 5}} \sum_{a=1}^{3}U^{e,*}_{R,{j a}} Y_{\nu,{a}}  Z_{{k 3}}^{H}  
  \Big\} 
  P_L
  \nonumber\\
  && 
- \,i \frac{1}{\sqrt{2}} \Big\{
- \sum_{a,b=1}^{3}Y^*_{e,{a b}} U_{L,{j a}}^{e}  Z_{{k 3 + b}}^{H}  U_{R,{i 5}}^{e} 
+ g_2 \sum_{a=1}^{3}U_{R,{i a}}^{e} Z_{{k 3 + a}}^{H}  U_{L,{j 4}}^{e} 
\nonumber\\
&&
+  \sum_{a,b=1}^{3}Y^*_{e,{a b}} U_{L,{j a}}^{e}  U_{R,{i b}}^{e} Z_{{k 1}}^{H}   
  + g_2 U_{R,{i 5}}^{e} U_{L,{j 4}}^{e} Z_{{k 1}}^{H} 
 + g_2 U_{R,{i 4}}^{e} U_{L,{j 5}}^{e} Z_{{k 2}}^{H} 
\nonumber\\
&&
  + \lambda^* U_{R,{i 5}}^{e} U_{L,{j 5}}^{e} Z_{{k 3}}^{H}
 -\sum_{a=1}^{3}Y^*_{\nu,{a}} U_{R,{i a}}^{e}  U_{L,{j 5}}^{e} Z_{{k 3}}^{H} 
  \Big\} 
P_R
\ .
\label{eq:coupling_even_sn->ll}
\eea
}
%%%%%%%%%%%%%%%%%%%%%%%%%%%%%%%%%%%%%%%%%%%%%%%%%%%%%%%%%%%%%%%%%
%%%%%%%%%%%%%%%%%%%%%%%%%%%%%%%%%%%%%%%%%%%%%%%%%%%%%%%%%%%%%%%%

\vspace{0.25cm}

\noindent
%%%%%%%%%%%%%%%%%%%%%%%%%%%%%%%%%%%%%%%%%%%%%%%%%%%%%%%%%%%%%%%%%
%%%%%%%%%%%%%%%%%%%%%%%%%%%%%%%%%%%%%%%%%%%%%%%%%%%%%%%%%%%%%%%%
%\subsection
{\bf One Pseudoscalar Higgs-Two Lepton--Interaction}
\label{SubSection:Coupling-All}
%%%%%%%%%%%%%%%%%%%%%%%%%%%%%%%%%%%%%%%%%%%%%%%%%%%%%%%%%%%%%%%%%
%%%%%%%%%%%%%%%%%%%%%%%%%%%%%%%%%%%%%%%%%%%%%%%%%%%%%%%%%%%%%%%%

%%%%%%%%%%%%%%%%%%%%%%%%%%%%%%%%%%%%%%%%%%%%%%%%%%%%%%%%%%%%%%%%%
%%%%%%%%%%%%%%%%%%%%%%%%%%%%%%%%%%%%%%%%%%%%%%%%%%%%%%%%%%%%%%%%
{\footnotesize
\bea 
&& \frac{1}{\sqrt{2}} \Big\{ 
  - U^{e,*}_{R,{j 5}} \sum_{a,b=1}^{3}U^{e,*}_{L,{i a}} Y_{e,{a b}}  Z_{{k 3 + b}}^{A}
 - g_2 U^{e,*}_{L,{i 4}} \sum_{a=1}^{3}U^{e,*}_{R,{j a}} Z_{{k 3 + a}}^{A}
\nonumber\\
&&
+\sum_{a,b=1}^{3}U^{e,*}_{R,{j b}} U^{e,*}_{L,{i a}} Y_{e,{a b}}   Z_{{k 1}}^{A} 
- g_2 U^{e,*}_{L,{i 4}} U^{e,*}_{R,{j 5}} Z_{{k 1}}^{A}  
- g_2 U^{e,*}_{R,{j 4}} U^{e,*}_{L,{i 5}} Z_{{k 2}}^{A}
\nonumber\\
&&
 - \lambda U^{e,*}_{R,{j 5}} U^{e,*}_{L,{i 5}} Z_{{k 3}}^{A}  
  + U^{e,*}_{L,{i 5}} \sum_{a=1}^{3}U^{e,*}_{R,{j a}} Y_{\nu,{a}}  Z_{{k 3}}^{A}  
 \Big\}
 P_L
 \nonumber\\
  && 
- \,\frac{1}{\sqrt{2}} \Big\{
 - \sum_{a,b=1}^{3}Y^*_{e,{a b}} U_{L,{j a}}^{e} Z_{{k 3 + b}}^{A}  U_{R,{i 5}}^{e} 
- g_2 \sum_{a=1}^{3}U_{R,{i a}}^{e} Z_{{k 3 + a}}^{A} U_{L,{j 4}}^{e} 
\nonumber\\
&&
 +\sum_{a,b=1}^{3}Y^*_{e,{a b}} U_{L,{j a}}^{e}  U_{R,{i b}}^{e} Z_{{k 1}}^{A} 
 - g_2 U_{R,{i 5}}^{e} U_{L,{j 4}}^{e} Z_{{k 1}}^{A} 
 - g_2 U_{R,{i 4}}^{e} U_{L,{j 5}}^{e} Z_{{k 2}}^{A} 
\nonumber\\
&&
 -\lambda^* U_{R,{i 5}}^{e} U_{L,{j 5}}^{e} Z_{{k 3}}^{A}
  + \sum_{a=1}^{3}Y^*_{\nu,{a}} U_{R,{i a}}^{e}  U_{L,{j 5}}^{e} Z_{{k 3}}^{A} 
  \Big\}
P_R
 \ .
\label{eq:coupling_odd_sn->ll}
\eea}
%%%%%%%%%%%%%%%%%%%%%%%%%%%%%%%%%%%%%%%%%%%%%%%%%%%%%%%%%%%%%%%%%
%%%%%%%%%%%%%%%%%%%%%%%%%%%%%%%%%%%%%%%%%%%%%%%%%%%%%%%%%%%%%%%%

\vspace{0.25cm}

\noindent
%%%%%%%%%%%%%%%%%%%%%%%%%%%%%%%%%%%%%%%%%%%%%%%%%%%%%%%%%%%%%%%%%
%%%%%%%%%%%%%%%%%%%%%%%%%%%%%%%%%%%%%%%%%%%%%%%%%%%%%%%%%%%%%%%%
%\subsection
{\bf One Higgs-Two Neutrino--Interaction}
\label{SubSection:Coupling-Hnn}
%%%%%%%%%%%%%%%%%%%%%%%%%%%%%%%%%%%%%%%%%%%%%%%%%%%%%%%%%%%%%%%%%
%%%%%%%%%%%%%%%%%%%%%%%%%%%%%%%%%%%%%%%%%%%%%%%%%%%%%%%%%%%%%%%%

%%%%%%%%%%%%%%%%%%%%%%%%%%%%%%%%%%%%%%%%%%%%%%%%%%%%%%%%%%%%%%%%%
%%%%%%%%%%%%%%%%%%%%%%%%%%%%%%%%%%%%%%%%%%%%%%%%%%%%%%%%%%%%%%%%
{\footnotesize
\bea 
 &&\frac{i}{2} 
\Big\{
g_1 (U^{V,*}_{i 4} \sum_{a=1}^{3}U^{V,*}_{j a}  
 + U^{V,*}_{j 4}  \sum_{a=1}^{3}U^{V,*}_{i a}) Z_{{k 3 + a}}^{H} 
  - g_2 (U^{V,*}_{i 5} \sum_{a=1}^{3}U^{V,*}_{j a} 
 +  U^{V,*}_{j 5} \sum_{a=1}^{3}U^{V,*}_{i a}) Z_{{k 3 + a}}^{H} 
 \nonumber\\
 &&
 - \sqrt{2} (U^{V,*}_{i 8} U^{V,*}_{j 7} + U^{V,*}_{i 7} U^{V,*}_{j 8})
 \sum_{a=1}^{3}Y_{\nu,{a}} Z_{{k 3 + a}}^{H}  
  +\sqrt{2} \lambda (U^{V,*}_{i 8} U^{V,*}_{j 7}
+ U^{V,*}_{i 7} U^{V,*}_{j 8}  ) Z_{{k 1}}^{H} 
  \nonumber\\
  &&
   + g_1 (U^{V,*}_{j 4} U^{V,*}_{i 6}
 +U^{V,*}_{i 4} U^{V,*}_{j 6}) Z_{{k 1}}^{H}
 - g_2 (U^{V,*}_{i 5} U^{V,*}_{j 6}
+U^{V,*}_{j 5} U^{V,*}_{i 6}) Z_{{k 1}}^{H} 
 \nonumber\\
 &&
  -g_1 (U^{V,*}_{j 4}  U^{V,*}_{i 7} + U^{V,*}_{i 4} U^{V,*}_{j 7}) Z_{{k 2}}^{H} 
+g_2 (U^{V,*}_{i 5} U^{V,*}_{j 7}
+ U^{V,*}_{j 5}  U^{V,*}_{i 7}) Z_{{k 2}}^{H} 
\nonumber \\ 
 &&    
 - \sqrt{2} ( U^{V,*}_{j 8} \sum_{a=1}^{3}U^{V,*}_{i a} 
+ U^{V,*}_{i 8} \sum_{a=1}^{3}U^{V,*}_{j a}) Y_{\nu,{a}}  Z_{{k 2}}^{H} 
 +\sqrt{2} \lambda (U^{V,*}_{i 8} U^{V,*}_{j 6}
 +U^{V,*}_{i 6} U^{V,*}_{j 8} ) Z_{{k 2}}^{H} 
\nonumber \\ 
 &&
 -2 \sqrt{2} \kappa U^{V,*}_{i 8} U^{V,*}_{j 8} Z_{{k 3}}^{H} 
 +\sqrt{2} \lambda (U^{V,*}_{i 7} U^{V,*}_{j 6} Z_{{k 3}}^{H} 
 + U^{V,*}_{i 6} U^{V,*}_{j 7}) Z_{{k 3}}^{H}
 \nonumber \\ 
 && 
 - \sqrt{2} (U^{V,*}_{j 7} \sum_{a=1}^{3}U^{V,*}_{i a} Y_{\nu,{a}} 
 + U^{V,*}_{i 7} \sum_{a=1}^{3}U^{V,*}_{j a} Y_{\nu,{a}} ) Z_{{k 3}}^{H} \Big\}
 P_L 
\nonumber\\
&&
+ \,
\frac{i}{2} \Big\{
g_1 \sum_{a=1}^{3}Z_{{k 3 + a}}^{H} (U_{{j a}}^{V} U_{{i 4}}^{V} 
 + U_{{i a}}^{V}  U_{{j 4}}^{V} )
-g_2\sum_{a=1}^{3}Z_{{k 3 + a}}^{H}( U_{{j a}}^{V} U_{{i 5}}^{V} 
+  U_{{i a}}^{V}  U_{{j 5}}^{V} )
\nonumber\\
&&
 - \sqrt{2} \sum_{a=1}^{3}Y^*_{\nu,{a}} Z_{{k 3 + a}}^{H} (
U_{{i 8}}^{V} U_{{j 7}}^{V} + U_{{i 7}}^{V} U_{{j 8}}^{V} )
 +\sqrt{2} \lambda^* Z_{{k 1}}^{H} (U_{{i 8}}^{V} U_{{j 7}}^{V} 
 + U_{{i 7}}^{V} U_{{j 8}}^{V} )
 \nonumber\\
 &&
 +g_1 Z_{{k 1}}^{H} (U_{{i 6}}^{V} U_{{j 4}}^{V}
  +  U_{{i 4}}^{V} U_{{j 6}}^{V})
 - g_2 Z_{{k 1}}^{H} (U_{{i 5}}^{V} U_{{j 6}}^{V} 
+U_{{i 6}}^{V} U_{{j 5}}^{V} )
\nonumber\\
&&
- g_1 Z_{{k 2}}^{H} (U_{{i 7}}^{V} U_{{j 4}}^{V}
 +  U_{{i 4}}^{V} U_{{j 7}}^{V} )
 +g_2 Z_{{k 2}}^{H} (
U_{{i 5}}^{V} U_{{j 7}}^{V}+
U_{{i 7}}^{V} U_{{j 5}}^{V} )
\nonumber\\
&&
  - \sqrt{2} \sum_{a=1}^{3}Y^*_{\nu,{a}} (U_{{j a}}^{V}  
Z_{{k 2}}^{H} U_{{i 8}}^{V} 
  + U_{{i a}}^{V}  Z_{{k 2}}^{H} U_{{j 8}}^{V} )
  +\sqrt{2} \lambda^* Z_{{k 2}}^{H} (U_{{i 8}}^{V} U_{{j 6}}^{V}
+Z_{{k 2}}^{H} U_{{i 6}}^{V} U_{{j 8}}^{V})
  \nonumber\\
  &&
  -2 \sqrt{2} \kappa^* Z_{{k 3}}^{H} U_{{i 8}}^{V} U_{{j 8}}^{V} 
  +\sqrt{2} \lambda^* Z_{{k 3}}^{H} (U_{{i 7}}^{V} U_{{j 6}}^{V}
  + Z_{{k 3}}^{H} U_{{i 6}}^{V} U_{{j 7}}^{V}  )
  \nonumber\\
  &&
    - \sqrt{2} \sum_{a=1}^{3}Y^*_{\nu,{a}} (U_{{j a}}^{V}  
   Z_{{k 3}}^{H} U_{{i 7}}^{V} 
 +U_{{i a}}^{V}  Z_{{k 3}}^{H} U_{{j 7}}^{V} )
\Big\}
P_R\ .
\label{eq:coupling_even_sn->nn}
\eea }
%%%%%%%%%%%%%%%%%%%%%%%%%%%%%%%%%%%%%%%%%%%%%%%%%%%%%%%%%%%%%%%%%
%%%%%%%%%%%%%%%%%%%%%%%%%%%%%%%%%%%%%%%%%%%%%%%%%%%%%%%%%%%%%%%%

\vspace{0.25cm}

\noindent
%%%%%%%%%%%%%%%%%%%%%%%%%%%%%%%%%%%%%%%%%%%%%%%%%%%%%%%%%%%%%%%%%
%%%%%%%%%%%%%%%%%%%%%%%%%%%%%%%%%%%%%%%%%%%%%%%%%%%%%%%%%%%%%%%%
%\subsection
{\bf One Pseudoscalar Higgs-Two Neutrino--Interaction}
\label{SubSection:Coupling-Ann}
{\footnotesize
\bea 
 &&\frac{1}{2} 
\Big\{
g_1(U^{V,*}_{i 4} \sum_{a=1}^{3}U^{V,*}_{j a} 
+U^{V,*}_{j 4} \sum_{a=1}^{3}U^{V,*}_{i a}) Z_{{k 3 + a}}^{A} 
 - g_2( U^{V,*}_{i 5} \sum_{a=1}^{3}U^{V,*}_{j a}  
+  U^{V,*}_{j 5} \sum_{a=1}^{3}U^{V,*}_{i a}) Z_{{k 3 + a}}^{A}
\nonumber\\
&&
 +\sqrt{2} (U^{V,*}_{i 8} U^{V,*}_{j 7}+U^{V,*}_{i 7} U^{V,*}_{j 8}) 
 \sum_{a=1}^{3}Y_{\nu,{a}} Z_{{k 3 + a}}^{A}  
  - \sqrt{2}\lambda (U^{V,*}_{i 8} U^{V,*}_{j 7}
 +  U^{V,*}_{i 7} U^{V,*}_{j 8})   Z_{{k 1}}^{A}
\nonumber \\ 
 && 
 + g_1 (U^{V,*}_{i 4} U^{V,*}_{j 6} + U^{V,*}_{j 4} U^{V,*}_{i 6}) Z_{{k 1}}^{A}  
 - g_2 (U^{V,*}_{i 5} U^{V,*}_{j 6}
+ U^{V,*}_{j 5} U^{V,*}_{i 6}) Z_{{k 1}}^{A} 
\nonumber \\ 
 &&
 - g_1(U^{V,*}_{i 4} U^{V,*}_{j 7} 
  + U^{V,*}_{j 4} U^{V,*}_{i 7})  Z_{{k 2}}^{A} 
  + g_2 (U^{V,*}_{i 5} U^{V,*}_{j 7} 
  + U^{V,*}_{j 5}  U^{V,*}_{i 7}) Z_{{k 2}}^{A} 
 \nonumber\\
 &&
 +\sqrt{2}(U^{V,*}_{j 8} \sum_{a=1}^{3}U^{V,*}_{i a}  
 + U^{V,*}_{i 8} \sum_{a=1}^{3}U^{V,*}_{j a}) Y_{\nu,{a}}  Z_{{k 2}}^{A} 
  - \sqrt{2} \lambda  (U^{V,*}_{i 8} U^{V,*}_{j 6} 
 + U^{V,*}_{i 6} U^{V,*}_{j 8})Z_{{k 2}}^{A} 
 \nonumber \\ 
 &&
  -2 \sqrt{2} \kappa U^{V,*}_{i 8} U^{V,*}_{j 8} Z_{{k 3}}^{A}
 +\sqrt{2} \lambda(U^{V,*}_{i 7} U^{V,*}_{j 6} 
 + U^{V,*}_{i 6} U^{V,*}_{j 7})  Z_{{k 3}}^{A}
 \nonumber\\
 &&
- \sqrt{2} (U^{V,*}_{j 7} \sum_{a=1}^{3}U^{V,*}_{i a} Y_{\nu,{a}}  
  +  U^{V,*}_{i 7} \sum_{a=1}^{3}U^{V,*}_{j a} Y_{\nu,{a}})  Z_{{k 3}}^{A} 
\Big\}
 P_L 
\nonumber\\
&&
- \,
\frac{1}{2} 
\Big\{
g_1\sum_{a=1}^{3}Z_{{k 3 + a}}^{A}(U_{{j a}}^{V}  
   U_{{i 4}}^{V}
+ U_{{i a}}^{V}  U_{{j 4}}^{V}) 
-g_2\sum_{a=1}^{3}Z_{{k 3 + a}}^{A}(
   U_{{j a}}^{V}  
  U_{{i 5}}^{V} 
   + U_{{i a}}^{V}  U_{{j 5}}^{V} )
  \nonumber\\
  &&
  + \sqrt{2}\sum_{a=1}^{3}Y^*_{\nu,{a}} Z_{{k 3 + a}}^{A}  ( 
  U_{{i 8}}^{V} U_{{j 7}}^{V}+
  U_{{i 7}}^{V} U_{{j 8}}^{V} )
  -\sqrt{2} \lambda^* Z_{{k 1}}^{A} (U_{{i 8}}^{V} U_{{j 7}}^{V} 
  + U_{{i 7}}^{V} U_{{j 8}}^{V})
  \nonumber\\
  &&
 + g_1 Z_{{k 1}}^{A} (
 + U_{{i 4}}^{V} U_{{j 6}}^{V}
+U_{{i 6}}^{V} U_{{j 4}}^{V})
  -g_2 Z_{{k 1}}^{A} (
U_{{i 5}}^{V} U_{{j 6}}^{V}
+U_{{i 6}}^{V} U_{{j 5}}^{V})
\nonumber\\
&&
-g_1 Z_{{k 2}}^{A} (U_{{i 4}}^{V} U_{{j 7}}^{V}+
U_{{i 7}}^{V} U_{{j 4}}^{V} )
 + g_2 Z_{{k 2}}^{A} (U_{{i 5}}^{V} U_{{j 7}}^{V}+
U_{{i 7}}^{V} U_{{j 5}}^{V} )
 \nonumber\\
 &&
  +\sqrt{2} \sum_{a=1}^{3}Y^*_{\nu,{a}} (
 U_{{i a}}^{V}  Z_{{k 2}}^{A} U_{{j 8}}^{V}
+U_{{j a}}^{V} Z_{{k 2}}^{A} U_{{i 8}}^{V}  
)
- \sqrt{2} \lambda^* Z_{{k 2}}^{A} (U_{{i 8}}^{V} U_{{j 6}}^{V}
 + U_{{i 6}}^{V} U_{{j 8}}^{V})
  \nonumber \\ 
 && 
  -2 \sqrt{2} \kappa^* Z_{{k 3}}^{A} U_{{i 8}}^{V} U_{{j 8}}^{V}
+ \sqrt{2} \lambda^*Z_{{k 3}}^{A}(  U_{{i 7}}^{V} U_{{j 6}}^{V} 
+ U_{{i 6}}^{V} U_{{j 7}}^{V}) 
\nonumber\\
&&
 -\sqrt{2} \sum_{a=1}^{3}Y^*_{\nu,{a}} (U_{{i a}}^{V}  Z_{{k 3}}^{A} U_{{j 7}}^{V} 
 +U_{{j a}}^{V}  
  Z_{{k 3}}^{A} U_{{i 7}}^{V} )
 \Big\}
P_R\ .
\label{eq:coupling_odd_sn->nn} 
\eea 
}

%%%%%%%%%%%%%%%%%%%%%%%%%%%%%%%%%%%%%%%%%%%%%%%%%%%%%%%%%%%%%%%%%%%%%%%%%%%
%%%%%%%%%%%%%%%%%%%%%%%%%%%%%%%%%%%%%%%%%%%%%%%%%%%%%%%%%%%%%%%%%%%%%%%%%%%

%%%%%%%%%%%%%%%%%%%%%%%%%%%%%%%%%%%%%%%%%%%%%%%%%%%%%%%%%%%%%%%%%%%%%%%%%%%%
%%%%%%%%%%%%%%%        bibliography              %%%%%%%%%%%%%%%%%%%%%%%%%%%
%%%%%%%%%%%%%%%%%%%%%%%%%%%%%%%%%%%%%%%%%%%%%%%%%%%%%%%%%%%%%%%%%%%%%%%%%%%%
\bibliographystyle{JHEP}
\bibliography{munussm_v7}
%%%%%%%%%%%%%%%%%%%%%%%%%%%%%%%%%%%%%%%%%%%%%%%%%%%%%%%%%%%%%%%%%
%%%%%%%%%%%%%%%%%%%%%%%%%%%%%%%%%%%%%%%%%%%%%%%%%%%%%%%%%%%%%%%%
\end{document}

%% file: defsV2.tex
\usepackage[T1]{fontenc}
\usepackage{epsfig}
\usepackage{latexsym}
\usepackage{graphicx}
\usepackage{amsmath}
\usepackage{amsfonts}   % if you want the fonts
\usepackage{amssymb}    % if you want extra symbols
\usepackage{float}
\usepackage{bm}
\usepackage{url}
\usepackage{hyperref} 
\usepackage[nodisplayskipstretch]{setspace}
\def\tanb{\tan\beta}

\def\lsim{\raise0.3ex\hbox{$\;<$\kern-0.75em\raise-1.1ex\hbox{$\sim\;$}}}
\def\gsim{\raise0.3ex\hbox{$\;>$\kern-0.75em\raise-1.1ex\hbox{$\sim\;$}}}
\newcommand{\captions}{\sf\caption}
\def    \beq            {\begin{equation}}
\def    \eeq            {\end{equation}}
\def    \bea           {\begin{eqnarray}}
\def    \eea           {\end{eqnarray}}

\def \mn{\mu\nu{\rm SSM}}

\def\g2{{\rm GeV}^2}
\def\sw2{sin^2 \theta_w}

\def\a^tau{\alpha_{\tau}}

\def\beq{\begin{equation}}
\def\eeq{\end{equation}}
\def\beqa{\begin{eqnarray}}
\def\eeqa{\end{eqnarray}}

\newcommand{\gev}{\,\textrm{GeV}}

\newcommand{\newc}{\newcommand}
\newc\BR{BR}
\newc{\akappa}{A_{\kappa} }
\newc\deltagmtwo{\delta (g-2)_{\mu}} 
\newc\deltaamu{\Delta a_{\mu}}

\def\anti{\overline}

\def\la{\lambda}

\def\rpv{{R}_{p} \hspace{-0.4cm}\slash\hspace{0.2cm}}

\newc{\haa}{BR\(h_1\to a_1 a_1\)}
\newc{\abb}{BR\(a_1\to b\anti{b}\)}
\newc{\hbb}{BR\(h_1\to b\anti{b}\)}
\newc{\abund}{\Omega h^2}
\newc\bsgamma{b\rightarrow s \gamma }
\newc\bxsgamma{\overline{B}\rightarrow X_{s}\gamma}
\newc\brbsgamma{\BR(\overline{B}\rightarrow X_s\gamma)}

%% file: sneutrinoL-LSP-V60.bbl
\providecommand{\href}[2]{#2}\begingroup\raggedright\begin{thebibliography}{100}

\bibitem{LopezFogliani:2005yw}
D.~E. L\'opez-Fogliani and C.~Mu\~noz, \emph{{Proposal for a supersymmetric
  standard model}},
  \href{http://dx.doi.org/10.1103/PhysRevLett.97.041801}{\emph{Phys. Rev.
  Lett.} {\bf 97} (2006) 041801},
  [\href{http://arxiv.org/abs/hep-ph/0508297}{{\tt hep-ph/0508297}}].

\bibitem{Escudero:2008jg}
N.~Escudero, D.~E. L\'opez-Fogliani, C.~Mu\~noz and R.~R. de~Austri,
  \emph{{Analysis of the parameter space and spectrum of the $\mu\nu$SSM}},
  \href{http://dx.doi.org/10.1088/1126-6708/2008/12/099}{\emph{JHEP} {\bf 12}
  (2008) 099}, [\href{http://arxiv.org/abs/0810.1507}{{\tt 0810.1507}}].

\bibitem{Munoz:2009an}
C.~Mu\~noz, \emph{{Phenomenology of a New Supersymmetric Standard Model: The
  $\mu\nu$SSM}},  in \emph{{Proceedings of the 17th International Conference on
  Supersymmetry and the Unification of Fundamental Interactions (SUSY09):
  Boston, USA, June 5-10, 2009, AIP Conf. Proc.}}, vol.~1200, p.~413, 2010.
\newblock \href{http://arxiv.org/abs/0909.5140}{{\tt 0909.5140}}.

\bibitem{Munoz:2016vaa}
C.~Mu\~noz, \emph{{Searching for SUSY and decaying gravitino DM at the LHC and
  Fermi-LAT with the $\mu\nu$SSM}},  in \emph{{Proceedings of the 11th
  International Workshop on the Dark Side of the Universe (DSU 2015): Kyoto,
  Japan, December 14-18, 2015, PoS}}, 2016.
\newblock \href{http://arxiv.org/abs/1608.07912}{{\tt 1608.07912}}.

\bibitem{Martin:1997ns}
S.~P. Martin, \emph{{A Supersymmetry primer}},
  \href{http://dx.doi.org/10.1142/9789812839657_0001}{\emph{Adv. Ser. Direct.
  High Energy Phys. 18 (1998) 1} (1997) },
  [\href{http://arxiv.org/abs/hep-ph/9709356}{{\tt hep-ph/9709356}}].

\bibitem{Kim:1983dt}
J.~E. Kim and H.~P. Nilles, \emph{{The $\mu$ problem and the strong CP
  Problem}}, \href{http://dx.doi.org/10.1016/0370-2693(84)91890-2}{\emph{Phys.
  Lett.} {\bf B138} (1984) 150}.

\bibitem{Ghosh:2008yh}
P.~Ghosh and S.~Roy, \emph{{Neutrino masses and mixing, lightest neutralino
  decays and a solution to the $\mu$ problem in supersymmetry}},
  \href{http://dx.doi.org/10.1088/1126-6708/2009/04/069}{\emph{JHEP} {\bf 04}
  (2009) 069}, [\href{http://arxiv.org/abs/0812.0084}{{\tt 0812.0084}}].

\bibitem{Bartl:2009an}
A.~Bartl, M.~Hirsch, A.~Vicente, S.~Liebler and W.~Porod, \emph{{LHC
  phenomenology of the $\mu\nu$SSM}},
  \href{http://dx.doi.org/10.1088/1126-6708/2009/05/120}{\emph{JHEP} {\bf 05}
  (2009) 120}, [\href{http://arxiv.org/abs/0903.3596}{{\tt 0903.3596}}].

\bibitem{Fidalgo:2009dm}
J.~Fidalgo, D.~E. L\'opez-Fogliani, C.~Mu\~noz and R.~Ruiz~de Austri,
  \emph{{Neutrino physics and spontaneous CP violation in the $\mu\nu$SSM}},
  \href{http://dx.doi.org/10.1088/1126-6708/2009/08/105}{\emph{JHEP} {\bf 08}
  (2009) 105}, [\href{http://arxiv.org/abs/0904.3112}{{\tt 0904.3112}}].

\bibitem{Ghosh:2010zi}
P.~Ghosh, P.~Dey, B.~Mukhopadhyaya and S.~Roy, \emph{{Radiative contribution to
  neutrino masses and mixing in $\mu\nu$SSM}},
  \href{http://dx.doi.org/10.1007/JHEP05(2010)087}{\emph{JHEP} {\bf 05} (2010)
  087}, [\href{http://arxiv.org/abs/1002.2705}{{\tt 1002.2705}}].

\bibitem{LopezFogliani:2010bf}
D.~E. L\'opez-Fogliani, \emph{{The seesaw mechanism in the $\mu\nu$SSM}},  in
  \emph{CTP International Conference on Neutrino Physics in the LHC Era, Luxor,
  Egypt, November 15-19, 2009}.
\newblock \href{http://arxiv.org/abs/1004.0884}{{\tt 1004.0884}}.

\bibitem{Ghosh:2010ig}
P.~Ghosh, \emph{{Neutrino masses and mixing in $\mu\nu$SSM}},
  \href{http://dx.doi.org/10.1088/1742-6596/259/1/012063}{\emph{J. Phys. Conf.
  Ser.} {\bf 259} (2010) 012063}, [\href{http://arxiv.org/abs/1010.2578}{{\tt
  1010.2578}}].

\bibitem{Bandyopadhyay:2010cu}
P.~Bandyopadhyay, P.~Ghosh and S.~Roy, \emph{{Unusual Higgs boson signal in
  R-parity violating nonminimal supersymmetric models at the LHC}},
  \href{http://dx.doi.org/10.1103/PhysRevD.84.115022}{\emph{Phys. Rev.} {\bf
  D84} (2011) 115022}, [\href{http://arxiv.org/abs/1012.5762}{{\tt
  1012.5762}}].

\bibitem{Fidalgo:2011ky}
J.~Fidalgo, D.~E. L\'opez-Fogliani, C.~Mu\~noz and R.~Ruiz~de Austri,
  \emph{{The Higgs sector of the $\mu\nu$SSM and collider physics}},
  \href{http://dx.doi.org/10.1007/JHEP10(2011)020}{\emph{JHEP} {\bf 10} (2011)
  020}, [\href{http://arxiv.org/abs/1107.4614}{{\tt 1107.4614}}].

\bibitem{Ghosh:2012pq}
P.~Ghosh, D.~E. L\'opez-Fogliani, V.~A. Mitsou, C.~Mu\~noz and R.~Ruiz~de
  Austri, \emph{{Probing the $\mu$-from-$\nu$ supersymmetric standard model
  with displaced multileptons from the decay of a Higgs boson at the LHC}},
  \href{http://dx.doi.org/10.1103/PhysRevD.88.015009}{\emph{Phys. Rev.} {\bf
  D88} (2013) 015009}, [\href{http://arxiv.org/abs/1211.3177}{{\tt
  1211.3177}}].

\bibitem{Ghosh:2014rha}
P.~Ghosh, D.~E. L\'opez-Fogliani, V.~A. Mitsou, C.~Mu\~noz and R.~R. de~Austri,
  \emph{{Hunting physics beyond the standard model with unusual $W^\pm$ and $Z$
  decays}}, \href{http://dx.doi.org/10.1103/PhysRevD.91.035020}{\emph{Phys.
  Rev.} {\bf D91} (2015) 035020}, [\href{http://arxiv.org/abs/1403.3675}{{\tt
  1403.3675}}].

\bibitem{Ghosh:2014ida}
P.~Ghosh, D.~E. L\'opez-Fogliani, V.~A. Mitsou, C.~Mu\~noz and R.~Ruiz~de
  Austri, \emph{{Probing the $\mu\nu$SSM with light scalars, pseudoscalars and
  neutralinos from the decay of a SM-like Higgs boson at the LHC}},
  \href{http://dx.doi.org/10.1007/JHEP11(2014)102}{\emph{JHEP} {\bf 11} (2014)
  102}, [\href{http://arxiv.org/abs/1410.2070}{{\tt 1410.2070}}].

\bibitem{Fidalgo:2011tm}
J.~Fidalgo and C.~Mu\~noz, \emph{{The $\mu\nu$SSM with an Extra U(1)}},
  \href{http://dx.doi.org/10.1007/JHEP04(2012)090}{\emph{JHEP} {\bf 04} (2012)
  090}, [\href{http://arxiv.org/abs/1111.2836}{{\tt 1111.2836}}].

\bibitem{Ibanez:1983kw}
L.~E. Ibanez, \emph{{The Scalar Neutrinos as the Lightest Supersymmetric
  Particles and Cosmology}},
  \href{http://dx.doi.org/10.1016/0370-2693(84)90221-1}{\emph{Phys. Lett.} {\bf
  137B} (1984) 160}.

\bibitem{Hagelin:1984wv}
J.~S. Hagelin, G.~L. Kane and S.~Raby, \emph{{Perhaps Scalar Neutrinos Are the
  Lightest Supersymmetric Partners}},
  \href{http://dx.doi.org/10.1016/0550-3213(84)90064-6}{\emph{Nucl. Phys.} {\bf
  B241} (1984) 638}.

\bibitem{Falk:1994es}
T.~Falk, K.~A. Olive and M.~Srednicki, \emph{{Heavy sneutrinos as dark
  matter}}, \href{http://dx.doi.org/10.1016/0370-2693(94)90639-4}{\emph{Phys.
  Lett.} {\bf B339} (1994) 248},
  [\href{http://arxiv.org/abs/hep-ph/9409270}{{\tt hep-ph/9409270}}].

\bibitem{Arina:2007tm}
C.~Arina and N.~Fornengo, \emph{{Sneutrino cold dark matter, a new analysis:
  Relic abundance and detection rates}},
  \href{http://dx.doi.org/10.1088/1126-6708/2007/11/029}{\emph{JHEP} {\bf 11}
  (2007) 029}, [\href{http://arxiv.org/abs/0709.4477}{{\tt 0709.4477}}].

\bibitem{Hall:1997ah}
L.~J. Hall, T.~Moroi and H.~Murayama, \emph{{Sneutrino cold dark matter with
  lepton number violation}},
  \href{http://dx.doi.org/10.1016/S0370-2693(98)00196-8}{\emph{Phys. Lett.}
  {\bf B424} (1998) 305}, [\href{http://arxiv.org/abs/hep-ph/9712515}{{\tt
  hep-ph/9712515}}].

\bibitem{Chala:2017jgg}
M.~Chala, A.~Delgado, G.~Nardini and M.~Quiros, \emph{{A light sneutrino
  rescues the light stop}},
  \href{http://dx.doi.org/10.1007/JHEP04(2017)097}{\emph{JHEP} {\bf 04} (2017)
  097}, [\href{http://arxiv.org/abs/1702.07359}{{\tt 1702.07359}}].

\bibitem{Abdallah:2003xe}
{\scshape DELPHI} collaboration, J.~Abdallah et~al., \emph{{Searches for
  supersymmetric particles in e+ e- collisions up to 208-GeV and interpretation
  of the results within the MSSM}},
  \href{http://dx.doi.org/10.1140/epjc/s2003-01355-5}{\emph{Eur. Phys. J.} {\bf
  C31} (2003) 421}, [\href{http://arxiv.org/abs/hep-ex/0311019}{{\tt
  hep-ex/0311019}}].

\bibitem{Borgani:1996ag}
S.~Borgani, A.~Masiero and M.~Yamaguchi, \emph{{Light gravitinos as mixed dark
  matter}}, \href{http://dx.doi.org/10.1016/0370-2693(96)00956-2}{\emph{Phys.
  Lett.} {\bf B386} (1996) 189},
  [\href{http://arxiv.org/abs/hep-ph/9605222}{{\tt hep-ph/9605222}}].

\bibitem{Takayama:2000uz}
F.~Takayama and M.~Yamaguchi, \emph{{Gravitino dark matter without R-parity}},
  \href{http://dx.doi.org/10.1016/S0370-2693(00)00726-7}{\emph{Phys. Lett.}
  {\bf B485} (2000) 388}, [\href{http://arxiv.org/abs/hep-ph/0005214}{{\tt
  hep-ph/0005214}}].

\bibitem{Chung:2010cd}
D.~J.~H. Chung and A.~J. Long, \emph{{Electroweak Phase Transition in the
  $\mu\nu$SSM}},
  \href{http://dx.doi.org/10.1103/PhysRevD.81.123531}{\emph{Phys. Rev.} {\bf
  D81} (2010) 123531}, [\href{http://arxiv.org/abs/1004.0942}{{\tt
  1004.0942}}].

\bibitem{Choi:2009ng}
K.-Y. Choi, D.~E. L\'opez-Fogliani, C.~Mu\~noz and R.~R. de~Austri,
  \emph{{Gamma-ray detection from gravitino dark matter decay in the
  $\mu\nu$SSM}},
  \href{http://dx.doi.org/10.1088/1475-7516/2010/03/028}{\emph{JCAP} {\bf 03}
  (2010) 028}, [\href{http://arxiv.org/abs/0906.3681}{{\tt 0906.3681}}].

\bibitem{GomezVargas:2011ph}
G.~A. G\'omez-Vargas, M.~Fornasa, F.~Zandanel, A.~J. Cuesta, C.~Mu\~noz,
  F.~Prada et~al., \emph{{CLUES on Fermi-LAT prospects for the extragalactic
  detection of $\mu\nu$SSM gravitino dark matter}},
  \href{http://dx.doi.org/10.1088/1475-7516/2012/02/001}{\emph{JCAP} {\bf 02}
  (2012) 001}, [\href{http://arxiv.org/abs/1110.3305}{{\tt 1110.3305}}].

\bibitem{Albert:2014hwa}
A.~Albert, G.~A. G\'omez-Vargas, M.~Grefe, C.~Mu\~noz, C.~Weniger, E.~Bloom
  et~al., \emph{{Search for 100 MeV to 10 GeV $\gamma$-ray lines in the
  Fermi-LAT data and implications for gravitino dark matter in $\mu\nu$SSM}},
  \href{http://dx.doi.org/10.1088/1475-7516/2014/10/023}{\emph{JCAP} {\bf 10}
  (2014) 023}, [\href{http://arxiv.org/abs/1406.3430}{{\tt 1406.3430}}].

\bibitem{Gomez-Vargas:2016ocf}
G.~A. G\'omez-Vargas, D.~E. L\'opez-Fogliani, C.~Mu\~noz, A.~D. Perez and
  R.~Ruiz~de Austri, \emph{{Search for sharp and smooth spectral signatures of
  $\mu\nu$SSM gravitino dark matter with Fermi-LAT}},
  \href{http://dx.doi.org/10.1088/1475-7516/2017/03/047}{\emph{JCAP} {\bf 1703}
  (2017) 047}, [\href{http://arxiv.org/abs/1608.08640}{{\tt 1608.08640}}].

\bibitem{Barbier:2004ez}
R.~Barbier et~al., \emph{{R-parity violating supersymmetry}},
  \href{http://dx.doi.org/10.1016/j.physrep.2005.08.006}{\emph{Phys. Rept.}
  {\bf 420} (2005) 1}, [\href{http://arxiv.org/abs/hep-ph/0406039}{{\tt
  hep-ph/0406039}}].

\bibitem{Ellwanger:2009dp}
U.~Ellwanger, C.~Hugonie and A.~M. Teixeira, \emph{{The next-to-minimal
  supersymmetric standard model}},
  \href{http://dx.doi.org/10.1016/j.physrep.2010.07.001}{\emph{Phys. Rept.}
  {\bf 496} (2010) 1}, [\href{http://arxiv.org/abs/0910.1785}{{\tt
  0910.1785}}].

\bibitem{Gonzalez-Garcia:2015qrr}
M.~C. Gonzalez-Garcia, M.~Maltoni and T.~Schwetz, \emph{{Global Analyses of
  Neutrino Oscillation Experiments}},
  \href{http://dx.doi.org/10.1016/j.nuclphysb.2016.02.033}{\emph{Nucl. Phys.}
  {\bf B908} (2016) 199--217}, [\href{http://arxiv.org/abs/1512.06856}{{\tt
  1512.06856}}].

\bibitem{Forero:2014bxa}
D.~V. Forero, M.~Tortola and J.~W.~F. Valle, \emph{{Neutrino oscillations
  refitted}}, \href{http://dx.doi.org/10.1103/PhysRevD.90.093006}{\emph{Phys.
  Rev.} {\bf D90} (2014) 093006}, [\href{http://arxiv.org/abs/1405.7540}{{\tt
  1405.7540}}].

\bibitem{Capozzi:2013csa}
F.~Capozzi, G.~L. Fogli, E.~Lisi, A.~Marrone, D.~Montanino and A.~Palazzo,
  \emph{{Status of three-neutrino oscillation parameters, circa 2013}},
  \href{http://dx.doi.org/10.1103/PhysRevD.89.093018}{\emph{Phys. Rev.} {\bf
  D89} (2014) 093018}, [\href{http://arxiv.org/abs/1312.2878}{{\tt
  1312.2878}}].

\bibitem{Staub:2008uz}
F.~Staub, \emph{{SARAH}},  \href{http://arxiv.org/abs/0806.0538}{{\tt
  0806.0538}}.

\bibitem{Staub:2011dp}
F.~Staub, T.~Ohl, W.~Porod and C.~Speckner, \emph{{A Tool Box for Implementing
  Supersymmetric Models}},
  \href{http://dx.doi.org/10.1016/j.cpc.2012.04.013}{\emph{Comput. Phys.
  Commun.} {\bf 183} (2012) 2165--2206},
  [\href{http://arxiv.org/abs/1109.5147}{{\tt 1109.5147}}].

\bibitem{Staub:2013tta}
F.~Staub, \emph{{SARAH 4 : A tool for (not only SUSY) model builders}},
  \href{http://dx.doi.org/10.1016/j.cpc.2014.02.018}{\emph{Comput. Phys.
  Commun.} {\bf 185} (2014) 1773}, [\href{http://arxiv.org/abs/1309.7223}{{\tt
  1309.7223}}].

\bibitem{Porod:2003um}
W.~Porod, \emph{{SPheno, a program for calculating supersymmetric spectra, SUSY
  particle decays and SUSY particle production at e+ e- colliders}},
  \href{http://dx.doi.org/10.1016/S0010-4655(03)00222-4}{\emph{Comput. Phys.
  Commun.} {\bf 153} (2003) 275},
  [\href{http://arxiv.org/abs/hep-ph/0301101}{{\tt hep-ph/0301101}}].

\bibitem{Porod:2011nf}
W.~Porod and F.~Staub, \emph{{SPheno 3.1: Extensions including flavour,
  CP-phases and models beyond the MSSM}},
  \href{http://dx.doi.org/10.1016/j.cpc.2012.05.021}{\emph{Comput. Phys.
  Commun.} {\bf 183} (2012) 2458}, [\href{http://arxiv.org/abs/1104.1573}{{\tt
  1104.1573}}].

\bibitem{Alwall:2014hca}
J.~Alwall, R.~Frederix, S.~Frixione, V.~Hirschi, F.~Maltoni, O.~Mattelaer
  et~al., \emph{{The automated computation of tree-level and next-to-leading
  order differential cross sections, and their matching to parton shower
  simulations}}, \href{http://dx.doi.org/10.1007/JHEP07(2014)079}{\emph{JHEP}
  {\bf 07} (2014) 079}, [\href{http://arxiv.org/abs/1405.0301}{{\tt
  1405.0301}}].

\bibitem{Sjostrand:2006za}
T.~Sjostrand, S.~Mrenna and P.~Z. Skands, \emph{{PYTHIA 6.4 physics and
  manual}}, \href{http://dx.doi.org/10.1088/1126-6708/2006/05/026}{\emph{JHEP}
  {\bf 05} (2006) 026}, [\href{http://arxiv.org/abs/hep-ph/0603175}{{\tt
  hep-ph/0603175}}].

\bibitem{preparation}
P.~Ghosh, E.~Kaptcha, D.~E. L\'opez-Fogliani, C.~Mu\~noz and R.~Ruiz~de Austri
  \href{http://arxiv.org/abs/in preparation}{{\tt in preparation}}.

\bibitem{Camargo-Molina:2013sta}
J.~E. Camargo-Molina, B.~O'Leary, W.~Porod and F.~Staub, \emph{{Stability of
  the CMSSM against sfermion VEVs}},
  \href{http://dx.doi.org/10.1007/JHEP12(2013)103}{\emph{JHEP} {\bf 12} (2013)
  103}, [\href{http://arxiv.org/abs/1309.7212}{{\tt 1309.7212}}].

\bibitem{Blinov:2013fta}
N.~Blinov and D.~E. Morrissey, \emph{{Vacuum Stability and the MSSM Higgs
  Mass}}, \href{http://dx.doi.org/10.1007/JHEP03(2014)106}{\emph{JHEP} {\bf 03}
  (2014) 106}, [\href{http://arxiv.org/abs/1310.4174}{{\tt 1310.4174}}].

\bibitem{Camargo-Molina:2014pwa}
J.~E. Camargo-Molina, B.~Garbrecht, B.~O'Leary, W.~Porod and F.~Staub,
  \emph{{Constraining the Natural MSSM through tunneling to color-breaking
  vacua at zero and non-zero temperature}},
  \href{http://dx.doi.org/10.1016/j.physletb.2014.08.036}{\emph{Phys. Lett.}
  {\bf B737} (2014) 156}, [\href{http://arxiv.org/abs/1405.7376}{{\tt
  1405.7376}}].

\bibitem{Bobrowski:2014dla}
M.~Bobrowski, G.~Chalons, W.~G. Hollik and U.~Nierste, \emph{{Vacuum stability
  of the effective Higgs potential in the Minimal Supersymmetric Standard
  Model}}, \href{http://dx.doi.org/10.1103/PhysRevD.90.035025,
  10.1103/PhysRevD.92.059901}{\emph{Phys. Rev.} {\bf D90} (2014) 035025},
  [\href{http://arxiv.org/abs/1407.2814}{{\tt 1407.2814}}].

\bibitem{Chattopadhyay:2014gfa}
U.~Chattopadhyay and A.~Dey, \emph{{Exploring MSSM for Charge and Color
  Breaking and Other Constraints in the Context of Higgs@125 GeV}},
  \href{http://dx.doi.org/10.1007/JHEP11(2014)161}{\emph{JHEP} {\bf 11} (2014)
  161}, [\href{http://arxiv.org/abs/1409.0611}{{\tt 1409.0611}}].

\bibitem{Hollik:2016dcm}
W.~G. Hollik, \emph{{A new view on vacuum stability in the MSSM}},
  \href{http://dx.doi.org/10.1007/JHEP08(2016)126}{\emph{JHEP} {\bf 08} (2016)
  126}, [\href{http://arxiv.org/abs/1606.08356}{{\tt 1606.08356}}].

\bibitem{Beuria:2016cdk}
J.~Beuria, U.~Chattopadhyay, A.~Datta and A.~Dey, \emph{{Exploring viable vacua
  of the Z$_{3}$-symmetric NMSSM}},
  \href{http://dx.doi.org/10.1007/JHEP04(2017)024}{\emph{JHEP} {\bf 04} (2017)
  024}, [\href{http://arxiv.org/abs/1612.06803}{{\tt 1612.06803}}].

\bibitem{Casas:1995pd}
J.~A. Casas, A.~Lleyda and C.~Munoz, \emph{{Strong constraints on the parameter
  space of the MSSM from charge and color breaking minima}},
  \href{http://dx.doi.org/10.1016/0550-3213(96)00194-0}{\emph{Nucl. Phys.} {\bf
  B471} (1996) 3}, [\href{http://arxiv.org/abs/hep-ph/9507294}{{\tt
  hep-ph/9507294}}].

\bibitem{Ade:2015xua}
{\scshape Planck} collaboration, P.~A.~R. Ade et~al., \emph{{Planck 2015
  results. XIII. Cosmological parameters}},
  \href{http://dx.doi.org/10.1051/0004-6361/201525830}{\emph{Astron.
  Astrophys.} {\bf 594} (2016) A13},
  [\href{http://arxiv.org/abs/1502.01589}{{\tt 1502.01589}}].

\bibitem{An:2015rpe}
{\scshape Daya Bay} collaboration, F.~P. An et~al., \emph{{New measurement of
  antineutrino oscillation with the full detector configuration at Daya Bay}},
  \href{http://dx.doi.org/10.1103/PhysRevLett.115.111802}{\emph{Phys. Rev.
  Lett.} {\bf 115} (2015) 111802}, [\href{http://arxiv.org/abs/1505.03456}{{\tt
  1505.03456}}].

\bibitem{BarShalom:1998xz}
S.~Bar-Shalom, G.~Eilam, J.~Wudka and A.~Soni, \emph{{R-parity violation and
  uses of the rare decay $\widetilde \nu \to \gamma \gamma$ in hadron and
  photon colliders}},
  \href{http://dx.doi.org/10.1103/PhysRevD.59.035010}{\emph{Phys. Rev.} {\bf
  D59} (1999) 035010}, [\href{http://arxiv.org/abs/hep-ph/9809253}{{\tt
  hep-ph/9809253}}].

\bibitem{Dawson:1983fw}
S.~Dawson, E.~Eichten and C.~Quigg, \emph{{Search for Supersymmetric Particles
  in Hadron - Hadron Collisions}},
  \href{http://dx.doi.org/10.1103/PhysRevD.31.1581}{\emph{Phys. Rev.} {\bf D31}
  (1985) 1581}.

\bibitem{Eichten:1984eu}
E.~Eichten, I.~Hinchliffe, K.~D. Lane and C.~Quigg, \emph{{Super Collider
  Physics}}, \href{http://dx.doi.org/10.1103/RevModPhys.56.579,
  10.1103/RevModPhys.58.1065}{\emph{Rev. Mod. Phys.} {\bf 56} (1984) 579}.

\bibitem{delAguila:1990yw}
F.~del Aguila and L.~Ametller, \emph{{On the detectability of sleptons at large
  hadron colliders}},
  \href{http://dx.doi.org/10.1016/0370-2693(91)90336-O}{\emph{Phys. Lett.} {\bf
  B261} (1991) 326}.

\bibitem{Baer:1993ew}
H.~Baer, C.-h. Chen, F.~Paige and X.~Tata, \emph{{Detecting Sleptons at Hadron
  Colliders and Supercolliders}},
  \href{http://dx.doi.org/10.1103/PhysRevD.49.3283}{\emph{Phys. Rev.} {\bf D49}
  (1994) 3283}, [\href{http://arxiv.org/abs/hep-ph/9311248}{{\tt
  hep-ph/9311248}}].

\bibitem{Baer:1997nh}
H.~Baer, B.~W. Harris and M.~H. Reno, \emph{{Next-to-leading order slepton pair
  production at hadron colliders}},
  \href{http://dx.doi.org/10.1103/PhysRevD.57.5871}{\emph{Phys. Rev.} {\bf D57}
  (1998) 5871}, [\href{http://arxiv.org/abs/hep-ph/9712315}{{\tt
  hep-ph/9712315}}].

\bibitem{Bozzi:2004qq}
G.~Bozzi, B.~Fuks and M.~Klasen, \emph{{Slepton production in polarized hadron
  collisions}},
  \href{http://dx.doi.org/10.1016/j.physletb.2005.01.060}{\emph{Phys. Lett.}
  {\bf B609} (2005) 339--350}, [\href{http://arxiv.org/abs/hep-ph/0411318}{{\tt
  hep-ph/0411318}}].

\bibitem{Aaboud2016}
{\scshape ATLAS} collaboration, M.~Aaboud et~al., \emph{{Search for
  supersymmetry in a final state containing two photons and missing transverse
  momentum in $\sqrt{s}\ =\ 13\ TeV\ pp$ collisions at the LHC using the ATLAS
  detector}},
  \href{http://dx.doi.org/10.1140/epjc/s10052-016-4344-x}{\emph{Eur. Phys. J.}
  {\bf C76} (2016) 517}, [\href{http://arxiv.org/abs/1606.09150}{{\tt
  1606.09150}}].

\bibitem{Aad:2016tuk}
{\scshape ATLAS} collaboration, G.~Aad et~al., \emph{{Search for supersymmetry
  at $\sqrt{s}=13$ TeV in final states with jets and two same-sign leptons or
  three leptons with the ATLAS detector}},
  \href{http://dx.doi.org/10.1140/epjc/s10052-016-4095-8}{\emph{Eur. Phys. J.}
  {\bf C76} (2016) 259}, [\href{http://arxiv.org/abs/1602.09058}{{\tt
  1602.09058}}].

\bibitem{Khachatryan:2017qgo}
{\scshape CMS} collaboration, V.~Khachatryan et~al., \emph{{Search for
  supersymmetry with multiple charged leptons in proton-proton collisions at
  $\sqrt{ s} = $ 13 TeV}},  \href{http://arxiv.org/abs/1701.06940}{{\tt
  1701.06940}}.

\bibitem{Khachatryan:2016iqn}
{\scshape CMS} collaboration, V.~Khachatryan et~al., \emph{{Searches for
  $R$-parity-violating supersymmetry in $pp$ collisions at $\sqrt{s} =$ 8 TeV
  in final states with 0-4 leptons}},
  \href{http://dx.doi.org/10.1103/PhysRevD.94.112009}{\emph{Phys. Rev.} {\bf
  D94} (2016) 112009}, [\href{http://arxiv.org/abs/1606.08076}{{\tt
  1606.08076}}].

\bibitem{Khachatryan:2016ojf}
{\scshape CMS} collaboration, V.~Khachatryan et~al., \emph{{Search for
  supersymmetry in events with photons and missing transverse energy in pp
  collisions at 13 TeV}},
  \href{http://dx.doi.org/10.1016/j.physletb.2017.04.005}{\emph{Phys. Lett.}
  {\bf B769} (2017) 391--412}, [\href{http://arxiv.org/abs/1611.06604}{{\tt
  1611.06604}}].

\bibitem{ATLAS-CONF-2016-096}
{\scshape ATLAS} collaboration, \emph{{Search for supersymmetry with two and
  three leptons and missing transverse momentum in the final state at
  $\sqrt{s}=$ 13 TeV with the ATLAS detector}},  Tech. Rep.
  ATLAS-CONF-2016-096, CERN, Geneva, Sep, 2016.

\bibitem{Aaboud:2016uro}
{\scshape ATLAS} collaboration, M.~Aaboud et~al., \emph{{Search for new
  phenomena in events with a photon and missing transverse momentum in $pp$
  collisions at $\sqrt{s}=13$ TeV with the ATLAS detector}},
  \href{http://dx.doi.org/10.1007/JHEP06(2016)059}{\emph{JHEP} {\bf 06} (2016)
  059}, [\href{http://arxiv.org/abs/1604.01306}{{\tt 1604.01306}}].

\bibitem{Aaboud:2016tnv}
{\scshape ATLAS} collaboration, M.~Aaboud et~al., \emph{{Search for new
  phenomena in final states with an energetic jet and large missing transverse
  momentum in $pp$ collisions at $\sqrt{s} =$ 13 TeV using the ATLAS
  detector}}, \href{http://dx.doi.org/10.1103/PhysRevD.94.032005}{\emph{Phys.
  Rev.} {\bf D94} (2016) 032005}, [\href{http://arxiv.org/abs/1604.07773}{{\tt
  1604.07773}}].

\bibitem{Aaboud:2016zdn}
{\scshape ATLAS} collaboration, M.~Aaboud et~al., \emph{{Search for squarks and
  gluinos in final states with jets and missing transverse momentum at
  $\sqrt{s} =$ 13 TeV with the ATLAS detector}},
  \href{http://dx.doi.org/10.1140/epjc/s10052-016-4184-8}{\emph{Eur. Phys. J.}
  {\bf C76} (2016) 392}, [\href{http://arxiv.org/abs/1605.03814}{{\tt
  1605.03814}}].

\bibitem{Aad:2016qqk}
{\scshape ATLAS} collaboration, G.~Aad et~al., \emph{{Search for gluinos in
  events with an isolated lepton, jets and missing transverse momentum at
  $\sqrt{s}$ = 13 Te V with the ATLAS detector}},
  \href{http://dx.doi.org/10.1140/epjc/s10052-016-4397-x}{\emph{Eur. Phys. J.}
  {\bf C76} (2016) 565}, [\href{http://arxiv.org/abs/1605.04285}{{\tt
  1605.04285}}].

\bibitem{Aad:2016eki}
{\scshape ATLAS} collaboration, G.~Aad et~al., \emph{{Search for pair
  production of gluinos decaying via stop and sbottom in events with $b$-jets
  and large missing transverse momentum in $pp$ collisions at $\sqrt{s} = 13$
  TeV with the ATLAS detector}},
  \href{http://dx.doi.org/10.1103/PhysRevD.94.032003}{\emph{Phys. Rev.} {\bf
  D94} (2016) 032003}, [\href{http://arxiv.org/abs/1605.09318}{{\tt
  1605.09318}}].

\bibitem{Aaboud:2016lwz}
{\scshape ATLAS} collaboration, M.~Aaboud et~al., \emph{{Search for top squarks
  in final states with one isolated lepton, jets, and missing transverse
  momentum in $\sqrt{s}=13$ TeV $pp$ collisions with the ATLAS detector}},
  \href{http://dx.doi.org/10.1103/PhysRevD.94.052009}{\emph{Phys. Rev.} {\bf
  D94} (2016) 052009}, [\href{http://arxiv.org/abs/1606.03903}{{\tt
  1606.03903}}].

\bibitem{ATLAS-CONF-2015-082}
\emph{{A search for Supersymmetry in events containing a leptonically decaying
  $Z$ boson, jets and missing transverse momentum in $\sqrt{s}=13~$TeV $pp$
  collisions with the ATLAS detector}},  Tech. Rep. ATLAS-CONF-2015-082, CERN,
  Geneva, Dec, 2015.

\bibitem{ATLAS-CONF-2016-013}
\emph{{Search for production of vector-like top quark pairs and of four top
  quarks in the lepton-plus-jets final state in $pp$ collisions at
  $\sqrt{s}=13$ TeV with the ATLAS detector}},  Tech. Rep. ATLAS-CONF-2016-013,
  CERN, Geneva, Mar, 2016.

\bibitem{ATLAS-CONF-2016-050}
{\scshape ATLAS} collaboration, \emph{{Search for top squarks in final states
  with one isolated lepton, jets, and missing transverse momentum in $\sqrt{s}$
  = 13 TeV pp collisions with the ATLAS detector}},  Tech. Rep.
  ATLAS-CONF-2016-050, CERN, Geneva, Aug, 2016.

\bibitem{ATLAS-CONF-2016-076}
{\scshape ATLAS} collaboration, \emph{{Search for direct top squark pair
  production and dark matter production in final states with two leptons in
  $\sqrt{s} = 13$ TeV $pp$ collisions using 13.3 fb$^{-1}$ of ATLAS data}},
  Tech. Rep. ATLAS-CONF-2016-076, CERN, Geneva, Aug, 2016.

\bibitem{Dercks:2016npn}
D.~Dercks, N.~Desai, J.~S. Kim, K.~Rolbiecki, J.~Tattersall and T.~Weber,
  \emph{{CheckMATE 2: From the model to the limit}},
  \href{http://dx.doi.org/10.1016/j.cpc.2017.08.021}{\emph{Comput. Phys.
  Commun.} {\bf 221} (2017) 383}, [\href{http://arxiv.org/abs/1611.09856}{{\tt
  1611.09856}}].

\bibitem{deFavereau:2013fsa}
{\scshape DELPHES 3} collaboration, J.~de~Favereau, C.~Delaere, P.~Demin,
  A.~Giammanco, V.~Lemaitre, A.~Mertens et~al., \emph{{DELPHES 3, A modular
  framework for fast simulation of a generic collider experiment}},
  \href{http://dx.doi.org/10.1007/JHEP02(2014)057}{\emph{JHEP} {\bf 02} (2014)
  057}, [\href{http://arxiv.org/abs/1307.6346}{{\tt 1307.6346}}].

\bibitem{Cacciari:2011ma}
M.~Cacciari, G.~P. Salam and G.~Soyez, \emph{{FastJet User Manual}},
  \href{http://dx.doi.org/10.1140/epjc/s10052-012-1896-2}{\emph{Eur. Phys. J.}
  {\bf C72} (2012) 1896}, [\href{http://arxiv.org/abs/1111.6097}{{\tt
  1111.6097}}].

\bibitem{Cacciari:2005hq}
M.~Cacciari and G.~P. Salam, \emph{{Dispelling the $N^{3}$ myth for the $k_t$
  jet-finder}},
  \href{http://dx.doi.org/10.1016/j.physletb.2006.08.037}{\emph{Phys. Lett.}
  {\bf B641} (2006) 57}, [\href{http://arxiv.org/abs/hep-ph/0512210}{{\tt
  hep-ph/0512210}}].

\bibitem{Cacciari:2008gp}
M.~Cacciari, G.~P. Salam and G.~Soyez, \emph{{The Anti-k(t) jet clustering
  algorithm}},
  \href{http://dx.doi.org/10.1088/1126-6708/2008/04/063}{\emph{JHEP} {\bf 04}
  (2008) 063}, [\href{http://arxiv.org/abs/0802.1189}{{\tt 0802.1189}}].

\bibitem{Read:2002hq}
A.~L. Read, \emph{{Presentation of search results: The CL(s) technique}},
  \href{http://dx.doi.org/10.1088/0954-3899/28/10/313}{\emph{J. Phys.} {\bf
  G28} (2002) 2693}.

\bibitem{Bechtle:2008jh}
P.~Bechtle, O.~Brein, S.~Heinemeyer, G.~Weiglein and K.~E. Williams,
  \emph{{HiggsBounds: Confronting Arbitrary Higgs Sectors with Exclusion Bounds
  from LEP and the Tevatron}},
  \href{http://dx.doi.org/10.1016/j.cpc.2009.09.003}{\emph{Comput. Phys.
  Commun.} {\bf 181} (2010) 138}, [\href{http://arxiv.org/abs/0811.4169}{{\tt
  0811.4169}}].

\bibitem{Bechtle:2011sb}
P.~Bechtle, O.~Brein, S.~Heinemeyer, G.~Weiglein and K.~E. Williams,
  \emph{{HiggsBounds 2.0.0: Confronting Neutral and Charged Higgs Sector
  Predictions with Exclusion Bounds from LEP and the Tevatron}},
  \href{http://dx.doi.org/10.1016/j.cpc.2011.07.015}{\emph{Comput. Phys.
  Commun.} {\bf 182} (2011) 2605}, [\href{http://arxiv.org/abs/1102.1898}{{\tt
  1102.1898}}].

\bibitem{Bechtle:2013gu}
P.~Bechtle, O.~Brein, S.~Heinemeyer, O.~Stal, T.~Stefaniak, G.~Weiglein et~al.,
  \emph{{Recent Developments in HiggsBounds and a Preview of HiggsSignals}},
  {\emph{PoS} {\bf CHARGED2012} (2012) 024},
  [\href{http://arxiv.org/abs/1301.2345}{{\tt 1301.2345}}].

\bibitem{Bechtle:2013wla}
P.~Bechtle, O.~Brein, S.~Heinemeyer, O.~Stal, T.~Stefaniak, G.~Weiglein et~al.,
  \emph{{$\mathsf{HiggsBounds}-4$: Improved Tests of Extended Higgs Sectors
  against Exclusion Bounds from LEP, the Tevatron and the LHC}},
  \href{http://dx.doi.org/10.1140/epjc/s10052-013-2693-2}{\emph{Eur. Phys. J.}
  {\bf C74} (2014) 2693}, [\href{http://arxiv.org/abs/1311.0055}{{\tt
  1311.0055}}].

\bibitem{Bechtle:2015pma}
P.~Bechtle, S.~Heinemeyer, O.~Stal, T.~Stefaniak and G.~Weiglein,
  \emph{{Applying Exclusion Likelihoods from LHC Searches to Extended Higgs
  Sectors}}, \href{http://dx.doi.org/10.1140/epjc/s10052-015-3650-z}{\emph{Eur.
  Phys. J.} {\bf C75} (2015) 421}, [\href{http://arxiv.org/abs/1507.06706}{{\tt
  1507.06706}}].

\bibitem{Chatrchyan:2012mea}
{\scshape CMS} collaboration, S.~Chatrchyan et~al., \emph{{Search for anomalous
  production of multilepton events in $pp$ collisions at $\sqrt{s}=7$ TeV}},
  \href{http://dx.doi.org/10.1007/JHEP06(2012)169}{\emph{JHEP} {\bf 06} (2012)
  169}, [\href{http://arxiv.org/abs/1204.5341}{{\tt 1204.5341}}].

\bibitem{Chatrchyan:2014aea}
{\scshape CMS} collaboration, S.~Chatrchyan et~al., \emph{{Search for anomalous
  production of events with three or more leptons in $pp$ collisions at
  $\sqrt{s} =$ 8 TeV}},
  \href{http://dx.doi.org/10.1103/PhysRevD.90.032006}{\emph{Phys. Rev.} {\bf
  D90} (2014) 032006}, [\href{http://arxiv.org/abs/1404.5801}{{\tt
  1404.5801}}].

\bibitem{Aaltonen:2010fv}
{\scshape CDF} collaboration, T.~Aaltonen et~al., \emph{{Search for R-parity
  Violating Decays of $\tau$ sneutrinos to $e\mu$, $\mu\tau$, and $e\tau$ pairs
  in $p\bar{p}$ Collisions at $\sqrt{s} = 1.96$ TeV}},
  \href{http://dx.doi.org/10.1103/PhysRevLett.105.191801}{\emph{Phys. Rev.
  Lett.} {\bf 105} (2010) 191801}, [\href{http://arxiv.org/abs/1004.3042}{{\tt
  1004.3042}}].

\bibitem{Abazov:2010km}
{\scshape D0} collaboration, V.~M. Abazov et~al., \emph{{Search for sneutrino
  production in emu final states in 5.3 fb$^{-1}$ of $p\bar{p}$ collisions at
  sqrt(s) =1.96 TeV}},
  \href{http://dx.doi.org/10.1103/PhysRevLett.105.191802}{\emph{Phys. Rev.
  Lett.} {\bf 105} (2010) 191802}, [\href{http://arxiv.org/abs/1007.4835}{{\tt
  1007.4835}}].

\bibitem{Achard:2001ek}
{\scshape L3} collaboration, P.~Achard et~al., \emph{{Search for R parity
  violating decays of supersymmetric particles in $e^{+} e^{-}$ collisions at
  LEP}}, \href{http://dx.doi.org/10.1016/S0370-2693(01)01367-3}{\emph{Phys.
  Lett.} {\bf B524} (2002) 65--80},
  [\href{http://arxiv.org/abs/hep-ex/0110057}{{\tt hep-ex/0110057}}].

\bibitem{Heister:2002kq}
{\scshape ALEPH} collaboration, A.~Heister et~al., \emph{{Search for R-parity
  violating production of single sneutrinos in $e^+ e^-$ collisions at
  $\sqrt{s}=189$ GeV to $209$ GeV}},
  \href{http://dx.doi.org/10.1007/s10052-002-1010-2}{\emph{Eur. Phys. J.} {\bf
  C25} (2002) 1--12}, [\href{http://arxiv.org/abs/hep-ex/0201013}{{\tt
  hep-ex/0201013}}].

\bibitem{Heister:2002jc}
{\scshape ALEPH} collaboration, A.~Heister et~al., \emph{{Search for
  supersymmetric particles with R parity violating decays in $e^{+} e^{-}$
  collisions at $\sqrt{s}$ up to $209$ GeV}},
  \href{http://dx.doi.org/10.1140/epjc/s2003-01311-5}{\emph{Eur. Phys. J.} {\bf
  C31} (2003) 1--16}, [\href{http://arxiv.org/abs/hep-ex/0210014}{{\tt
  hep-ex/0210014}}].

\bibitem{Abbiendi:2003rn}
{\scshape OPAL} collaboration, G.~Abbiendi et~al., \emph{{Search for R parity
  violating decays of scalar fermions at LEP}},
  \href{http://dx.doi.org/10.1140/epjc/s2004-01596-8}{\emph{Eur. Phys. J.} {\bf
  C33} (2004) 149--172}, [\href{http://arxiv.org/abs/hep-ex/0310054}{{\tt
  hep-ex/0310054}}].

\bibitem{Abdallah:2003xc}
{\scshape DELPHI} collaboration, J.~Abdallah et~al., \emph{{Search for
  supersymmetric particles assuming R-parity nonconservation in $e^+ e^-$
  collisions at $\sqrt{s}=192$ GeV to $208$ GeV}},
  \href{http://dx.doi.org/Erratum:, 10.1140/epjc/s2004-01881-6,
  10.1140/epjc/s2004-01976-0}{\emph{Eur. Phys. J.} {\bf C36} (2004) 1--23},
  [\href{http://arxiv.org/abs/hep-ex/0406009}{{\tt hep-ex/0406009}}].

\bibitem{CMS:2015neg}
{\scshape CMS} collaboration, \emph{{Search for Resonances Decaying to Dijet
  Final States at $\sqrt{s} = 8$ TeV with Scouting Data}},
  {\emph{CMS-PAS-EXO-14-005} (2015) }.

\bibitem{Aad:2014aqa}
{\scshape ATLAS} collaboration, G.~Aad et~al., \emph{{Search for new phenomena
  in the dijet mass distribution using $p-p$ collision data at $\sqrt{s}=8$ TeV
  with the ATLAS detector}},
  \href{http://dx.doi.org/10.1103/PhysRevD.91.052007}{\emph{Phys. Rev.} {\bf
  D91} (2015) 052007}, [\href{http://arxiv.org/abs/1407.1376}{{\tt
  1407.1376}}].

\bibitem{Khachatryan:2015dcf}
{\scshape CMS} collaboration, V.~Khachatryan et~al., \emph{{Search for narrow
  resonances decaying to dijets in proton-proton collisions at $\sqrt(s) =$ 13
  TeV}}, \href{http://dx.doi.org/10.1103/PhysRevLett.116.071801}{\emph{Phys.
  Rev. Lett.} {\bf 116} (2016) 071801},
  [\href{http://arxiv.org/abs/1512.01224}{{\tt 1512.01224}}].

\bibitem{ATLAS:2015nsi}
{\scshape ATLAS} collaboration, G.~Aad et~al., \emph{{Search for new phenomena
  in dijet mass and angular distributions from $pp$ collisions at $\sqrt{s}=$
  13 TeV with the ATLAS detector}},
  \href{http://dx.doi.org/10.1016/j.physletb.2016.01.032}{\emph{Phys. Lett.}
  {\bf B754} (2016) 302--322}, [\href{http://arxiv.org/abs/1512.01530}{{\tt
  1512.01530}}].

\bibitem{Aad:2015pfa}
{\scshape ATLAS} collaboration, G.~Aad et~al., \emph{{Search for a Heavy
  Neutral Particle Decaying to $e\mu$, $e\tau$, or $\mu\tau$ in $pp$ Collisions
  at $\sqrt{s}=8$ TeV with the ATLAS Detector}},
  \href{http://dx.doi.org/10.1103/PhysRevLett.115.031801}{\emph{Phys. Rev.
  Lett.} {\bf 115} (2015) 031801}, [\href{http://arxiv.org/abs/1503.04430}{{\tt
  1503.04430}}].

\bibitem{Khachatryan:2016ovq}
{\scshape CMS} collaboration, V.~Khachatryan et~al., \emph{{Search for lepton
  flavour violating decays of heavy resonances and quantum black holes to an
  e$\mu$ pair in proton-proton collisions at $\sqrt{s}$ = 8 TeV}},
  \href{http://dx.doi.org/10.1140/epjc/s10052-016-4149-y}{\emph{Eur. Phys. J.}
  {\bf C76} (2016) 317}, [\href{http://arxiv.org/abs/1604.05239}{{\tt
  1604.05239}}].

\bibitem{Fuks:2013lya}
B.~Fuks, M.~Klasen, D.~R. Lamprea and M.~Rothering, \emph{{Revisiting slepton
  pair production at the Large Hadron Collider}},
  \href{http://dx.doi.org/10.1007/JHEP01(2014)168}{\emph{JHEP} {\bf 01} (2014)
  168}, [\href{http://arxiv.org/abs/1310.2621}{{\tt 1310.2621}}].

\bibitem{pgs}
PGS,
  \emph{{http://conway.physics.ucdavis.edu/research/software/pgs/pgs4-general.htm}},
  .

\bibitem{Jadach:1990mz}
S.~Jadach, J.~H. Kuhn and Z.~Was, \emph{{TAUOLA: A Library of Monte Carlo
  programs to simulate decays of polarized tau leptons}},
  \href{http://dx.doi.org/10.1016/0010-4655(91)90038-M}{\emph{Comput. Phys.
  Commun.} {\bf 64} (1990) 275--299}.

\bibitem{Jadach:1993hs}
S.~Jadach, Z.~Was, R.~Decker and J.~H. Kuhn, \emph{{The tau decay library
  TAUOLA: Version 2.4}},
  \href{http://dx.doi.org/10.1016/0010-4655(93)90061-G}{\emph{Comput. Phys.
  Commun.} {\bf 76} (1993) 361--380}.

\bibitem{Aad:2014iza}
{\scshape ATLAS} collaboration, G.~Aad et~al., \emph{{Search for supersymmetry
  in events with four or more leptons in $\sqrt{s}$ = 8 TeV pp collisions with
  the ATLAS detector}},
  \href{http://dx.doi.org/10.1103/PhysRevD.90.052001}{\emph{Phys. Rev.} {\bf
  D90} (2014) 052001}, [\href{http://arxiv.org/abs/1405.5086}{{\tt
  1405.5086}}].

\bibitem{Aad:2015hea}
{\scshape ATLAS} collaboration, G.~Aad et~al., \emph{{Search for photonic
  signatures of gauge-mediated supersymmetry in 8 TeV pp collisions with the
  ATLAS detector}},
  \href{http://dx.doi.org/10.1103/PhysRevD.92.072001}{\emph{Phys. Rev.} {\bf
  D92} (2015) 072001}, [\href{http://arxiv.org/abs/1507.05493}{{\tt
  1507.05493}}].

\bibitem{prepa}
I.~Lara, D.~E. L\'opez-Fogliani, C.~Mu\~noz, N.~Nagata, H.~Otono and R.~Ruiz~de
  Austri \href{http://arxiv.org/abs/in preparation}{{\tt in preparation}}.

\bibitem{Lopez-Fogliani:2017qzj}
D.~E. L\'opez-Fogliani and C.~Mu\~noz, \emph{{On a reinterpretation of the
  Higgs field in supersymmetry and a proposal for new quarks}},
  \href{http://dx.doi.org/10.1016/j.physletb.2017.05.018}{\emph{Phys. Lett.}
  {\bf B771} (2017) 136}, [\href{http://arxiv.org/abs/1701.02652}{{\tt
  1701.02652}}].

\bibitem{Aguilar-Saavedra:2017giu}
J.~A. Aguilar-Saavedra, D.~E. L\'opez-Fogliani and C.~Mu\~noz, \emph{{Novel
  signatures for vector-like quarks}},
  \href{http://arxiv.org/abs/1705.02526}{{\tt 1705.02526}}.

\bibitem{Escudero:2005hk}
N.~Escudero, C.~Mu\~noz and A.~M. Teixeira, \emph{{FCNCs in supersymmetric
  multi-Higgs doublet models}},
  \href{http://dx.doi.org/10.1103/PhysRevD.73.055015}{\emph{Phys. Rev.} {\bf
  D73} (2006) 055015}, [\href{http://arxiv.org/abs/hep-ph/0512046}{{\tt
  hep-ph/0512046}}].

\bibitem{Escudero:2005ku}
N.~Escudero, C.~Mu\~noz and A.~M. Teixeira, \emph{{Phenomenological viability
  of orbifold models with three Higgs families}},
  \href{http://dx.doi.org/10.1088/1126-6708/2006/07/041}{\emph{JHEP} {\bf 07}
  (2006) 041}, [\href{http://arxiv.org/abs/hep-ph/0512301}{{\tt
  hep-ph/0512301}}].

\bibitem{Escudero:2007db}
N.~Escudero, C.~Mu\~noz and A.~M. Teixeira, \emph{{Lepton masses and mixings in
  orbifold models with three Higgs families}},
  \href{http://dx.doi.org/10.1088/1126-6708/2007/12/080}{\emph{JHEP} {\bf 12}
  (2007) 080}, [\href{http://arxiv.org/abs/0710.3672}{{\tt 0710.3672}}].

\bibitem{Chemtob:2004xr}
M.~Chemtob, \emph{{Phenomenological constraints on broken R parity symmetry in
  supersymmetry models}},
  \href{http://dx.doi.org/10.1016/j.ppnp.2004.06.001}{\emph{Prog. Part. Nucl.
  Phys.} {\bf 54} (2005) 71--191},
  [\href{http://arxiv.org/abs/hep-ph/0406029}{{\tt hep-ph/0406029}}].

\bibitem{Dreiner:2012mx}
H.~K. Dreiner, K.~Nickel, F.~Staub and A.~Vicente, \emph{{New bounds on
  trilinear R-parity violation from lepton flavor violating observables}},
  \href{http://dx.doi.org/10.1103/PhysRevD.86.015003}{\emph{Phys. Rev.} {\bf
  D86} (2012) 015003}, [\href{http://arxiv.org/abs/1204.5925}{{\tt
  1204.5925}}].

\bibitem{deCarlos:1996ecd}
B.~de~Carlos and P.~L. White, \emph{{R-parity violation effects through soft
  supersymmetry breaking terms and the renormalization group}},
  \href{http://dx.doi.org/10.1103/PhysRevD.54.3427}{\emph{Phys. Rev.} {\bf D54}
  (1996) 3427}, [\href{http://arxiv.org/abs/hep-ph/9602381}{{\tt
  hep-ph/9602381}}].

\bibitem{Gabbiani:1996hi}
F.~Gabbiani, E.~Gabrielli, A.~Masiero and L.~Silvestrini, \emph{{A Complete
  analysis of FCNC and CP constraints in general SUSY extensions of the
  standard model}},
  \href{http://dx.doi.org/10.1016/0550-3213(96)00390-2}{\emph{Nucl. Phys.} {\bf
  B477} (1996) 321}, [\href{http://arxiv.org/abs/hep-ph/9604387}{{\tt
  hep-ph/9604387}}].

\bibitem{Brignole:1997dp}
A.~Brignole, L.~E. Ibanez and C.~Munoz, \emph{{Soft supersymmetry breaking
  terms from supergravity and superstring models}},
  \href{http://dx.doi.org/10.1142/9789814307505_0004}{\emph{Adv. Ser. Direct.
  High Energy Phys.} {\bf 21} (2010) 244},
  [\href{http://arxiv.org/abs/hep-ph/9707209}{{\tt hep-ph/9707209}}].

\end{thebibliography}\endgroup
